\documentclass[12pt,english]{article}
\usepackage{graphicx}
\usepackage{epsfig}
\usepackage{xspace}
\usepackage{amsmath}
\usepackage{amssymb}
\usepackage{babel}
\newcommand{\dmsq}{\Delta m^{2}}

\newcommand{\nuebar}{\overline{\nu}_{e}}

\newcommand{\sstc}{$\sin ^{2} 2\theta_{13} \:$}

\def\adm2{\Delta{{m}^2_{{atm}}}}
\def\sdm2{\Delta{{m}^2_{{sol}}}}
\def\Dm2{\Delta{m}^2}
\def\t13{\theta_{{13}}}

\def\simgt{\lower.5ex\hbox{$\; \buildrel > \over \sim \;$}}
\newcommand{\quq}{\theta_{13}}

\newcommand{\nueb}{\ensuremath{\overline{\nu}_e}\xspace}
\newcommand{\Epos}{\ensuremath{E_{e^+}}\xspace}
\newcommand{\Evis}{\ensuremath{E_{\text{vis}}}\xspace}
\newcommand{\Enueb}{\ensuremath{E_{\nueb}}\xspace}
\newcommand{\NBins}{\ensuremath{N_{\text{bins}}}\xspace}

\newcommand{\capdef}{}
\newcommand{\mycaption}[2][\capdef]{\renewcommand{\capdef}{#2}%
        \caption[#1]{{\footnotesize #2}}}
\newcommand{\be}{\begin{equation}}
\newcommand{\ee}{\end{equation}}
\newcommand{\deltacp}{\delta_{\mathrm{CP}}}
\newcommand{\stheta}{\ensuremath{\sin^2 2 \theta_{13}}}
\newcommand{\meff}{\mbox{$\left| m_{ee} \right|$}}
\newcommand{\ie}{{\it i.e.}}
\newcommand{\cf}{{\it cf.}}

\newcommand{\Ref}{Reference}
\newcommand{\Refs}{References}

\begin{document}
\pagenumbering{roman}
\title{
\huge {
\bf Double Chooz:  A Search for the \\
Neutrino Mixing Angle $\theta_{13}$ 
} 
\nopagebreak
}
\author{
F.~Ardellier$^{19}$ \and
I.~Barabanov$^{10}$ \and
J.~C.~Barri\`ere$^{19}$ \and
F.~Bei{\ss}el$^{1}$ \and
S.~Berridge$^{23}$ \and
L.~Bezrukov$^{10}$ \and
A.~Bernstein$^{14}$ \and
T.~Bolton$^{12}$ \and
N.S.~Bowden$^{20}$ \and
Ch.~Buck$^{16}$
% ...
}
\date{20 June 2006 (\small revised 26 October 2006) }
\maketitle

\newpage
\begin{enumerate}
\item RWTH {\bf Aachen} University
\item University of { \bf Alabama}
\item { \bf Argonne} National Laboratory
\item {\bf AstroParticule} et Cosmologie Universit\'e Paris 7 (APC)
\item {\bf CIEMAT} Centro de Investigaciones Energ\'eticas, 
Medioambientales y
Tecnol\'ogicas Madrid
\item {\bf Drexel} University
\item INFN, Laboratori Nazionali del {\bf Gran} Sasso
\item Universit\"at {\bf Hamburg}
\item { \bf Illinois} Institute of Technology
\item { \bf Institute} for Nuclear Research RAS
\item { \bf Institute} for Physical Chemistry and Electrochemistry RAS
\item { \bf Kansas} State University
\item RRC { \bf Kurchatov} Institute
\item Lawrence { \bf Livermore} National Laboratory
\item { \bf Louisiana} State University
\item {\bf Max-Planck-Institut} f\"ur Kernphysik (Heidelberg)
\item INFN, Milano
\item University of {\bf Notre} Dame
\item CEA/DSM/DAPNIA {\bf Saclay}
\item {\bf Sandia } National Laboratories
\item {\bf SUBATECH } Nantes (IN2P3 - Universit\'e de Nantes - EMN)
\item Technische Universit\"at {\bf M\"unchen}
\item University of { \bf Tennessee}
\item Eberhard-Karls Universit\"at {\bf T\"ubingen}
\end{enumerate}

\newpage
\tableofcontents
\newpage

\pagenumbering{arabic}
\setcounter{page}{1}
\cleardoublepage
\section*{Executive Summary}
%
%
\par It is widely recognized that the potential of reactor anti-neutrino disappearance experiments 
has not been fully exploited. High precision neutrino oscillation measurements could be achieved 
by a multi-detector experiment, whereby the sensitivity of the experiment would be nearly 
unaffected by the dominant uncertainties related to neutrino production and interaction~\cite{bib:kr2det}.
This strategy has been adopted by the Double Chooz collaboration in order to search for a 
non-vanishing $\theta_{13}$ mixing angle, which might open the way to unveiling CP violation in 
the leptonic sector. 
Furthermore, the reactor experiment results are complementary to those from next generation neutrino 
accelerator experiments, as it has been demonstrated 
a few years ago~\cite{bib:minakata1, bib:huber}. 
%
%
\par The Double Chooz initiative started in the summer 2003 after reviewing a few possible sites
suitable to perform a new reactor neutrino experiment dedicated to  $\theta_{13}$ in France. 
The Chooz site was selected because of 
the availability of the underground neutrino laboratory (300~m.w.e) located at 1.05~km from 
the nuclear cores, funded and constructed by \'Electricit\'e de France (E.D.F.)~\cite{bib:edf} 
for the first experiment done at Chooz, end of the 90's~\cite{bib:chooz}. 
This site selection was done in parallel with other similar efforts in Brazil,  China, Japan, 
South-Korea, Russia, Taiwan, and the United States. There have been four international workshops mainly 
dedicated to the feasibility of new reactor neutrino experiments as well as for reviewing the 
potential of each site. All these initiatives have been described in the White Paper, 
``A New Nuclear Reactor Neutrino Experiment to Measure $\theta_{13}$''~\cite{bib:white}.
But since its publication the worldwide situation has changed and the projects still
being considered are Angra~\cite{bib:angra} in Brazil, Daya Bay~\cite{bib:daya} in China, 
Double Chooz in France (see~\cite{bib:loi,bib:choozusprop} and this proposal), KASKA~\cite{bib:kaska} 
in Japan and RENO~\cite{bib:reno} in South Korea. 
A recent comparison of the capabilities of these experiments can be found in~\cite{bib:icfa, bib:lasserresobel}.
Double Chooz is particularly attractive because it could limit $\sin^2(2\theta_{13})$ to ~0.022-030 
(for $\Delta{m}^2_{31}=3.5-2.5 \times 10^{-3}~eV^2$), within an unrivaled time scale and a modest cost. 
Installation of the experiment will start with the far detector located at the former 
experimental site of the CHOOZ experiment. This first phase of the experiment will allow us to exceed 
the CHOOZ sensitivity within a few months, to reach a limit of $\sin^2(2\theta_{13})<0.08$ if no oscillation 
signal is detected after 1,5 years of data taking. 
Then the collaboration will install an identical 
near detector 280~m from the Chooz nuclear cores, in a new neutrino laboratory in a 45~m deep shaft to be excavated. 
This second phase will considerably reduce the overall systematic uncertainties and allow the final
sensitivity of the experiment to be reached within 3 years of data taking. 

\par Double Chooz will also lay the
 foundation for future experiments through the development of innovative 
technologies (scintillators, detector inter-calibration, etc). 	
In addition, the International Atomic Energy Agency (IAEA) mandated by the United Nations Organization 
is  interested in the prospect applications of neutrino physics. The detection of the antineutrinos produced 
by a nuclear power plant would provide a real-time, remote, non-intrusive and impossible-to-fake means to address 
certain safeguards applications. The near detector of Double Chooz will perform a measurement of the 
antineutrino flux and energy spectrum with an unprecedented accuracy, which will enable the collaboration 
to achieve a feasibility study of the detection of antineutrinos for safeguards applications, and test the 
potential of neutrinos to detect various diversion scenarios. 
%
%
\par The Double Chooz collaboration is presently composed of 24 institutions from France, Germany, Italy, 
Russia, Spain and the United States. In this proposal, and during the funding request phase of the experiment, 
the detector design has been divided in work packages shared in the collaboration under the assumption that each
group is full funded. 
French groups are responsible for the detector mechanics. They are also in charge of the digitization and 
the data acquisition systems. 
German groups have  responsibility for 
the scintillators, purification, and fluid systems, as well as 
the inner muon veto, and the level-1 trigger. They are also involved in the calibration.
Spanish collaborators are involved in the inner detector 
photodetection and the related mechanics.
United States groups have taken the inner phototubes, the Front-End electronics, the calibration systems, 
and the slow-control. The Outer Veto is also under the responsibility of the United States groups.
Russian groups are involved in simulation, calibration, and scintillator developments.  
We note as well the participation of the Italian/Russian chemistry group of the INFN--LNGS 
laboratory to the loaded liquid scintillator effort.
Simulation and software developments are distributed throughout the collaboration. 
Near and Far detector infrastructures, as well as the technical coordination and detector integration
are managed by the French institutes. Finally, a few 
 other items are still under discussion. 
%
%
\par The Double Chooz experiment has been approved by the French scientific councils of CEA--DSM--DAPNIA 
and CNRS--IN2P3 since March 2004. Recently it has been reviewed again, leading to the 
definitive approval and funding from both agencies, starting from 2006.  
The Double Chooz experiment is supported by the Astroparticle Physics European Coordination (ApPEC) 
roadmap~\cite{bib:appec}.
In Germany the experiment receives funding through the Max-Planck Society with a substantial increase beginning 
in 2006. The funding of the university groups is currently in the final approval process at the Deutsche 
Forschungsgemeinschaft (DFG). 
In the United States, a new reactor neutrino experiment is one of the high priority recommendations of the 
APS neutrino study~\cite{bib:aps}. 
The Neutrino Scientific Assessment Group (NuSAG) and the High Energy Physics Advisory 
Panel (HEPAP) both recommended U.S. participation to Double Chooz~\cite{bib:nusag, bib:hepap}.
However, the High Energy Physics division at DOE turned down the Double Chooz R$\&$D request and 
has not acted on the construction request.  
Construction funding is currently being discussed at NSF (National Science Foundation). 
Russian participation has been funded by Russian Foundation of Basics Research. 
In Spain funding will be decided in June 2006. 
%
%
\par In conclusion, the design and work sharing presented in this proposal may change in the forthcoming 
months as new collaborators join the experiment. Since the major funding of Double Chooz is secured it is 
planned to start the far detector integration in 2007 in order to take data in 2008. 
The near detector will be installed after the completion of the shaft civil engineering construction in 2009.  
\subsubsection*{Acknowledgments}
Upgrade of the far site laboratory is done in cooperation with the CIDEN (Centre d' Ing\'enierie, 
D\'econstruction et Environnement Nucléaire) division of \'Electricit\'e de France (E.D.F.)~\cite{bib:ciden}. 
Study of the near site laboratory civil engineering is done in cooperation with the CNEN (Centre 
National d'Etudes Nucl\'eaires) division of \'Electricit\'e de France (E.D.F.). 
We also thank the staff of E.D.F. CHOOZ nuclear plant~\cite{bib:edfchooz} for their continuous support.
We are very grateful to the Conseil G\'en\'eral des Ardennes for providing us with the facilities 
for the experiment. We thank also the Mayor of Chooz and the local members of the parliament.
Special thanks to the technical staff of our laboratories for their excellent work in designing the detector.

%
\cleardoublepage
%
%
%
\section{Overview}
\subsection{Physics case and experimental context}
\par It has recently been widely recognized that a reactor 
antineutrino
disappearance experiment with two or more detectors is one of the most
cost-effective ways to extend our reach in sensitivity for the neutrino
mixing angle $\theta_{13}$ without ambiguities from CP violation and matter
effects~\cite{bib:white}. 
This is a proposal by an international collaboration of neutrino physicists
to modify the existing neutrino physics facility at the Chooz-B nuclear
power station in France\cite{bib:loi,bib:choozusprop}. The experiment, 
called Double Chooz, is planned
to reach a sensitivity 
to \sstc down to 0.03 over a three
year run from early 2009, with two detectors running simultaneously. 
This will cover roughly 85\% of the currently allowed region.  
The costs and time to first results for this critical parameter can be
minimized since our project takes advantage of an existing
laboratory.  

\par The Double Chooz reactor neutrino experiment~\cite{bib:loi}
offers the world particle physics community a 
relatively 
quick and inexpensive opportunity to measure the mixing angle
$\quq$ if it is not too small: 
$0.19 > \sin^2 2\quq > 0.03$.  The data taking will be divided in two phases: a first one with 
the Far detector only, and a second phase with both Near and Far detectors running simultaneously.
For $\dmsq \approx 2.5 \times 10^{-3}$eV$^2$, 
Double Chooz will be sensitive to 0.05 after 1.5 year of data taking in phase I, and
0.03 or better after 3 years of operation with two detectors.  
If $\quq$ is in this range,
long-baseline off-axis neutrino experiments will be able to measure 
matter effects and search for CP violation~\cite{bib:shav, bib:huber,
bib:minakata}.
\par
There have been four international workshops and a white paper which have
outlined the challenges and benefits of a new reactor experiment to
measure $\quq$~\cite{bib:white}.  
Projects currently being considered are ANGRA~\cite{bib:angra} in Brazil,
Daya Bay~\cite{bib:daya} in China, Double Chooz~\cite{bib:dc} in France, 
KASKA~\cite{bib:kaska} in Japan and
RENO~\cite{bib:reno} in South Korea.  
For a recent comparison of the
capabilities of these 
five  experiments, see References ~\cite{bib:icfa, bib:lasserresobel}.

Of these proposed experiments, Double Chooz has the opportunity
to obtain results first and quickly explore this important region of
neutrino parameter space.
The Reactor Neutrino White Paper~\cite{bib:white} identified civil construction costs as
approximately two-thirds of the cost of a new reactor neutrino experiment. 
By taking advantage of the Chooz site where a hall near the correct distance already
exists, tremendous savings are available, with regard not only to cost and
schedule but also 
in the areas of risk and contingency.  Further, a timely
measurement of $\quq$, available only from such an experiment, will be of
considerable value in planning the more expensive off-axis accelerator
projects, which are sensitive to matter effects and CP violation in the
neutrino sector, but only if $\quq$ is large enough.

\subsection{Experimental site: the Chooz nuclear reactors}
\par
\begin{figure}[htb]
\begin{center}
\includegraphics[width=\textwidth]{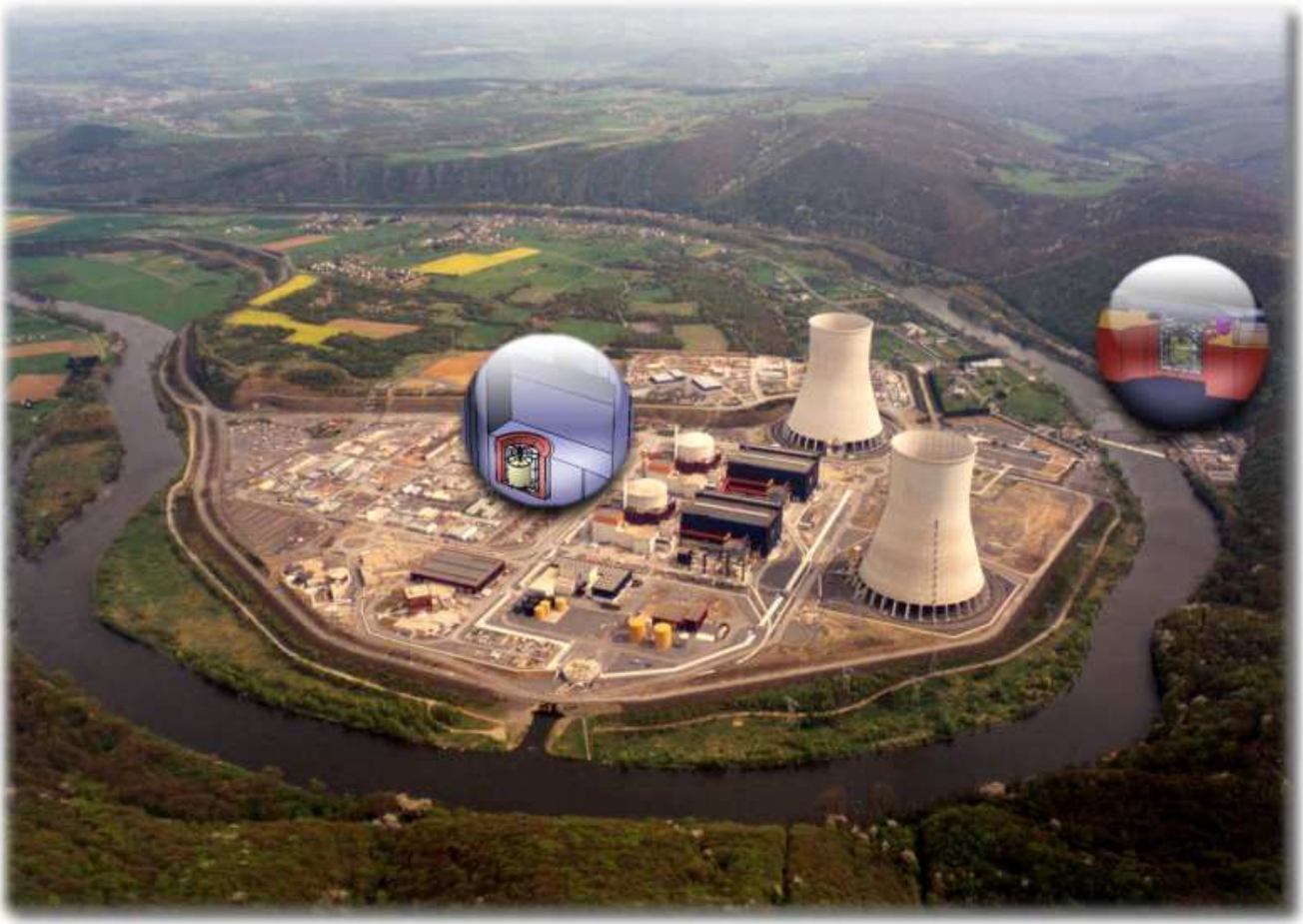}
\caption[Overview of the experiment site]{Overview of the experiment site.}
\label{fig:chooznearfarsites}
\end{center}
\end{figure}

\begin{figure}[htb]
\begin{center}
\includegraphics[width=\textwidth]{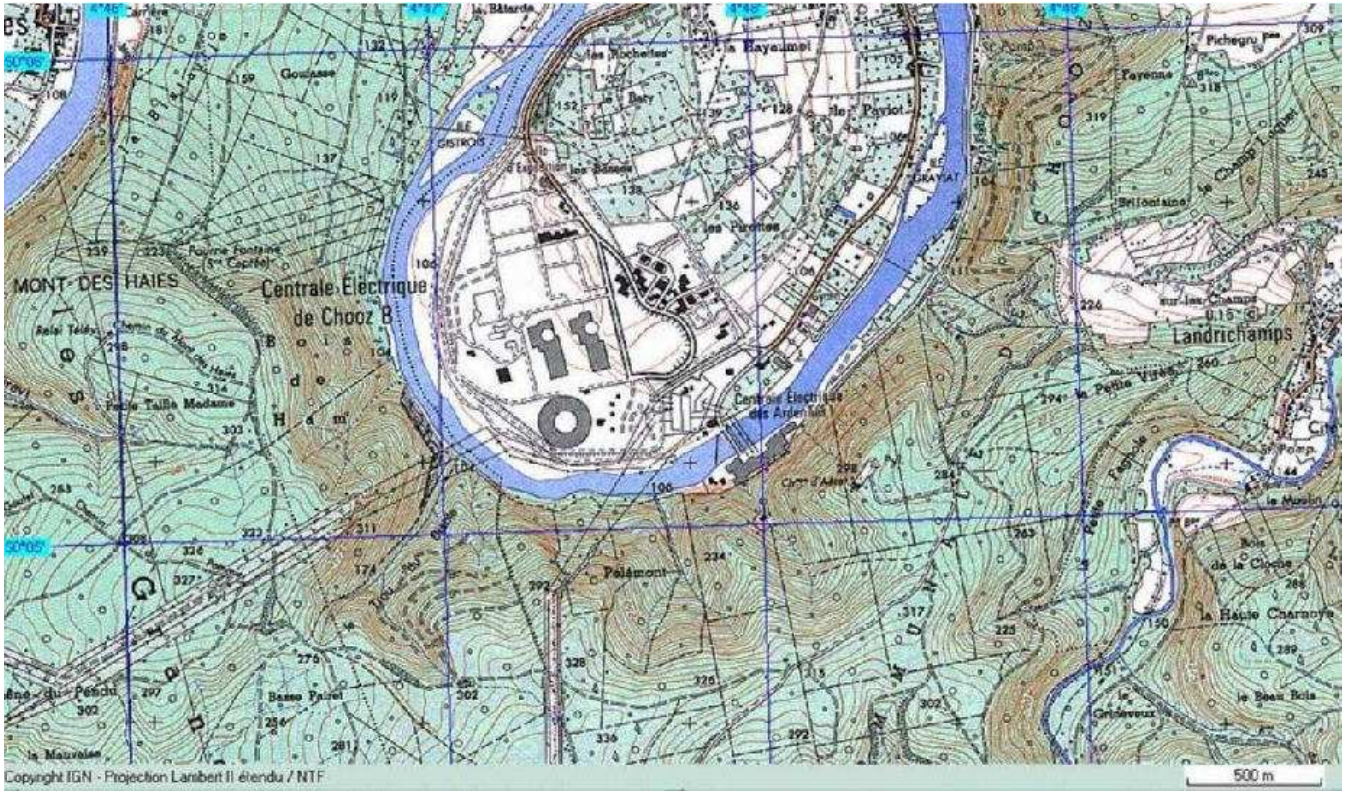}
\caption[Map of the experiment site]{Map of the experiment site. The
  two cores are separated by a distance of 140~meters. The far
  detector site is located 1.0 and 1.1~km from the two cores.}
\label{fig:chooznearfarsitemap}
\end{center}
\end{figure}

 The antineutrinos used in the experiment are those produced by the pair
of reactors located at the Chooz-B nuclear power station operated by
the French  company Electricit\'e de France (EdF) in partnership
with the Belgian utilities Electrabel S.A./N.V. and Soci\'et\'e Publique
d'Electricit\'e.  They are located in the Ardennes region, in the northeast of
France, very close to the Belgian border, in a meander of the Meuse
river (see Figures \ref{fig:chooznearfarsites} and \ref{fig:chooznearfarsitemap}).
At the Chooz site, there are two nuclear reactors. Both are of
the  most recent ``N4" type (4 steam generators), with a thermal power of
4.27~GW$_{\text{th}}$ each, and were recently upgraded from 
1.45~GW$_{\text{e}}$ to 1.5~GW$_{\text{e}}$.
Each reactor is off about one month per year.
These 
are pressurized water reactors (PWR) and are fed with
UOx type fuel. They are the most powerful reactor type in operation in
the world. One unusual characteristic of the N4 reactors is their
ability to vary their output from 30\% to 95\% of full power in
 less than 30~minutes, using the so-called gray control
 rods  in the reactor core. These rods are referred to as gray
 because  they absorb fewer free neutrons than conventional (``black")
 rods. One advantage is 
 greater thermal homogeneity.
A total of 
205 fuel assemblies are contained within each reactor core.
The entire reactor vessel is 
a cylinder 4.27~m tall and 3.47~m diameter.
The first reactor started 
full-power operation 
in May 1997,
and the second one in September of the same year.

\begin{figure}[htb]
\begin{center}
\includegraphics[width=\textwidth]{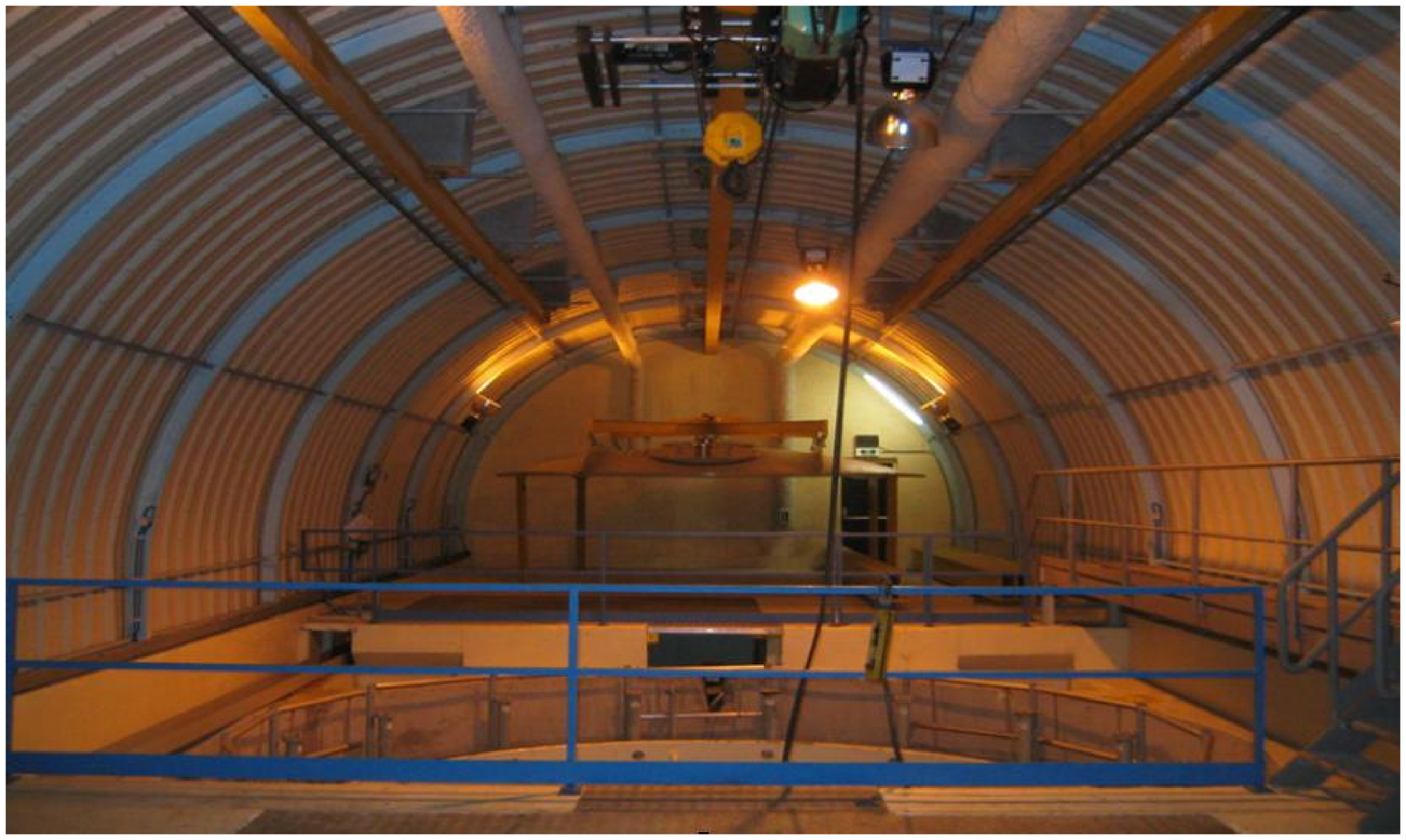}
\caption[Picture of the Double Chooz-far detector site]
{Picture of the Double Chooz-far detector site taken in September
 2003. The original CHOOZ laboratory hall constructed by EdF, 
 located close the old Chooz A underground power plant, is still in
 perfect condition and could be re-used without additional civil engineering 
 construction.}
\label{fig:choozfarfoto}
\end{center}
\end{figure}

%
The Double Chooz experiment will 
employ two almost identical detectors of medium size,
each containing 10.3 cubic meters of liquid scintillator target doped with
0.1\% of gadolinium (see Section~\ref{sec:scintillator}).  
The neutrino laboratory of the first CHOOZ experiment,\footnote{For clarity, the first reactor neutrino experiment conducted at the Chooz reactor is herein referred to in uppercase.} located 1.05~km
from the two cores of the Chooz nuclear plant, will be used again
(see Figure \ref{fig:choozfarfoto}). This is the main advantage of
this site compared  with other locations. 
We label this site the far detector site or {\bf Double Chooz-far}. 
A sketch of  the Double Chooz-far detector is shown in Figure~\ref{fig:choozfar}. 
The Double Chooz-far site is shielded by about 300~m.w.e.\ of 2.8~$\text{g/cm}^3$ rock. 
It is intended to start taking data at Double Chooz-far at the beginning of 2008. 

In order to cancel the systematic errors originating from the nuclear
reactors (lack of knowledge of the $\nuebar$ flux and spectrum), as well as to
reduce the set of systematic errors related to the detector and to the
event selection procedure, a second detector will be installed  close
to the nuclear cores. We label this detector site the near site or
{\bf Double Chooz-near}.
\begin{figure}[htb]
\begin{center}
\includegraphics[width=0.5\textwidth]{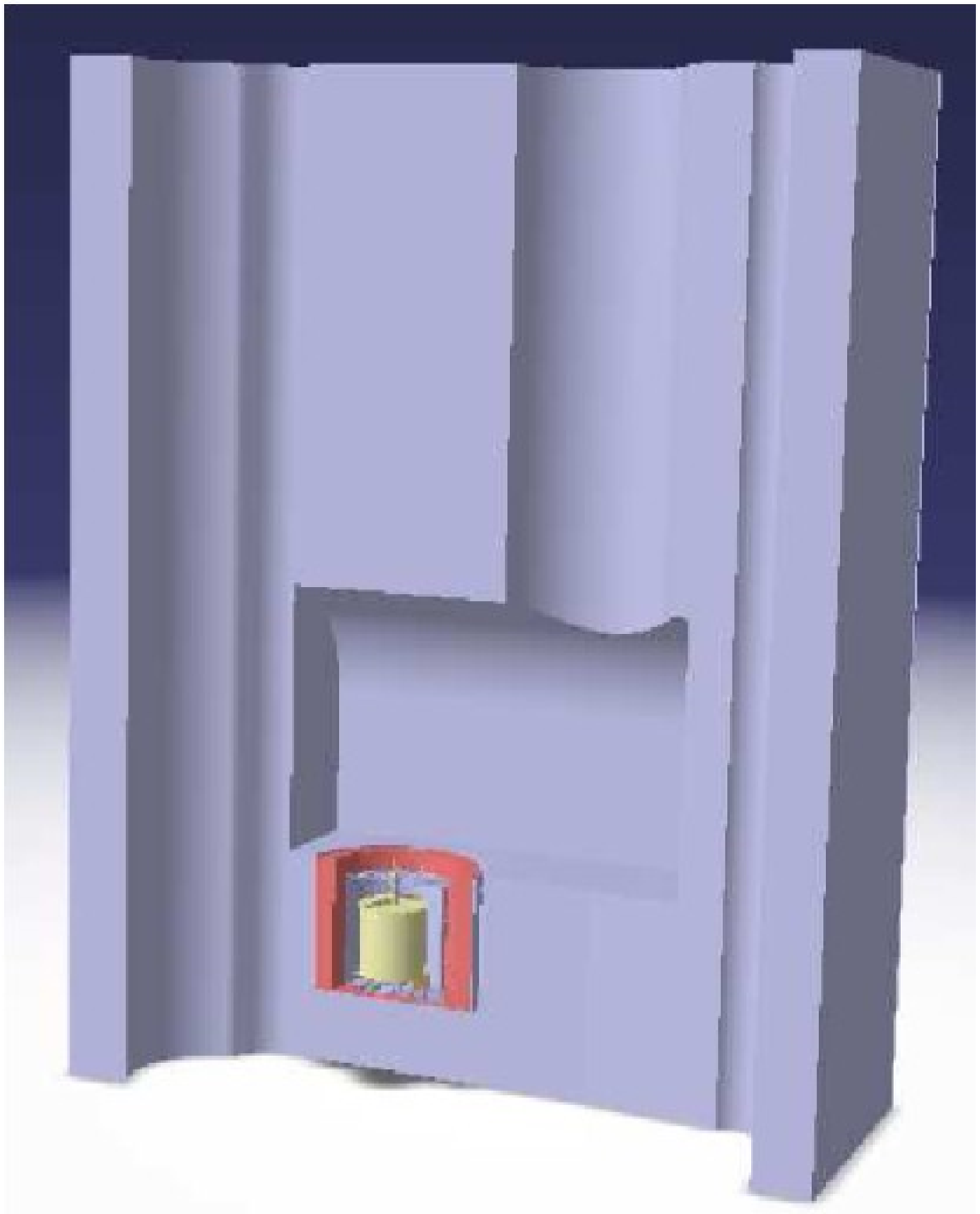}
\caption[3D rendering of the Double Chooz-near site]
{3D rendering of the Double Chooz-near site. The detector is located in a $\sim$45-meter-deep shaft, 
about 250--300~meters from the nuclear cores. Several civil engineering options are being 
studied to provide an overburden of 30~meters of rock (2.8~$\text{g/cm}^3$). 
Since more space will be available here than at the far site, 
we are studying the option of using low-radioactivity sand (70~cm) instead of the steel shielding used at Double Chooz far.}
\label{fig:choozneasite}
\end{center}
\end{figure}
An initial  study has been completed by the French electric 
power company EdF to determine the best combination of
location and overburden as well as the preliminary cost of the needed civil 
construction. This study 
suggested the feasibility of excavating a $\approx$40~m deep
shaft at a 250--300~m distance from the nuclear reactor cores, see Figure~\ref{fig:choozneasite}.  
The plan is to start taking data at Double Chooz-near at the end of 
2008 
or early in 2009.
%
\subsection{The new Double Chooz detector concept}
\begin{figure}[htb]
\begin{center}
\includegraphics[width=0.6\textwidth]{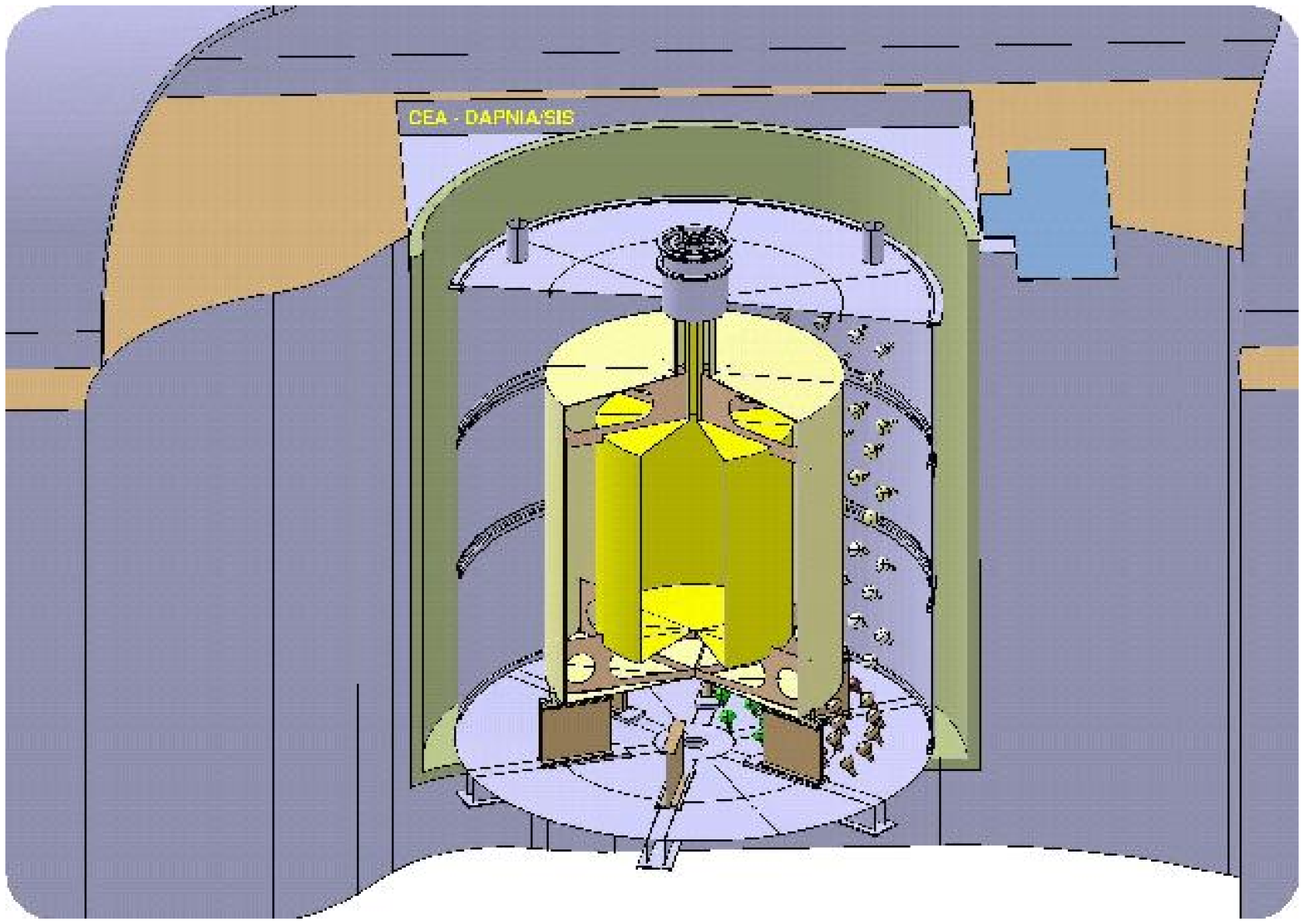}
\caption[Double Chooz-far detector]
{The Double Chooz-far detector, at the Chooz underground site.
The detector is located in the tank used for the CHOOZ experiment
(7 meters high and 7 meters in diameter) 
which is still available.
A total of 10.3~$\text{m}^3$ of a dodecane+PXE-based liquid scintillator doped
with gadolinium is contained in a transparent acrylic cylinder surrounded
by the $\gamma$-catcher region (22.6 m$^3$) and the buffer (114.2 m$^3$).
The design goal is to achieve  a light yield of about 200 pe/MeV which requires
an optical coverage of about 13\%, provided by the surrounding PMTs.
The PMTs are mounted on the cylindrical steel structure that 
optically isolates the outer part of the detector, used as a muon
veto (90 m$^3$), from the inner part.}
\label{fig:choozfar}
\end{center}
\end{figure}
\par The Double Chooz-far detector will consist of
a target cylinder of 115~cm radius 
and 246~cm height, providing a volume of 10.3~m$^3$.
The near and far detectors will be identical inside the PMT 
support structure. 
This will allow a relative normalization systematic error of 0.6\%.
However, due to the differing overburdens (70--80 vs.\ 300 m.w.e.), 
the outer shielding will not be identical, since the
cosmic ray background differs considerably between Double Chooz-near and -far. 
The overburden of the near detector has been chosen in order to keep
a high true-neutrino-signal to background ratio. A review of the backgrounds
at both sites is given in Section~\ref{sec:backgrounds}.

Starting from the center of the target the detector elements are as
follows (see Figure~\ref{fig:choozfar}):
\vspace{-0.25cm}
\subsubsection*{Target and $\gamma$-catcher}
Target and $\gamma$-catcher vessels will be built from acrylic plastic material, transparent
to UV and visible photons with wavelengths above 400 nm. 
The vessels are designed to contain the target
and $\gamma$-catcher aromatic liquids with long-term 
hermeticity (no leak for 10~years) and stability. 
The strongest constraint is the chemical compatibility between the vessel and the
scintillating liquids of the target and 
$\gamma$-catcher (chemical stability for a period of at least 5~years).
The $\gamma$-catcher vessel must also be chemically compatible
with the mineral oil of the buffer region,  which is however known to be a weaker constraint.
Target and $\gamma$-catcher vessels will be made of cast acrylic. 
The target vessel is a cylinder of 246~cm height, 230~cm diameter,
and 8~mm thickness. It contains a target volume of 10.3~m$^3$.
The $\gamma$-catcher is a 55-cm-thick buffer of nonloaded liquid scintillator (22.6~m$^3$) 
with the same light yield as the $\nuebar$ target.
This scintillating buffer around the target is necessary for efficiently
measuring 
the gammas from 
neutron capture on Gd and 
from positron annihilation, 
and to reject the background from fast neutrons (see Section~\ref{sec:backgrounds}).  
\vspace{-0.25cm}
\subsubsection*{Nonscintillating buffer}
A 105-cm-thick region of nonscintillating liquid (114.2 m$^3$) serves to
decrease the level of accidental background (mainly the contribution
from photomultiplier-tube radioactivity). This region is crucial to
keeping the singles rate below 10~Hz in the sensitive region
(target+$\gamma$-catcher).
\vspace{-0.25cm}
\subsubsection*{Buffer vessel and PMT support structure}
This vessel is made of 3-mm-thick stainless steel sheets
and stiffeners.
A total of 534 phototubes (8 inch) in a uniform array are mounted from the interior surface of the buffer vessel.
\vspace{-0.25cm}
\subsubsection*{Inner veto system}
A 50-cm-thick veto region filled with liquid scintillator for both the near and 
far detectors.
\vspace{-0.25cm}
\subsubsection*{Outer veto system}
A proportional-tube tracker  system will identify and locate
``near-miss" muons. This improves the muon rejection by a
factor~20 compared to that of the inner veto by itself.

\par The main uncertainties at CHOOZ came from the uncertainty in the
knowledge of the antineutrino flux coming from the reactor. 
This systematic error is made to vanish by the addition of a near detector.
The nonscintillating buffer will reduce the singles rates in each
detector by two orders of magnitude with respect to those in CHOOZ, which
had no such buffer.  The positron detection threshold will be about 500--700~keV, 
well below the 1.022~MeV physical threshold of the inverse beta decay reaction.
A full review of the detector systematics 
is given in Section~\ref{sec:systematics}.
\begin{table}[htbp]
\caption[Detectors] {Summary of the some parameters  of the
proposed Double Chooz experiment.}
\label{tab:detsum}
\begin{center}
\begin{tabular}{lrc}
\hline
Thermal power & 4.27 GW & each of 2 cores \\
Electric power & 1.5 GWe & each of 2 cores \\
$\nuebar$ target volume & 10.3 m$^3$ & Gd loaded LS (0.1\%) \\
$\gamma$-catcher thickness & 55 cm & Gd-free LS\\
Buffer thickness & 105~cm & non scintillating oil\\
Total liquid volume & $\sim$237~m$^3$ & \\
Number of phototubes per detector & 534 8{\tt "} & 13\% coverage \\
Far detector distance  & 1050~m & averaged\\
Near detector distance & 280~m & averaged\\
Far detector overburden & 300 m.w.e. & hill topology\\
Near detector overburden & 70--80 m.w.e. & shaft\\
$\nuebar$ far detector events (5 yr) & 75,000 & with a 60.5\% efficiency\\
$\nuebar$ near detector events (5 yr) & 789,000 & with a 43.7\% efficiency\\
Relative normalization error        & 0.5\% & \\
Effective bin-to-bin error       & 1\%   & background systematics \\
Running time with far detector only & 1--1.5 year & \\
Running time with far+near detector & 3 years & \\
$\sin^2(2\theta_{13})$ goal in 3 years with 2 detectors & 0.02--0.03 & (90\% CL) \\
\hline
\end{tabular}
\end{center}
\end{table}
\subsection{Time scale}
\par A summary of key detector parameters is given in Table~\ref{tab:detsum}. 
At the Chooz site, the laboratory previously used by the CHOOZ experiment is
vacant and available for use as a far site with minimal preparation. The ventilation
and electrical distribution systems have been upgraded and safety upgrades of the
gallery are
currently in progress.
The relationship between the members of the CHOOZ collaboration
and the reactor company, Electricit\'e de France (EdF), has been very cooperative
and cordial.  For the Double Chooz 
experiment, EdF has again shown the
same level of close cooperation and has signed a letter outlining that
cooperation.
\par The schedule for Double Chooz is aggressive in order to
take into account the great worldwide interest in $\quq$.  Construction of the far
detector 
will be completed by the end of 2007, and that 
of the near detector 
by the end of 2008 (with some uncertainty concerning
the schedule of the near laboratory construction).  Detector operation will be
for 4.5 years, starting with just the far detector, followed by
three years of operation with both detectors (2009--2011).
Important first results are possible with just the far detector because the 
luminosity (12 GW-ton-years) of the original 
CHOOZ experiment will be matched in just a few months.  
Using both detectors, Double Chooz will reach a $\sin^2(2\theta_{13})$ sensitivity of 0.06 in 2009
and 0.03 in 2011. Whether running any longer at that time makes sense will
depend on an evaluation of systematic errors and backgrounds
achieved, as well as the world situation regarding $\quq$.

%
\cleardoublepage
%
\section{Performance and Physics Reach}
\label{sec:signal}
\subsection{Signal}
In this section we provide a short summary of 
the antineutrino energy spectrum as expected 
in the Double Chooz detectors. We first introduce the 
energy spectrum parametrization and then 
include the effect of the oscillations. The goal is to give a good estimation of the sensitivity
 to small values of $\sin^2(2\theta_{13})$ in Section~\ref{cha:sensitivity}.
%
\label{sec:spectrum}
The antineutrino flux provided by the two nuclear cores of the Chooz power plant results from 
$\beta^-$ decay of the fission products of four main isotopes $^{235}$U, $^{239}$Pu, $^{241}$Pu 
and $^{238}$U. The overall \nueb spectrum is evaluated from measurements of $^{235}$U, $^{239}$Pu, 
$^{241}$Pu and theoretical prediction for $^{238}$U~\cite{Huber:2004xh}. We take an average 
burning cycle composition of 55.6$\%$ of $^{235}$U, 
32.6$\%$ of $^{239}$Pu, 7.1$\%$ of $^{238}$U 
and 4.7$\%$ of $^{241}$Pu. A ten percent burnup effect is taken 
into account in detailed simulations however. 
The \nueb number produced and the released energy per 
fission are shown in Table~\ref{tab:Nnueb_and_Enueb}.
\begin{table}[htpb]
\caption{\label{tab:Nnueb_and_Enueb}Total number of \nueb produced and energy released by 
fission above the threshold energy of 1.8~MeV~\cite{Huber:2004xh}.}
\centering\begin{tabular}{lrr}
\hline
& Number of \nueb & Energy released per\\
& per fission              & fission (in MeV)\\
\hline
$^{235}$U  & $1.92\pm 0.036$ & $201.7 \pm 0.6$ \\
$^{238}$U  & $2.38\pm 0.048$ & $205.0 \pm 0.9$ \\
$^{239}$Pu & $1.45\pm 0.030$ & $210.0 \pm 0.9$ \\
$^{241}$Pu & $1.83\pm 0.035$ & $212.4 \pm 1.0$ \\
\hline
\end{tabular} 
\end{table}
We note the west and east reactor cores as $R_W$ and $R_E$. 
The maximum operating thermal power of each core amounts to 4.27~GW. 
The far detector is located at a distance of $1114.6\pm 0.1$~m from 
$R_E$, and at 
$997.9\pm 0.1$~m from $R_W$, leading to $2.86$~$\nu$~events/h. 
A near detector preferred location has been chosen
 at 290.7~m from $R_E$ and 260.3~m from $R_W$, 
leading to $41.2$~events/h. 
At this location, both detectors receive the same neutrino flux ratio coming from both 
nuclear cores. This cancels systematic uncertainties related to the uncorrelated fluctuations of 
the thermal power of each reactor (see 
Section~\ref{cha:sensitivity} for more details). 

The Double Chooz Target contains $10.32~m^3$ of liquid scintillator 
(see Section~\ref{sec:scintillator}). 
Before addition of the small amount of fluors 
as well as the gadolinium compound, the scintillator can be 
described as 20\% of PXE ($C_{16}H_{18}$) and 80\% of dodecane ($C_{12}H_{26}$). 
This leads to $6.79~10^{29}$ free protons available 
for the detection reaction:
\begin{equation} 
\nueb + p\to e^+ + n\;.
\end{equation} 
We thus have a direct relation between positron energy and the incoming neutrino energy, 
taking into account neutron recoil, 
\begin{equation}
\Enueb = \frac{1}{2}\frac{2M_p \Epos+M_n^2-M_p^2-m_e^2}{M_p-\Epos+\sqrt{\Epos^2-m_e^2}\cos\theta_{e^+}}\;.
\end{equation}
We define the visible energy ($\Evis$) as the kinetic energy from the detected positron 
plus its annihilation on an electron. Introducing the notation $\Delta=M_n-M_p=1.293$~MeV, and assuming 
$\langle\cos\theta_{e^+}\rangle=0$ ($\theta_{e^+}$ is the angle between the neutrino and positron directions), 
we have the following expression 
\begin{equation}
\Evis = \Epos + m_e \simeq \Enueb - \Delta + m_e\;.
\end{equation} 
%
The antineutrino cross-section on proton expressed with the same variables is given by:
\begin{equation}
\sigma(\Enueb) = K\times(\Enueb-\Delta)\sqrt{(\Enueb-\Delta)^2-m_e^2}\;,
\end{equation}
where $K$ is directly extracted from the neutron lifetime
\begin{equation}
K=(9.559\pm0.009)\;10^{-44}\text{~cm}^2\,\text{MeV}^{-2}\;.
\end{equation}
%
%
For $\sin^2(2\theta_{13})$ above~0.001, oscillation terms 
with $\Delta{m}^2_{21}$ can be ignored, and we can write:
\begin{equation}
  \label{eq:survivalprobability}
  P_{ee}(\Enueb,L,\Delta{m}^2_{31},\theta_{13}) = 
  1-\sin^2(2\theta_{13})\sin^2\left(1.27\frac{\Delta{m}^2_{31}[10^{-3}\;
  \text{eV}^2]L[\text{km}]}{\Enueb[\text{MeV}]}\right)\;.
\end{equation}
We compute the number of events in detector $D$ from reactor $R$ in the i$^{\text{th}}$ antineutrino 
energy interval $[E_i;E_{i+1}]$ as
\begin{equation}
N_i^{R,D} = {\cal T}(R)\int_{E_i}^{E_{i+1}} \sigma(\Enueb)\,\phi_{R,D}(\Enueb)\,R_i(\Enueb)
P_{ee}(\Enueb,L(D,R),\Delta{m}^2_{31},\theta_{13})\;\text{d}\Enueb\;,
\label{eq:Ni}
\end{equation}
where the flux from reactor $R$ to detector $D$ has the following expression
\begin{equation}
\phi_{R,D}(\Enueb)=\frac{1}{4\pi L(D,R)^2}\sum_l N^{\text{fis},R}_l\phi_l(\Enueb)\;.
\end{equation}
We assume a polynomial parametrization of the four isotopes antineutrino energy spectra 
($l=^{235}$U, $^{239}$Pu, $^{241}$Pu or $^{238}$U):
\begin{equation}
\phi_l(\Enueb)=\exp\left(\sum_{k=0}^6a_{kl}\Enueb^k\right)\;.
\end{equation}
The $a_{kl}$ coefficients may be found in Reference~\cite{Huber:2004xh}, 
and $N^{\text{fis},R}_l$ is the 
number of fissions per second of the isotope~$l$, which can be 
extracted from~\cite{Bemporad:2001qy}.

\begin{center}
\begin{figure}[hbtp]
		\begin{tabular}{cc}
		\includegraphics[width=0.5\textwidth]{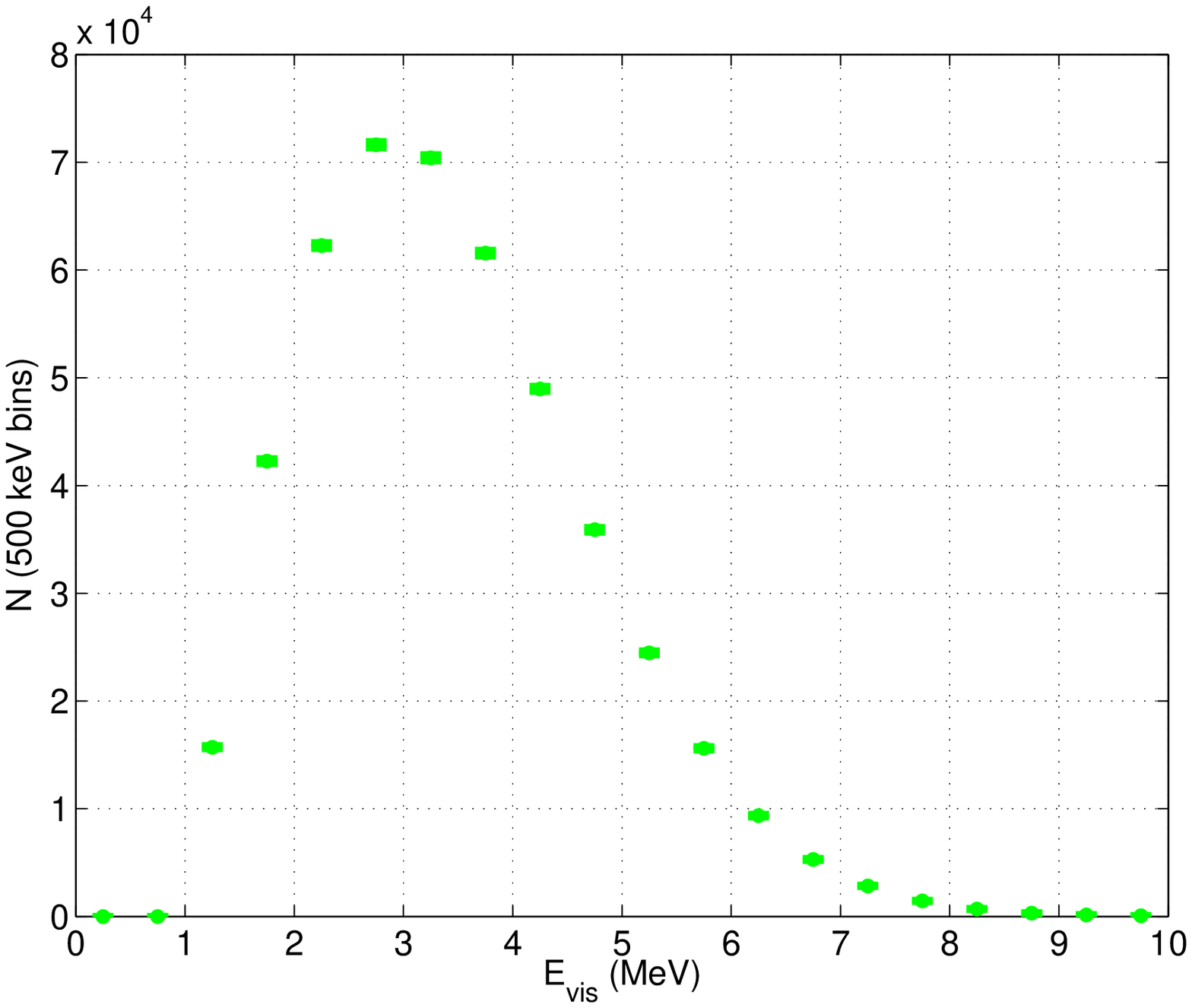} & \includegraphics[width=0.5\textwidth]{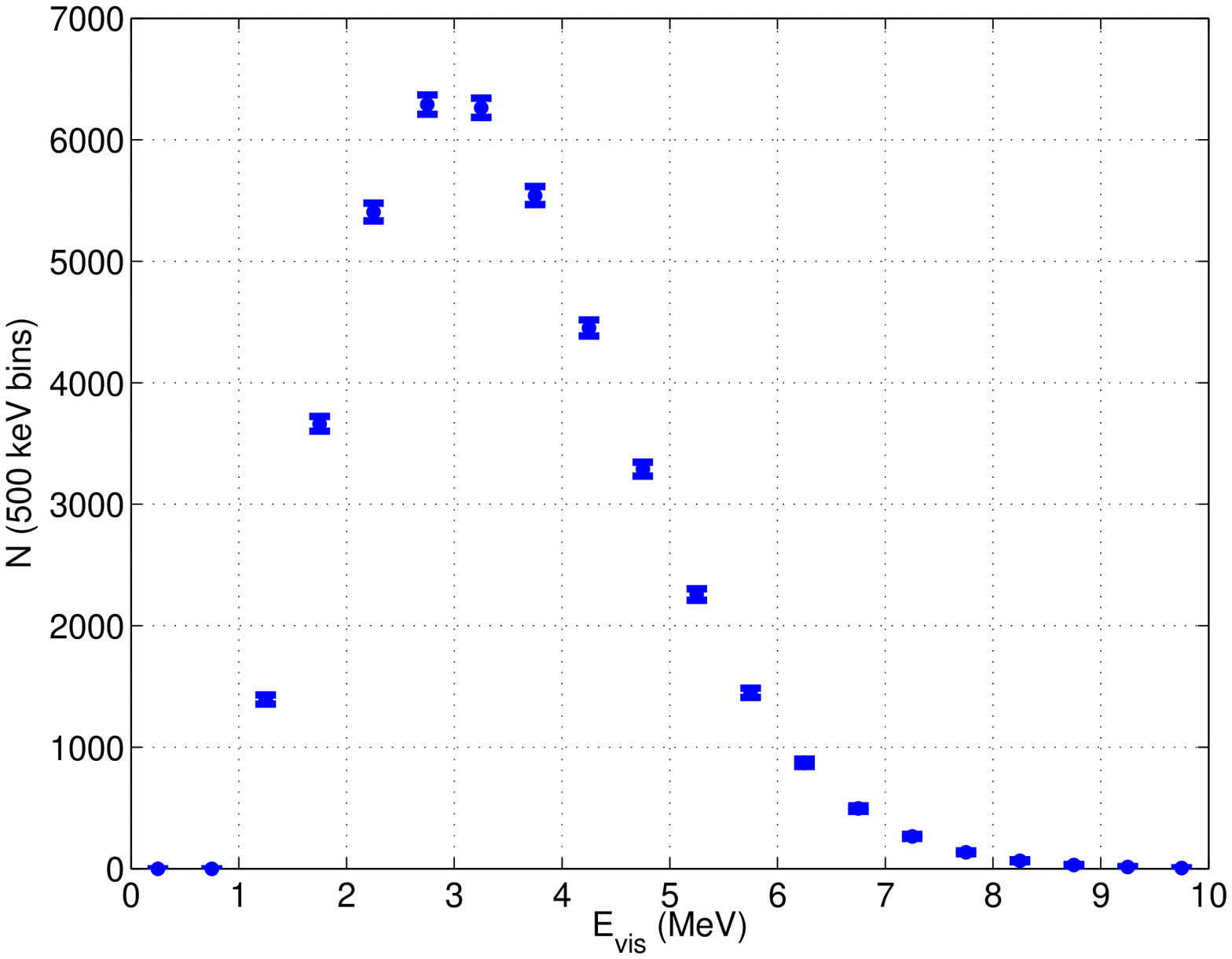}\\
		\includegraphics[width=0.5\textwidth]{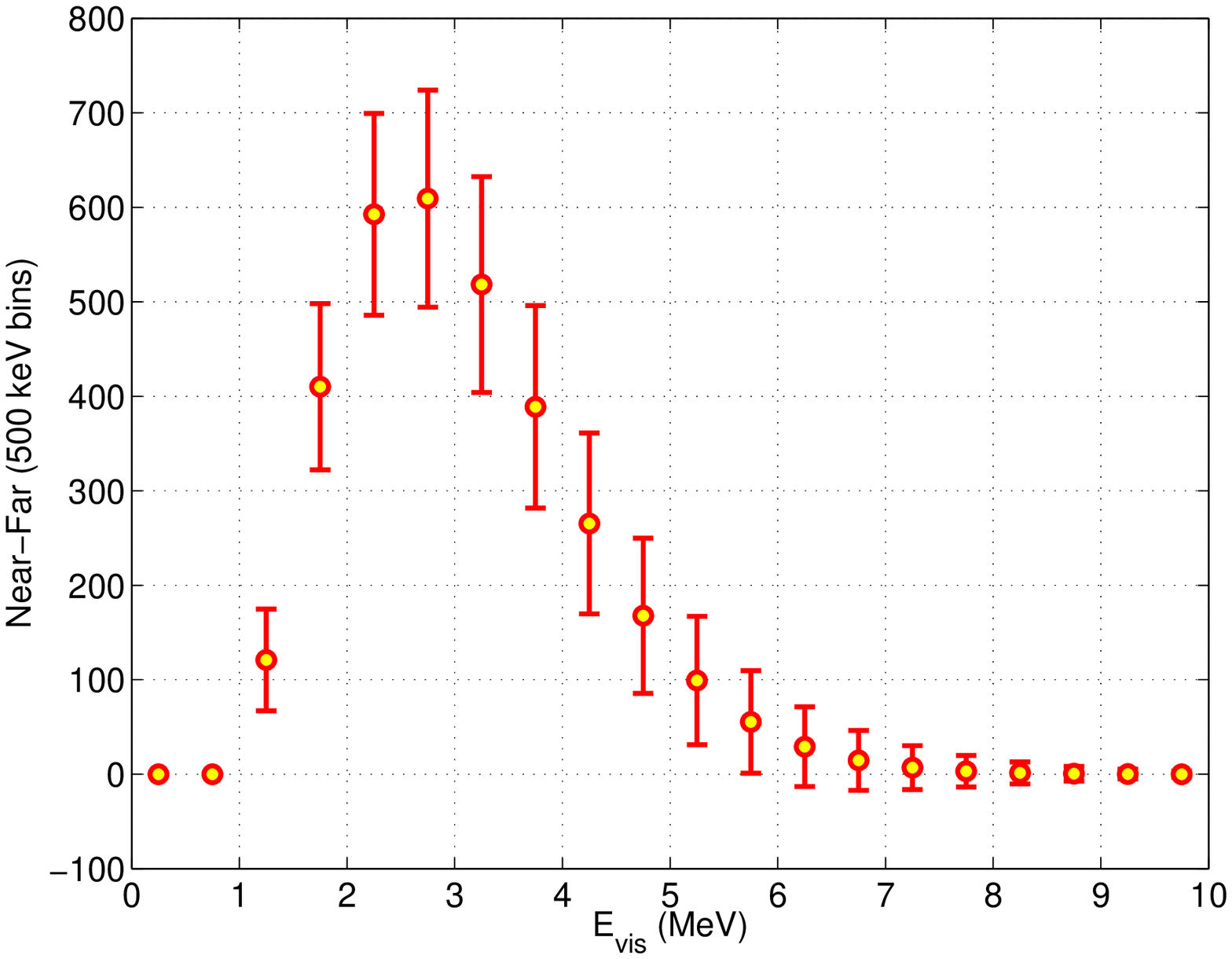} & \includegraphics[width=0.5\textwidth]{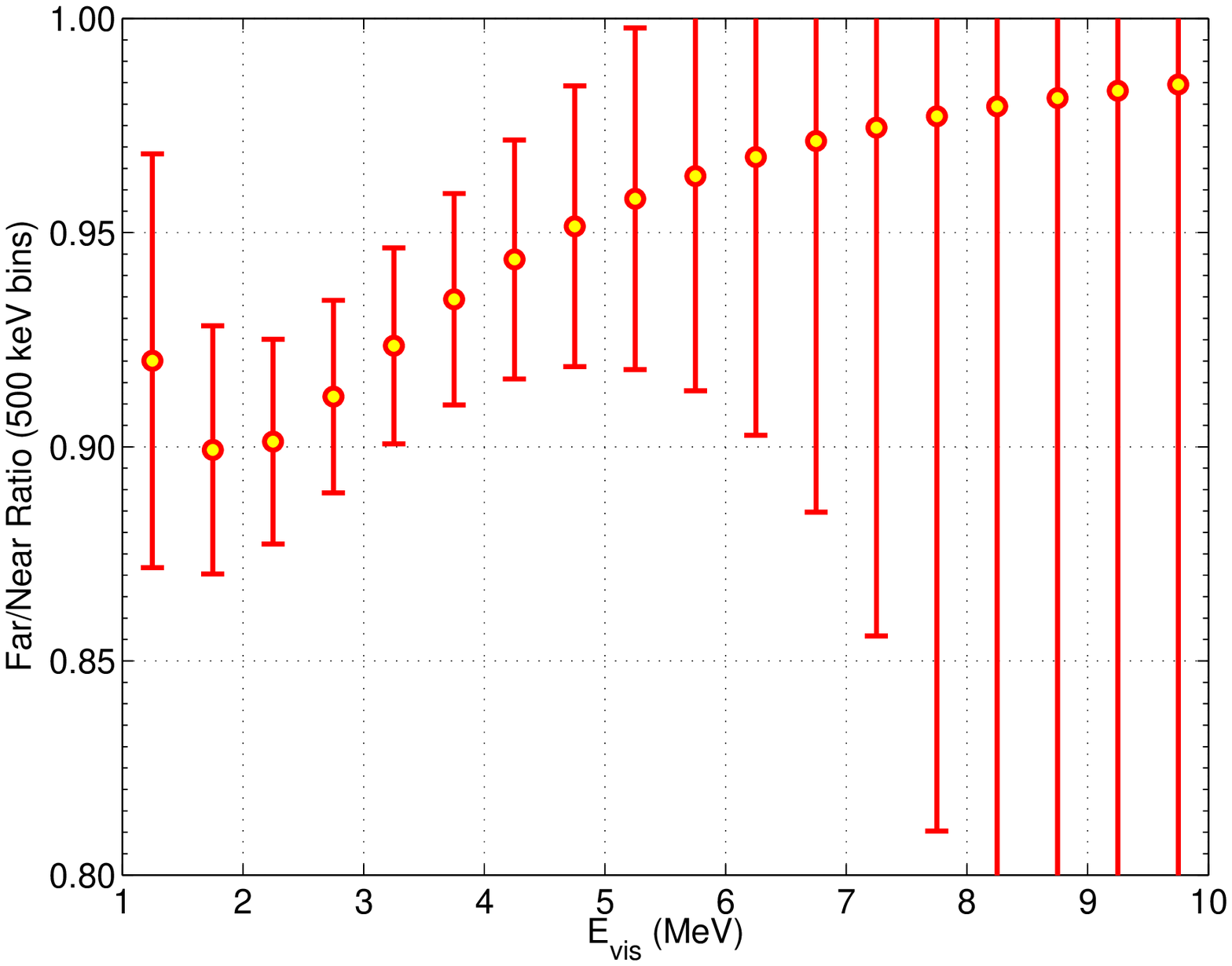}\\
		\end{tabular}
\label{fig:expsignal}
\caption{Expected signal assuming a true value of $\sin^2(2\theta_{13})=0.1$ and $\Delta{m}^2_{31}=2.5 \times 10^{-3}~eV^2$, for 3 years of data taking with both detectors. Top left: positron energy spectrum at the Near detector. Top right: positron energy spectrum at the Far detector. Bottom left: spectrum difference between the Near and Far detector (normalized to the Far detector statistics). Bottom right: Far to Near spectrum ratio. Errors bars quoted are only statistical.}
\end{figure}
\end{center}

The generic factor ${\cal T}(R)$ in equation~(\ref{eq:Ni}) takes into account the experiment life time and 
the global load factor of reactor $R$. The detector ($D$) response is expressed through $R_{i}(\Enueb)$, 
including the number of free protons, a dead time of 30$\%$ for the near detector due to its shallow depth, 
an energy resolution of 7$\%$ (at 1~MeV), and a detector efficiency of 80$\%$.
The global load factor of the Chooz power plant was $73.3\%$ and $81\%$ for $R_W$ (Chooz-B1), and 
$79.7\%$ and $76.2\%$ for $R_E$ (Chooz-B2), in 2003 and 2004 respectively. 
In the following we assume the average value of $78\%$. 
After 5 years the amount of data available would be 76,000~reactor
neutrino events in the far detector (efficiency included) and 800,000 events in the near detector, 
accounting for the estimated reactor and detector efficiencies (efficiency included). Neutrino rates and 
associated information are summarized
in Table~\ref{tab:neutrinorates}. 
\begin{table}[htb]
\begin{center}
\caption{$\nu$ rate expected in the Near and Far detectors, with and 
without reactor and detector efficiencies. 
Final distances for the Near detector have not been decided yet, but the values presented are good estimates 
within 30$\%$. The rate without efficiency is used for signal 
to background comparison. The integrated rate in the last line
includes detector efficiency, dead time, and reactor off periods averaged over a year.}
{\label{tab:neutrinorates}} 
\begin{tabular}{lrr}
\hline
 \multicolumn{1}{c}{Detector}       &         Near             &          Far              \\
\hline 
 Distance from West reactor  (m)    &       $290.7$            &   $1114.6\pm 0.1$        \\
 Distance from East reactor  (m)    &       $260.3$            &   $998.1\pm 0.1$          \\
 Detector Efficiency                 &       80$\%$             &   80$\%$                  \\
 Reactor Efficiency                  &       78$\%$             &   78$\%$                  \\
 Dead Time                          &       30$\%$             &   3$\%$                   \\      
 Rate without efficiency ($d^{-1}$) &       1011.5              &   68.8                    \\
 Rate with detector efficiency ($d^{-1}$) & 566.4              &   53.4                    \\
 Integrated rate ($y^{-1}$)         &       161,260             &   15,200                   \\
\hline
\end{tabular}
\end{center}
\end{table}
Figure~\ref{fig:expsignal} shows the positron energy spectrum expected at Chooz-Near and Chooz-Far for 3 years of data taking,
assuming a true value of $\sin^2(2\theta_{13})=0.1$ and $\Delta{m}^2_{31}=2.5 \times 10^{-3}~eV^2$.
\subsection{Experimental errors}
\label{sec:systematics}
Double Chooz will use two $\nueb$ detectors in order to cancel or decrease significantly 
the systematic uncertainties that would 
limit the sensitivity to $\theta_{13}$. 
For the sake of comparison, the total systematic error 
of the CHOOZ experiment amounted to $2.7\%$~\cite{bib:chooz}. 
This error was dominated by the reactor antineutrino flux and spectrum 
uncertainties, that amounted to 1.9$\%$.
At neutrino luminosities addressed by Double Chooz (less than a few hundred thousand events
in the far detector) the relative normalization between the two detectors is the most important 
source of error and must be carefully controlled. The goal of Double 
Chooz is to reduce this uncertainty to 0.6$\%$. 
Double Chooz systematic uncertainties were already reviewed in the Double Chooz Letter of Intent (LOI)~\cite{bib:loi}. 
We discuss them briefly, focusing on the work done to confirm the LOI~\cite{bib:loi}. 
It is worth noting that we do not assume any major innovation to decrease the systematic error 
below one percent. Considering the 
CHOOZ results, we identified the uncertainties that cancel
(some of them only partially) by using two identical detectors 
and derived the predicted systematics 
for Double Chooz. However, two cases are discussed in 
more detail below because they needed some 
dedicated R$\&$D in the collaboration: 
the dead time measurement on page~\pageref{'dead'}, and
the Target volume measurement on page~\pageref{'volume'}.
\subsubsection{Reactor induced uncertainties}
\label{sec:sysReactor}
Since the neutrino flux is isotropic, all the reactor induced systematic uncertainties cancel but the 
solid angle. A summary of the reactor induced systematics is given in Table~\ref{tab:systdc_reactor}.
\begin{table}[htb]
\begin{center}
\caption[Reactor induced systematics]
{\label{tab:systdc_reactor} Reactor induced systematics} 
\begin{tabular}{lrr}
\hline
Reactor power             &         CHOOZ          &   Double Chooz     \\
\hline
Reactor power             &         $\sim$2$\%$   &   negligible     \\
Energy per fission        &         0.6$\%$       &   negligible     \\
$\nueb$/fission           &         0.2$\%$       &   negligible     \\
Neutrino cross section    &         0.1$\%$       &   negligible     \\
\hline
\end{tabular}
\end{center}
\end{table}
\subsubsection*{Solid angle}
%
and each core have been measured with a precision of $10$~cm.
%
The distances between the center of the detectors and 
the reactor core center have to be
precisely known. The far detector-cores distance have been measured
with a precision of 10~cm. Such a precision is 
achieved using geometric measurements at
the near site. Considering a near detector at about 250-300~m, this leads to a
systematic error of 0.06\%.

Two other effects at this baseline distance could add up: a statistical fluctuation of
the neutrino detected number whose mean 
may vary; a displacement of the production
``barycenter'' inside cores.

Concerning the first effect, a simple Monte Carlo simulation 
shows us the distance traveled
by neutrinos between production and detection closely follows a normal
distribution, and the most important point is 
that its skewness is very low. A study using the 
central limit theorem tells us the mean won't vary by more than
0.01\%.

\begin{figure}[htbp]
\centering
\includegraphics[width=.6\textwidth]{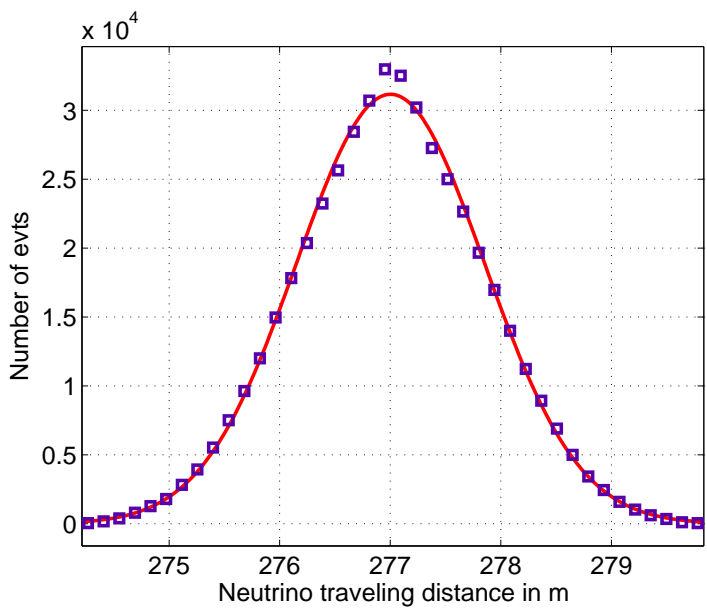}
\caption{Neutrino traveling distance histogram for near detector (274~m away, 40~m
depth). The traveling distance distribution is centered around 277~m, with no bias
(skewness around $10^{-3}$). In this Monte Carlo simulation we assumed a uniform
neutrino production inside a typical cylindrical reactor core.}
\label{fig:nu_travelling_distance}
\end{figure}

Regarding the second effect, the ``barycenter'' of the neutrino emission in the reactor
core have to be monitored with a similar precision (a level of 5~cm was achieved in
the Bugey experiment~\cite{bib:bugey}). Such an effect would lead to an uncertainty of 0.03\%.
Actually it was shown in the Bugey experiment that the barycenter moves by tens of
centimeters vertically, but is very stable laterally. A 10~cm vertical displacement would
lead to a distance variation of about 1~cm (at 300~m and 40~m depth) which is
negligible. In conclusion, determining precisely the barycenter of each reactor core and 
the center of each detector is sufficient
to take into account flux uncertainties from finite size effects.

\paragraph*{Core power uncertainties:}

\begin{figure}[htbp]
\centering
\includegraphics[width=.45\textwidth]{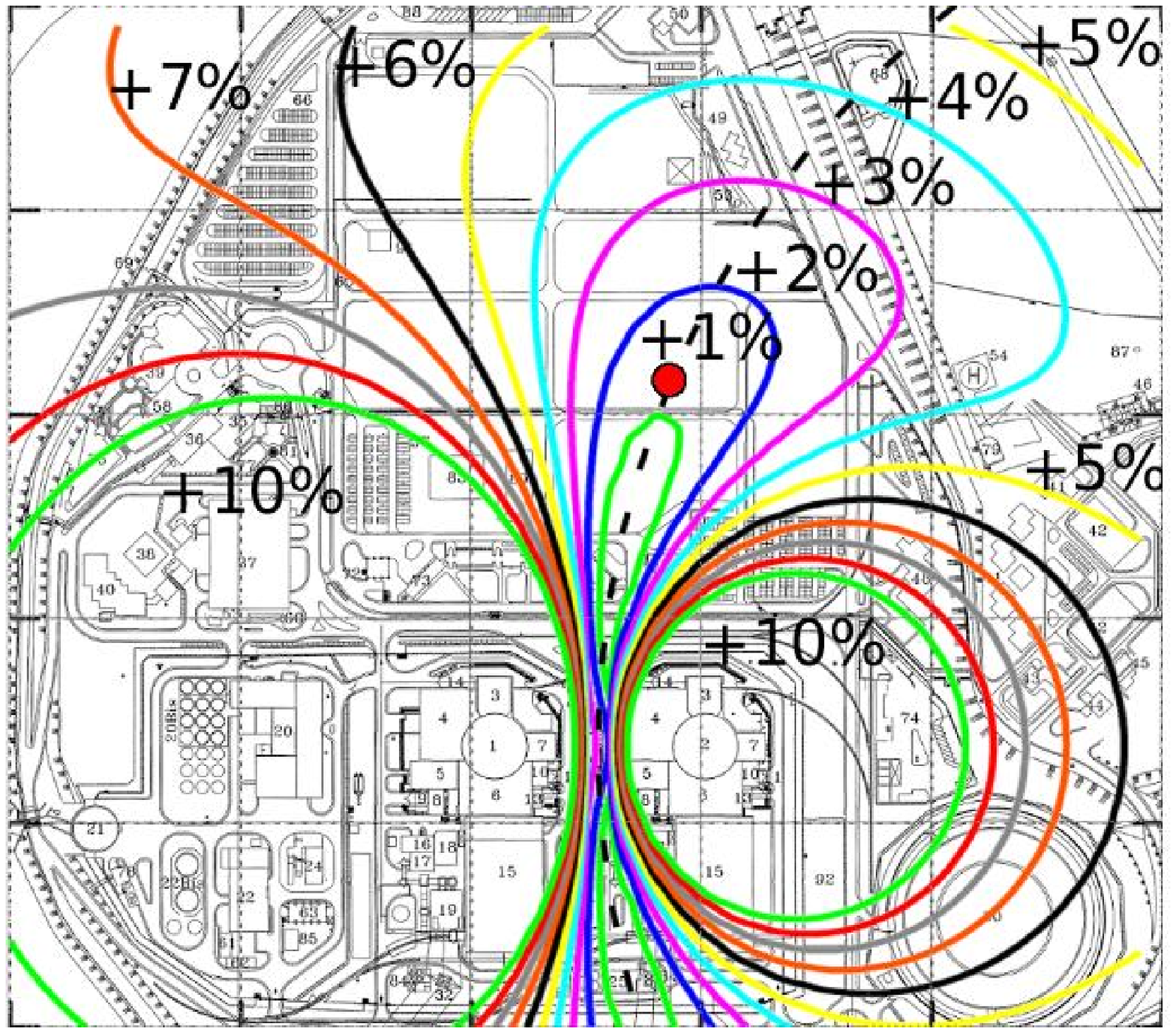}\hfill
\includegraphics[width=.45\textwidth]{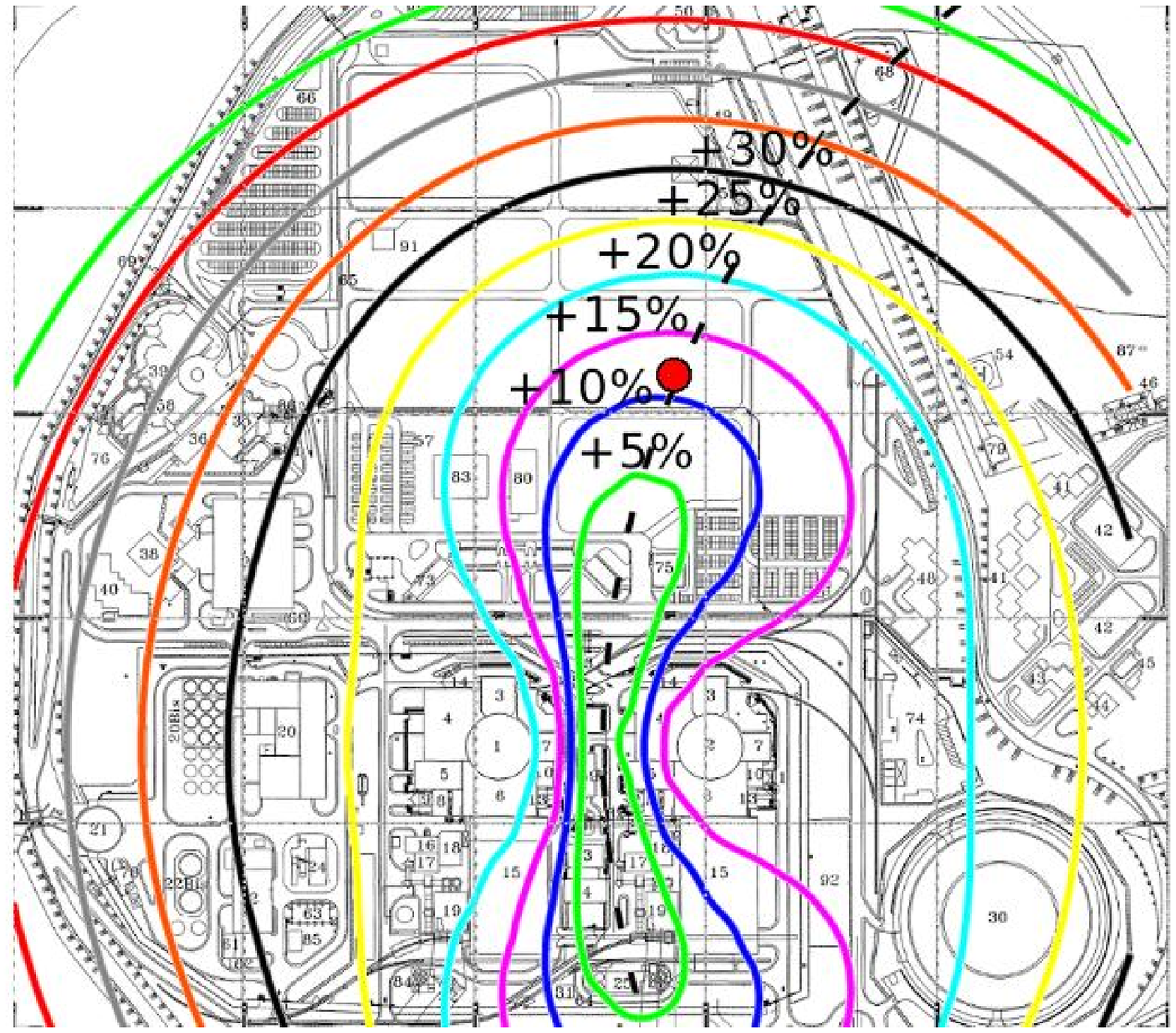}
\caption{Iso-sensitivity contours as a function of the near detector location
superimposed on the power plant viewgraph. Buildings~1 and~2 
respectively are the
west and east reactor cores. For the left figure, we assumed a 3\% uncorrelated
uncertainty on the power of both cores and $\sigma_{\text{abs}}=2.0\%$,
$\sigma_{\text{rel}}=0.6\%$, $\sigma_{\Delta{m}^2}=20\%$. Each superimposed colored
curve indicates a +1\% relative degradation on $\sin^2(2\theta_{13})$ sensitivity with
respect to the best sensitivity inside this plane. On the right figure we assumed the
same uncertainties plus $\sigma_{\text{shp}}=2.0\%$. Here, each superimposed colored
curve indicates a +5\% relative degradation on $\sin^2(2\theta_{13})$ sensitivity with
respect to the best sensitivity inside this plane. On both figures, the red circle shows
the chosen location of the near detector.\label{fig:iso_contours_map}
}
\end{figure}

The thermal power of each core may not be known with a precision better than 1 to 3\%.
Specific reactor core power uncertainties have been included in a
$\chi^2$ statistical study.
Results are illustrated on Figure~\ref{fig:iso_contours_map}. From
this study we observe that even a large 3\% uncorrelated uncertainty 
on each core power
does not significantly degrade the $\sin^2(2\theta_{13})$ sensitivity (left 
figure), the
main degrading effect coming from reactor spectrum shape uncertainty (right 
figure),
since installing the near detector farther from cores results
 in a poorer knowledge of
the reactor spectrum shape. In order to eliminate core power uncertainties
and other possible core uncertainties, the best choice is to place the near
detector on the same flux ratio line as for the far detector. The 
best near detector location is marked by a red dot (solid circle)
on Figure~\ref{fig:iso_contours_map}
 (260.3~m away from the east reactor
and 290.7~m away from the west reactor).

\subsubsection{Spent Fuel signal at Chooz}
\label{sec:spentfuel}
About every 8~months the reactor is stopped for unloading of used fuel 
and reloading with a
fresh fuel enriched
at 3.45$\%$ in $^{235}U$. During this stop only one quarter of the fuel 
assembly is exchanged, while the
others remains in the core although relocated following a precise 
design.
The operation of the N4-reactor is such that the total amount of energy 
extracted per
ton of enriched Uranium is
  45~GW.day/ton = 3.89 $1O^{15}$~J/ton or around 1.2~$1O^{26}$ fissions. 
This operation
produced in particular
long lived fissions products which can be evaluated by the cumulative 
fission yields
and checked for some of them by analysis
of the spent fuel ; nevertheless the uncertainty on the exact amount 
can be at the level of 20\%. Most of these fission products are beta 
emitters, but if we restrict
ourself to those producing
neutrinos above the inverse beta decay threshold (1.8 MeV) with an 
half-life of more than a day, only 4~fissions products
remains :
\begin{table}[htb]
\begin{center}
\caption[Spent Fuel signal at Chooz]
{\label{tab:spentfuel} Radioelements stored in the spent fuel pool on 
site. BR stands for branching ratio.}
\begin{tabular}{lrrrr}
\hline
Isotope                       &         $T_{1/2}$        & Mass (kg)   & $\beta$-End point   &  BR $> 1.8$ MeV  \\
\hline
Ba$^{140}$/La$^{140}$         &         12.752 d       & 6.15         & 3.762     & 4.8  \\
Ce$^{144}$/Pr$^{144}$         &         284.893 d      & 5.44         & 2.997     & 100. \\
Ru$^{106}$/Rh$^{106}$         &         373.59 d       & 3.06         & 3.678     & 86. \\
Sr$^{90}$/Y$^{90}$            &         28.79 y        & 3.81         & 2.280     & 100. \\
\hline
\end{tabular}
\end{center}
\end{table}
All these fissions products are in pairs: the first one is responsible 
of the long
lifetime and the second one
produced much more quickly the beta decay with the end point indicated 
in the table. The
amount indicated above
corresponds to the total production of a N4-reactor when the 120~tons 
of enriched UO$_2$
has produced the required energy ;
hence only one quarter is added every 8~months in the pool closed to 
the reactor core.
The pools at Chooz stores all the spent fuel since the start of 
operation of the reactor. These pools
are located in a building closer
from the near site by about 30~m compared to the nuclear core. The 
evaluation of the
signal induced by these fission product
  can be made using the informations in the table and the cross section.
The signal due to the Ba$^{140}$/La$^{140}$ is weaken by the small 
branching ratio of energetic antineutrinos and decay quickly. On the 
longer term
Sr$^{90}$/Y$^{90}$ produced a more
or less constant background less than 0.5~interaction per day for both 
pools.

The signal due to spent fuel accumulated in the past years averaged over
the year compared to the signal directly produced by the two cores is 
around 0.1\%.
But it is concentrated in the lower part of the positron spectrum ($<~1.7$~MeV) 
where it represent an excess close to 1
\subsubsection{Detector induced uncertainties}
\label{sec:sysDetector}
The uncertainty from detector hardware and 
the selection cuts has to be reduced by
a factor of three from the CHOOZ experiment. A new detector 
design provides us with such an improvement, as well 
as a reduction of the accidental backgrounds. A steel shield
will be used instead of  the sand that was 
used in CHOOZ. A non scintillating region, called the ``Buffer", 
is created to reduce the single rate in the Target+$\gamma$-catcher. An 
efficient Inner Veto 
is designed for muon tagging and fast neutron background rejection. 
An outer muon tracker
 will allow us to further reduce the cosmogenic backgrounds. 
Possible detector-related
contributions to systematic error are considered in what follows.
\subsubsection*{Gadolinium concentration}
The concentration of Gd will be identical in both detectors since Target liquids will be 
produced in a single batch. It can be measured from the neutron capture time 
at a level of 0.3$\%$~\cite{bib:chooz} or by 
analyzing the neutron multiplicity 
from a Californium source (see Section~\ref{sec:calibration}). 
The major concern is a 
non-uniform Gd concentration (e.g., a top/bottom asymmetry).
This can be studied with a z-axis Californium calibration. 
Comparison  between the near and far detectors, with the same set of 
calibration sources, will allow us to detect less than 0.3$\%$ time capture differences. 
\subsubsection*{spill in - spill out}
A neutrino event is defined by the neutron capture on Gd 
after positron annihilation.  It is possible
to have a neutrino interacting in 
the Target region but a neutron capture in the $\gamma$-catcher region, 
called ``spill out", 
and the n capture on Gd for an event interacting in the $\gamma$-catcher, called 
``spill in".
In CHOOZ the spill-in effect accounted for 4$\%$ of the events, and the
 spill-out 2$\%$ (the quoted error was about 40\%). It led
to a $\sim$1\% uncertainty (from Monte-Carlo simulation). 
However, the effect cancels if the two detectors have a similar Target 
(a few millimeter geometrical effect has  
a negligible effect on the results).
\subsubsection*{Target proton measurement}
\label{'volume'}
The Target liquid will be prepared in a large single batch of about 16 tons.
 Thus, the uncertainty on the scintillator density ($\sim$0.01$\%$)
as well as chemical composition ($\sim$1$\%$) disappears from the relative
normalization error. 

The mass of both Targets has to be known with an uncertainty of 0.1$\%$. 
In 2004-2005, we made investigations on several different methods 
to ensure accurate volume measurements. 
Two particular methods have been designed and tested:
MPI-Heidelberg suggested to use pH-metry 
and showed that the volume determination can reach 
0.2$\%$ of uncertainty (with water).
The CEA-DAPNIA proposed a different method: a 
weight measurement. For the experiment, the idea 
is to weigh an intermediate vessel before and 
after the filling of the Target. 
The difference between the two measurements gives 
the weight of liquid in the detector. 
In order to reduce systematics, the 
same intermediate vessel, equipped with the weight sensors 
has to be used for both the far and near detectors.
We performed tests on the 
weight sensors. We first checked their repeatability, 
weighing a 1.5~ton mass 20~times. The standard deviation was 0.15~kg.
Then, we made a cross-check between a 
flow-meter and the weight sensors filling a 2~cubic meter 
stainless-steel vessel. The flow meter 
data sheet indicates a global accuracy of 0.5$\%$ which can 
be reduced to 0.3$\%$ with local calibration. 
(This calibration uses a weight measurement). 
After the calibration of the flow meter 
and several trials with both sensors, we obtained compatible 
values of volume with a (one sigma) dispersion of 0.12$\%$. 
We plan to make an 
absolute calibration of the weight sensor at 
Laboratoire National d'Essais in Paris.
We are working on a simple method 
to ensure the local value of gravity is the same at the far and 
near sites. 

To achieve that, we have accurately weighed a 100 g standard at both
the near and far sites. Any weight difference below 10 mg means that
this gravity variation can be neglected. We have measured differences
smaller than 3 mg. (The precision scale we have used was calibrated
by measuring the same standard on different sites close to Chooz, at
different altitudes and thus different values of g.)
As was written before, the result of these measurements is a weight.
To compute the volume, we also have to perform a density measurement.
For that, we checked the availability of a densitometer which can
perform measurements with an accuracy of $10^{-4}$ g/cm$^3$.  This
device uses a measurement of the sound velocity and it is not affected
by variations of gravity. A temperature control and stabilizer at the
level of one degree Celsius will be installed in the detector.
\begin{table}[htb]
\begin{center}
\caption{Systematic uncertainties related to the Target 
free H measurement. This error will be less than~0.2$\%$ 
since a single batch will be prepared for both
detector Target. The ``Single Detector" case is  important for the 
non proliferation component of Double Chooz, however. 
The H/C determination depends on the scintillator solvent 
composition. Dodecane is favored over mineral 
oil whose composition is less well
 defined (1$\%$). In the 
 case of a single detector we estimate a systematic error contribution of 
$<$0.55$\%$.} \label{tab:systdc_volume}
\begin{tabular}{lrr}
\hline
               & \multicolumn{1}{c}{Single Detector}     & \multicolumn{1}{c}{Double Chooz}  \\
\hline
Volume         &                           0.2$\%$      &   0.2$\%$                        \\
Ratio H/C      &                           $<$ 0.5$\%$  &   0                               \\
Density        &                           0.01$\%$     &   0                               \\
Gravimetry (g) &                           0.005$\%$    &  0.002$\%$                       \\
\hline
Total          &                           $<$0.55$\%$   &   ~0.2$\%$                        \\
\hline
\end{tabular}
\end{center}
\end{table}
\subsubsection*{Dead time measurement}
\label{sec:sysDeadtime}
\label{'dead'}
As explained in Section~\ref{sec:daqReadout}, the digitization and
acquisition systems do not introduce dead time. The only two sources
of dead time are the first level trigger (hardware) and the offline cuts:
\begin{enumerate}
\item the first level trigger necessarily introduces dead time: when an
event comes very shortly after a first one, it cannot be detected;
\item offline, a short veto will be applied after each event, to reject
pile-up, and, more importantly, a 500 $\mu$s veto after each muon.
\end{enumerate}

The veto times will always be longer than the electronics dead time and care
will be taken that they always include it. Therefore, the veto times account
for the total detector dead time. This time can be measured with the
precision of the main clock.

The main clock is distributed to the Waveform Digitizers by the hardware
trigger system. The stability of this clock is still not known but a
precision of 10$^{-4}$ (9s/d) is achieved in modest devices. The frequency
and instability of the clock will be measured with a GPS clock.

Since the uncertainty on dead time contributes directly
to the uncertainty of the neutrino flux measurement and the dead time
itself differs strongly between the two detectors, a cross check is
performed by the mean of a random fake event generator which produces
light pulses in the inner volume of the detectors and records them.
It is then possible to count how many have been rejected or missed.

Those methods of dead time measurements  will be intensively tested
with the Far detector. Simulation of the Near detector higher rate
could be tested using the random fake event generator.

\subsubsection{Uncertainties in the efficiency determination}
$\nueb$ events manifest themselves as two local energy depositions of more than 500~keV 
in less than 200~$\mu$s. Delayed energy should be consistent with a neutron capture on Gd.  
In the CHOOZ experiment, the selection cut uncertainty
 was 1.5$\%$, coming from a set of 7 
analysis cuts to extract the $\nueb$ candidates~\cite{bib:chooz}. 
The Double Chooz detector design differs slightly from CHOOZ since a non scintillating Buffer region 
shields the active region (Target and $\gamma$-catcher) from the phototube intrinsic radioactivity. 
This allows us to reduce the number of selection cuts, 
while keeping a small accidental background contamination 
(a few percent of the signal). A summary of the systematic errors associated with $\nueb$ event
selection cuts is given in Table~\ref{tab:systdc_cut}.
\subsubsection*{Identification of the prompt positron signal}
The positron induced by the antineutrino interaction in the Target has a very short track 
in the detector. It annihilates and creates two back-to-back 511~keV gammas. 
One of the gammas may leave the Target, but most of the time it will deposit a large fraction 
of its energy in the Target+$\gamma$-catcher sensitive volume.
Thanks to the addition of a Buffer volume between phototubes and the active detection region, 
the energy threshold will be between 500 and 700~keV. 
Using the expected energy resolution
of 7$\%$ at 1~MeV, the fraction of neutrino events leading to a positron below the threshold 
will be smaller than 0.03$\%$. Thus almost no $\nueb$ candidate will be rejected by 
a prompt event energy cut since the antineutrino induced positron deposits at least 1~MeV. 
The systematic error associated is negligible.
\subsubsection*{Identification of the neutron delayed signal}
Neutrons induced by neutrino interactions can travel about 5~cm before being 
captured. The thermal neutron is captured either on hydrogen or on gadolinium. (We 
neglect here carbon captures.) The gammas emitted after the capture on Gd can escape in
some cases the Target+$\gamma$-catcher volume (when a high 
energy gamma is emitted in the gamma shower). 
Thanks to the two detector concept, the error on the the absolute knowledge of the gamma 
spectrum from a Gd neutron capture disappears. 
The selection cut that identifies the neutron will be set at about 6~MeV (an optimization is in progress).
An error of $\sim$100~keV in the energy scale leads to a 0.2$\%$ effect on the neutron selection.
\subsubsection*{Neutron time capture on Gd}
The analytical behavior describing the neutron capture time on Gd is not known. The shape is rather
exponential, however. Furthermore there is no reason that Gd capture differs significantly from 
neutron capture on light nuclei 
which follows a Gamow-like behavior.
With a single detector at CHOOZ, the systematics 
error was 0.4$\%$, but a relative comparison between 
detectors with identical scintillator relies only on the control of the electronic time cuts. 
A relative comparison of the electronics of the 
two detectors will lead to an effect less than 0.1$\%$.
\begin{table}[htb]
\begin{center}
\caption[Summary of the uncertainties associated with the data reduction]
{Summary of the uncertainties associated with the 
data reduction (selection cuts). 
CHOOZ values have been taken from~\cite{bib:chooz}.}
\label{tab:systdc_cut}
\begin{tabular}{lrr}
  \hline
                   &  CHOOZ              & Double Chooz \\
  \hline
  selection cut    &   \multicolumn{2}{c}{rel. error $(\%)$} \\
  \hline
  e$^+$ capture energy containment &  0.8 &  0.1 \\
  neutron capture                  &  1.0 &  0.2 \\ 
  n capture energy containment     &  0.4 &  0.2 \\
  neutron delay                    &  0.4 &  0.1 \\ 
  combined                         &  1.4 &  $\sim$0.3 \\
    \hline
\end{tabular}
\end{center}
\end{table}
\subsubsection{Summary of the systematic uncertainty cancellations}
Table~\ref{tab:systematicsoverview} summarizes the identified systematic errors
that are currently being considered for the Double Chooz experiment.
\begin{table}[htb]
\begin{center}
\caption{Total systematic error on the normalization between the detectors}
\label{tab:systematicsoverview}
\begin{tabular}{llrrr}
\hline 
 \multicolumn{3}{c}{}                                                               & CHOOZ    & Double Chooz \\
\hline
 Reactor & & Solid Angle        & ---          & 0.06$\%$          \\
\hline
Detector &  H nuclei in Target                                   & Volume             & 0.3$\%$      & 0.2$\%$          \\
         &                                                     & Fiducial Volume    & 0.2$\%$      & 0            \\
         &                                                     & Density            &              & 0.1$\%$          \\
         &                                                     & H/C                & 0.8$\%$      & 0            \\
\hline
Detector & Electronics                                         & 
Dead Time          & ---          & 0$\%$          \\     
\hline
 Particle& Positron                                            & Escape             & 0.1$\%$      & 0            \\
 Identification&                                               & Capture            & 0            & 0            \\
         &   & Energy Cut & 0.8$\%$      & 0.2$\%$          \\
\hline
Particle& Neutron                                              & Escape             & 1.0$\%$      & 0            \\
 Identification&                                               & Capture ($\%$~Gd)  & 0.85$\%$     & 0.3$\%$          \\
         &                                                     & Identification Cut & 0.4$\%$      & 0.1$\%$          \\
\hline
Particle & Antineutrino                                        & Time Cut           & 0.4$\%$      & 0.1$\%$          \\
 Identification&                                               & Distance Cut       & 0.3$\%$      & 0            \\
         &                                                     & Unicity            & 0.5$\%$      & 0            \\
\hline
Total    &                                                     &                    & 1.5$\%$      & 0.5$\%$          \\   
\hline
\end{tabular}
\end{center}
\end{table}
\subsection{Backgrounds}
\label{sec:backgrounds}
There are backgrounds from primordial and man-made radioactivity, 
as well as backgrounds induced by cosmic ray interactions. 
Previous experiments provide quantitative guidance
on how to design an experiment which, in 
principle, allows us to measure $\nuebar$'s 
with a high signal to noise ratio.
\subsubsection{Accidental backgrounds}
\label{sec:bkg_acc}
Naturally occurring radioactivity can create accidental as well as correlated backgrounds. 
Selection of high purity materials for detector construction and passive 
shielding provide an
 efficient handle against this type of background. 
Gamma, beta and neutron signals in the inner detector or in the rock may generate 
accidental background events which mimic prompt $\nuebar$ interaction signal (positron-like).
The delayed background (neutron-like) comes mainly from neutron capture on Gd.
However, part of those neutron-like events could be due to bremsstrahlung photons radiated 
from cosmic muons which traverse the rock surrounding the detector. 
The neutron like background rate had been measured at the far site in the CHOOZ detector, 
at the level of $45\pm 2/h$ (after cuts)~\cite{bib:chooz}. 
Based on that measurement and the new target volume,
we assume a neutron rate of 83/h in Double Chooz. 
However, such a high rate remained unexplained in CHOOZ, and current simulation 
is not reliable in that it gives 
a much smaller rate~(see Section~\ref{sec:bkg_nlike} for details).  
The neutron-like rate is taken at $0.023 Hz$ in the Target and can thus be neglected 
as a relevant contribution for
the positron-like events since phototubes contribute at the level 
of the Hertz.

The accidental background rate $b_{acc}$ is given by $b_{acc}= b_p V_p b_d V_d \tau_d $. 
Here $b_p$ and $b_d$ are the specific background rates (in units of $Hz \, m^{-3}$) 
for the prompt and the delayed events, respectively. 
The time window for the coincidence is given by $\tau_d \sim 100 \,\, \mu s$, 
$V_p=V_{T} + V_{GC}=32.89 \,\, m^3$ is the volume considered for the positron-like event 
(no use of a distance cut), $V_{d}=V_{T}=10.32 \,\, m^3$ is the volume accounted for 
the neutron-like event. 
We note $R_p=b_p V_p$ the total prompt positron-like events, and $R_p=b_d V_d$ the total delayed 
neutron-like events. 
A good estimate of the daily accidental background without distance cut is given by
%
\begin{equation}
b_{acc} \sim 0.2 \times \frac{R_p}{1 \,\, s^{-1}} \times \frac{R_d}{83 \,\, h^{-1}} ~day^{-1} \,\, .
\end{equation}
%
If we now require that the accidental background rate from all material but phototubes is less than $1\%$ 
of the neutrino signal (69/d and ~1012/d at the far and near site respectively), we get the constraints 
$R^{far}_p < 10$ Hz and $R^{near}_p < 14$ Hz. As a guideline, we require each detector element 
contribution to be less than 0.2 Hz (6 detectors components and 3 isotopes).
\subsubsection*{Inner detector materials and phototubes}
The results presented here have been obtained by using the
DCGLG4sim simulation, the adaptation to Double Chooz
of the GLG4sim GEANT4-based simulation (see Section~\ref{sec:glg4}).
The inner detector simulated here includes three perfectly cylindrical
volumes: the inner Target, the $\gamma$-catcher and the Buffer.
The three regions are filled with liquids, whose chemical and physical
properties are those of the base-case currently considered (for Target
and $\gamma$-catcher, PXE/Dodecane mixture with 20$\%$ and 80$\%$ volume
ratio, plus addition of fluors and -for the Target only- of a Gd-complex
to a Gd concentration of 1 g/l; pure dodecane for the Buffer). However,
all optical properties have been switched off for the calculation
presented here, since the focus here is on the fraction of energy
deposited in the sensitive volumes, rather than on the exact detector
response. The Target and $\gamma$-catcher vessels are made of acrylics
and have a thickness of 8 mm and 12 mm, respectively. 
The $^{40}\textrm{K}$, $^{238}\textrm{U}$ and $^{232}\textrm{Th}$ nuclides 
have been generated uniformly in several parts of the detector: Target scintillator, 
acrylics of the Target vessel, $\gamma$-catcher scintillator, $\gamma$-catcher acrylics, 
Buffer liquid, stainless steel Buffer tank. Phototubes have been simulated separately.
The simulation gives the spectra of the energy deposited in the sensitive
volumes. We then extract the fraction of events depositing an energy in the Target plus 
$\gamma$-catcher volumes above a 500~keV threshold and we define the maximum concentration
of $^{40}K$, $^{238}U$, $^{232}Th$ allowed for each detector component.
%
%
\begin{table}[htb]
\caption{Allowed concentration (g/g) of $^{40}K$, $^{238}U$, and $^{232}Th$ for the main components 
of the Double Chooz detector (with a safety margin). All values above $\sim 10^{-11}$ g/g can be measured 
through gamma spectroscopy at underground laboratories. For 
the case of Target and $\gamma$-catcher liquid scintillator, 
other counting methods will have to be used to reach a sensitivity at the level $10^{-13}$ g/g.}
{\label{tab:accidentals}} 
\begin{center}
\begin{tabular}{lrrrr}
\hline
                 & $^{40}K$   & $^{238}U$ & $^{232}Th$   &  $^{60}Co$  \\
                 & g/g        & g/g       & g/g          & mBq/Kg      \\      
\hline
Target LS        & $10^{-10}$ & $10^{-13}$  & $10^{-13}$ & ---  \\
Target Acrylics  & $10^{-8}$  & $10^{-11}$  & $10^{-11}$ & ---  \\
GC LS            & $10^{-10}$ & $10^{-13}$  & $10^{-13}$ & ---  \\
GC Acrylics      & $10^{-8}$  & $10^{-11}$  & $10^{-11}$ & ---  \\
Buffer Oil       & ---        & $10^{-12}$  & $10^{-12}$ & ---  \\
Buffer Vessel    & ---        & $10^{-9}$   & $10^{-9}$  & $15$ \\
Veto LS          & ---        & $10^{-10}$  & $10^{-10}$ & ---  \\
\hline
\end{tabular}
\end{center}
\end{table}
Phototubes will be selected according to their radiopurity, and their 
contribution to the single rate is expected to dominate, at the level of a few Hertz. 
Several phototube candidates are being currently investigated, from ETL, Hamamatsu and Photonis. 
This is discussed in details in Section~\ref{sec:photo}. 
Estimates can be given taking into account the whole set of measurements for each phototube 
candidates. In that case the expected single rate (positron-like) above 500~keV is expected to be
 between 9 and 14~Hz for ETL and Hamamatsu.  With extra care, this could be lowered to 4-10~Hz. 
This translates to 1-1.4 and 0.4-1 accidental events per day respectively.

We remind here that accidental background (rate and spectrum) can be measured {\it in situ} with
a precision better than $10\%$.
\subsubsection*{Steel shielding thickness optimization}
\label{sec:kbg_rocks}
 The Double Chooz inert steel shield has been designed to minimize
 the incidence of the gamma-rays from the rock surrounding the
 detector into the target volumes (referring to the so-called Target
 and $\gamma$-catcher volumes).
 This element grants Double Chooz with a critical reduction on the
 uncorrelated background, as compared to the CHOOZ experiment.
 Simulation studies based on GEANT3 and GEANT4 have been performed
 to infer the optimal thickness of the shield.
 The contribution of the K, U and Th were modulated by the
 available knowledge gathered by the CHOOZ collaboration 
 ~\cite{bib:chooz} (and references therein) 
about the activity of the relevant types of rock found in the detector cavity.
 The goal of the optimization is to keep the rate of rock gammas
with energy deposition above the prompt energy threshold
 (set at $0.5$~MeV) lower than the expected activity from the PMTs,
 which has been estimated to be $\mathcal O$($5$~Hz); 
see Section~\ref{sec:bkg_acc}. 
 The studies have shown that a $17$~cm shield would suffice.
 In such a scenario, the overall rate of rock gammas is expected to
 be $<2$~Hz, dominated by $0.98$~Hz due to $^{208}$Tl (Th chain) and
 $0.86$~Hz due to $^{40}$K.
 Figure~\ref{fig:steelshielding3D} shows the current baseline
 engineering model for the Double Chooz shield.
\subsubsection{Simulation of muon background}
\label{sec:musim}
To properly simulate the detector response to cosmic muons, it is
necessary to use angular resolved muon distributions and energy
spectra for both detector locations. 
The muon propagation tool MUSIC~\cite{bib:music} 
(see Section~\ref{sec:music}) was used for this purpose, in
combination with GEANT4 and FLUKA based simulations.
\subsubsection*{Far site}
\begin{figure}[htbp]
\centering
\includegraphics[width=.6\textwidth]{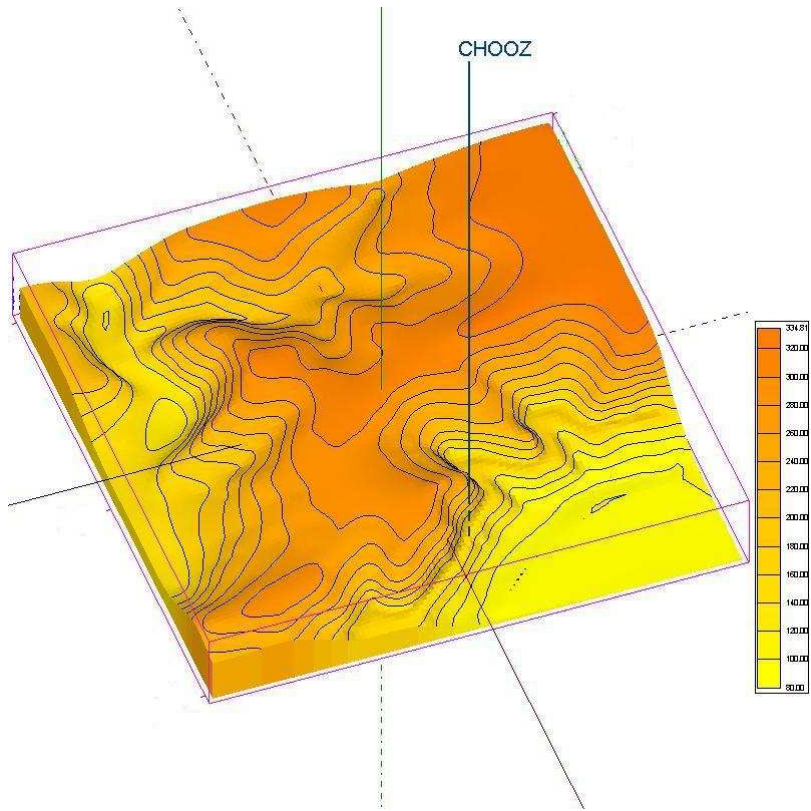}
\caption{Hills surrounding the Chooz sites.\label{fig:hill}
}
\end{figure}

\begin{figure}[htbp]
\centering
\includegraphics[width=.7\textwidth]{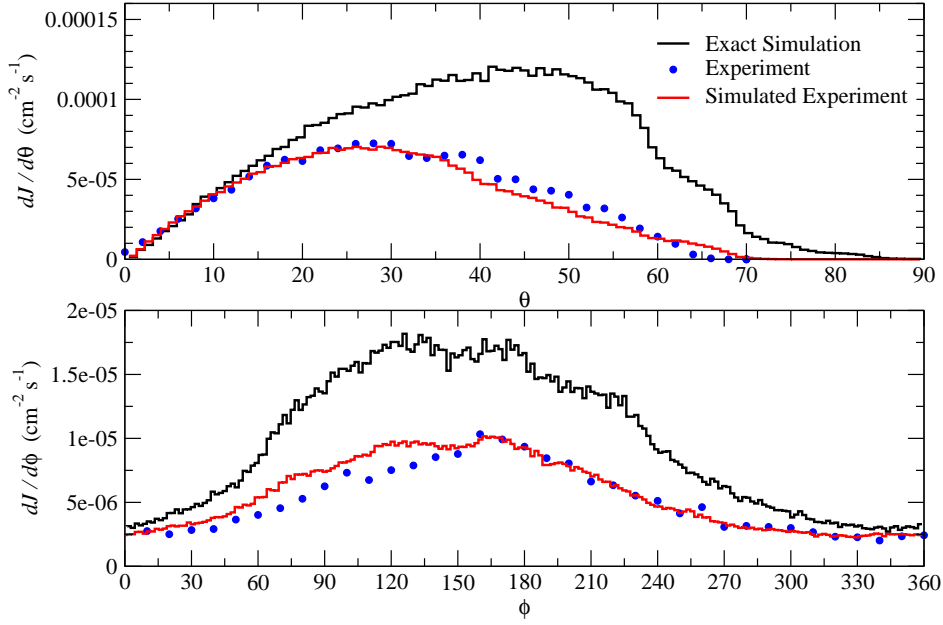}
\caption{Comparison of angular dependence 
        of muons for the far detector \cite{bib:tang}.
The experiment refers to data from an RPC setup which was used
to characterize the $\mu$ flux during the original CHOOZ
experiment.  The exact simulation shows the expected $\mu$ rates
for the CHOOZ site as derived from the new MUSIC propagation
simulation.  When the results from the exact simulation are
combined with the known acceptances of the RPC setup, the
resulting simulated experiment is in good agreement with
the measured data.\label{fig:iv_angular}
}
\end{figure}

\begin{figure}[htbp]
\centering
\includegraphics[width=.7\textwidth]{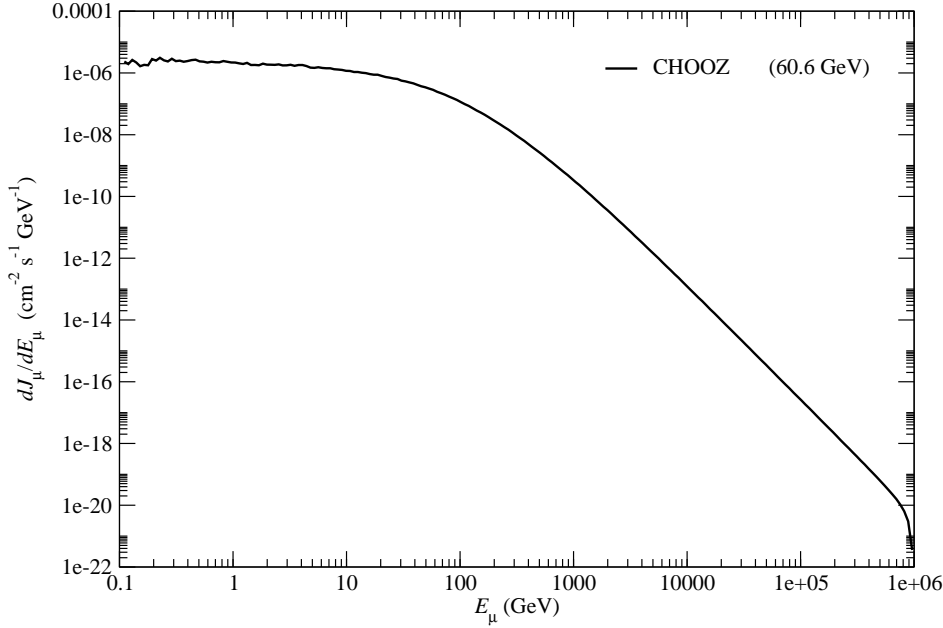}
\caption{Energy spectrum of cosmic muons at the far detector.
\label{fig:iv_far_mu_energy}
}
\end{figure}

\begin{figure}[htbp]
\centering
\includegraphics[width=.7\textwidth]{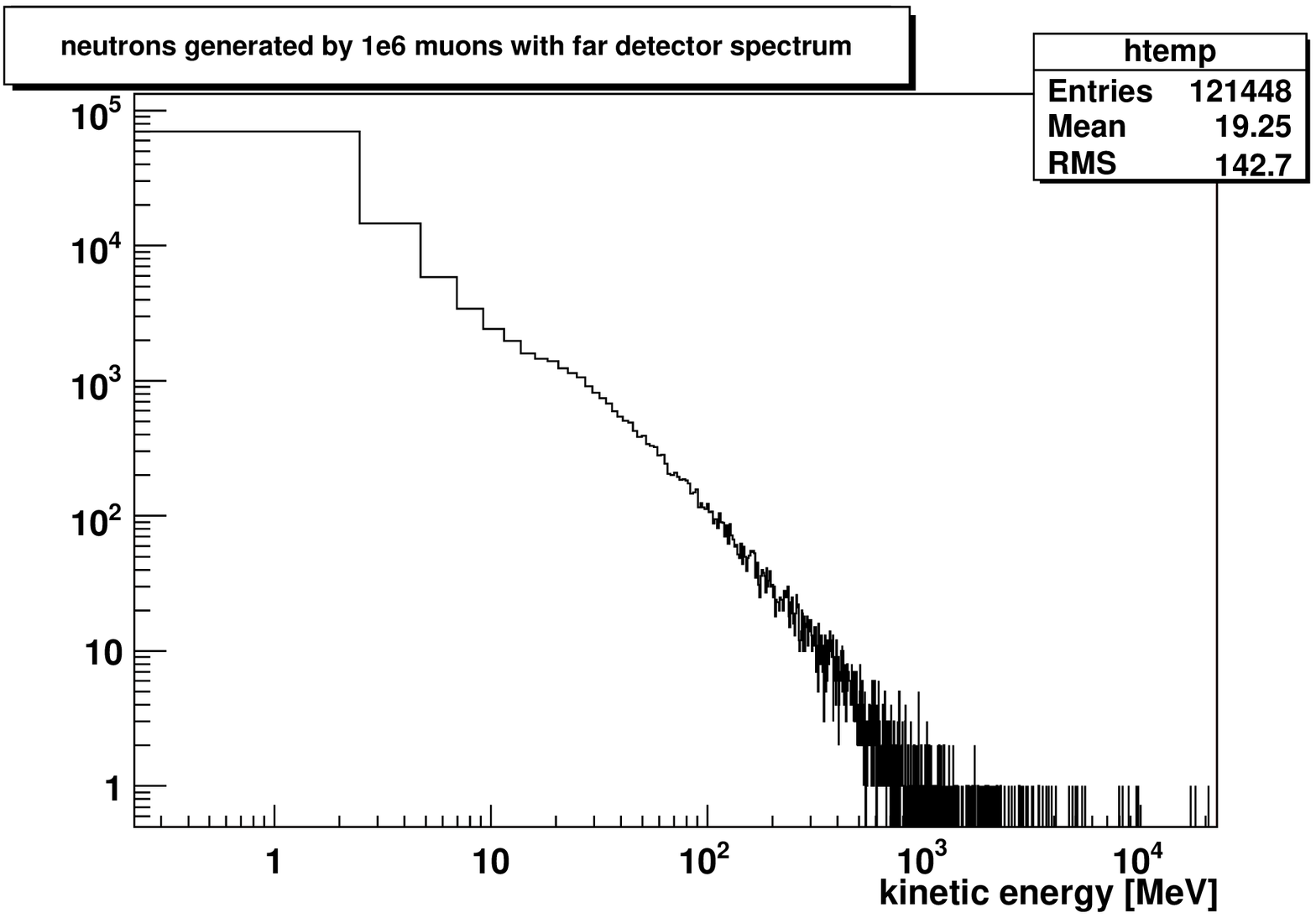}
\caption{Neutron energy distribution as generated by the far detector muon 
spectrum in detector and surrounding rock.
\label{fig:iv_far_neutrons}
}
\end{figure}

A measurement of the angular distribution of muons at the Double Chooz
far site was performed prior to the Chooz
experiment~\cite{bib:giannini}. No measurement of the energy spectrum,
however, is available.
A modification of the MUSIC code was used to create an independent
spectrum and angular distribution by propagating 
surface muons through rock~\cite{bib:tang}. 
Details on the rock composition, as measured by chemical analysis of
several samples in~\cite{bib:giannini}, were taken into account. 
The strong angular dependence of the muon flux is justified by the
large variation of the rock overburden with the incoming muon
direction: this required the creation of a digitized topographical map
of the Chooz hill profile, shown in Figure~\ref{fig:hill}.
The simulation predicts a muon flux of $(6.2)\cdot10^{-5}\,
{\rm{cm}^{-2}\,\rm{s}^{-1}}$, (about 5 Hz through the target region)
slightly higher than the
approximate measured value quoted in \cite{bib:chooz}.
The calculated angular distributions of the muons are shown in 
Figure~\ref{fig:iv_angular}. The simulation agrees well with the
measured data, once the acceptance of the experimental
apparatus is taken into account.
The simulated energy spectrum of the muons 
is presented in Figure~\ref{fig:iv_far_mu_energy}. The
mean muon energy at the far detector location will be about 61~GeV.
With these muon distributions, a secondary neutron spectrum was
generated with the DCGLG4sim software, as shown in plot
\ref{fig:iv_far_neutrons}. This neutron distribution is used as  a
starting point for dedicated neutron simulations (as in
Section~\ref{sec:photo}). 
\subsubsection*{Near site}
During the design phase, the calculation of muon background at the
near detector site will be essential for the definition of the near
lab design.

\begin{figure}[htbp]
\centering
\includegraphics[width=.6\textwidth]{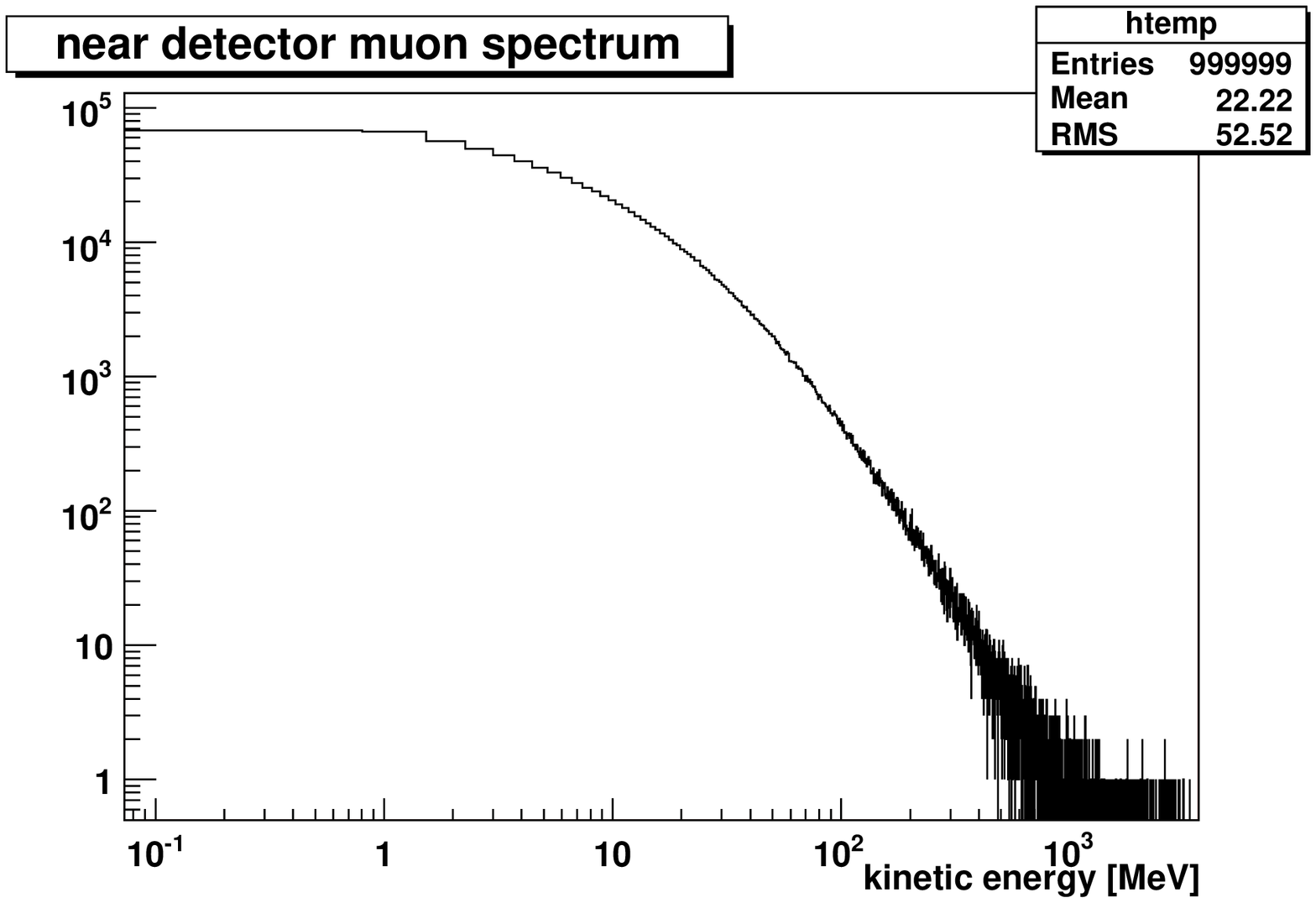}
\caption{Preliminary energy spectrum of cosmic muons for the 
near detector, simulated with a flat topography.
\label{fig:iv_near_mu_energy}
}
\end{figure}

A calculation of the muon energy spectrum, assuming a flat topography of
the rock overburden, was performed with the MUSIC code. The calculated
energy spectrum is shown in Figure~\ref{fig:iv_near_mu_energy}: the
mean energy is 30 GeV, and the flux  $5.9\cdot10^{-4}\,
{\rm{cm}^{-2}\,\rm{s}^{-1}}$ (about 55 Hz through the target).

\subsubsection{Neutron-like background}
\label{sec:bkg_nlike}
All energy deposit over 6 MeV isolated in time from other deposits 
are candidate single neutrons. 
In the first CHOOZ experiment, events occurred with a rate
of about $10^{-2}$ Hz. The origin of these energy deposits is unknown.
The radial distribution of the reconstructed vertices, decreasing 
by two orders of magnitude from the outer wall to the center, suggests
an origin outside the detector.

The hypothesis has been made that neutron-like events could be due to
bremsstrahlung photons radiated from cosmic muons which traverse the rock 
surrounding the detector (``near-miss" muons).

A detailed Monte Carlo study was carried out to test this hypothesis.
Cosmic muon samples were generated according to the measured angular 
distributions~\cite{bib:giannini} at the site and to an energy spectrum 
appropriate for an overburden of 300 m.w.e.~\cite{pdg}.
The interaction with the rock and in a detector with similar geometry 
to CHOOZ were simulated with the GEANT4 package.

\begin{figure}[htb]
\centerline{\epsfig{file=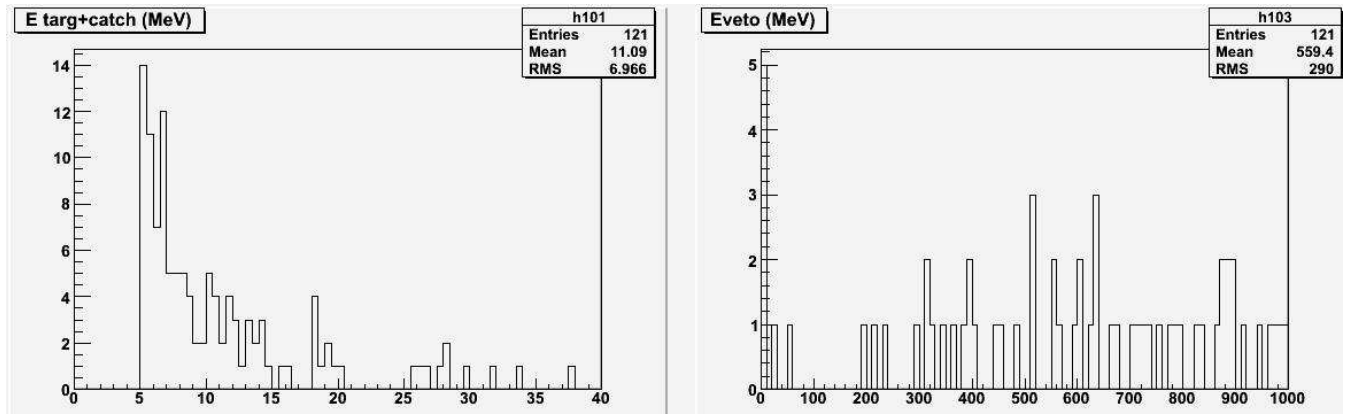,width=\linewidth}}
\caption{Distribution of the energy deposited by photons originated
from near-miss muons in the Target (left) and in the veto (right).}
\label{fig:egam}
\end{figure}

The bulk of the photons entering the detector were found to have originated 
from delta and knock-on electrons produced by  
muons in the rock. 
The spectrum of energy deposits in the Target due to such photons
is shown in Figure~\ref{fig:egam}(left): it reproduces very well the shape
observed in Chooz.
The probability to have such a signal in the Target is at the level of 
$10^{-4}$ per muon; taking into account the measured muon rate
at the site, 0.4 Hz/(m$^2$), and the surface area 
 relevant for the muon interactions, a rate of events in the
Target of about of 0.006 Hz is found, which is the same
order of magnitude
as the measured one.

However, most of these signals in the Target are accompanied by 
significant energy deposition in the veto, as shown in 
Figure~\ref{fig:egam}(right), due to the showering of the photons.
Using a threshold around 200 MeV,
only about 10$\%$ of the events would not be tagged in the veto,
while no signal is present in the neutron-like background.

In conclusion, detailed simulation of photons from ``near-miss"
muons explains only 1/10 of the neutron-like energy deposits
observed in CHOOZ.
We don't know yet the origin of these energy deposits. It will be
an item of study in Double Chooz.

\subsubsection{Fast neutron background}
\label{sec:bkg_fastn}
The fast neutron background was the main single
 source of background in the CHOOZ experiment~\cite{bib:chooz}. 
This background was also studied in other neutrino experiments such as Bugey, Karmen and Palo Verde.
These fast neutrons are created by near miss muons interacting 
around the detector. Their large interaction length allows them to cross over the full detector. 
Some of them happen to be slowed down in 
the central scintillators, and eventually be captured in the target. 
The recoil proton can mimic the positron (the deposited energy is higher, but part of it can be unseen), 
and the neutron capture is obviously correlated as a neutrino event. Some secondary neutrons can also contribute.
The energy shape of these events is known from previous experiments to be very smooth. In the  CHOOZ paper 
(Figure~46 of~\cite{bib:chooz}) a rather flat 
spectrum is shown, with some saturation effect even creating a 
slow rise. Since there is no reactor 
neutrino over some 15~MeV, this background can be permanently monitored in 
the 15 - 50 MeV range.
The total correlated background in the CHOOZ 
experiment was published at 1.01 event/day, and found to be in good agreement
 with the extrapolation of the measured rate over 15 MeV.
To prepare the new experiment, the Munich  group\cite{bib:grieb}
first tried to reproduce these numbers by simulation.  
31 hours of data taking in the CHOOZ experiment were 
simulated, and 1 background event observed. This is 0.8/day, 
corresponding to an upper limit 
of 3.0 event/day at 90$\%$ confidence level. 
This is in good agreement with the 1 event/day quoted 
in Reference~\cite{bib:chooz}.

Having validated the Monte Carlo, the Double Chooz detector 
was simulated. The program was run with an overburden 
of 100 m.w.e., and then a scaling done to the 
actual overburden of 300~m.w.e. for the far site and 80 m.w.e. for the near 
one. 43 hours of data taking were simulated, 
and 1 event observed. This leads to expect 0.15 event/day at the far site 
(less than 0.6 event/day at 90$\%$
 confidence level) and 3.3 event/day at the Near detector.  
This is significantly less than that in the CHOOZ experiment, 
as expected due to improvements in the detector design. Using steel 
instead of sand increases the neutron path toward 
the target by about one attenuation length, from which a factor 
3 decrease can be predicted, leading to 0.3 event/day. 
This number is in good agreement with the simulation result quoted above. 
All this work was done without using the muon tracker.
An additional background rejection of a factor of 15 is expected,
making this fast neutron background very small. 
Another handle on rejecting fast neutrons is that some would make
and energy deposit in the Inner Veto.
\begin{table}[htb]
\begin{center}
\caption{Summary of the fast neutron background prediction in Double Chooz Near and Far detectors. This can be compared 
with the data of the CHOOZ experiment at the Far site as well as its new simulation.}
{\label{tab:fastn}} 
\begin{tabular}{lrrr}
\hline
Event/day            &   Data                         & \multicolumn{2}{c}{Simulation}               \\
                     &                                &   Mean           &  90$\%$ C.L. Upper Limit  \\
\hline
CHOOZ                &   $1 .01 \pm 0.04 \pm 0.1$     &   0.8            &  3.0                      \\
\hline
Double Chooz Far     &     ---                        &   0.2           &  0.8                      \\
Double Chooz Near    &     ---                        &   1.3            &  5.0                      \\
\hline
\end{tabular}
\end{center}
\end{table}
In the Double Chooz experiment, some Pulse Shape 
Discrimination (PSD) will be possible, 
allowing us to separate the recoil proton 
signal, which is more highly
ionizing and slower, from the neutron one. 
This will be used, not to reject the noise (which would create 
systematic errors), but to tag and monitor 
the background, and measure its energy shape. Since it will 
also be measured above 15 MeV and below 1 MeV 
(once the accidentals events are subtracted), a three-fold fit will be possible.
\subsubsection{Muon capture}
\label{sec:bkg_mucapt}
%
%
Muon capture contributes to the background by creating gammas, 
neutrons, and cosmogenic
nuclei.
Cosmogenics nuclei are described on page~\pageref{'li9'}
and gamma backgrounds have been discussed previously in 
this section.
We focus here on neutrons from such a capture.
Fast neutrons are a potential source of correlated background: they 
may propagate though
the detector
and reach the $\gamma$-catcher or even the Target,  creating recoil 
protons that can mimic the
positron inverse beta decay signal, and later on the capture on Gd.
If the recoil protons are not seen, the neutron capture can be 
randomly associated with
a gamma,
creating an accidental background event.
The rate of this background can be neglected compared to other 
accidental
backgrounds in the detector.

\par The main issue concerning the capture of muons on nuclei in 
the detector is
the presence
of dead materials along the muons track.  We review below the 
potentially critical detector elements.
The corresponding background is summarized in the Table~\ref{tab:muoncapture},
which gives an estimate accurate to an order of magnitude.

\subsubsection*{Target and $\gamma$-catcher}
Inside the Target and the $\gamma$-catcher, muon tracks as well as 
neutron capture  are
seen due to the huge amount of energy deposit. 
 However, the  acrylic supports linking the 
Target and the $\gamma$-catcher
vessels are dead materials. The corresponding muon capture 
background is estimated to 0.1 event/day,
and the rejections are shown in the rightmost column of the table.

\subsubsection*{Buffer}
The volume of the Buffer is large (114 m$^3$). Its contribution is 
dominated by the last
layer, near the $\gamma$-catcher.
Following a standard rule of thumb, 4~tons of mineral oil  close to 
the $\gamma$-catcher,
corresponding to one
attenuation length (15~cm),  will dominate.
In the scintillator liquid,  
 less than 10$\%$ of neutrons are captured on carbon.  These neutrons have 
to cross the $\gamma$-catcher to reach
the Target and are vetoed (the efficiency  is  99.9$\%$)

\subsubsection*{Buffer vessel}
The Buffer and the stainless steel Buffer vessel  are  not 
instrumented (or weakly so, since Cerenkov
 light from muon tracks in the Buffer could be detected). All stopping 
muons in these
volumes are captured by iron
 nuclei. Neutrons produced in this 3~mm thick stainless steel vessel 
have to travel
through the Buffer and the $\gamma$-catcher to reach the Target (a total 
path of about
150~cm).
The total rejection from the propagation is estimated to be as large
as $10^5$. The Inner Veto rejection is also applied. 
The estimated background is 0.03 event/day.

\subsubsection*{Steel shielding}
The detector shielding is a large amount of iron (about 306 tons),
which we consider as dominated by the last attenuation length.
This shielding will be covered by the Outer Veto detector providing a 
rejection factor
of about 20 (applied in the table).
The resulting background is 0.05 event/d.  This number scaled to the 
far detector is
0.002/d
(30 times less cosmic ray muons).

The first CHOOZ experiment provides an experimental confirmation to 
this estimate.  The 70~cm sand layer at the outermost part of the detector was 
producing fast neutrons
from muon capture,
roughly as much as the iron shielding of the new experiment. 
Furthermore, in the new
detector the distance
from the steel shielding to the Target region is enlarged by 15~cm. 
Thus, according to
our estimate, in the first experiment
the sand muon capture background was about 0.15 event/d.
This is well within the published total correlated background (1~event/day).

\subsubsection*{Chimney}
The detector chimney is also  a dead volume, at least above the Buffer 
vessel.
Muons going down through the chimney are not seen by the vetoes if the 
chimney is not
instrumented.
Thanks to an effort to minimize the chimney diameter, this solid angle 
is small.
Since the liquid inside the chimney is scintillating, and further 
since the chimney is made
of transparent acrylics
up to the Buffer vessel,  the muon track and capture in the chimney 
would be seen by the
inner phototubes.
We consider here that only a neutron capture at the Buffer vessel 
level can contribute
to the background at a level of 0.1 event/day.
(This background would disappear if the liquid level is kept below the 
Buffer vessel).
\subsubsection*{Muon capture summary}
In this study, neutron propagation was estimated taking into account its 
energy spectrum
and the corresponding
attenuation length. The neutron energy from muon capture is peaked 
toward low energy:
 90$\%$ of
the produced neutrons have an energy lower than 10~MeV and the 
corresponding attenuation
length makes them
unable to cross the $\gamma$-catcher and reach the Target.
In Table~\ref{tab:muoncapture} higher energy neutrons were propagated using an 
attenuation length of 15~cm,
which corresponds to 20~MeV neutrons. For a reasonable variation of 
the energy
spectrum and attenuation
 length, it was checked that the result is stable within an order of 
magnitude.
\begin{table}
\begin{center}
\caption{Fast neutron background from muon captures in the near 
detector (overburden 80~m.w.e). Appropriate masses of material within one
interaction length of each component were used in the calculation.
}

\label{tab:muoncapture}
\begin{tabular}{lrrrrrr}
 \hline
                    & Steel              & Buffer           &  Buffer      & Chimney     & Target      \\
                    & shielding          &  structure       & oil          &             & structure   \\
\hline
mass (tons)         & 14                 &  7               & 4            &  0.005      &  0.1        \\
\hline
Capture rate        & 70/s               & 40/s             & 0.5/s        &$0.6\,10^{-3}$/s  & 0.01/s \\
neutron rate        & 80/s               & 50/s             & 0.8/s        & 10$^{-3}$/s & 0.015/s     \\
Target rate (no Veto)  &1/d              & 5/d              & 160/d        &  0.1/d      & 100/d       \\
Target rate (Veto)  & 0.05/d             & 0.005/d          & 0.16/d       &   0.1/d     &  0.1/d      \\
\hline
\end{tabular}
\end{center}
\end{table}

In conclusion, the muon capture correlated background is well below 
1\% of the
signal at the near and far detectors.
The accidental background originating from these neutrons is small.  
With less
than 1 neutron per day together with a
few Hertz of gammas within 100~$\mu$s, the background is less than $10^{-3}$ 
events per day, which is negligible.

%
%
\subsubsection{Cosmogenic correlated background: Lithium 9}
\label{sec:bkg_li9}
The cosmogenically-produced isotope $^{9}Li \:$ can be a serious problem for reactor
experiments, since the beta decay has a branching ratio of about 50\%
into states that de-excite via neutron emission. This means that the electron plus neutron
combination can mimic a reactor neutrino signal. The 178 ms half life could 
also make it difficult to veto since it would lead to an unacceptable
deadtime. In addition, the best measurements
of cosmogenic production come from large
 depths, where the hard muon flux makes extrapolation
to shallow depths uncertain.

\par Fortunately, Double Chooz has the great advantage of having the data from the
original CHOOZ experiment to draw on (especially 138 days of Reactor Off
data). This allows us to extract the expected rate of $^{9}Li \:$ at our Far
Lab based on using data rather than MC or extrapolation 
from large depths. The expected 
rate is 1-2\% of the signal (before veto). Thus, this is a background we can handle by veto
and measurement, as shown below.

Detecting the initial muon is a necessary part of being able to measure and subtract
the background due to $^{9}Li \:$. Both the Near and Far Detectors will have 50-cm
thick, $4\pi$ scintillating muon vetoes in order to tag incoming events with
high efficiency\footnote{The Near Detector veto may be thicker than 60 cm}.
In addition, each lab will have a muon tracker on top. 
Thus entering muons will be efficiently detected. 
In addition, the current plan for the electronics has included the concept of
``muon electronics'' specifically designed to measure the muon energy
deposit over a much larger dynamic range than that planned for the waveform
digitizers being designed for the low-energy events.

In the KamLAND experiment, it was found that a cut of $10^{6}$ photoelectrons
was effective in tagging muon-initiated showers that were 
important in $^{9}Li \:$
production. The $10^{6}$ photoelectron cut in KamLAND
has been simulated by using the KamLAND detector simulation program, KLG4sim.
The simulation predicts that about $7.7 \pm 0.4$\% of cosmic ray muons will
be cut by a cut at $10^{6}$ p.e. (which corresponds to roughly 3.3 GeV
of deposited energy). This is to be compared with 6\% obtained by an
analysis of one day of KamLAND data - a reasonably good agreement.
The {\it same} simulation
package adapted for Double Chooz depth (300 m.w.e.) and geometry predicts
that a similar cut (slightly above the peak of the energy loss
distribution, about 2.8 GeV) would cut only
$1.3\%$  in Double Chooz (Far). With a muon rate of ~25 Hz, this would
correspond to a veto rate of only one every 3 seconds.
\subsubsection*{What can we learn from the CHOOZ data}
\label{'li9'}
\par We have extracted the $^{9}Li \:$ 
production rate in the CHOOZ experiment in three different ways: 
(1) total background
rate, (2) spectral fit to official CHOOZ data between 2.8 and 10.0 MeV,
and (3) a more recent spectral fit to extended CHOOZ data between 2.8 and
30.0 MeV.

CHOOZ had a measured accidental coincidence background of $0.42 \pm 0.05$
event/day and a background from fast neutrons of $1.01 \pm 0.11$ event/day
during the period when most of their Reactor On data was taken. The
corresponding background rate from Reactor Off data in this same period
was $1.4 \pm 0.24$ event/day, which is consistent with the sum of the two
individual rates. Based on the 0.11 event/day uncertainty in
extrapolating the proton recoil spectrum below 10 MeV, there could
be as much as 0.2 event/day of $^{9}Li \:$ hidden under the reactor signal.
This analysis depends critically on systematic uncertainties which are
difficult to quantify. A better technique is to fit the shape of 
the prompt energy spectrum.

\begin{figure}
\centerline{\rotatebox[]{0}
{\scalebox{0.6}{\includegraphics[width=0.7\textwidth]{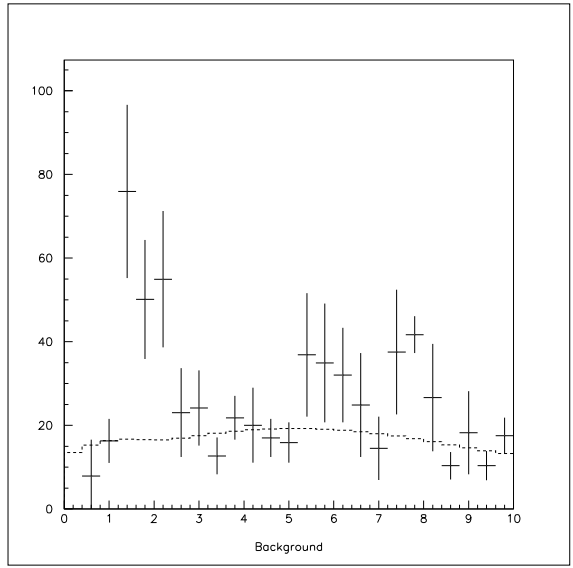}}}}
\caption{\label{F:Li9_fig1}
Official CHOOZ scaled Reactor Off visible energy spectrum
fit with a combination
of flat and $^{9}Li \:$-shaped background. The fit is only above 2.8 MeV to avoid
the accidentals below that energy. The best fit for $^{9}Li \:$ is $0.6 \pm 0.4$
event/day.}
\end{figure}

A spectral fit to the official CHOOZ Reactor Off data is shown in 
figure~\ref{F:Li9_fig1}. The fit is to a flat proton recoil visible energy spectrum
and a $^{9}Li \:$-shaped decay spectrum that includes the effect of the broad lines
on the endpoint. The fit is only done above 2.8 MeV to avoid the peak due
to accidental coincidences below that energy. The best fit for the $^{9}Li \:$ rate
is $0.6 \pm 0.4$ event/day, although the rates of the two backgrounds are 
highly correlated.

\begin{figure}
\centerline{\rotatebox[]{0}
{\scalebox{0.6}{\includegraphics[width=0.7\textwidth]{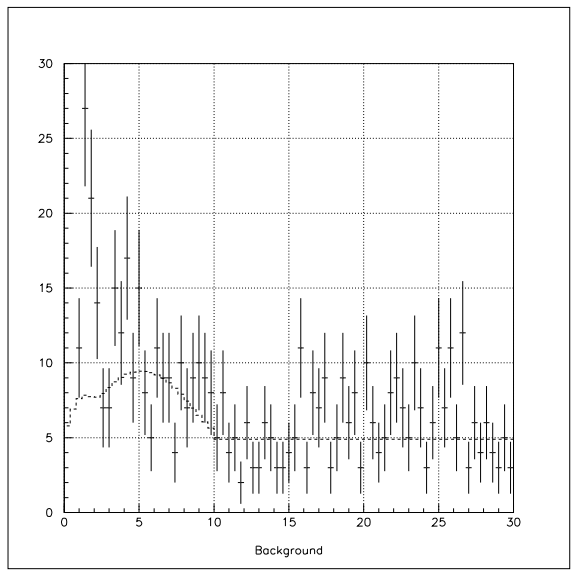}}}}
\caption{\label{F:Li9_fig2}
Unofficial CHOOZ Reactor Off visible energy spectrum
fit with a combination
of flat and $^{9}Li \:$-shaped background. The fit is only above 2.8 MeV to avoid
the accidentals below that energy. The best fit for $^{9}Li \:$ is $0.7 \pm 0.2$
event/day.}
\end{figure}

The degeneracy between the two backgrounds can be broken by fitting the
background above 10 MeV, beyond the $^{9}Li \:$ endpoint. Such data is not (yet)
officially available by consensus from the CHOOZ collaboration, but an
unofficial spectrum can be fit based on roughly 114 days of reactor
off data with energy up to 30 MeV, as shown in their published scatter plot.
This is shown in Figure~\ref{F:Li9_fig2}.
Now the degeneracy is almost completely broken, and the measured rate is
$0.7 \pm 0.2$ event/day.

\subsubsection*{Extrapolation from CHOOZ to Double Chooz}
The three methods
using CHOOZ data imply a $^{9}Li \:$ rate for Double Chooz in the range 0.5 to
1.5 event/day, or about 1-2$\%$ of the expected signal. This is a rate
that can be handled either by veto, measurement to an accuracy of 10\%,
or (most likely) both. With the expected muon rate through the Target
of 25 Hz, we could afford to veto these muons for up to
400 ms (factor of ten reduction) before incurring substantial dead time.
Thus $^{9}Li \:$ in the Far Detector can be dealt with in multiple ways.

In the Near Detector, the extrapolated rate using an $E^{0.75}$ dependence
gives a rate of 0.9\% of the expected signal. The muon rates are
expected to be too high to effectively veto $^{9}Li \:$, however, a better than 
20\% measurement of this rate can be done with the expected One Reactor
Off time required for refueling (~one month per year per reactor), or with
two weeks of Two Reactor Off time. Thus, this background can be subtracted
to the required precision.
\subsubsection{Background subtraction error}
In this section we summarize the backgrounds estimated in the Double Chooz detectors. 
We also refer to the background measured in CHOOZ~\cite{bib:chooz}. The 
results
are presented in Table~\ref{tab:back}.
Accidental backgrounds have been estimated through simulation (see Section~\ref{sec:bkg_acc}). 
For Double Chooz we separate the contribution of the detector materials from the 
phototubes. The radiopurity of our PMT candidates have been used for this estimation. 
We only report the total accidental background measured in CHOOZ.
In the case of the correlated background we also 
split the different contributions:
fast neutron (see Section~\ref{sec:bkg_fastn}), muon capture (see Section~\ref{sec:bkg_mucapt}), 
and $^9$Li production (see Section~\ref{sec:bkg_li9}). 
We report the total correlated background measured in CHOOZ, 
and we also present an attempt to extract the different 
contributions of fast neutron and $^9$Li from 
a new analysis of the CHOOZ data ($^9$Li induced background 
had not been measured in CHOOZ).

To estimate the 
background at the Near detector site different 
methods are used: a full simulation 
at 100 m.w.e. for the fast neutron background (see Section~\ref{sec:bkg_fastn}), and a rescaling for the 
accidental backgrounds, the muon capture and the $^9$Li backgrounds. We scale the background rate 
using the ratio of muon flux times the mean 
muon energy to the power 0.75, between 300 m.w.e. and 80 m.w.e.,
 leading to a factor of 7 to a good approximation.
Taking into account all those backgrounds and a 0.6$\%$ relative systematics error leads to a sensitivity 
of $\sin^2(2\theta_{13})<0.027$ (90$\%$~C.L.) after 3 years of data taking with two detectors 
(at $\Delta{m}^2_{31}=2.5\times10^{-3}~\text{eV}^2$). 
\begin{table}[htb]
\begin{center}
\caption{Summary of the background subtraction error at the Far and Near detector (preliminary). 
Background rate and shape with their corresponding uncertainties 
are used for the calculation 
of the sensitivity. 
The systematics correspond to our best estimate of the error associated 
with each particular background (this can be used as 
a ``background systematic error").  The uncertainty on the background
rate is larger than the systematic error because we haven't yet chosen
certain materials, though when we do, the error will be quite small.\label{tab:back}}
\begin{tabular}{llrrrrrr}
\hline
Detector     & Site &              & \multicolumn{5}{c}{Background} \\
\hline
             &      &              & \multicolumn{2}{c}{Accidental}    & \multicolumn{3}{c}{Correlated}  \\
             &      &              &     Materials   &    PMTs    & Fast n           & $\mu$-Capture  & $^9$Li \\
\hline
CHOOZ        &      & Rate ($d^{-1}$)&      ---      &       ---  &      ---         &     ---        & $0.6 \pm 0.4$  \\
(24~$\nu$/d) &      & Rate ($d^{-1}$)& \multicolumn{2}{c}{$0.42 \pm 0.05$}& \multicolumn{3}{c}{$1 .01 \pm 0.04(stat) \pm 0.1(sys)$} \\
             & Far  & bkg/$\nu$      & \multicolumn{2}{c}{1.6$\%$}        & \multicolumn{3}{c}{4$\%$}                   \\ 
             &      & Systematics  & \multicolumn{2}{c}{0.2$\%$}        & \multicolumn{3}{c}{0.4$\%$}                   \\
\hline
Double Chooz &      & Rate ($d^{-1}$)&  $ 0.5 \pm 0.3$& $ 1.5 \pm 0.8$  
& $ 0.2 \pm 0.2$ & $<0.1$ & $1.4 \pm 0.5$    \\
(69~$\nu$/d) & Far  & bkg/$\nu$      &  0.7$\%$        & 2.2$\%$       
& 0.2$\%$          & $<$0.1$\%$        & 1.4$\%$        \\
             &      & Systematics    &  $<$0.1$\%$       & $<$0.1$\%$      
& 0.2$\%$          & $<$0.1$\%$        & 0.7$\%$        \\

\hline
Double Chooz &      & Rate ($d^{-1}$)& $ 5 \pm 3$ & $ 17 \pm 9$  & 
$1.3 \pm 1.3$   & $ 0.4$ & $ 9 \pm 5$    \\
(1012~$\nu$/d)& Near & bkg/$\nu$      &   0.5$\%$      & 1.7$\%$         
& 0.13$\%$         & $<$0.1$\%$      & 1$\%$         \\
             &      & Systematics    &   $<$0.1$\%$      & $<$0.1$\%$         
& 0.2$\%$          & $<$0.1$\%$      & 0.2$\%$         \\
\hline
\end{tabular}
\end{center}
\end{table}
%
\subsection{Sensitivity and discovery potential}
\label{cha:sensitivity}
We used a $\chi^2$ pull approach in order 
to compute the $\sin^2(2\theta_{13})$ sensitivity. This method 
allows us to introduce and assess the impact of systematic errors  
with ease and to study the interplay between them by 
switching them on/off. This type of $\chi^2$ has a 
generic structure: a sum over terms like: 
$$\left(\frac{\text{Data}-\text{Prediction}-\text{Systematics effects}}
{\text{Uncorrelated error}}\right)^2$$ and another sum term 
which weights the systematic effects by the accuracy with which we 
know them: $$\left(\frac{\text{Systematic effect amplitude}}
{\text{Systematic effect knowledge}}\right)^2\;.$$ This overall 
$\chi^2$ is minimized with respect to all systematic amplitudes. 
In the case of Double Chooz, this leads to the 
following $\chi^2$ formula~\cite{bib:mention}:
\begin{equation}
\label{eq:chipulldef}
\chi^2=\min_{\{\alpha^D_{i,k}\}}\left\{
  \sum_{D=N,F}\sum_{i=1}^{N_{\text{bins}}}
\left[\left(
  \frac{O_i^D - T_i^D - \sum_{k=1}^K\alpha_{i,k}^D S_{i,k}^D}{U_i^D}
\right)^2
+\sum_{k=1}^Kc_{i,k}^D\left(\frac{\alpha_{i,k}^D}{\sigma_k^D}\right)^2\right]\right\}.
\end{equation}
where there are N$_{\text{bins}}$ energy bins.
This $\chi^2$ includes all the spectral 
shape and normalization from both 
detectors (D = N,F). The (K = 5)
systematic amplitudes included, $\alpha_{i,k}^D$, 
are gathered in Table~\ref{tab:chi2paramtable}.
\begin{table}[htpb]
\begin{center}
\caption{Table of systematic parameters used in the $\chi^2$.}
\scalebox{1}{\begin{tabular}[t]{@{\hspace{4mm}}l@{\hspace{4mm}}c@{\hspace{4mm}}c@{\hspace{4mm}}c@{\hspace{4mm}}c@{\hspace{4mm}}c@{\hspace{4mm}}}
 \hline
  Error type& $k$ & $c_{i,k}^D$ & $\alpha_{i,k}^D$ & $S_{i,k}^D$ & $\sigma_k^D$\\
  \hline
  Global normalization &  1  & $1/2\NBins$ & $\alpha_{\text{abs}}$ &
  $T_i^D$ & \hfill $\sigma_{\text{abs}}=2.0\%$\\
  Relative normalization &  2  & $1/\NBins$ & $\alpha_{\text{rel}}^D$ &
  $T_i^D$ & \hfill $\sigma_{\text{rel}}=0.6\%$\\
  Spectrum shape &  3  & $1/2$ & $\alpha_{i,{\text{shp}}}$ &
  $T_i^D$ & \hfill $\sigma_{\text{shp}}=2.0\%$\\
  Energy scale & 4 & $1/\NBins$ & $\alpha_{i,\text{scl}}^D$ & $\left.\frac{\text{d}N_{i}^D}{\text{d}\alpha_{i,\text{scl}}}\right|_{\alpha_{i,\text{scl}}^D=0}$ & \hfill $\sigma_{\text{scl}}=0.5\%$ \\
  $\Delta{m}^2_{31}$ knowledge &  5  & $1/2\NBins$ & $\alpha_{\Delta{m}^2_{31}}$ &
  --- & \hfill $\sigma_{\Delta{m}^2_{31}}=\; 10-20\% $ \\
  \hline
\end{tabular}}
\label{tab:chi2paramtable}
\end{center}
\end{table}
The current knowledge of the antineutrino reactor spectrum, 
although already well known is limited, due to production/detection 
uncertainties, to about 2$\%$ of the precision on the rate 
(the number of \nueb in the full range between 1.8 and 10~MeV) and 
the shape (number of \nueb per energy interval in the same range). 
These two limitations are included in the $\chi^2$ through 
$\alpha_{i,1}^D=\alpha_{\text{abs}}$ and 
$\alpha_{i,3}^D=\alpha_{i,\text{shp}}$ which do not depend on 
$D$, the detector index, since they have an identical effect in 
both detectors. We also introduced a relative normalization error 
between the two detectors through $\alpha_{i,2}^D=\alpha_{\text{rel}}$. 
Moreover we took into account the fact that 
the $\Delta{m}^2_{31}$ is not known at a precision better 
than 20$\%$ as a systematic effect (at the time of this 
proposal, but it should be known with a better precision of 
10$\%$ with MINOS data)\cite{bib:minos}.

We define our $\sin^2(2\theta_{13})$ sensitivity limit as the lowest 
value of $\sin^2(2\theta_{13})$ we can obtain with an 
experiment at a given confidence level if $\quq = 0$.
We compute the expected number of events $O_i^D$ as 
$N_i^D(\Delta{m}^2_{31},\theta_{13}=0)\sum_{R_1,R_2}N_i^{R,D}(\Delta{m}^2_{31},0)\;.$ 
(see Equation~\ref{eq:Ni}). The theoretical prediction on the
number 
of events, $T_i^D$, is computed by the same formula 
but with a value of $\theta_{13}$. $S_{i,k}^D$ 
describes the $1\,\sigma$ systematic impact on the number of 
events moderated by the amplitude $\alpha_{i,k}^D$.

We also include background rejection uncertainties, $\sigma_{\text{bkg}}$, in both detectors through the uncorrelated errors $U_i^D$:
\begin{equation}
U_i^D = \sqrt{O_i^D + B_i^D(1+\sigma_{\text{bkg}}^2 B_i^D)}
\end{equation}

Table~\ref{tab:sensitivityDm2} displays the sensitivity for different values of $\Delta{m}^2_{31}$. According to the last MINOS and Super-Kamiokande results~\cite{SK_atm_nu2002,Fukuda:2002pe, bib:minos}, the preferred values lie between 2.5 and 3. times $10^{-3}~eV^2$ (the latter value being the best fit value of MINOS). According to those new results, the optimum distance for a new reactor neutrino experiment is around 1.3~km, thus the Double Chooz far detector (1.05~km) is well located to efficiently search for a non vanishing $\theta_{13}$ mixing angle.
\begin{table}[htb]
\begin{center}
\caption[Sensitivity and discovery potential dependence on $\Delta{m}^2_{31}$]
{\label{tab:sensitivityDm2} Sensitivity and discovery potential 
dependence on $\Delta{m}^2_{31}$, for 
3 years of data taking with both detectors.} 
\begin{tabular}{lrrrrrrrrr}
\hline
$\Delta{m}^2_{31}$ ($10^{-3}~eV^2$) &  1.8   & 2.0   & 2.2   & 2.4   & 2.6   & 2.8   & 3.0   & 3.2   & 3.4  \\
\hline
Sensitivity (90$\%$ C.L. )          &  0.043 & 0.037 & 0.032 & 0.030 
& 0.027 & 0.025 & 0.024 & 0.023 & 0.022 \\
Discovery potential (3$\sigma$)     &  0.078 & 0.067 & 0.060 & 0.054 & 
0.050 & 0.046 & 0.044 & 0.042 & 0.041 \\
\hline
\end{tabular}
\end{center}
\end{table}
\begin{figure}[htb]
\centering
\includegraphics[width=.7\textwidth]{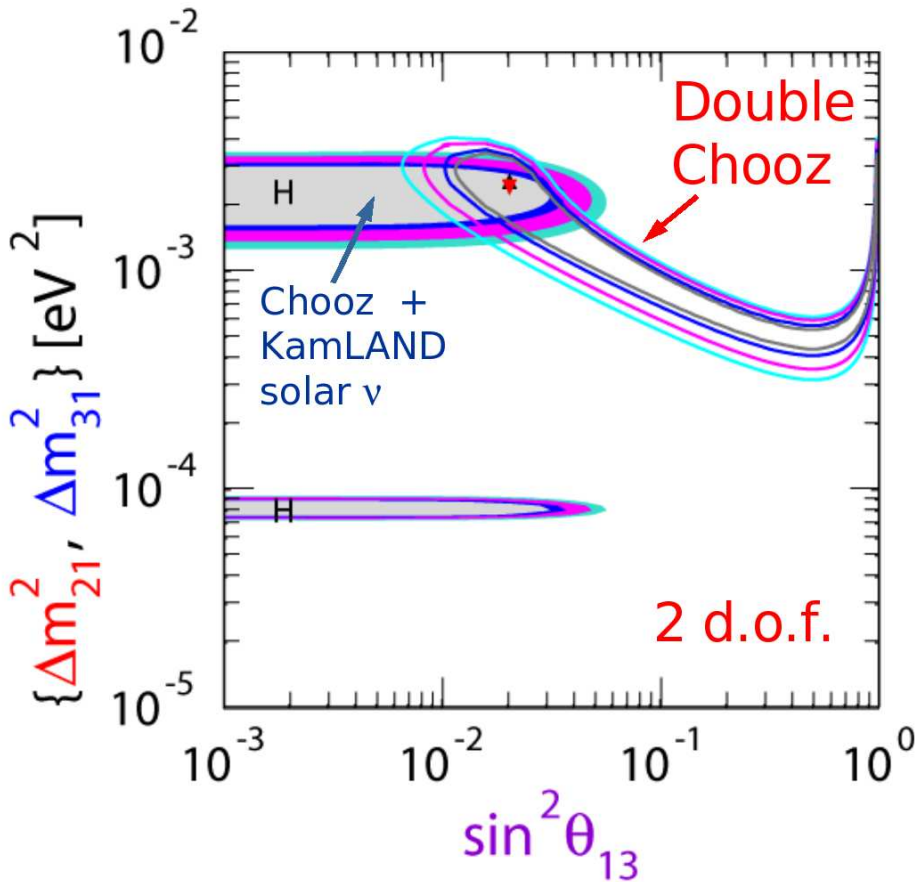}
\caption{Double Chooz sensitivity contours (gray $1\,\sigma$, 
blue 90$\%$, magenta $2\,\sigma$, cyan $3\,\sigma$ C.L.) 
in $(\sin^2\theta_{13},\Delta{m}^2_{31})$ plane generated for 
$\sin^2\theta_{13}=0.02$ and $\Delta{m}^2_{31}=2.5\times10^{-3}~\text{eV}^2$ 
(which is also the best fit) with 1$\%$ background 
in near and far detectors, systematics included in 
Table~\ref{tab:chi2paramtable} and parametrization described in 
Section~\ref{sec:signal}, Section~\ref{sec:spectrum} except 
that we relaxed the constraint on $\Delta{m}_{31}^2$ in 
this figure and computed contours with 2 degrees 
of freedom after 3~years of data taking. Also shown here are
the current contours from global analysis with the same 
color convention~\cite{Valle:2005ai}. Double Chooz is able 
to provide a 20$\%$ precision measurement of $\theta_{13}$ 
as long as this parameter is not too low.}
\label{fig:contours}
\end{figure}
\begin{figure}[htb]
\centering
\includegraphics[width=.8\textwidth]{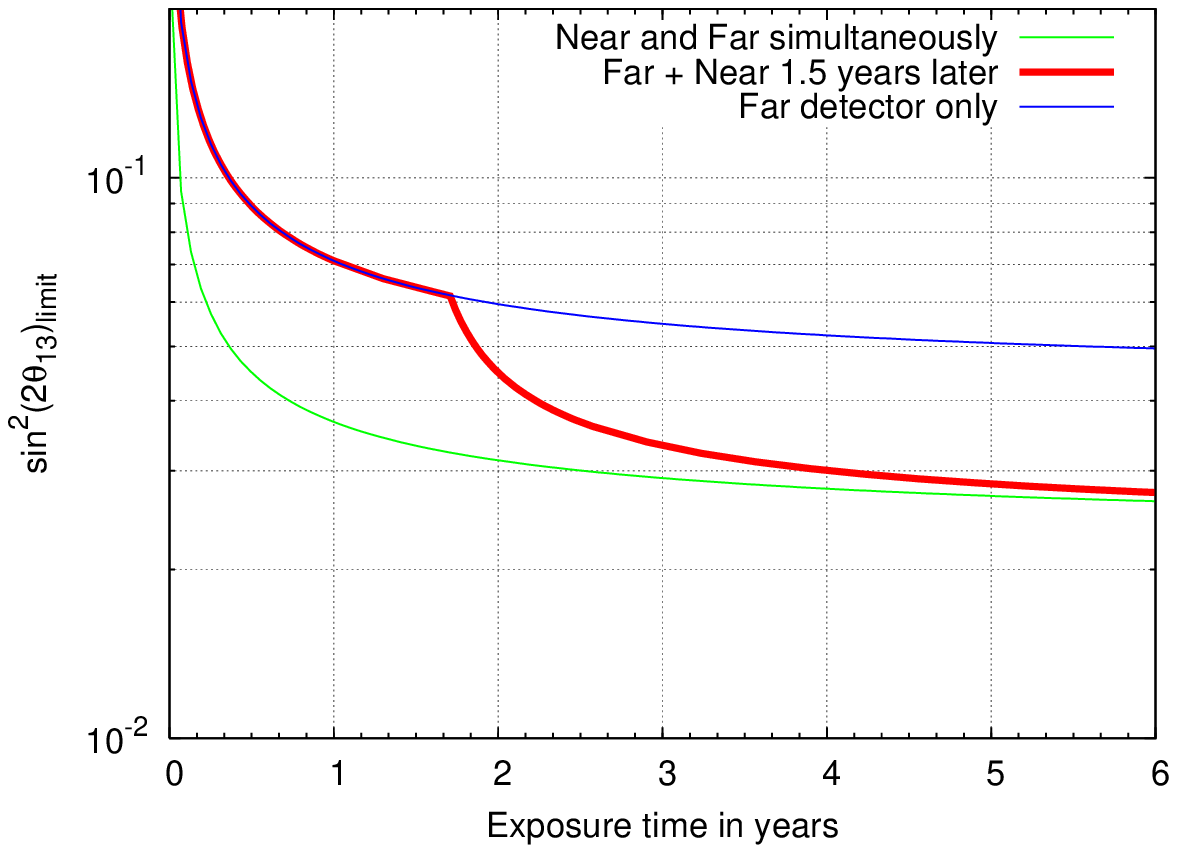}
\caption{$\sin^2(2\theta_{13})$ sensitivity limit for the detectors installation scheduled scenario}
\label{fig:sensitivity}
\end{figure}

\begin{figure}[htb]
\centering
\includegraphics[angle=-90, width=.8\textwidth]{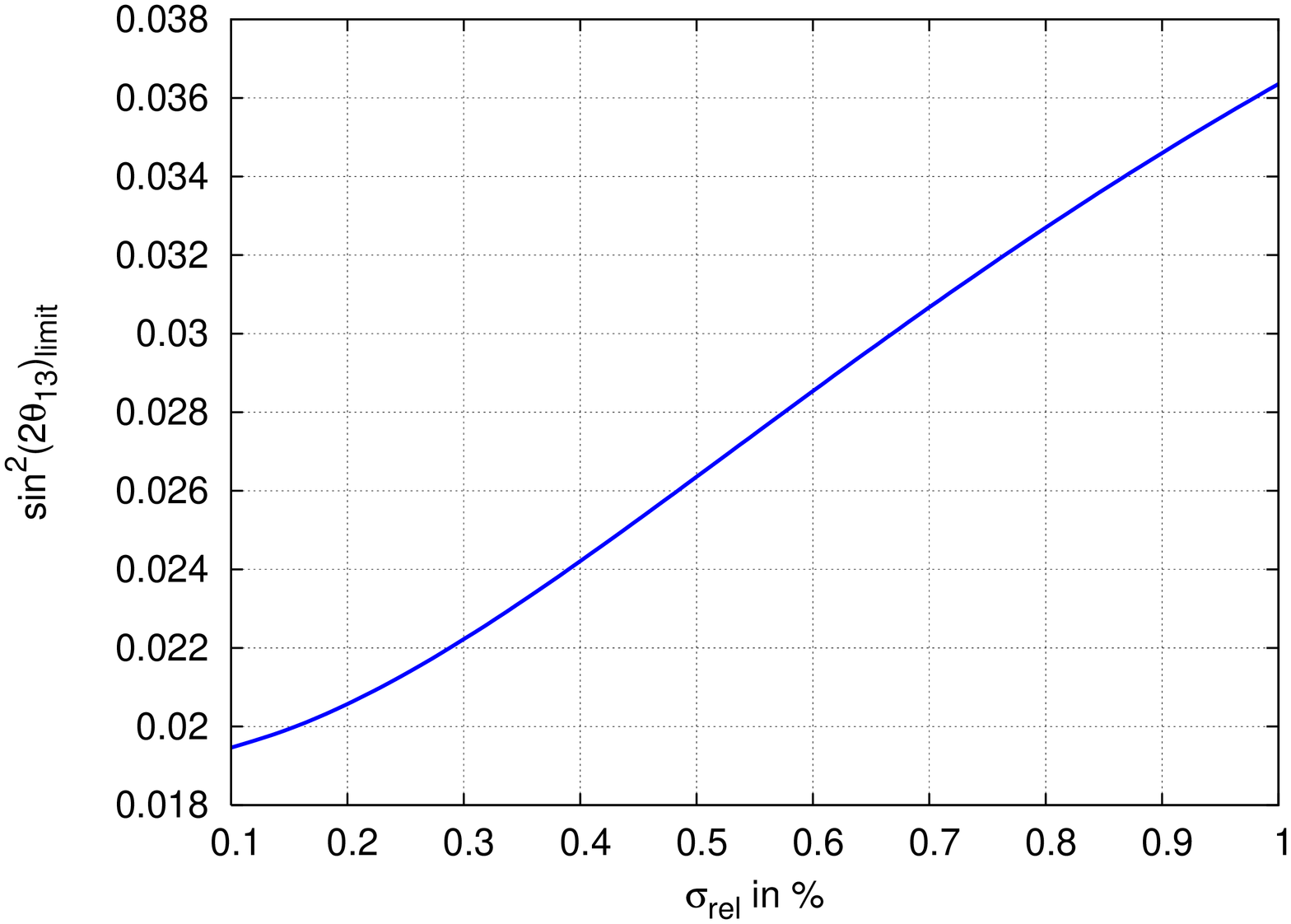}
\caption{$\sin^2(2\theta_{13})$ sensitivity (limit at 90\% C.L. if no signal) 
evolution with respect to the relative normalization error, from 0.1\% to 1\%, 
between the Near and Far detectors.}
\label{fig:senssigmarel}
\end{figure}

Figure~\ref{fig:senssigmarel} shows the variation of the sensitivity (limit at 90\% C.L. if no signal) 
with respect to the relative normalization error between 
the Near and Far detectors. Even though we foresee 
a relative error of 0.5\% 
(see Table~\ref{tab:systematicsoverview}), we used 0.6\% to 
quote our results. \\

\begin{figure}[htb]
\centering
\includegraphics[angle=-90, width=.8\textwidth]{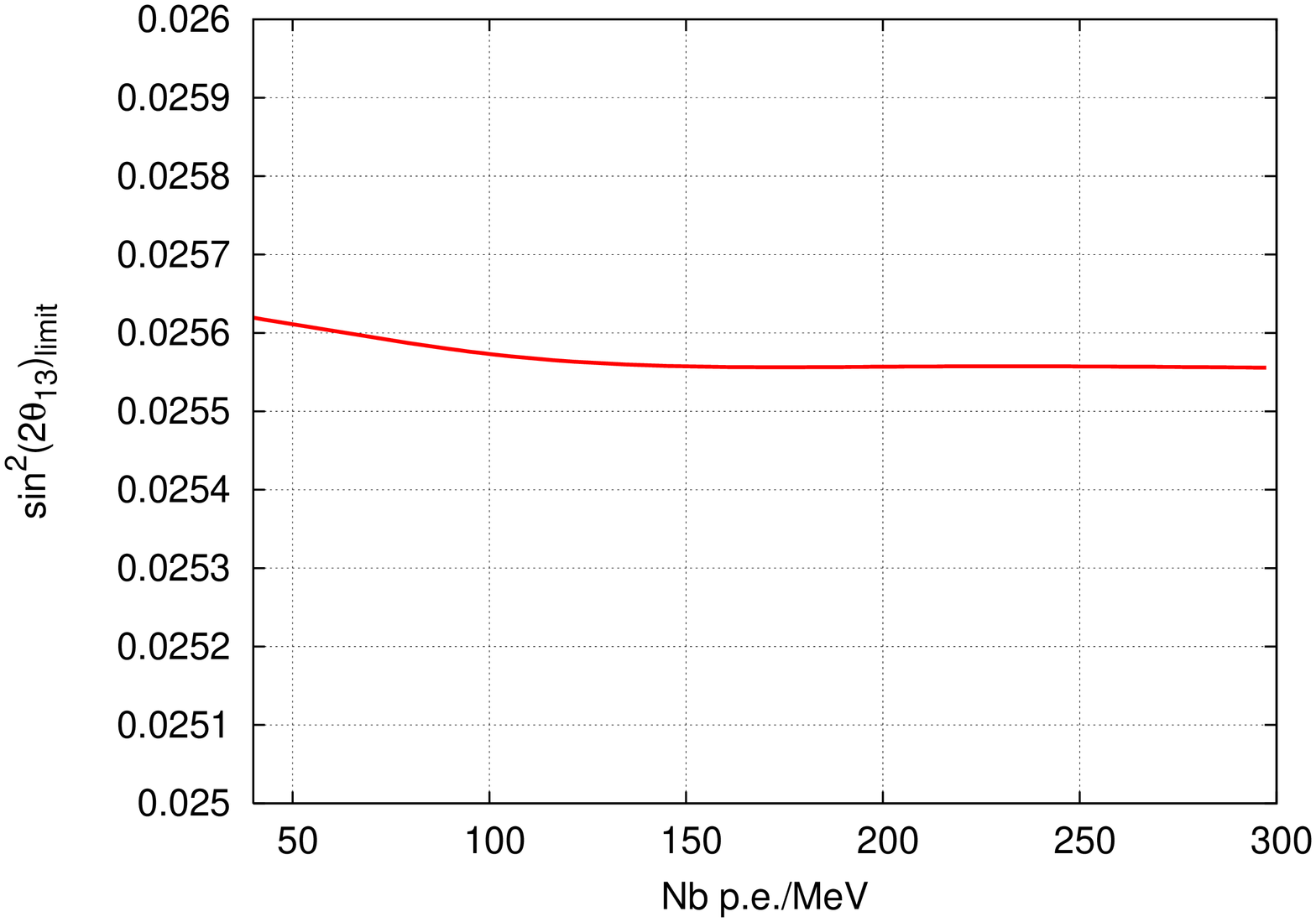}
\caption{$\sin^2(2\theta_{13})$ sensitivity (limit at 90\% C.L. if no signal) 
evolution with respect to the energy resolution in photoelectrons (p.e.) per MeV. 
The Double Chooz predicted value is between 200 and 300 p.e./MeV. The full backgrounds 
have not been included in this study.}
\label{fig:sensresolution}
\end{figure}

Figure~\ref{fig:sensresolution} displays the variation of the sensitivity (limit at 90\% C.L. if no signal) 
with respect to the energy resolution. We plan to use about 500 PMTs that will lead to 200 to 300 
photoelectrons for an energy deposition of 1~MeV at 
the center of the detector (see~Section~\ref{sec:photo}). We can see on 
the figure that the resolution is not a critical parameter (above 100 p.e./MeV); this is because we do look for an 
oscillation signal whose shape extend on the MeV scale. A good energy resolution is important for the understanding 
of the tail of the low energy spectrum, however.   
\subsection{Complementarity with the Superbeam program}

\begin{figure}[t!]
\begin{center}
\includegraphics[width=1\textwidth]{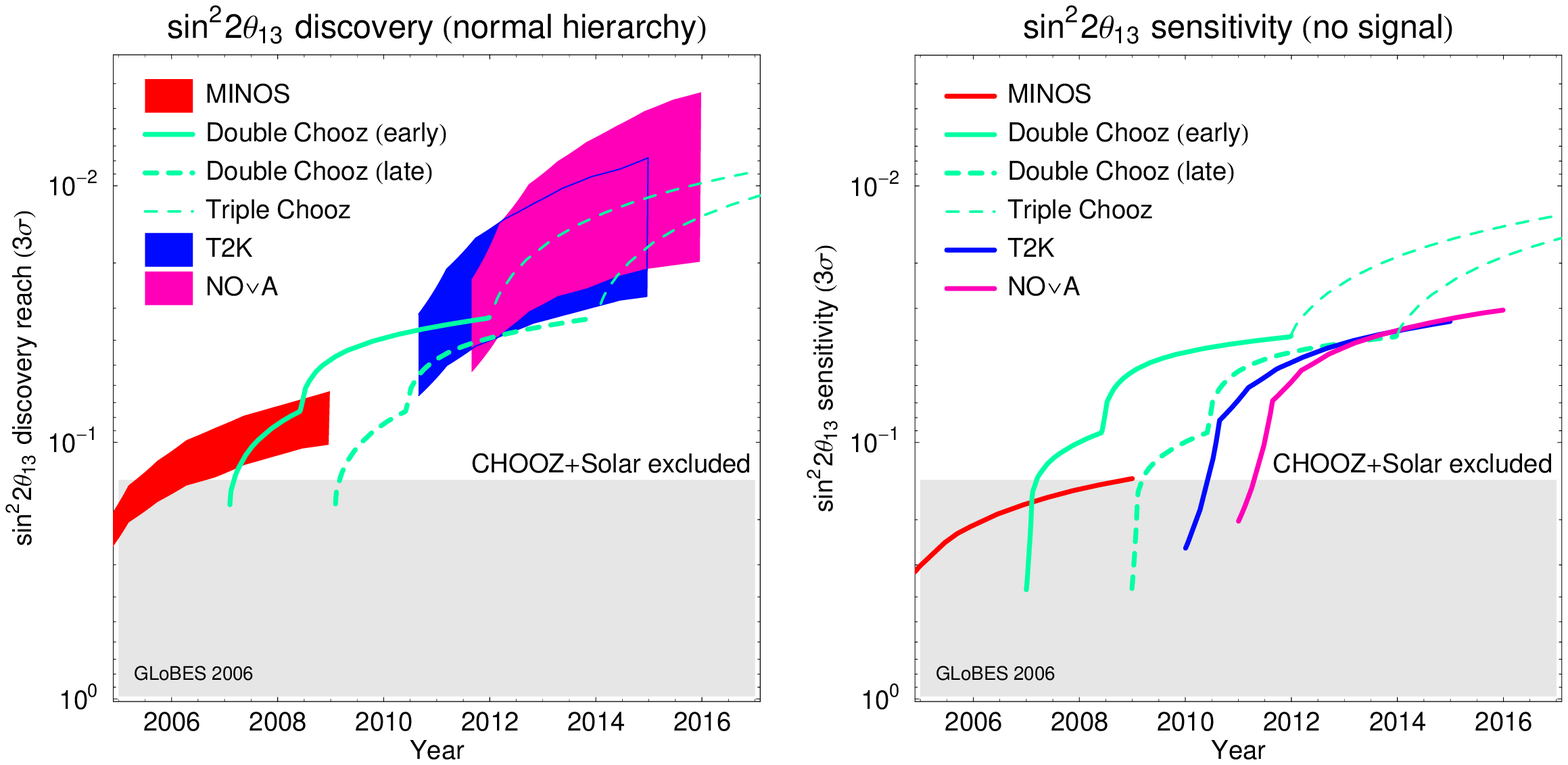}
\end{center}
\mycaption{\label{fig:evolution} A possible evolution of the
   $\stheta$ discovery potential (left) and $\stheta$ sensitivity/exclusion
   limit (right) at 3$\sigma$ as a function of time including statistics, 
   systematics, and correlations ($3 \sigma$). The bands reflect 
   for the neutrino beam experiments the dependence on the unknown 
   value of $\deltacp$, \ie, the actual sensitivity evolution will lie
   in between the best case (upper) and worst (lower) curve depending 
   on the value of $\deltacp$ chosen by nature. All experiments are 
   assumed to be operated five years and the beam experiments are
   operated with neutrino running only. The full detector mass is 
   assumed to be available right from the beginning for the beam 
   experiments, \ie, the starting times are chosen accordingly.
   Double Chooz is assumed to start data taking with the near 
   detector 1.5~years after the far detector, where two possible far 
   detector starting times are shown. In addition, the possible upgrade
   to Triple Chooz is included after five years of data taking.
   Though the starting times of the experiments have been chosen as 
   close as possible to those stated in the respective LOIs, they have 
   to be interpreted with care. A normal mass hierarchy is assumed for 
   this plot and for an inverted hierarchy, the accelerator-based
   sensitivities are expected to be slightly worse. The calculations
   (including time evolution) are based on simulations with
   GLoBES~\cite{Huber:2004ka}, which were performed in
   \Refs~\cite{Huber:2002mx,Huber:2002rs,Huber:2004xh,Huber:2004ka}
   for the beams and in ref.~\cite{Huber:2006vr} for Double Chooz
   and Triple Chooz. Figure from \Ref~\cite{Huber:2006vr} (similar to the ones in
   \Refs~\cite{Albrow:2005kw,PDNOD}).}
\end{figure}

In order to discuss the role of Double Chooz in the global context, we
show in Figure~\ref{fig:evolution} a possible evolution of the $\stheta$
discovery potential (left) and $\stheta$ sensitivity limit (right) as
function of time. In the left panel of Figure~\ref{fig:evolution}, we assume
that $\stheta$ is finite and that a certain unknown value of
$\deltacp$ exists.  The bands in the figure reflect the dependence on
the unknown value of $\deltacp$, \ie, the actual discovery reach will lie
in between the best case (upper) and worst case (lower) curve, depending on
the value of $\deltacp$ realized in nature. In addition, the curves
for the beam experiments shift somewhat to the worse for the inverted
mass hierarchy.  The right panel of the figure shows the $\stheta$
limit which can be obtained for the hypothesis $\stheta=0$, \ie, no
signal. Since particular parameter combinations can easily mimic
$\stheta=0$ in the case of the neutrino beams, their final $\stheta$
sensitivity limit is spoiled by correlations (especially with
$\deltacp$) compared to Double Chooz\footnote{Note that we define the
  $\stheta$ sensitivity limit as the largest value of $\stheta$ which
  fits (the true) $\stheta=0$ at the given confidence level.
  Therefore, this definition has no dependence on the true value of
  $\deltacp$, and the fit $\deltacp$ is marginalized over (\cf, App.~C
  of \Ref~\cite{Huber:2004xh}).}.  The two panels of Figure~\ref{fig:evolution}
very nicely illustrate the complementarity of beam and reactor
experiments: Beams are sensitive to $\deltacp$ (and the mass hierarchy
for long enough baselines), reactor experiments are not. On the other
hand, reactor experiments allow for a ``clean'' measurement of
$\stheta$ without being affected by correlations.

There are a number of important observations which can be gleaned
from Figure~\ref{fig:evolution}. First of all, assume that Double Chooz starts
as planned (solid Double Chooz curves). Then it will quickly exceed
the $\stheta$ discovery reach of MINOS and CNGS, especially after the
near detector is online (left panel). It will be the most sensitive
experiment until at least 2011 and its $\stheta$ discovery potential is
remarkable. In some scenarios, like inverted mass
hierarchy and specific values of $\delta_{\mathrm{CP}}$, the
reactor measurement would have the best discovery
potential. Note, that even the far detector of Double Chooz alone
would improve the current bounds on $\stheta$ considerably down to
0.04 after 4 years and 0.03 after 10 years at the 90\% confidence level. 
The information gained by Double Chooz can also be used for a
fine-tuning of the running strategy of second generation superbeams
with anti-neutrinos. If a finite value of $\sin^22\theta_{13}$ were
established at Double Chooz, the superbeam experiments could possibly
avoid the time consuming (due to lower cross sections) anti-neutrino
running and gain more statistics with neutrinos.  The breaking of
parameter correlations and degeneracies could in this case be achieved
by the synergy with the Double Chooz experiment.

The Chooz reactor complex even allows for a very interesting upgrade,
called Triple Chooz~\cite{Huber:2006vr}. There exists another
underground cavern at roughly the same distance from the reactor cores
as the Double Chooz far detector. A 200~t detector could be constructed
there without requiring significant civil engineering efforts.
This upgrade would in principle be equivalent to the Reactor-II setup
described in Reference~\cite{bib:huber}. Figure~\ref{fig:evolution} 
shows that it could play a
leading role, since its sensitivity is unrivaled by any of the first
generation beam experiments for the next decade and even the
discovery potential is excellent and covers more than 1/2 of the
region superbeams can access. In the case of a value of $\stheta$ not
too far below the current CHOOZ bound, this might even lead to the
possibility to restrict the CP parameter space at superbeams
for large enough luminosities. The advantage offered by this staged approach
compared to other reactor projects which would start right away with a
very large detector is that Double Chooz will serve as a testbed for the
upgrade. Thus the systematics will be well enough understood to allow a
reliable sensitivity prediction for Triple Chooz.
\begin{figure}[t]
  \begin{center}
    \includegraphics[width=0.75\textwidth]{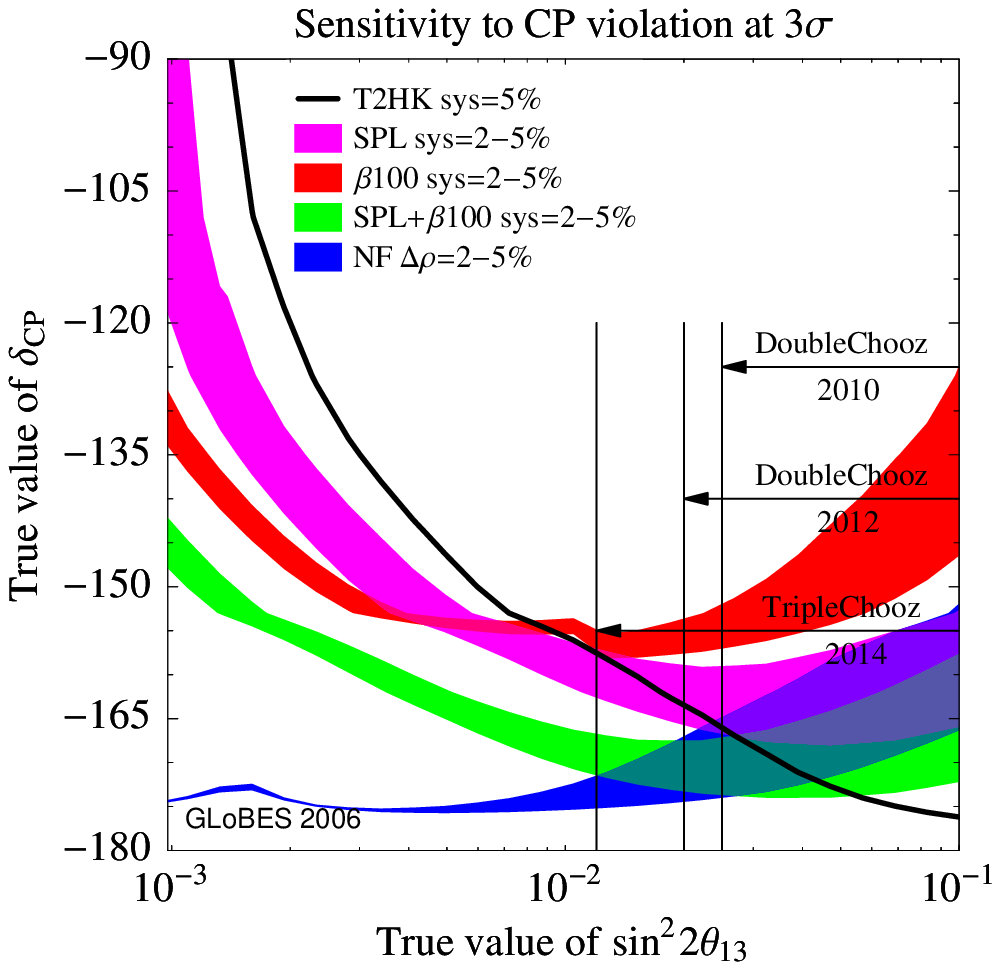} \hspace{0.5 cm}
  \end{center}
  \mycaption{\label{fig:beams} $\sin^22\theta_{13}$ sensitivity limits
    at $90\%$ C.L. of Double and Triple Chooz in comparison to the
    $3\,\sigma$ discovery reaches for CP violation of various, second
    generation beam experiments. Sensitivity to CP violation is
    defined, for a given point in the $\theta_{13}$-$\delta$-plane, by
    being able to exclude $\delta=0$ and $\delta=\pi$ at the given
    confidence level. All curves have been prepared with
    GLoBES~\cite{Huber:2004ka}.  Degeneracies and correlations are
    fully taken into account. For all setups the appropriate
    disappearance channels have been included. The beta beam is
    lacking muon neutrino disappearance, which is replaced by a 
    10\% precision on $\Delta m_{31}^2$ (corresponding to the 
    T2K disappearance information). In all cases systematics between neutrinos,
    anti-neutrinos, appearance and disappearance is uncorrelated. For
    all setups with a water Cerenkov detector the systematics applies
    both to background and signal, uncorrelated.  The neutrino factory
    assumes $3.1\cdot10^{20}$ $\mu^+$ decays per year for 10 years and
    $3.1\cdot10^{20}$ $\mu^-$ decays for 10 years. It has one detector
    with $m=100\,\mathrm{kt}$ at $3000\,\mathrm{km}$ and another
    detector with $30\,\mathrm{kt}$ at $7000\,\mathrm{km}$. The
    density errors between the two baselines are uncorrelated. The
    systematics are 0.1\% on the signal and 20\% on the background,
    uncorrelated. The detector threshold and the other parameters are
    taken from~\cite{Huber:2002mx} and approximate the results
    of~\cite{Cervera:2000vy}.  The beta beam assumes $5.8\cdot10^{18}$
    He decays per year for five years and $2.2\cdot 10^{18}$ Ne decays
    per year for five years. The detector mass is $500\,\mathrm{kt}$.
    The detector description and the glb-file is
    from~\cite{Campagne:2005jh}.  The SPL setup is taken
    from~\cite{Mezzetto:2005ae}, and the detector mass is
    $500\,\mathrm{kt}$.  The T2HK setup is taken
    from~\cite{Huber:2002mx} and closely follows the
    LOI~\cite{Itow:2001ee}. The detector mass is $1000\,\mathrm{kt}$
    and it runs with $4\,\mathrm{MW}$ beam power, 6 years with
    anti-neutrinos and 2 years with neutrinos. The systematic error on
    both background and signal is 5\%. Figure from \Ref~\cite{Huber:2006vr}.}
\end{figure}

Figure~\ref{fig:beams} shows the sensitivity to CP violation at
$3\,\sigma$ C.L.\  ($\Delta\chi^2=9$) for various second generation
beam experiments.  In this figure two regimes can be clearly
distinguished: very large $\sin^22\theta_{13}\geq0.01$ and very small
$\sin^22\theta_{13}\leq0.01$. At large $\theta_{13}$ the sensitivity
to CP violation is basically completely determined by factors like
systematic errors or matter density uncertainty. Thus the question of
the optimal technology cannot be answered with confidence at the
moment, since for most of the controlling factors the exact magnitude
can only be estimated. The technology decision for large
$\theta_{13}$, therefore, requires considerable R\&D. On the other
hand, in the case of small $\theta_{13}$ the optimal technology seems
to be a neutrino factory\footnote{Not shown in Figure~\ref{fig:beams}
  is the $\gamma=350$ beta beam~\cite{Burguet-Castell:2005pa}. Such a higher gamma beta beam could also 
  play the role of a neutrino factory~\cite{Huber:2005jk,Burguet-Castell:2003vv}.} 
quite independently from any of the
above mentioned factors. The branching point between the two regimes
is around $\sin^22\theta_{13}\sim0.01$ which coincides with the
sensitivities obtainable at the Chooz reactor complex. Moreover, the
information from Double Chooz would be available around 2010 which is
precisely the envisaged time frame for the submission of a proposal
for those second generation neutrino beam facilities. Thus the Double
Chooz results are of central importance for the long term strategy of
beam-based neutrino physics.

Double Chooz is also important for the next generations of 
searches for neutrino-less double beta decay. The amplitude for  
this process depends on the so-called 
effective mass, 
\be
\meff \equiv \left| \sum_i m_i \, U_{ei}^2 \right| 
\mbox{ with } 
m_{ee} = |m_{ee}^{(1)}| + |m_{ee}^{(2)}| \, e^{2i\alpha} + 
|m_{ee}^{(3)}| \, e^{2i\beta}~, 
\label{eq:meff1}
\ee
where $m_i$ is the mass of the $i^{th}$ neutrino state, 
the sum is over all light neutrino mass states and $\alpha, \beta$ are 
the two Majorana phases. For an overview on the current situation 
of neutrino-less double beta decay, see \cite{Aalseth:2004hb} and 
references therein. 

\begin{figure}[tbh]
\begin{center}
\epsfig{file=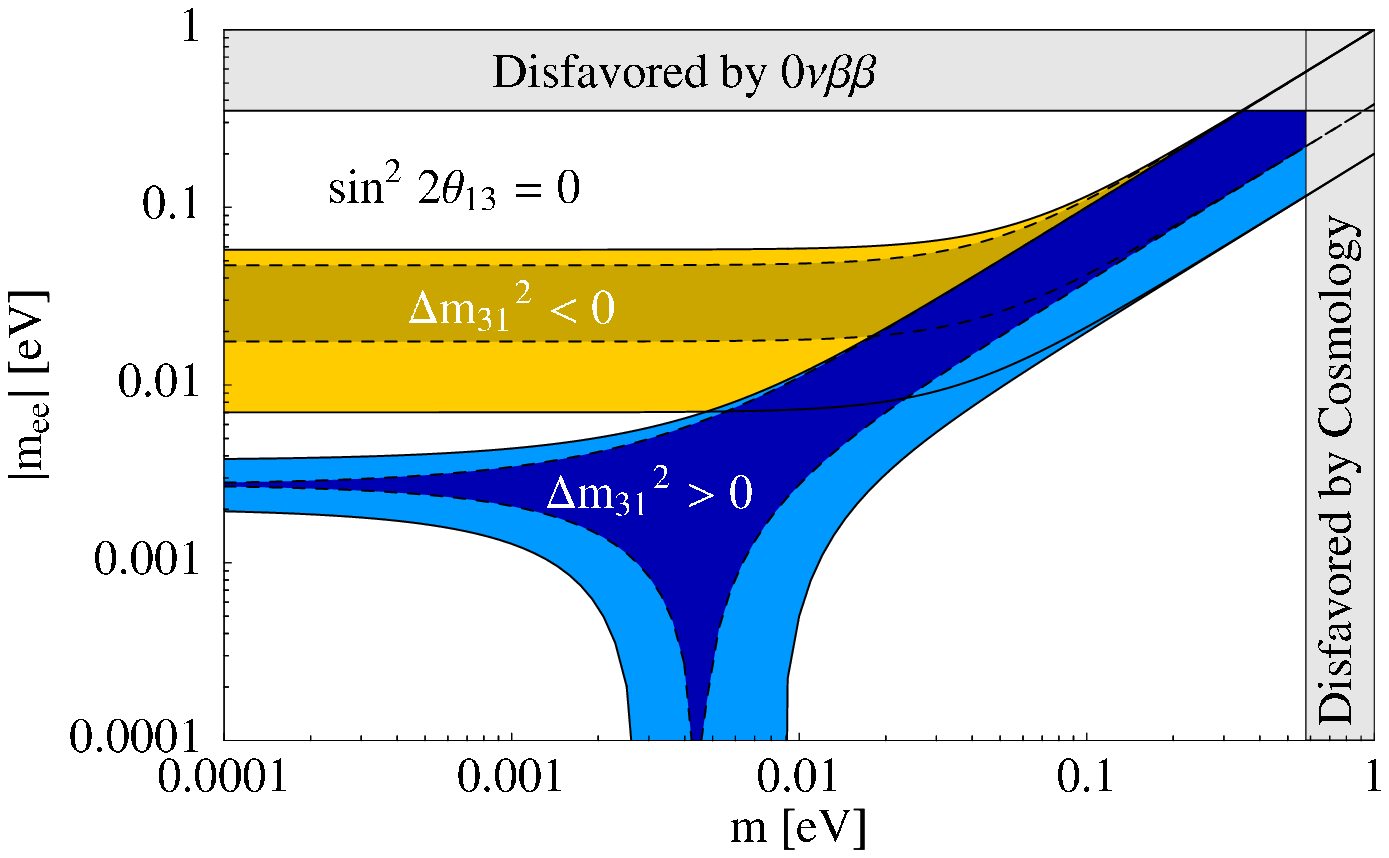,width=8cm,height=6cm}
\epsfig{file=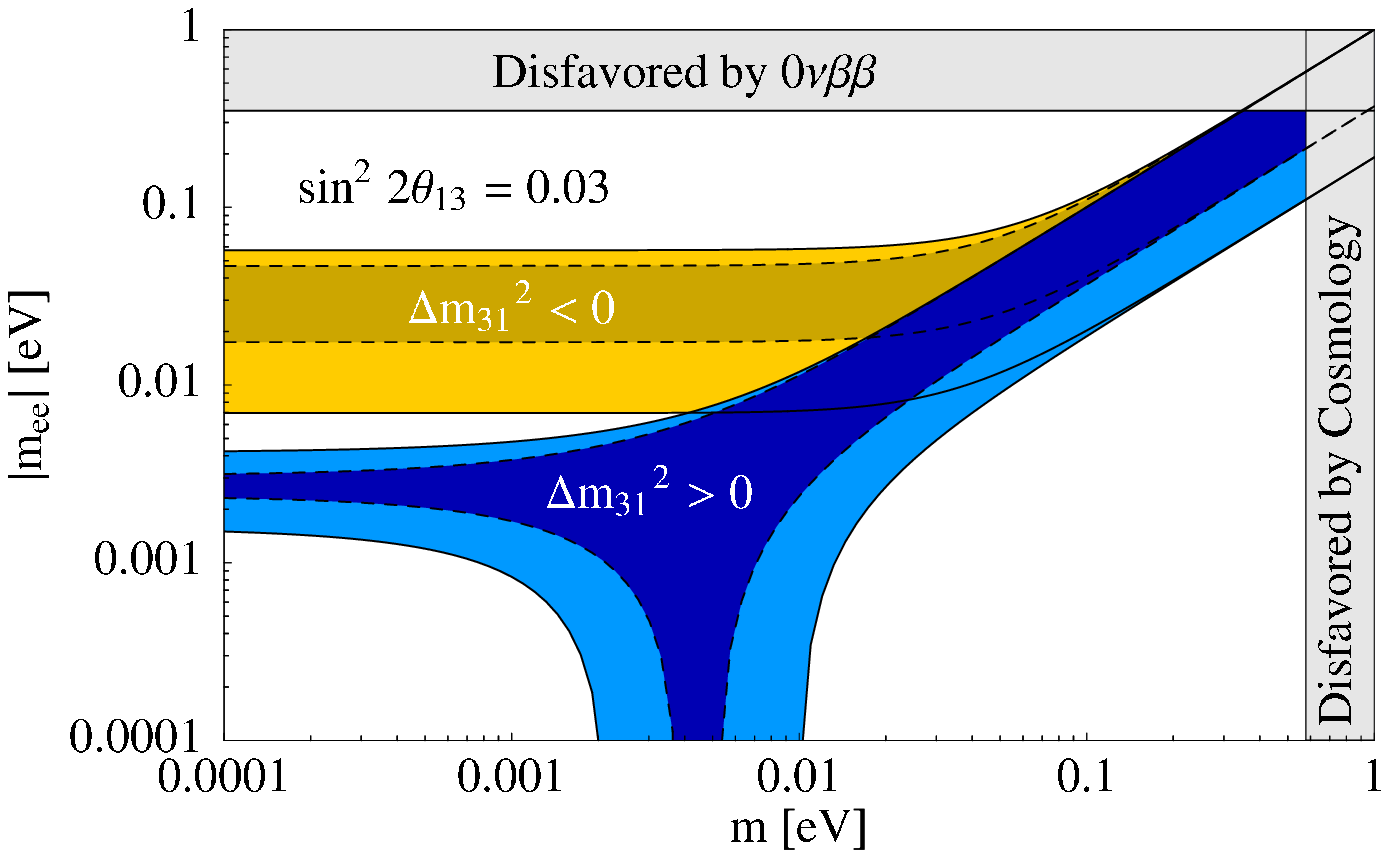,width=8cm,height=6cm}
\epsfig{file=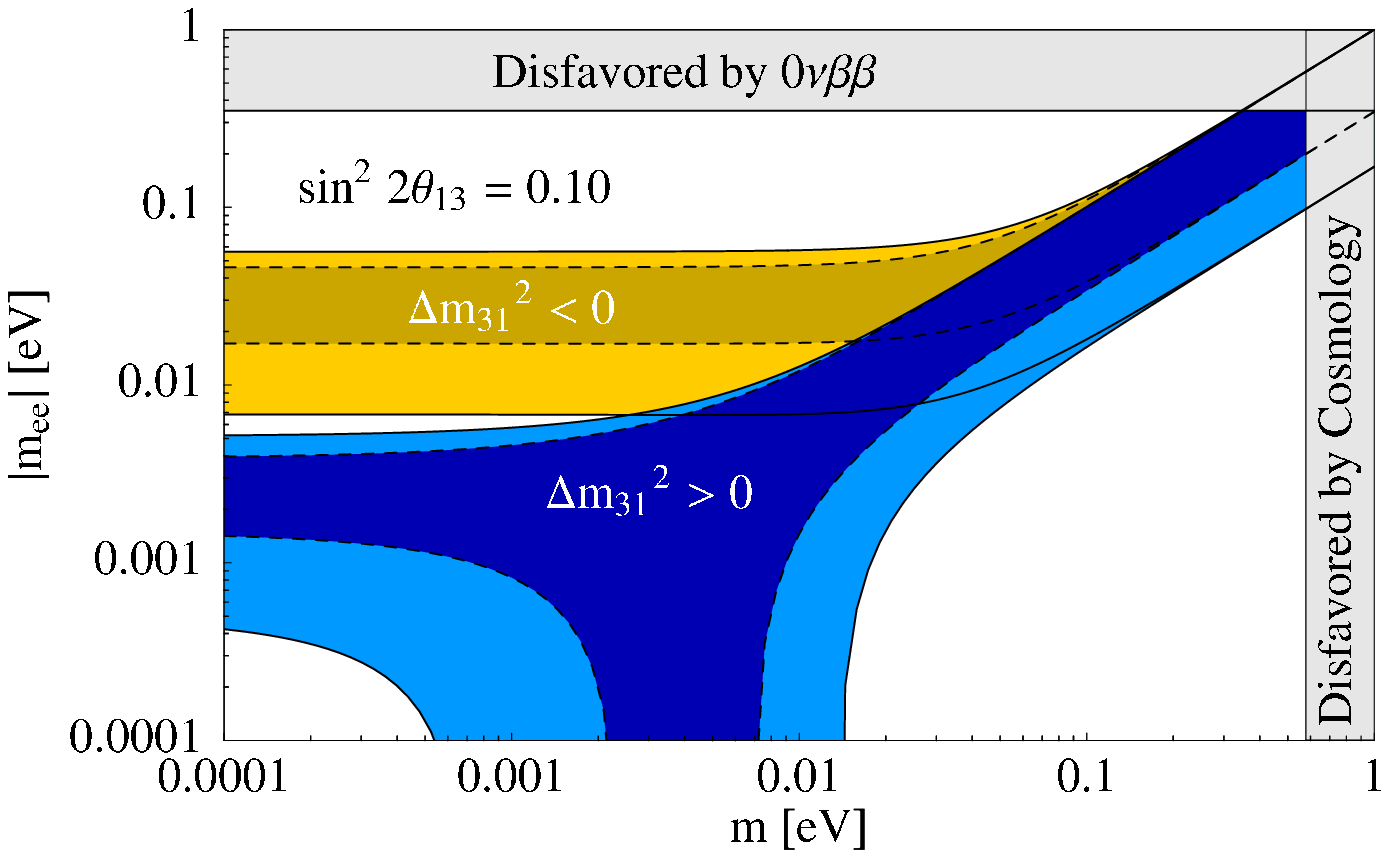,width=8cm,height=6cm}
\epsfig{file=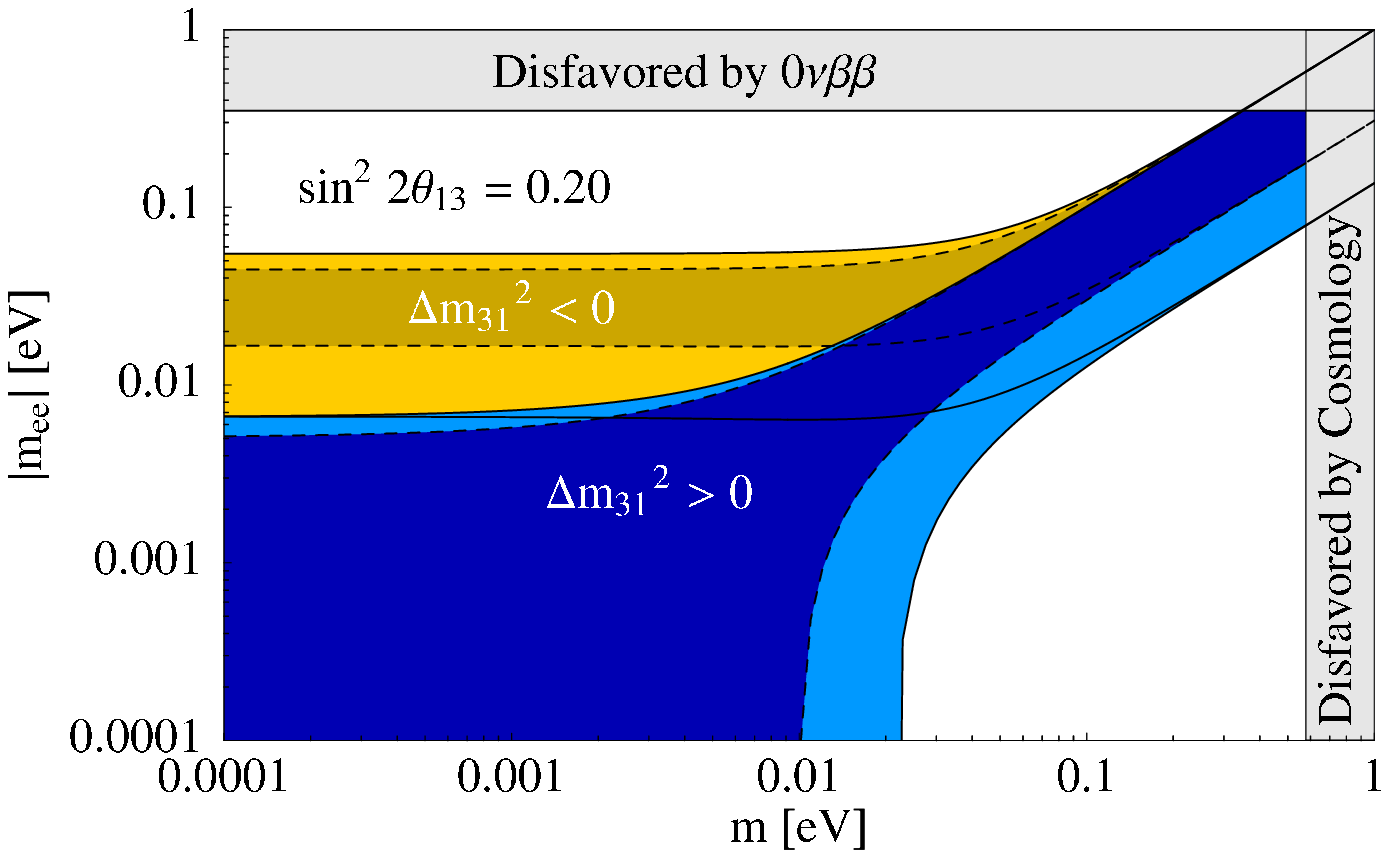,width=8cm,height=6cm}
\caption{\label{fig:usual_boring_plot}The effective mass (in eV) 
for the normal and inverted ordering as a function of the smallest 
neutrino mass (in eV) for different values of $\sin^2 2 \theta_{13}$. 
The prediction for the 
best-fit values of the oscillation parameters and for 
the $3\sigma$ ranges is given. A typical bound from cosmology 
and the current limit on the effective mass 
are indicated. From \cite{Lindner:2005kr}.}
\end{center}
\end{figure}

Assuming a certain value of $\theta_{13}$ and 
predicting the effective mass as a function of the smallest neutrino mass
\cite{Klapdor-Kleingrothaus:2000gr,Bilenky:2001rz,Feruglio:2002af}, 
one obtains two bands for $|m_{ee}|$, corresponding to a positive or negative sign of the 
atmospheric $\Delta m^2$ (see Figure~\ref{fig:usual_boring_plot}). This 
allows one to distinguish the 
normal from the inverted mass ordering 
\cite{Pascoli:2005zb,Choubey:2005rq,deGouvea:2005hj}. 
However, this simplified picture gets changed once the
current knowledge on $\theta_{13}$ is taken into account 
\cite{Lindner:2005kr}. 

As can be seen from Figure\ \ref{fig:usual_boring_plot}, 
the gap between the cases $\Delta m^2_{31} > 0$ 
and $\Delta m^2_{31} < 0$ depends on $\theta_{13}$. 
For todays largest possible values 
of $\theta_{13}$, the normal and inverted hierarchy regions overlap, 
making the distinction of normal and inverted hierarchy impossible. 
The width of the ``chimney'' for very small values of \meff~also depends 
strongly on $\theta_{13}$, in the sense that it becomes narrower 
with decreasing $\theta_{13}$. 
For very small $\theta_{13}$, a vanishing \meff~would correspond to rather 
specifically chosen parameter values. 

All these aspects demonstrate how the outcome of Double Chooz affects the
future program for neutrino-less double beta decay. 

\subsection{Other physics}
\subsubsection{Study of the directionality}
The near detector of Double Chooz will detect on the order of
one million events per year, which opens exciting possibilities.
One topic is the possibility to determine the direction of the
neutrino from the forward scattering of the neutron in the 
inverse beta decay reaction. Successive scatterings of the 
neutron leads to a broad distribution of capture locations 
and it is almost as likely that the neutron is captured in 
backward direction as in forward direction. This means that 
the direction of a neutrino source cannot be determined
on an event by event basis. However, the neutron has a slightly
larger probability to be captured in forward direction, which
can be measured with sufficient statistics and resolution. 
This effect was first 
seen in the Bugey experiment\cite{bib:bugey}
and even better in the CHOOZ experiment\cite{bib:chooz}. 

The very high statistics of the near detector of Double Chooz 
will allow a detailed exploration of the directionality effect. 
The near detector will record events when both reactor cores 
are on, when one or the other core is off, and it will also 
have data with both reactors off. A comparison of these data 
sets, in combination with a modeling of the expected 
event distributions will allow us to understand and test 
directionality much better. These studies should ultimately 
clarify if directionality can be used in future experiments. 
Possible applications could emerge in astrophysics, reactor 
physics or in the context of geo-neutrinos. 
\subsubsection{Mass-varying neutrinos}
Mass-varying neutrinos are an interesting idea which could be
tested by reactor experiments.
The concept of mass-varying neutrinos has been introduced by
imposing a relation between neutrinos and the dark energy of the
Universe~\cite{Hung:2000yg,Gu:2003er,Fardon:2003eh,Peccei:2004sz}
through a scalar field, the acceleron. Including the possibility of
acceleron couplings to matter fields implies that the neutrino
oscillation parameters in vacuum/air and a medium could be very
different~\cite{Kaplan:2004dq}. Since reactor experiments
do have different paths in vacuum and air, or the material
along the path may be altered at relatively moderate effort,
the direct test of mass-varying neutrinos may be possible
using reactor experiments. In \Ref~\cite{Schwetz:2005fy},
a different parameterization of $\theta_{13}$ and $\Delta m_{31}^2$
was used for air and matter. This approach allows for arbitrary
effects different in air and matter, since any Hermitian
addition to the Hamiltonian in matter can be described by
a re-diagonalization leading to a different effective
mixing angle and mass squared difference in matter.
If only one experiment (in air or matter) measures these parameters,
no conclusion about mass-varying neutrinos can be drawn.
Only by the combination of different experiments in air and
vacuum, constraints to mass-varying neutrinos can be obtained.
In particular, since the Double Chooz baseline is
partly in air (roughly 50\%), Double Chooz will be a key component
to such measurements to be combined with beam or reactor data in
matter. Note that all proposals for other reactor or beam
experiments have baselines practically only in matter. Extremely
high sensitivities can be finally obtained by physically moving the
material between near and far detector, because one does not only
have {\em identical} detectors then, but also {\em the same}
detectors in two consecutive phases of the experiment in order to
cancel systematics almost completely. For a more extensive discussion,
see \Ref~\cite{Schwetz:2005fy}.

%
\cleardoublepage
%
%
\section{Detector Structures, Materials and Radiopurity}
\label{sec:mechanics}
\subsection{Detector dimensions}
\begin{figure}[htb]
\centering
\includegraphics[width=0.8\textwidth]{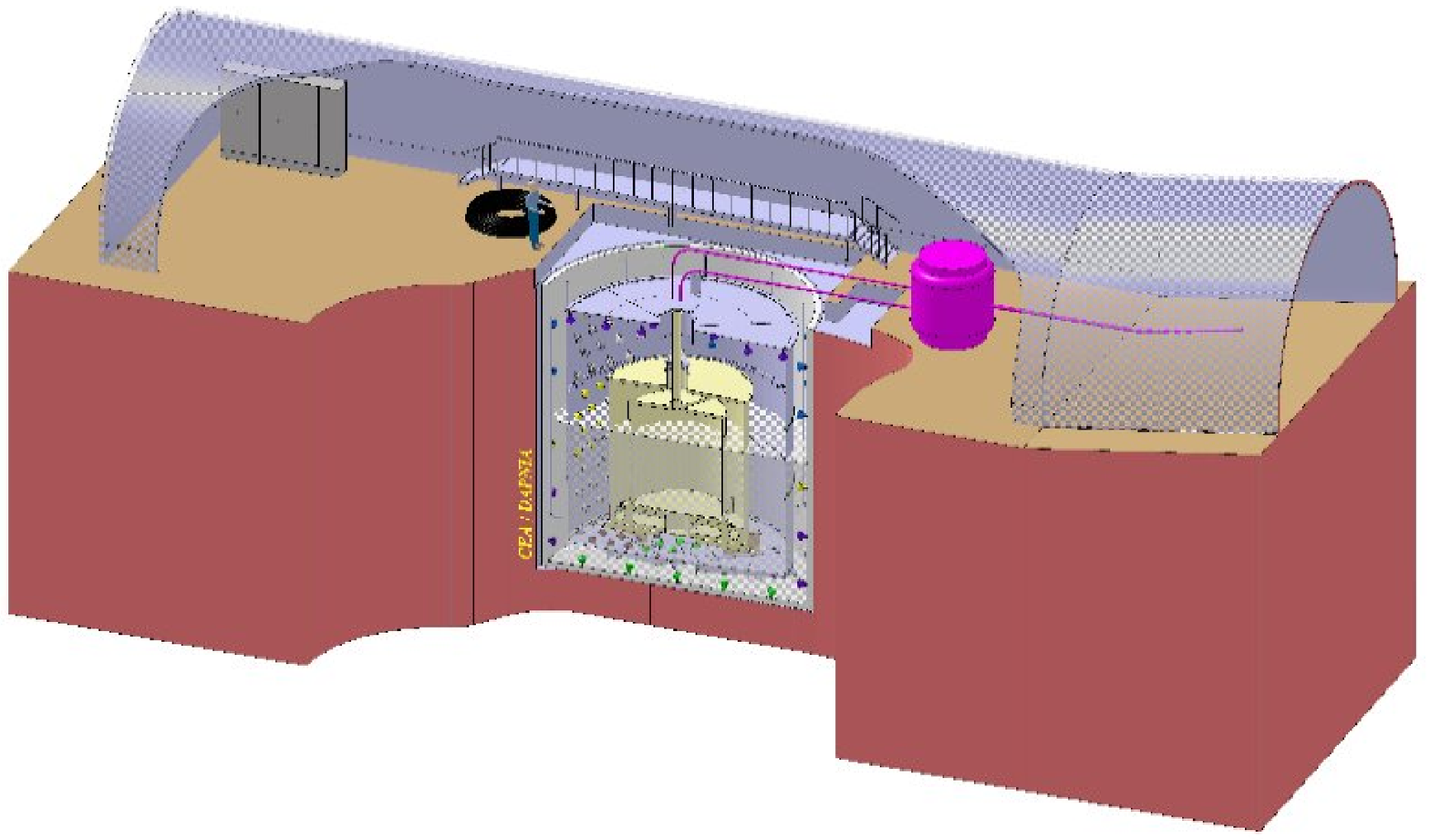}
\caption{Viewgraph of the Far detector integrated in the the neutrino laboratory located at 
the end of the "Marini\`ere" gallery, 1~km from the CHOOZ-B nuclear power station cores.}
\label{fig:detectorsynoptic}
\end{figure}

\begin{figure}[htbp]
\centering
\includegraphics[width=\textwidth]{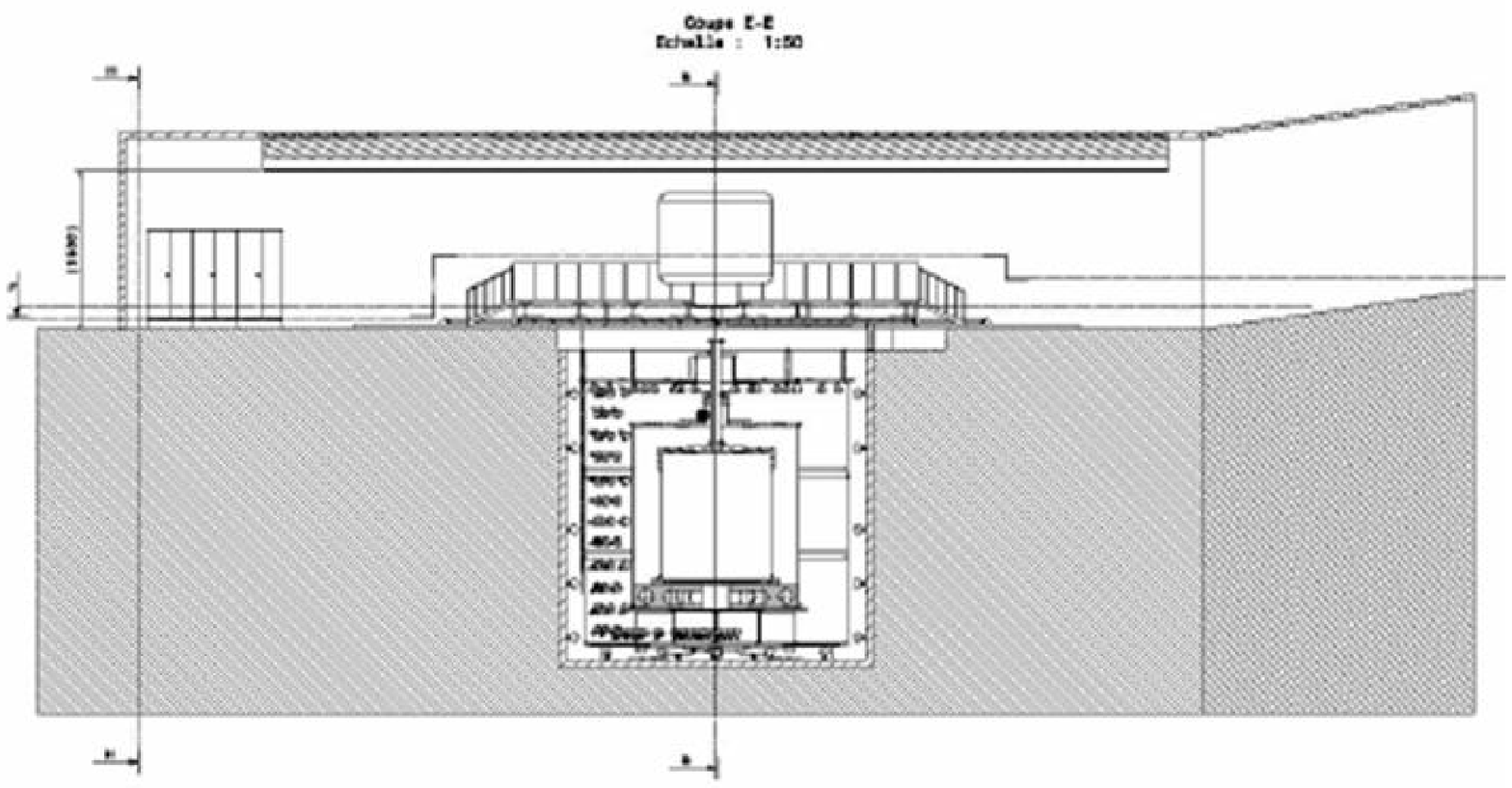}
\caption{Side view of the Far detector in the neutrino laboratory located at the end of 
the "Marini\`ere" gallery. The slope at the right side is the entrance of the laboratory.}
\label{fig:detector_side}
\end{figure}

\begin{figure}[htbp]
\centering
\includegraphics[width=\textwidth]{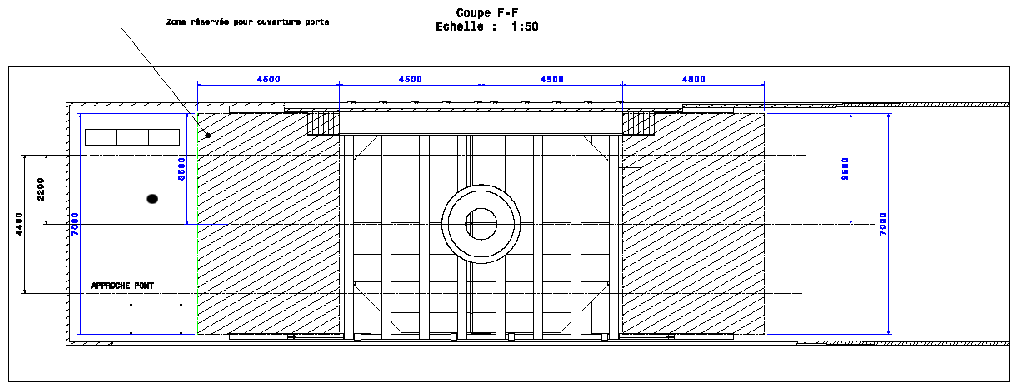}
\caption{Top view of the the neutrino laboratory located at the end of 
the "Marini\`ere" gallery. Entrance is at the right side. We can see the bridge crossing the 
detector that will have to be modified in order to be used as a safety issue in case of 
incident in the neighboring area (CHOOZ-A).}
\label{fig:detector_top}
\end{figure}

Detector dimensions are summarized in Tables~\ref{tab:innerdimensions},~\ref{tab:regiondimensions},~\ref{tab:chimneyimensions}.
Viewgraph of the Far detector integrated in the the neutrino laboratory are shown in Figures~\ref{fig:detectorsynoptic},
~\ref{fig:detector_side},~\ref{fig:detector_top}.
%
%
%
%
\begin{table}[htpb]
\caption{\label{tab:innerdimensions} Dimensions of the mechanical structure of the detector.}
\centering\begin{tabular}{ccccccc}
\hline
Inner     & Inner         & Inner      & Thickness  & Filled   & Volume    & Mass     \\
Detector  & Diameter (mm) & Height (mm) & (mm)      &   with   & 
(${\rm m}^3$)   & (tons)   \\
\hline
Target    & 2300          & 2458       & 8          &  Gd-LS   & 10.3      & 0.35     \\
$\gamma$-catcher &  3392     & 3574       & 12(-15) &  LS      & 22.6      & 1.1-1.4  \\  
PMTs      & ---           & ---        & ---        &  ---     & ---       & ---      \\
Buffer    & 5516          & 5674       & 3          &  Oil     & 114.2     & 7.7      \\
Veto      & 6590          & 6640$\pm100$ & 10       &  Oil     & 90        & 20       \\
Shielding & 6610          & 6660$\pm100$ & 170      &  Steel   & ---       & 300      \\
Pit       & 6950          & 7000       & ---        &  ---     & ---       & ---      \\
\hline
\end{tabular} 
\end{table}
\begin{table}[htpb]
\caption{\label{tab:regiondimensions} Thickness of the four detector regions filled with liquids}
\centering\begin{tabular}{ccccc}
\hline
Region                   &     Target    & $\gamma$-catcher  &   Buffer   &   Veto  \\       
\hline
Radial Thickness (mm)    &     1150      & 550           &   1050     &   500   \\
Vertical Thickness (mm)  &     2458      & 550           &   1050     &   500 below and 600 above   \\  
\hline
\end{tabular} 
\end{table}
\begin{figure}[htbp]
\centering
\includegraphics[width=0.9\textwidth]{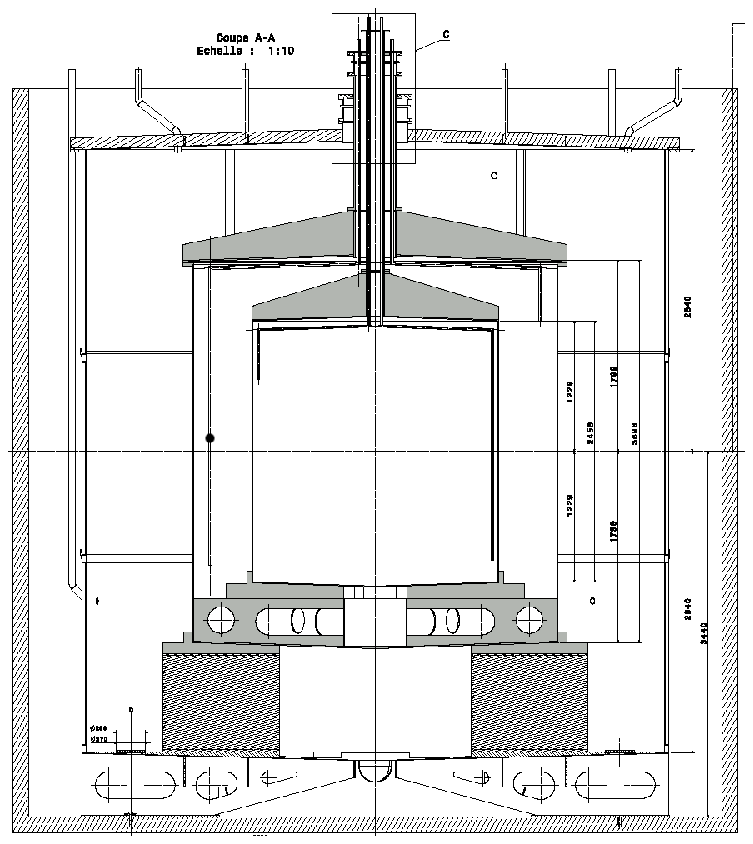}
\caption{Detail drawing of the Far detector. From the innermost 
region, one has: the neutrino
Target enclosed in a acrylic vessel (8~mm) filled with 
a Gd-doped liquid scintillator, the $\gamma$-catcher  
region enclosed in another acrylic vessel (12-15~mm) 
and filled with an undoped liquid scintillator, the
Buffer region filled with non scintillating oil 
enclosed in a stainless steel vessel (3~mm) supporting also
 the phototubes (534), the Inner Veto enclosed in 
a steel wall (10~mm) and filled with scintillating oil, 
and the steel shielding (170~mm). The top of the 
steel shielding as well as the outer muon veto are not
 represented.}
\label{fig:detectordrawing}
\end{figure}
\begin{figure}[htb]
\centering
\includegraphics[width=0.7\textwidth]{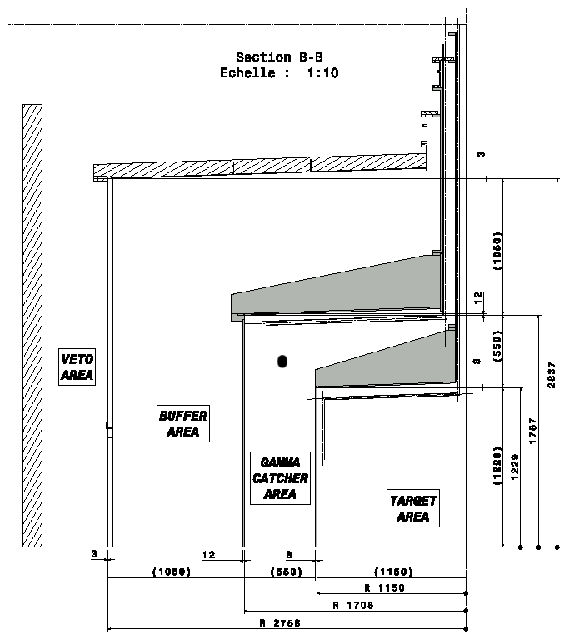}
\caption{Side view of the detector regions. Effective thickness of the 
$\gamma$-catcher  region is
550~mm. Effective thickness of the Buffer region 
is 1050~mm. Thickness of the Veto region is 
500~mm (600~mm on the top part of the Veto).}
\label{fig:detector_regions}
\end{figure}
%
%
%
%
\begin{table}[htpb]
\caption{\label{tab:chimneyimensions} Dimensions of the detector chimneys. The effective radius is defined as 
the radius available for the calibration systems. It corresponds to the radius of the chimney minus the size 
taken by the filling tubes.}
\centering\begin{tabular}{cccccc}
\hline
  Chimney      & Effective     & Thickness  & Height & Filled          \\
               & space    (mm) & (mm)       & (mm)   &   with          \\
\hline
Target         & 50 (radius)   &  8         &        & LS+N$_2$         \\
$\gamma$-catcher   & 40 (gap)      &  12-15     &        & LS+N$_2$     \\
\hline
\end{tabular} 
\end{table}
\subsection{Double acrylic vessel}
\par Target and $\gamma$-catcher  vessels will be built with acrylic plastic material, transparent 
to photons with wavelengths above 400 nm. Both vessels are designed to contain the Target 
and $\gamma$-catcher  aromatic liquids with a long term tightness (no leak leak for 10~years) 
and stability. The strongest constraint is the chemical compatibility between the vessel and the 
scintillating liquids of the Target and $\gamma$-catcher , for at least 5~years. 
We tolerate neither a modification of the liquid properties (scintillation, absorbency) nor a 
degradation of the acrylic material (breaking or crazing of more than 
a few percent of the acrylic surface area). 
The $\gamma$-catcher  vessel must also be chemically compatible 
with the mineral oil of the Buffer region, 
which is known to be a weaker constrain.
Material compatibility R$\&$D has been pursued 
within the collaboration since 2003, based on knowledge 
obtained from the LENS R$\&$D program~\cite{bib:lens}. Acrylic materials immersed in liquids of interest have been 
studied under mechanical stress up to 30~MPa 
and at different temperature to accelerate the aging processes.
According to these studies, partially realized in 
collaboration with the Roehm acrylic company in Germany~\cite{Roehm}, 
the maximum Von Mises stress tolerated in Double Chooz has been 
set to 5~MPa. 

\par Mechanically, the double vessels have to be strong enough to ensure identical shapes between Near 
and Far target vessels. No deformations of more than 5~mm are allowed during the running phase. This small 
geometrical difference between the two target acrylic vessels 
eliminate any measurable difference 
of the spill in/out effect between the near and far detectors. 
We note here that the number of free protons inside the Target vessel has to be measured at better 
than 0.2$\%$. Thus, according to the 5~mm geometrical tolerance given by the plastic manufacturer, 
the volume difference between both Target vessel could be as large 
as 0.6~$\%$ (60 liters). In consequence,
 a weight based measurement method has been developed and tested to control the target 
content (mass) at the required precision~(see page~\ref{'volume'}). 

\par The mechanical structures of the double acrylic vessel have been studied with the finite element simulation 
software CASTEM~2000~\cite{Castem2000}. The vessel has
 been designed to account for the transportation, 
integration, filling and running phases. 
The goal of the mechanical simulation was to define the structure of the vessel within the distortion 
and stress tolerances. During this work a few critical cases have been identified (see below).

\par In this simulation, some optimizations can be tested
to improve the signal to noise ratio. 
For example, the Target vessel thickness and the acrylic structural materials 
between the Target and $\gamma$-catcher 
can be minimized to reduce dead materials within 
the active volume, as well as accidental background 
originating from radio-impurities. Dead materials can
 lead to untagged muon capture, see 
section~\ref{sec:bkg_mucapt}. 
The $\gamma$-catcher  vessel, however, is not considered a dead 
material, but a part of the  non scintillating 
Buffer. Its thickness is only constrained by its radioimpurities content. 

\par
\begin{figure}[htb]
\centering
\includegraphics[width=0.5\textwidth]{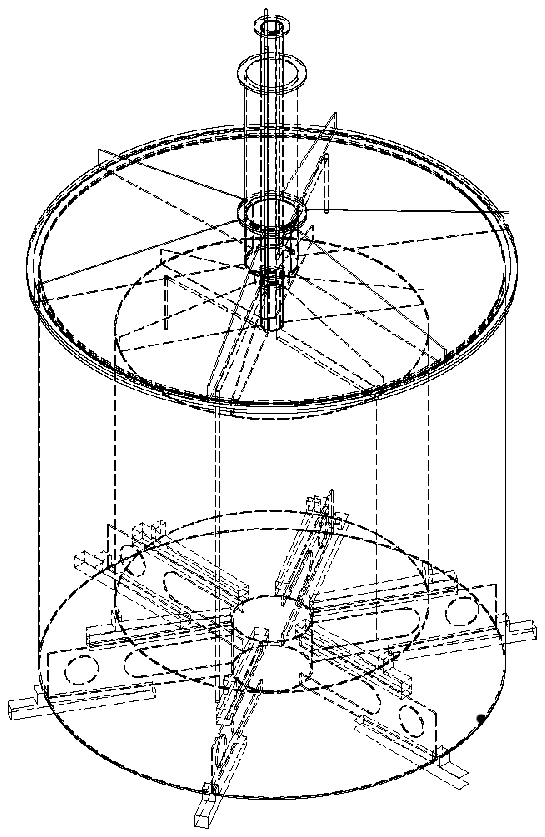}
\caption{3D view of the double acrylic vessel. The Target vessel is a cylinder of 2474~mm height, 
2316~mm diameter, and 8~mm thick. It weights 350~kg, and contains a volume 
of 10.3~$m^3$. 
The $\gamma$-catcher  vessel is a cylinder surrounding the Target of 3422~mm height, 3374~mm diameter 
(external dimensions), and 12~mm thick, and contains a volume of 22.6~$m^3$}
\label{fig:AcrylicVessel3D}
\end{figure}

 Considering all previous requirements the Target and 
$\gamma$-catcher  vessels will be made of 
cast acrylic from the Roehm Company. 
The assembly will be done by gluing pieces 
according to the manufacturer's expertise. 
The Target vessel is a cylinder of 2474~mm height, 2316~mm diameter (external dimensions), 
and 8~mm thick. It weights 350~kg, and contains a volume of 10.3~${\rm m}^3$ 
(without the chimney). 
The $\gamma$-catcher  vessel is a cylinder surrounding the Target of 3422~mm height, 3374~mm diameter 
(external dimensions), and 12~mm thick. It weights 1100~kg, and contains 
a volume of 22.6~${\rm m}^3$ 
(without the chimney). The Target chimney diameter 
will be less than 900~mm. Drawings of the Target 
and the $\gamma$-catcher  are shown in Figure~\ref{fig:AcrylicVessel3D}.
\par Simulation has shown that the transportation phase is hazardous
 for a double acrylic vessel which has been completely assembled by glue. 
Vibrations generated by the suspension system during ground transportation
could be significant if the full double vessel construction was completed
at the manufacturer's point.
Calculation of the structural brittleness of the acrylic 
gives a maximum acceptable acceleration of $11~g$. 
To avoid any resonance problem and completely decouple from the
suspension system, the eigen frequency of 
the structure must be greater than 10~Hz.
The computation shows that the first eigen frequencies of 
our double vessel are close to 8~Hz.
A first simple solution would be to add acrylic stiffeners to the structure. 
Eigen frequencies will then be raised to more than 13~Hz, above the critical region. 
This problem could also be solve, without changing the baseline design, by transporting the Target 
and $\gamma$-catcher  vessels separately, integrate them 
in the pit, and glue the $\gamma$-catcher  top lid 
and the Chimney in the neutrino laboratory. 
There will be no annealing after the vessels are made,
since the oven is far from Chooz and the 
transportation would induce intolerable stress. 

\par As a result of these studies, it has been decided that the 
integration of the double acrylic vessel will be done in several steps: 
a) the Target vessel will be entirely built, annealed and 
checked for tightness at the manufacturer;
b) The chimney will be glued in the Chooz far site laboratory;
c) the $\gamma$-catcher  vessel will be built without top lid--it will 
be glued later in the pit of the detector. 
For the final step, air conditions in the neutrino 
laboratory (in and just around the pit) have to be well controlled. 
During the 24~hours of the polymerization of the glue, the temperature has to be kept above 25 degrees centigrade
 and the relative humidity at 40$\%$, whereas nominal conditions in the tunnel are 14 degrees 
centigrade and 100$\%$ respectively. 
In both Chooz neutrino laboratories, the assembly in the 
pit will be handled with the ceiling crane and 
specifically developed on tools~(see Figure~\ref{fig:acrylictool}). 
\begin{figure}[htb]
\centering
\caption{Dedicated tool to handle the acrylics vessel from the construction till the integration 
in the detector. A full mechanical study has been performed to guarantee a tolerable stress 
level during all the phases of the life of the acrylic vessels.}
\label{fig:acrylictool}
\end{figure}
The $\gamma$-catcher  vessel 
will arrive mounted on a supporting structure used
to minimize the deformations, driven in the tunnel, and 
rotated above the pit prior to its final
 insertion in the Buffer vessel~(phototubes will already
be partially mounted). 
The Target vessel will then be inserted into the $\gamma$-catcher.
After a first cleaning on site the top lid of the 
$\gamma$-catcher  and the chimney will be glued by 
technicians from the manufacturer.

\par In Double Chooz all regions within the 
Buffer vessel have to be filled simultaneously. 
The filling phase generates constraints related to the 
differences in height of liquid.
According to the mechanical simulation, if we neglect density variations, the 
difference in height 
acceptable is 30~cm. 

\par During the running phase static loads induced by slight liquid density 
differences, 
put stress in the acrylic vessels. Simulations 
have been done to assess the stress levels 
with respect to density differences. The 
following table is the summary of the slight densities 
differences acceptable in each zone.
\begin{table}[htpb]
\caption{\label{tab:acrylics_density} Summary of stress and   effects 
induced by density differences between the Target, $\gamma$-catcher  and 
buffer liquids.}
\centering\begin{tabular}{ccccc}
\hline
Density difference & Dead load & 0$\%$ & 1$\%$ (T $\&$ GC) & 1$\%$ (GC $\&$ 
Buffer) \\
\hline
Max. distortion in the top lid (mm) &  -2 & -1 & -1 & -2 \\
Max. distortion in the bottom (mm) & -1 & -0.7 & -2.8 & 2.7 \\
Max. stress (MPa) & 1.2 & 0.9 & 1.2 & 1.0 \\
\hline
\end{tabular} 
\end{table}
\subsection{Stainless steel Buffer vessel}
\par The Buffer vessel surrounds the $\gamma$-catcher 
 region, 1050~mm away from the double acrylic vessel.
It will be built with stainless steel 304L, 
and has the following requirements: a) to be tight to mineral 
oil over a long term 
(10 years); b) to be chemically compatible with the 
mineral oil of the Buffer region and the scintillating 
oil of the Inner Veto region; c) to be strong 
enough to support the five hundred photomultiplier tubes
 (positioning precision is given at about 1~cm); d) to
be as light as possible to reduce backgrounds. 
\par
\begin{figure}[htb]
\centering
\includegraphics[width=.5\textwidth]{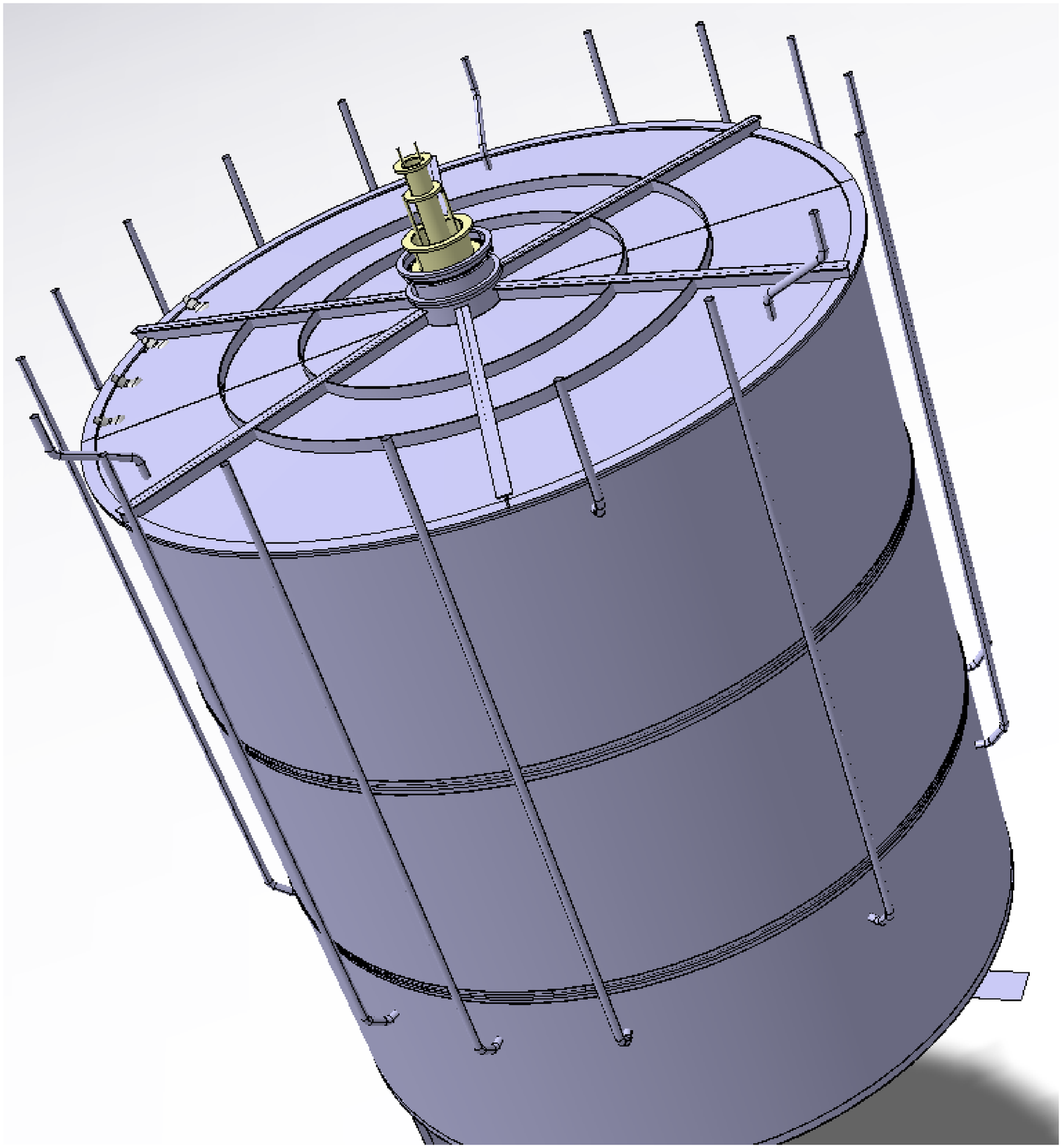}
\caption{\label{fig:buffervessel3D}
3D view of the stainless steel Buffer vessel. This is a cylinder of 5680~mm height, 
5522~mm diameter, and 3~mm thick (with 
stiffeners to guarantee the rigidity). 
It weights 7.7~tons, and contains a volume of 114.2~${\rm m}^3$. The cable 
phototubes go out of the detector 
through stainless steel pipes (welded) that run along the Inner Veto wall.}
\end{figure}

 The Buffer vessel is a cylinder of 5680~mm height, 
5522~mm diameter (external dimensions), 
and 3~mm thick. It weights 7.7~tons and contains a 
volume of 114.2~${\rm m}^3$ (without the chimney). 
A drawing of the Buffer is shown in Figure~\ref{fig:buffervessel3D}. The vessel will be made of a 
stainless steel structure covered by steel sheets 3~mm thick. The structure has been studied with
the finite element method of the CASTEM 2000 software~\cite{Castem2000}. 
The thickness constraint has been obtained in the critical case of dead load. 
Risk of buckling have also been considered. \\
The density of the Veto oil could be slightly different from the density of the Buffer oil. 
A mechanical simulation of the filling phase 
leads to an acceptable difference of~3$\%$. 
\begin{figure}[htbp]
\centering
\includegraphics[width=.7\textwidth]{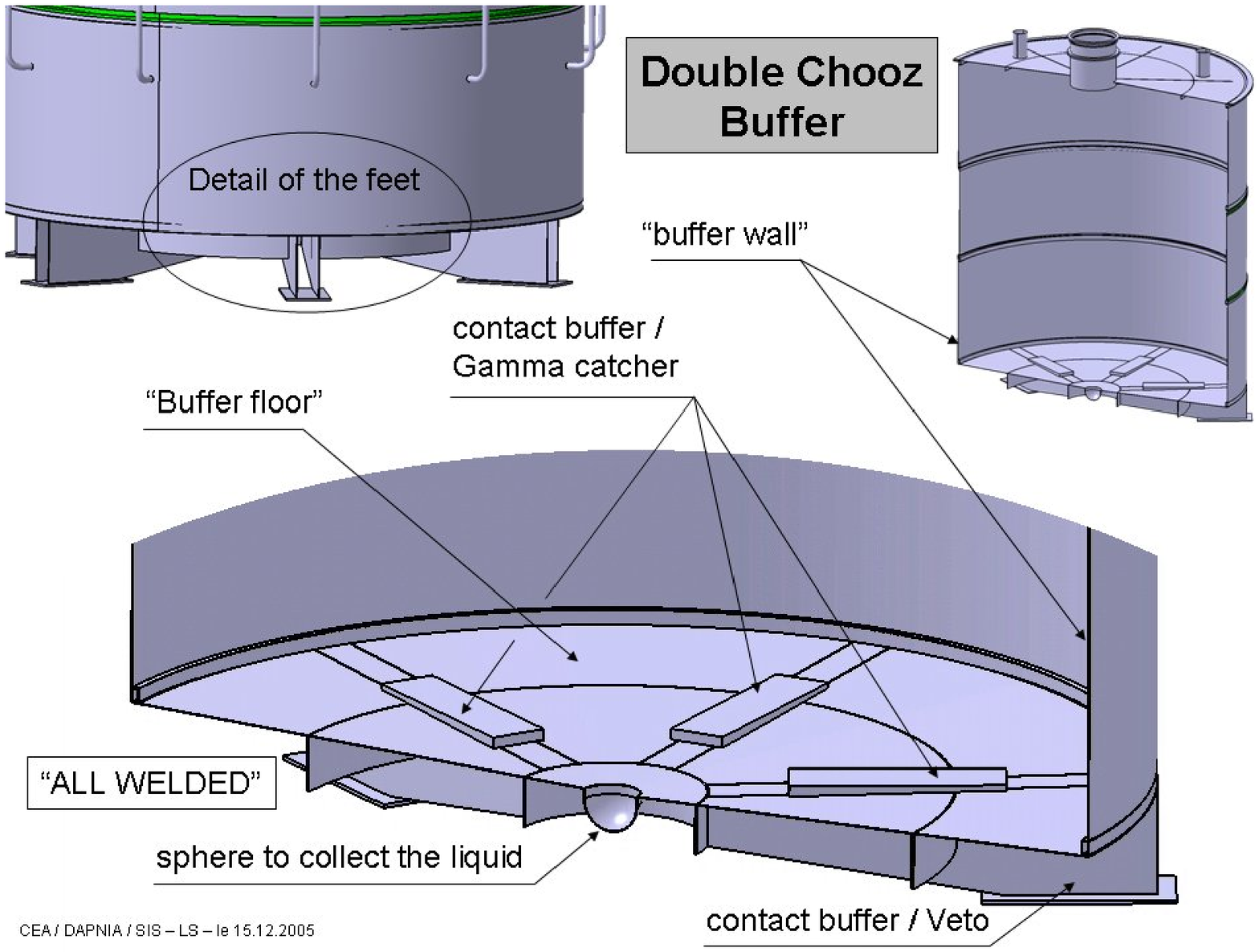}
\caption{Details of the bottom lid of the Buffer vessel that is supported by six feet.}
\label{fig:buffervessel}
\end{figure}
\par The Buffer vessel will be machined in several pieces by
 industry and transported 
to the neutrino laboratories. It consists of six half rings of stainless steel sheet, and two half 
bottom and top lids. All pieces will be pickled and passivated 
at the company. 
Half rings and lids will be initially welded beside the pit.
The Buffer vessel structure will be erected by welding 3 rings of stainless steel sheets in the pit.
Tubes for the inner photomultipliers cable paths will be welded afterward. Platform and dedicated tools
 will be developed and realized to facilitate the mounting as well as to guarantee a good cylindricity 
of the  vessel. Special care will be required because part of the welding will be done after the 
installation of the Tivek sheets (See Section~\ref{sec:innerveto_mechanics})
and phototubes. Leakage of welds will be systematically checked 
through the sweating method. The Buffer vessel will then be cleaned with
de-ionized water and weak nitric acid. 
Inside the Buffer we decided to keep the original stainless steel surface 
, providing a 40$\%$ reflectivity, as our baseline solution (see section~\ref{sec:photo}).

\par At the Far detector site, the integration constraints are given by the size of the access tunnel: 
about 3.5~meters of diameter 
(4,3~m width and 4~m high), 
the height under the crane of 3.5~meters, the crane capacity of 5~tons.
In addition, working space available in front and behind 
the pit is limited. At the Near detector site, we don't foresee more
stringent space constraints.
\subsection{PMT supporting structure}
\begin{figure}[htbp]
\centering
\includegraphics[width=.7\textwidth]{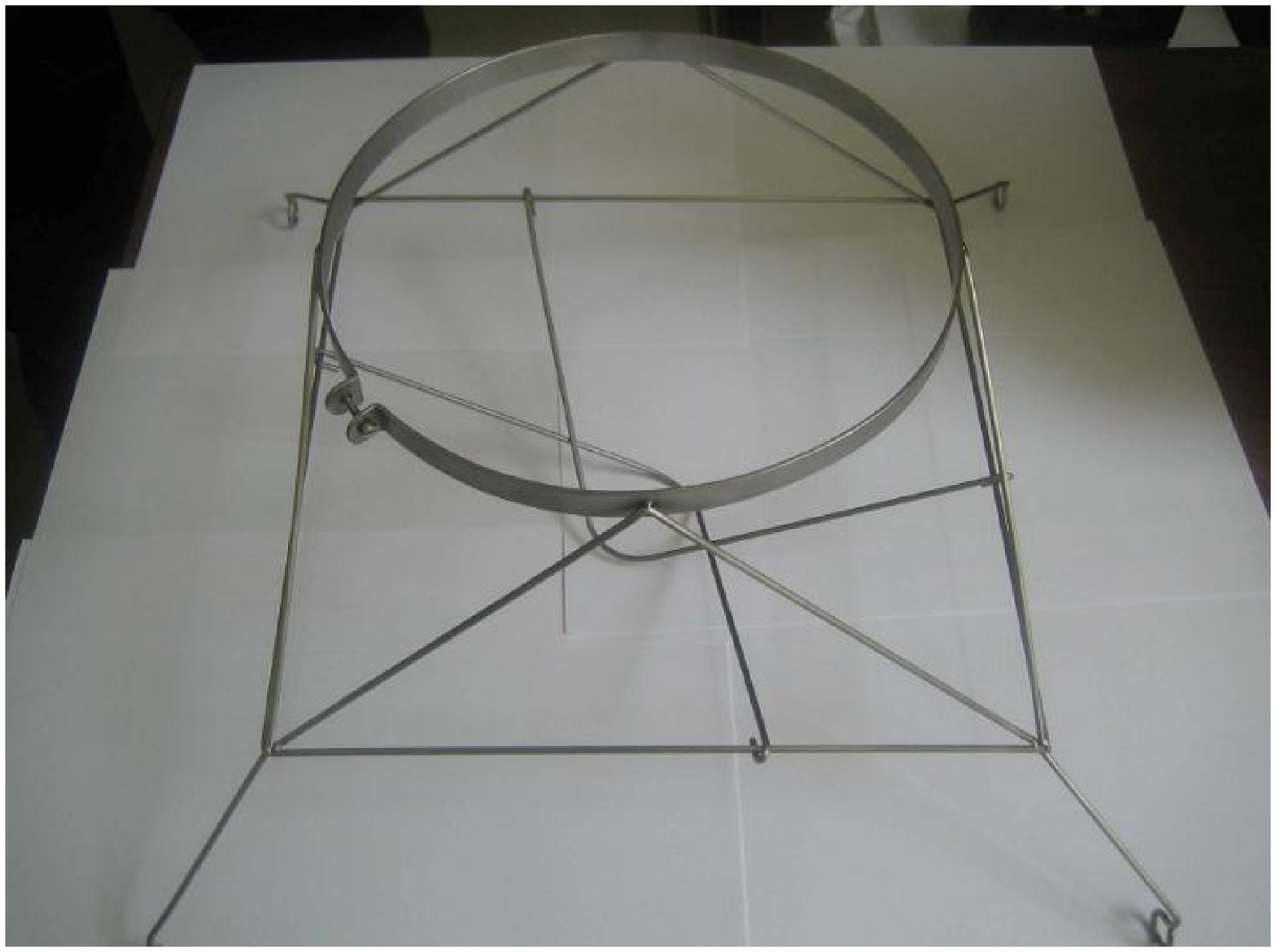}
\caption{Mockup of the PMT supporting structure done at the CIEMAT.}
\label{fig:pmtstructure}
\end{figure}

Each photomultiplier will have an independent mechanical support. As shown later (see
section~\ref{sec:514}, and Figure~\ref{PMTsupport}, we are planning to use light weight 
mountings similar to the ones developed for the Mini-Boone experiment. Base material of these 
mechanical supports is stainless steel wire of 2.5 mm diameter (see Figure~\ref{fig:pmtstructure}).

The inner face of the stainless steel buffer should be prepared to hold in place, in
their assigned locations, the set of photomultipliers. The location will be defined by
bar ribs, welded to the buffer vessel inner wall during the assembly process. The
photomultiplier mechanical supports will be simply bolted to the ribs, and the
photomultiplier location may be surveyed after mounting.   
\subsection{Inner Veto Mechanics}
\label{sec:innerveto_mechanics}
\subsection*{Materials and dimensions}
The Inner Veto is contained in the cylindrical setup of the detector. In case of the far detector, this means that the dimensions of the Inner Veto have to be chosen to fit in the existing pit in the far laboratory. The veto consists of a steel tank directly inside of the steel shielding, creating a cylindrical shell surrounding the Buffer vessel. The Buffer vessel serves at the same time as support structure for the inner PMTs. 
%
%
Horizontally, the veto region is defined by inner 
dimensions of the steel cylinder next to the steel 
shielding and the outer dimensions of the Buffer vessel. The 
resulting radial veto thickness is 50 cm between Buffer and shielding. 

A stainless steel support for the inner vessels consisting of six girders below the Buffer tank is mounted inside the veto which has an overall height of 50 cm in this region. Above the Buffer vessel, the gap between Buffer and lid is 60 cm. A central chimney with a radius of 15 cm is needed to allow access to the inner vessels at the top. Additionally, several pipes will cross the veto. These pipes are foreseen to feed the cables of the inner PMTs through the veto region up to the top lid of the Buffer.
The pipes are attached to the Buffer vessel in the lower third of the 
vessel. Muons that propagate along one of the pipes 
are potentially critical, because they might not be detected 
unless secondary particles enter the scintillator. For this 
reason, care has to be taken not to allow 
a path along one of the pipes without at least 
a part crossing the scintillator liquid.
\par The remaining space will be filled with 
a medium light-yield liquid scintillator based on mineral 
oil. All welds of the steel tank, the Buffer vessel and 
the pipes for cabling will have to be conducted in an oil-tight 
manner. A pipe to the bottom of the veto is used for 
filling. The overall volume to be filled with liquid 
is approximately 87 m$^{3}$ (90.5 m$^{3}$ minus PMT volumes).
\subsection{Steel Shielding}
\label{sec:steelshielding}
\par
\begin{figure}[htbp]
\centering
\includegraphics[width=\textwidth]{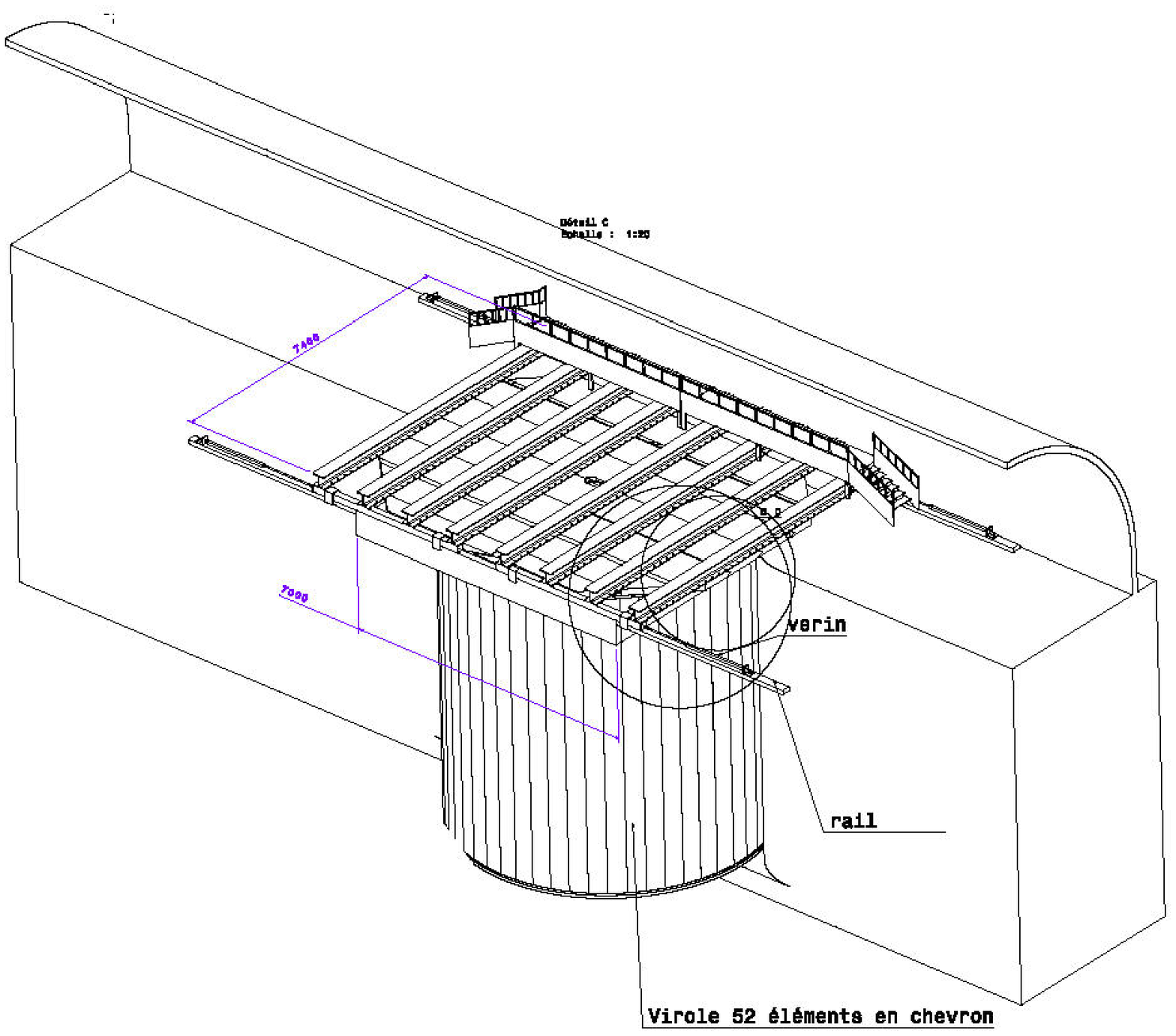}
\caption{3D overview of the Steel shielding. The top part of the steel shielding that opens 
in two halves to access the inner detector.}
\label{fig:steelshielding3D}
\end{figure}

 The detector outer vessel will be a steel tank surrounding the Veto region, 
500-600~mm away from the Buffer vessel. 
It is a cylinder of 6966~mm height, 6966~mm diameter 
(external dimensions), 
and 10~mm thick, weighing 20~tons.
This tank will contain all other liquids 
in case of any internal mechanical failure. It will be closed 
by a top lid coupled with a nitrogen blanket system to prevent
 oxygen contamination of the liquids. 
Outside this vessel a 170~mm thick low
 activity steel shielding will protect the detector from the natural 
radioactivity of the rocks around the pit (see 
Figure~\ref{fig:steelshielding3D}). 
The thickness was determined by a full detector simulation 
of the accidental backgrounds from 
the rocks (see section~\ref{sec:kbg_rocks}).

\par
\begin{figure}[htbp]
\centering
\includegraphics[width=.2\textwidth]{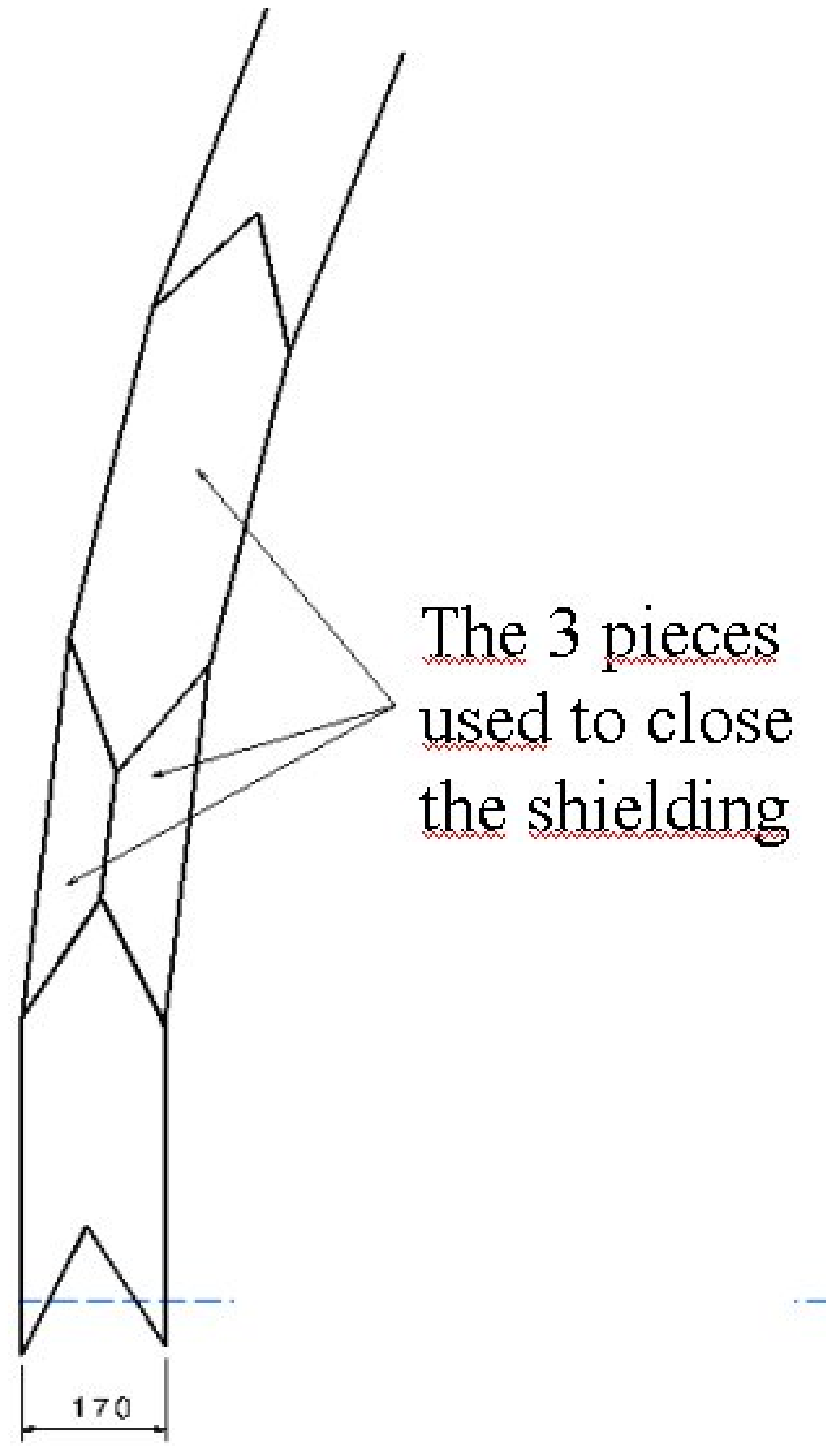}
\caption{Shape of the shielding bars to prevent gamma ray leaks through air gaps.}
\label{fig:steelshielding_60d}
\end{figure}

 The shielding will be made of steel S235JRG2 (old French name: E24-2; American name: A283C). 
Taking into account the limited size of the access tunnel and the limitations of the crane 
(see Figures~\ref{fig:detector_side} and ~\ref{fig:detector_top}), the shielding will be made of 
52~identical board-shaped pieces, plus 2~ additional pieces to close the system. 
%
%
Each side piece is a bar of 7000~mm long, 
400~mm wide and 170.2~mm thick (see Figure~\ref{fig:steelshielding3D}). 
The interface between neighboring pieces will be machined with a 60~degrees rafter shape 
(see~Figure~\ref{fig:steelshielding_60d}). This shape is very efficient to avoid gamma ray leaks through the 
shielding cracks between neighboring pieces. 
(The thickness "seen" by any gamma ray is always more than 
170~mm independent of the incident angle). 
The bottom part of the shielding will also be cut into pieces. 
The top shielding will be assembled in two halves.  A rail moving system 
will allow it to move to 
the sides of the pit, and thus to access the inner detector~(see Figure~\ref{fig:steelshielding3D}). 
%
%
\par The 10~mm thick steel tank integration is also constrained by the limited space in the access 
tunnel at the far site. The bottom of the tank will be made of 3 parts and the cylindrical side will 
be cut in 8 to 12 elements. All weldings will be done in the neutrino laboratories. 
The top part of the side tank will couple to a flange to fit with the lid. 
The internal surface of the tank will be coated with a highly reflecting white paint 
(currently being tested in the 1/5~ Double Chooz mock-up in Saclay). 
The cleaning procedure is currently being investigated 
(under test with the mock-up). 
A liquid recirculating system in the veto will be needed to have a good temperature
 homogeneity in the inner detector. The temperature difference between both detectors will
 be kept within 3 degrees. 
The top lid will be made of a permanently installed 
part housing PMTs cables feed through, and a central 
movable cylindrical part.
The lid inner surface 
will be painted with the same reflecting paint, whereas the external surface will be coated with an epoxy 
protecting paint. The lid will be fixed on the tank flange, and its tightness provided by a Teflon gasket. 
It will be equipped of 3 or 5 flanges (1 
at the center and one on each side to access 
$\gamma$-catcher  and Buffer). 
These flanges will be located on the closing line between the two shielding halves.
The near detector will be very similar to the far one. The differences would be in the shielding
(replacing iron by sand is under study) and in the elementary pieces definition 
(the handling constraints will be different). 
%
%

%
\subsection{Simulation of the detector geometry}
At present, the geometry of each detector assembly implemented in
DCGLG4sim is a slightly simplified version of the envisaged final
detector geometry. Work is underway to further improve the realism
of the simulation. All sub-volumes of the inner detector are perfect
coaxial cylinders located inside a pit, also cylindrical. The latter
is the only already existing part of the assembly, hence its dimension
are an important constraint. Table \ref{tab:geo_parameters} summarizes
the implemented geometry. 

\begin{table}
\caption{{\small \label{tab:geo_parameters}Geometrical parameters of the
simulated inner detector. All units in mm}}
\begin{center}{\small }\begin{tabular}{l|c|c|c|c|c}
{\small Region}&
{\small Radius}&
{\small Height}&
{\small Thickness}&
{\small Vert. translation}&
{\small Comment}\tabularnewline
&
{\small (mm)}&
{\small (mm)}&
{\small (mm)}&
{\small (mm)}&
\tabularnewline
\hline
\hline 
{\small Pit}&
{\small 3475}&
{\small 7000}&
{\small 0}&
{\small 0}&
{\small Already existing}\tabularnewline
\hline 
{\small Shielding}&
{\small 3475}&
{\small 7200}&
{\small 0}&
{\small + 100}&
{\small Top 200 mm above pit level}\tabularnewline
\hline 
{\small Veto}&
{\small 3265}&
{\small 6780}&
{\small 10}&
{\small 0}&
{\small 200 mm shielding + 10 mm tank}\tabularnewline
\hline 
{\small Veto chimney}&
{\small 155}&
{\small 500}&
&
&
{\small Same thickness as Veto}\tabularnewline
\hline 
{\small Buffer}&
{\small 2758}&
{\small 5674}&
{\small 3}&
{\small -50}&
{\small 100 mm thicker veto on top}\tabularnewline
\hline 
{\small Buffer chimney}&
{\small 150}&
{\small 1000}&
&
&
{\small Same thickness as Buffer}\tabularnewline
\hline 
{\small $\gamma$-catcher}&
{\small 1720}&
{\small 3574}&
{\small 12}&
{\small 0}&
{\small Vessel acts as buffer extension}\tabularnewline
\hline 
{\small $\gamma$-catcher chimney}&
{\small 113}&
{\small 1992}&
&
&
{\small Same thickness as $\gamma$-catcher}\tabularnewline
\hline 
{\small Target}&
{\small 1150}&
{\small 2458}&
{\small 8}&
{\small 0}&
{\small Fiducial volume = $10.0\,\textrm{m}^{3}$}\tabularnewline
\hline 
{\small Target chimney }&
{\small 75}&
{\small 2550}&
&
&
{\small Same thickness as target}\tabularnewline
\end{tabular}\end{center}
\end{table}

\begin{figure}
\begin{center}\includegraphics[%
  width=0.60\columnwidth,
  keepaspectratio]{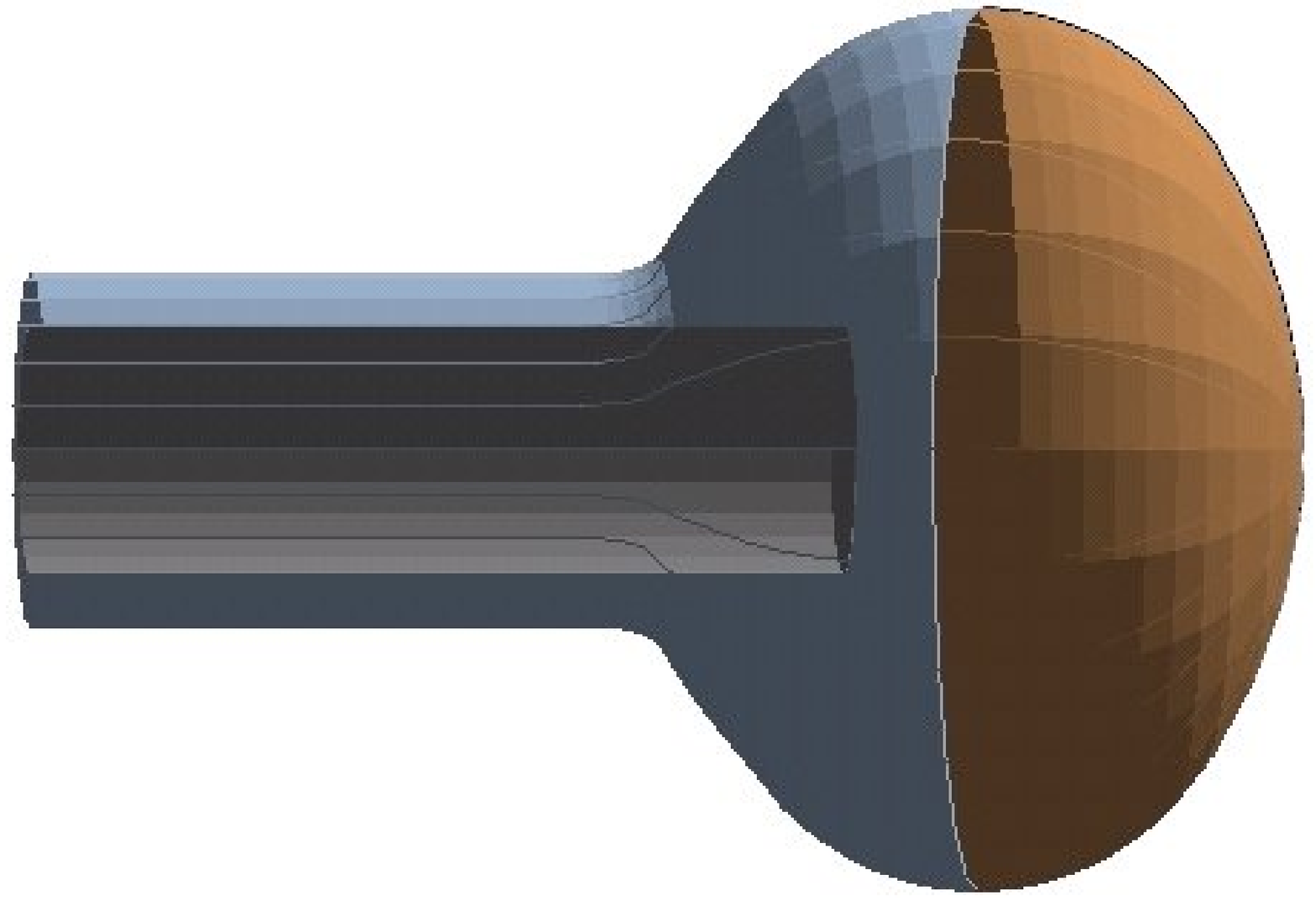}\end{center}

\caption{{\small \label{fig:Generic-PMT}Generic $8^{\prime\prime}$ PMT implemented
in DCGLG4sim. The shape can be controlled by the user by appropriate
choice of the relevant geometrical parameters.}}
\end{figure}

Some detector dimensions have slightly 
changed in the optimization process, 
but it does not change the results presented. A total of 
534 PMTs are arranged in 12 
rings of 30 PMTs on the lateral surface of the buffer tank, and in 5 rings of 30, 24, 18, 12 and 3
on the top and bottom caps. The geometry of each PMT is accurately
described within GLG4sim through the \emph{GLG4TorusStack} class,
which allows the definition of any generic shape (see Figure~\ref{fig:Generic-PMT}
for an example).

 The PMTs can be tilted at any angle, e.g. pointing toward the center
of the target (this configuration is preferred, as it assures the
best light output), or normal to the Buffer tank. Number and distribution
of the PMTs are not yet locked up for the design of the final detector
and the simulation will provide useful comparative data as regards
this issue.

\begin{figure}
\begin{center}\includegraphics[%
  width=\columnwidth,
  keepaspectratio]{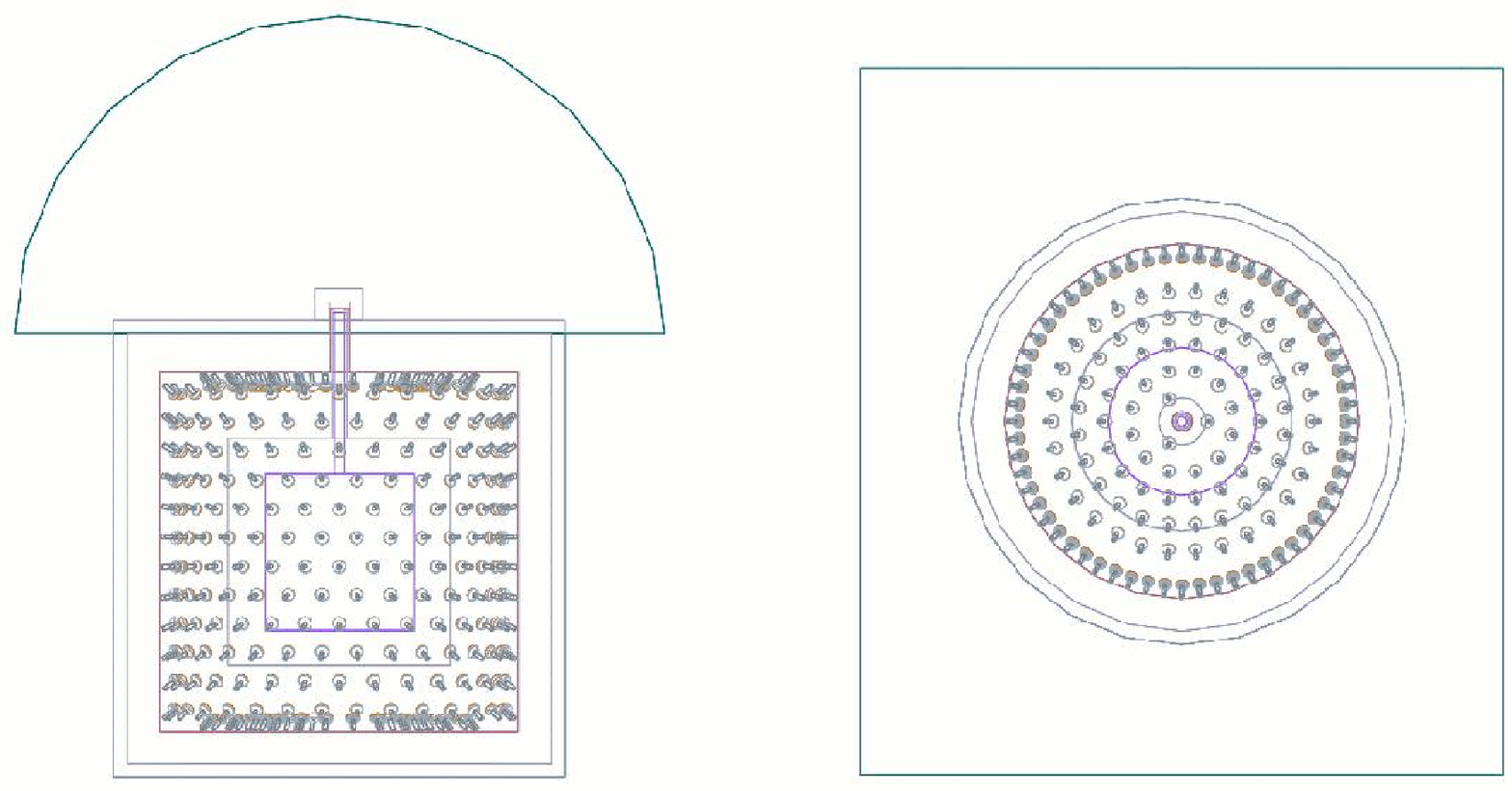}\end{center}

\caption{{\label{fig:detector_layout}Side and top view of the Double
Chooz detector layout as simulated with GEANT4.}}
\end{figure}

The detector layout as implemented in DCGLG4sim is 
shown in Figure~\ref{fig:detector_layout}.
As can be seen from the figure, the Monte Carlo representation of
the detector is already quite elaborate and faithful, including up
to a simplified schema of the chimney, described as a system of coaxial
cylinders -each one associated with each of the detector sub-volumes
and laid on top of it- with identical properties to the parent volume.

DCGLG4sim is conceived in such a way to retrieve all input parameters
of the geometry from external data-base files. This allows to readily
change the dimensions of detector sub-units and test the impact on
the performance.

\subsection{Simulation of the detector materials}
The volumes displayed in Figure~\ref{fig:detector_layout} are filled
with materials whose properties are already very close to those of
the final detector. The rock around the tunnel is defined as $\textrm{SiO}_{2}$,
with density $\rho=2.7\,\textrm{g/cm}^{3}$. The shielding material
and as well the Buffer tank are implemented as stainless steel, with
$\rho=7.87\,\textrm{g/cm}^{3}$ and elemental composition by mass:
71 \% Fe, 19 \% Cr, 10 \% Ni. The $\gamma$-catcher  and target vessels
are made of acrylics $\rho=1.14\,\textrm{g/l}$, a polymer whose basic
monomer is {[}$\textrm{-CH}_{2}-\textrm{CH}_{3}-\textrm{COOCH}_{3}-\textrm{C}-${]}.

The four different liquids filling respectively the Target, $\gamma$-catcher,
Buffer and Veto volumes have been implemented as mixtures of basic
components. GEANT4 does not keep track of the molecular composition
of materials defined as mixtures, as only the atomic abundances of
the final mix are believed to be of relevance. However, since most
of the properties of the resulting liquids do depend on the molecular
species present in their formulation, and their concentration, we
have introduced a sort of chemical approach to their definition, where
global variables store the complete information about the liquid composition.
This allows at the same time to keep all the detector liquids flexible
and user-configured, and to calculate at run-time the optical properties
of the detector. A micro-physical approach is implemented, based on
the known/measured properties of the basic components and the underlying
optical model of the system (see section~\ref{sec:opticalsimulation}). A comprehensive set
of primary materials has been defined, which are classified as \emph{aromatics}
(PXE, PC, ...), \emph{oils} (dodecane, mineral oil, ...), \emph{fluors}
(PPO, BPO, ...), \emph{wavelength-shifters} (bis-MSB, ...), and \emph{Gd-Compounds}
(Gd-dpm, Gd-acac, Gd-carboxylates). The Monte Carlo user is free to
build a detector with any combination of the above ingredients, by
specifying content and respective concentrations. The densities are
calculated accordingly, as well as the resulting optical properties. 
In Table \ref{tab:liquids_composition}
the typical liquids simulated are reported. Minor modifications are
expected with respect to the final design of Double Chooz.%
\begin{table}

\caption{{\small \label{tab:liquids_composition}Typical composition of the
Double Chooz liquids implemented in the Monte Carlo. 
$C_{12}H|{24}$ stands for Dodecane, and $C_{16}H_{18}$ stands for PXE.}}

\begin{center}{\small }\begin{tabular}{c|c|c|c|c|c||c}
&
{\small Aromatic}&
{\small Oil}&
{\small Fluor}&
{\small WLS}&
{\small Gd-Compound}&
{\small Density}\tabularnewline
\hline
\hline 
{\small Target}&
{\small $C_{16}H_{18}$ (20 \%)}&
{\small $C_{12}H_{24}$ (80 \%)}&
{\small PPO (6 g/l)}&
{\small bis-MSB (50 mg/l)}&
{\small Gd-dpm (1 g/l)}&
{\small 0.800}\tabularnewline
\hline 
{\small $\gamma$-catcher}&
{\small $C_{16}H_{18}$ (20 \%)}&
{\small $C_{12}H_{24}$ (80 \%)}&
{\small PPO (3 g/l)}&
{\small bis-MSB (50 mg/l)}&
{\small /}&
{\small 0.798}\tabularnewline
\hline 
{\small Buffer}&
{\small /}&
{\small Mineral Oil}&
{\small /}&
{\small /}&
{\small /}&
{\small 0.820}\tabularnewline
\hline 
{\small Veto}&
{\small /}&
{\small Mineral Oil}&
{\small PPO (6 g/l)}&
{\small bis-MSB (10 mg/l)}&
{\small /}&
{\small 0.821}\tabularnewline
\end{tabular}\end{center}
\end{table}
\subsection{Outer Veto}
Due to the shallow depth at the near detector, a high rate of muons is 
expected.  Since the primary background signal for this measurement will be
initiated by cosmic muons, an additional outer veto system is required.
This will be used to help identify muons which could cause neutrons or
other cosmogenic backgrounds and allow them to be eliminated from the data
set.  In some sense, the outer veto provides redundancy for the inner veto in
tagging background associated coincidences, but such a redundancy is crucial
to making a confident measurement of the background.  Comparison of a single
measurement with a full simulation would not provide such confidence 
because the cross sections for muon spallation products are not accurately 
measured.  In addition, the outer veto will provide three other benefits:
1) it will achieve a tracking resolution not possible with the inner
veto alone, allowing the differentiation between a muon which passed through
the target and one which merely passed near the target, 2) it will 
well measure those muons which only clip the corners of the inner veto.  
Such muons are especially 
dangerous because the inner veto efficiency will be low for these.
3)  It will provide sufficient granularity to detect high multiplicities
that would allow identification of high-energy showering muons.  Evidence
from KamLAND suggests that such muons have a high-likelihood of creating
the long-lived radioactive isotopes such as $^9$Li or $^8$He.
The overall goal is to provide a system with greater than 99\% 
efficiency that covers sufficient area to reduce the untagged flux of
neutrons to less than 0.005 Hz.  Reduction to this level should ensure that
the correlated background rate is well bellow 1\% of the signal.

\subsubsection*{Module Design}

The Outer Veto system design is based on the use of individual modules of 
gas filled long-tube wire proportional chambers.  Each chamber will be 
constructed from 2-inch diameter aluminum tubes with gas tight PVC end plugs
which also serve to hold a tensioned gold-tungsten wire which is strung down 
the length of the tube.  The modules will be 
constructed in a 8 wide by 3 deep close-packed configuration.   A gas system
will provide a low rate continuous flow of Ar/CO$_2$.  Electronic connection 
to the module will be made via two threaded brass posts for each individual 
chamber (tube): one in contact with the crimped wire and the other providing 
the return current connection to the outer tube.

\subsubsection*{Implementation at Near and Far Detectors}

The modules mentioned above will be assembled into square, flat panels by 
placing modules 
side-by-side and then placing two of these groups together orthogonally.  
At the near detector laboratory, the limited overburden will require a 
significant coverage to eliminate all angles of incoming muons.  As
a result, the intended design will use panels made up of 13 side-by-side 
5.3 meter long modules.  This will create a panel unit that is a 5.3 m square.
The top of the detector will be covered by 4 of the 5.3 m square panels.
Six more of these panels will be placed vertically in a hexagon around the side
of the detector.  The minimum radius of the hexagon will be 4.5 meters.
Since the outer wall of the inner veto system will be at 3.4 meters radius,
this allows approximately 1 meter of passive neutron absorber (foreseen to be 
low radioactivity sand) to be inserted between them.  

At the far detector, geometric constraints of the already existing laboratory
will not allow a similar design for the outer veto to be installed.  However,
the greater overburden reduces the rate of high angle muons to below our
need for identification.  As a result, a similar quality of rejection
can be achieved by constructing two 6.8m square panels which will cover
only the top of the main detector.  These panels will be constructed by
using 17 side-by-side 6.8 meter long modules.  

\subsection{Radiopurity Assessment}
\begin{figure}
\begin{center}
\psfig{file=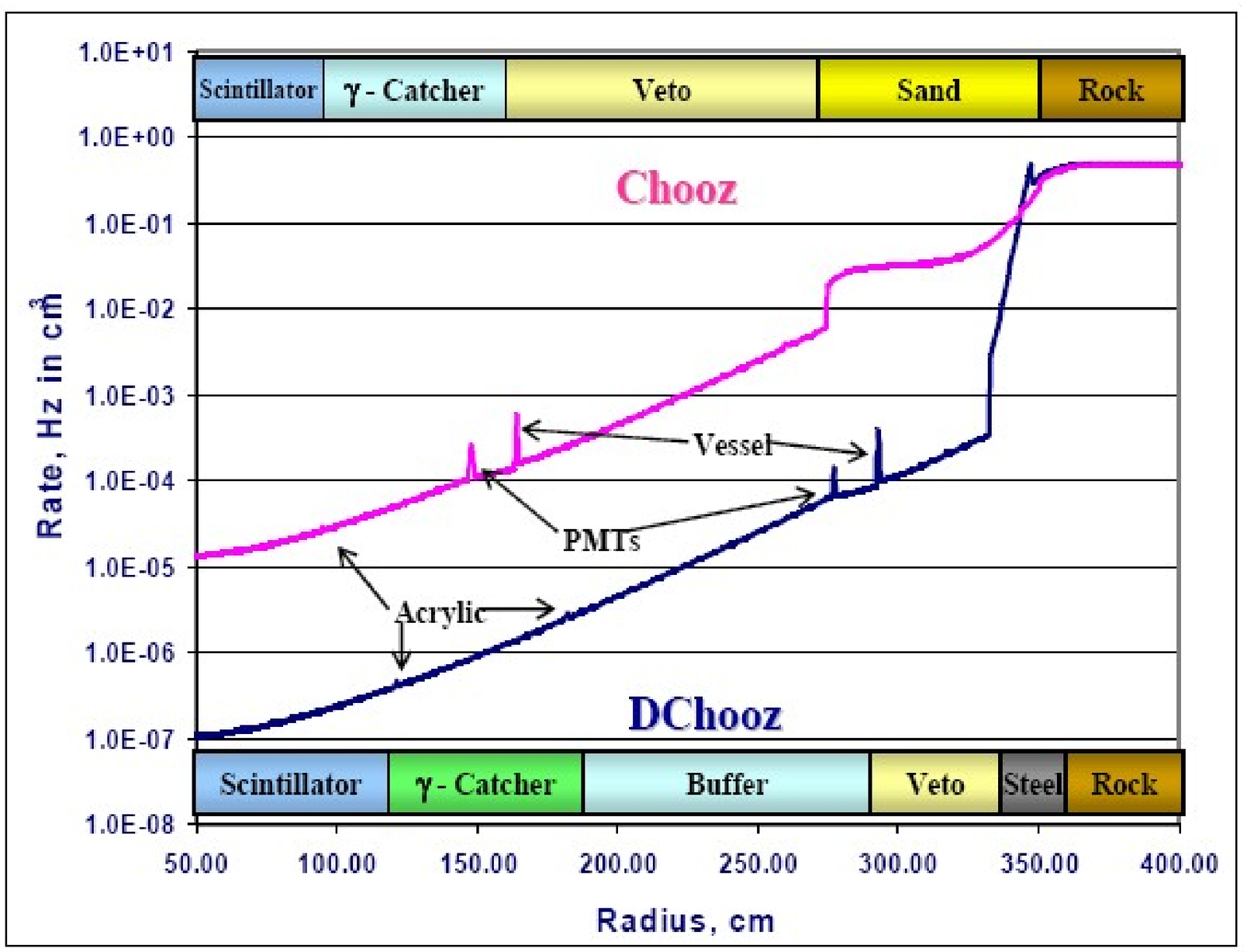,width=0.5\textwidth,clip}
\caption{\label{fig:singles}Comparison of the predicted singles rates in CHOOZ and
Double Chooz. The figure illustrate a simple model where both CHOOZ and Double Chooz
are considered to be spherical instead of Cylindrical. The single rate comes only from
gamma, which flux is analytically computed according to radiopurity concentration
of the different region and materials. The main differences between CHOOZ and Double Chooz
 arise with the addition of a non scintillating Buffer close to the PMTS, and the replacement
of 70 cm of sand by 17 cm of steel to shield the detector from external gammas.}
\end{center}
\end{figure}

%
%
The CHOOZ detector had a background singles rate of around 130~Hz, which became
65 Hz above 1.3~MeV after all data cuts. Couples with a ``neutrino-like''
event rate of 45~Hz, this meant that about 30\% of their correlated background
was due to random coincidences. We wish to improve upon that number for Double
Chooz, even though we have a larger detector by a factor of two.
To this end, we have designed a shielding plan that improves on the singles
rate in the target by more than an order of magnitude while using the same size detector
cavity. A calculation of the singles rates in CHOOZ and Double Chooz is
shown in Figure~\ref{fig:singles}.
This plot is based on a simple analytical model designed to calculate the rate of
single gamma hits for detectors with complex multi-layer geometries. Several
approximations are made for this estimate:
\begin{itemize}
\item To simplify only 1 MeV monoenergetic gammas are simulated. 
This simple simulation assumed 1 MeV gamma rays since these have the lowest attenuation
coefficient. Thus the results represent an upper bound on the number of gamma rays that
reach the inner layers of the detector. Furthermore, the mass attenuation coefficients
vary slowly in this energy region, meaning that the results will not be much different
for higher energy gamma rays of concern, e.g. 2.6 MeV.
\item Gamma production in materials is proportional to concentrations of
uranium, thorium and potassium.
\item Coefficients of proportionality were taken from measurements of gamma
production from PMT glasses with various levels of radioactivity.
\end{itemize}
The attenuation coefficients were checked by modeling the
KamLAND detector, where single rates are known very well.
This analytical model allows us to see the effects of the shield design
interactively, such that coarse optimization can be done rapidly and adjusted
later by detailed GEANT simulations.
Radioactive contaminants used for this model are given in Table~\ref{tab:singles}. 
They are typical values and not difficult to achieve
with reasonable care.
\begin{table}
\begin{center}
\begin{tabular}{l|l|l|l} \hline
 &                Th, ppb         & U, ppb          & K-40, ppb  \\ \hline
Target          & 0.000001        & 0.000001        & 0.000001  \\ \hline
Acrylic         & 0.01            & 0.01            & 0.01 \\ \hline
PMT             & 26              & 28              & 10 \\ \hline
Catcher & 0.000001        & 0.000001        & 0.000001 \\ \hline
Buffer          & 0.000001        & 0.000001        & 0.000001 \\ \hline
Veto            & 0.000001        & 0.000001        & 0.000001 \\ \hline
Steel           & 3.3             & 2.3             & 0.54 \\ \hline
Rock            & 5000            & 2000            & 1638 \\ \hline
\end{tabular}
\caption{\protect{\label{tab:singles}}
Assumptions for calculation of singles rates.}
\end{center}
\end{table}
Figure~\ref{fig:singles} shows the activity per unit volume starting from the
rock for both CHOOZ and Double Chooz. The largest difference comes from
replacing the sand with a steel shield, which drops the activity rate
considerably. Note that the PMTs have also been moved back from the target.
In the end, the predicted activity per unit volume is much lower than CHOOZ, almost two
orders of magnitude at 1 MeV.
\subsubsection*{Activity of Materials}
A survey of the literature has been done to get an idea of the radiopurity
levels available in the materials used in Double Chooz. Representative values
for each material are presented in Table~\ref{T:activities} (Note: measured
PMT activity is discussed in the photodetection section). These data do not
represent the total survey sample, but are included to show what variations
have been seen.
\begin{table}
\begin{center}
\begin{tabular}{r|r|r|l} \hline
U(ppb) & Th(ppb) & K(ppm) & source \\
\multicolumn{3}{c}{Steel} & \\ \hline
0.7 & 0.7 & 1.7 & KamLAND \\ \hline
0.3 & 0.7 & 0.85 & KamLAND \\ \hline
5.0 & 14. & 10. & SNO \\ \hline
\multicolumn{3}{c}{Acrylic} & \\ \hline
0.008 & 0.05 & 0.07 & KamLAND \\ \hline
0.004 & 0.008 & & SNO \\ \hline
0.001 & 0.001 & 0.00012 & Chooz \\ \hline
\multicolumn{3}{c}{Chooz Rock} & \\ \hline
14000. & 5000. & 2000. & Chooz \\ \hline
\end{tabular}
\caption{\label{T:activities}
Typical activities of various materials from a literature search.}
\end{center}
\end{table}
Cobalt-60 is also a radioactive component of the stainless steels. The constraint for the 
Double Chooz Buffer vessel is about 15 mBq/kg.
\subsubsection*{Check with a Detailed MC}
GEANT3 and GEANT4 based simulations have been built for estimating the
expected singles rate spectrum for the activities shown in Table~\ref{T:activities}.
In addition, an event generator has been written that
includes all the known gamma lines from 40-K, 208-Tl, and 214-Bi decay. These
are the major radioactive species that could produce events above our
threshold of around 0.7 MeV. Table~\ref{T:activities} shows the activity expected from
various detector components. The ``high'' and ``low'' designations refer to the highest
and lowest radiopurities found in the literature, which gives an idea of the spread of
possible values.\\
\begin{table}
\begin{center}
\begin{tabular}{l|c|c|c|c} \hline
 & \multicolumn{2}{c}{High} & \multicolumn{2}{c}{Low} \\
component & $> 0.5$ MeV & $>1.0$ MeV & $> 0.5$ MeV & $> 1.0$ MeV \\ \hline
PMT & 5.0 & 2.9 & 2.1 & 1.2 \\
target tank & 0.11 & 0.085 & 0.0034 & 0.0029 \\
GC tank & 0.17 & 0.12 & 0.0062 & 0.0037 \\
buffer tank & 1.8 & 0.93 & 0.67 & 0.25 \\
veto and shield & 1.7 & 1.0 & 0.11 & 0.063 \\
Chooz rock & 1.4 & 0.86 & 1.4  & 0.86 \\ \hline
total & 10. & 5.9 &  4.3 & 2.4 \\ \hline
\end{tabular}
\label{T:singlerate}
\caption{Expected rate (Hz) of single gammas above 0.5 and 1.0 MeV striking the scintillator. 
These results are based on
a GEANT3 simulation using the material activities given above, with high and low indicating the 
activity value used for the  simulation}
\end{center}
\end{table}
Thus we expect the background rate in Double Chooz at 1 MeV to be a factor of roughly twenty less than Chooz at 1.3 MeV.
Assuming a neutron-like event rate of 90 per hour (based on the Chooz experience), an average value for the internal activity,
and a 100 $\mu$s coincidence window,this translates into a random coincidence rate of about one per day at Double Chooz Far.\\
\subsubsection*{Radiopurity Monitoring}
To ensure that the ambitious sensitivity goal of the Double Chooz experiment
 is attained, great care must be taken to keep the background trigger rate at
or below the design level of a few Hz. This in turn requires that great care
 be taken in material selection, so that the intrinsic radioactivity of
detector components does not produce too high a background rate. We will
undertake a comprehensive radioactivity survey of the materials that will
be used in the Double Chooz antineutrino detectors to ensure that all
components will meet or exceed the required level of radiopurity. This
survey will also prove useful to other types of highly sensitive
experiments, including those searching for neutrino-less double beta
decay and dark matter candidates.
Activity measurements will be made at the Oroville Low Background Counting
Facility. This facility, operated by Lawrence Berkeley National Lab,
 is located in the powerhouse of the Oroville Dam in Northern California, under 180 m of 
rock cover (~600 m.w.e.). It has three Ge spectrometers: a 115\% n-type, an 80\% p-type,
and a 30\% p-type. Sensitivities of 50 parts-per trillion (PPT) for U and
daughters, 200 PPT for Th and daughters, and 100 parts-per-billion for K
are realized at the Oroville site.
Others measurements are being done in the Heidelberg (MPIK) low background counting facility, 
as well as in the very low background Gran Sasso Germanium detectors (scintillator and 
other liquid components, acrylics, \ldots).
The French underground laboratory of Modane is also providing access to its low 
background counting facility (Germanium detectors) for measurements of the inner 
materials of the detector, such as the acrylics and the stainless steel.

Sandia National Laboratories will manage this survey, collecting the results from the various 
groups building detector components, coordinating counting with LBNL, and analyzing and summarizing
the results. All materials to be used in the Double Chooz will have to be
validated in this way before being built into the detectors. 
%

%


%
\cleardoublepage
\section{Scintillator, Fluid Handling and Purification}
\label{sec:scintillator}
%
%

%
\subsection{Detector liquids and fluid handling}
In the Double Chooz detector design there are three separated
volumes filled with liquids inside the stainless steel buffer
tank. The inner volume is the $\bar{\nu}_e$-target and it will
contain 10.3~m$^3$ liquid scintillator loaded with gadolinium
(Gd-LS) at a concentration of approximately 1~g/$\ell$. The acrylic
vessel holding the target volume is surrounded by the
$\gamma$-catcher with a volume of 21.5~m$^3$. This volume is
filled with a Gd-free scintillator. Finally, there is a volume of
about 100~m$^3$ outside the $\gamma$-catcher containing a
non-scintillating buffer liquid. At the inner wall of this volume
the PMTs will be mounted. The densities of the liquids should be
similar in all of the three volumes in order to avoid strong buoyancy forces
in the detector.

It is foreseen to do the purification and the mixing of the liquid
scintillators off-site, both for the target region and as well as for the
$\gamma$-catcher. The scintillators will then be
shipped to the Chooz site in transport tanks. On-site the
tanks will be hooked up to dedicated filling systems. All three
volumes have to be filled simultaneously.

Material compatibility tests were made and will be made especially
for the Gd-LS. All materials in contact with the Gd-LS in the
detector or the liquid handling systems have to be checked
thoroughly. To test the material compatibility with acrylic a
mock up of the Double Chooz detectors was built (1/5~th scale). The
inner volume of this mock up contains about 110~$\ell$ of Gd-LS
(we did not tested the final scintillator recipe which was not yet available 
on the 100 liter scale). Smaller acrylic vessels with a capacity of 
approximately 30~$\ell$ are also available for long-term stability and 
material compatibility testing.

\subsection{Target}
Double Chooz will use a Gd-loaded liquid scintillator with 1~g/$\ell$
Gd-loading, as already pointed out.
The Gd-scintillator for both detectors should be
produced together as ``single batch'' to assure identical proton
per volume concentrations in both detectors, and to assure that if
there are any aging effects, they are more likely to be the same. 
As scintillator
solvent it is planned to use a PXE (phenyl-xylylethane)/dodecane
mixture at a volume ratio of 20:80. The admixture of the dodecane
reduces the light yield, but it improves the chemical
compatibility with the acrylic and increases the number of free
protons in the target. Besides technical requirements the solvent
mixture was selected due to safety considerations. In particular,
both components have high flash points (PXE: fp 145$^\circ$C,
dodecane: fp 74$^\circ$C). The scintillation yield of the unloaded
PXE based scintillator was measured as a function of dodecane
concentration. A scintillation yield of 78\% with respect to pure
PXE is observed at a volume fraction of 80\% dodecane. Pure
phenyldodecane or pseudocumene based mixtures could be used as
scintillator solvent as alternatives to the PXE/dodecane mixture.

Metal loading of liquid scintillators have been comprehensively
studied at MPIK and LNGS/INR for several years. Research with
gadolinium loaded scintillators in both institutes indicates that
suitable scintillators can be produced. Two scintillator
formulations have been investigated, one based on carboxylic acids and
the other on Gd-$\beta$-diketonates. Both systems show good
performance and are viable candidate liquid scintillators for the
$\bar{\nu}_e$-target.

\subsection*{Beta-diketonate (BDK) Gd-LS:}
The studies of the synthesis and properties of beta-diketonates of
rare elements and their relevant chemistry, especially stability at
high temperatures, is illustrated in References~\cite{Har92, Har85}. First
results of Gd-betadiketonate loaded liquid scintillators have been
reported in Reference~\cite{Mun}.  Promising results were already achieved
with the simplest beta-diketone molecule, acetylacetone (Hacac)
that was also used to produce a highly In-loaded scintillator
(5\%~wt. In) in the framework of the LENS (Low Energy Solar
Neutrino Spectroscopy) R\&D phase. With Gd(acac)$_3$ dissolved in
PXE high light yields and reasonable attenuation lengths were
achieved at a Gd-loading of 1~g/$\ell$. A Gd-loaded solution in pure
PXE was produced and successfully tested for stability for more
than 18 months.

However, Gd(acac)$_3$ does not dissolve at the required
concentration in the preferred solvent for Double Chooz
(PXE/dodecane at a ratio of 20/80) or in PC based solvents,
therefore other beta-diketone molecules were tested. The best
choice seems to be dipivaloylmethane (Hdpm). The Gd(dpm)$_3$
molecule is known to be more stable than the Gd(acac)$_3$ and can
be purified by sublimation.

The observed light yield for the Gd-BDK system corresponds to
approximately 80\% of the unloaded scintillator mixture at a
Gd-loading of 1~g/$\ell$. As primary and secondary fluor PPO (6~g/$\ell$)
and bis-MSB (approximately~20~mg/$\ell$) were used in these tests. The achieved
light yields are regarded to be sufficient for the requirements of
the Double Chooz experiment. Improvements concerning the light
yields appear to be possible by optimization of the fluor choice
and concentration. Attenuation lengths of more than 10~m were
measured for Gd(dpm)$_3$ solutions around 430~nm, the region of
the largest scintillator emission, at the required Gd-concentration of
1~g/$\ell$. Long-term stability tests with Gd(dpm)$_3$ based
scintillators gave promising results over several months. No
significant degradation was observed at room temperature for the
time period of the measurement of more than ten months.

About 500~g of Gd(dpm)$_3$ corresponding to more than 100~g of Gd
were produced at MPIK in an optimized synthesis. This batch
will be utilized to do long-term stability and material compatibility
tests of the full scintillator mixture on the 100~$\ell$ scale.

\subsection*{Carboxylate (CBX) Gd-LS:}
The chemical preparation of Gd loaded carboxylic acid based
scintillators (single acid, pH controlled) has been established
and demonstrated to be sound \cite{Cat1,Cat2,Dan,Mun2}. The Gd
scintillator can be synthesized by adding a crystalline material
or by a solvent extraction technique in a two phase system. Proper
control of pH during the synthesis is crucial.

\begin{figure}
\begin{center}
\includegraphics[angle=0, width=0.6\textwidth]{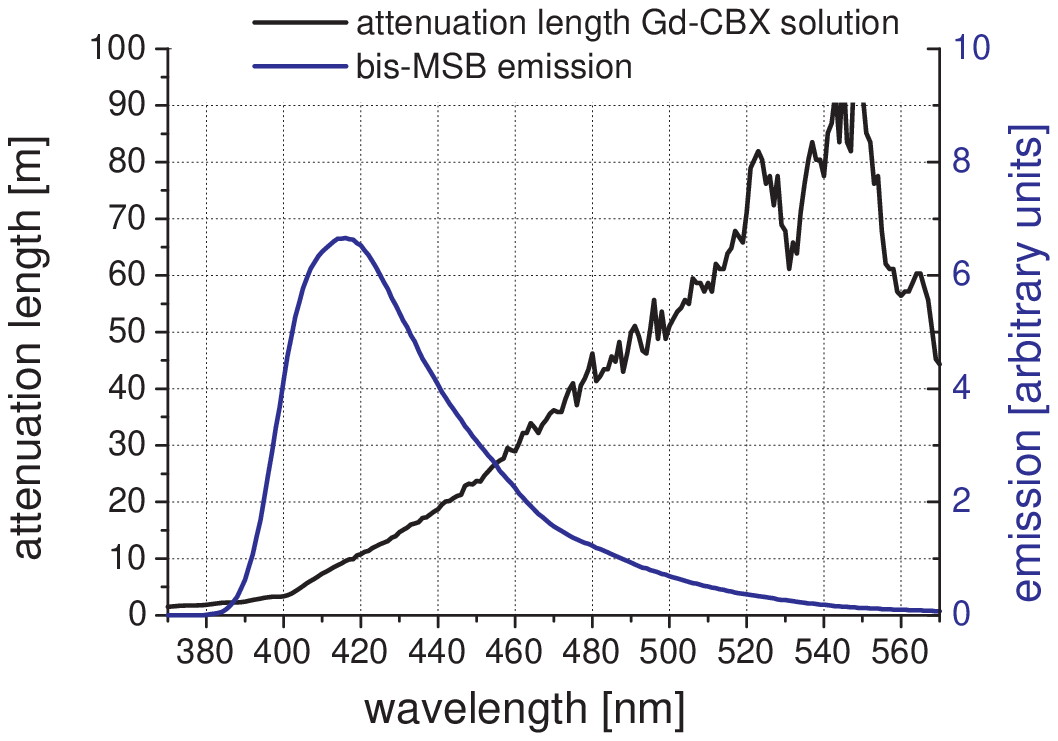}
\caption[CBX]{\label{CBXatt}
\it The plot shows the wavelength dependence of the
attenuation length of a Gd-carboxylate scintillator (1~g/$\ell$ Gd)
without wavelength shifters along with the emission spectrum of 
scintillator containing bis-MSB as secondary wavelength shifter.}
\end{center}
\end{figure}

A variety of Gd-CBX scintillators have been produced, varying the
Carbon number of the acid and using several solvent mixtures. The
Gd-CBX systems have excellent optical properties. The light yields
are above 80\% compared to the unloaded scintillator version and
attenuation lengths of about 10~m at 430~nm in the full
scintillator mixture were achieved. Figure \ref{CBXatt} shows the
attenuation length of a Gd-CBX solution before the addition of
fluors. The fluor absorption is not as critical as the absorbency
of the other components since the light can be re-emitted at higher
wavelengths and is therefore not necessarily lost. Along with the
attenuation length the scintillator emission is plotted in the
same figure. It is essentially the emission spectrum of the
secondary wavelength shifter bis-MSB. The spectrum was measured in
a triangular cell using a geometry that includes short distance
self-absorption and reemission of photons.

Long-term stability tests at room temperature and at elevated
temperatures of various CBX-versions were performed for test
periods of several months up to close to two years. Some
degradation in the attenuation length was observed for the
CBX-versions without any additives on time-scales of a few months
and more. The changes were observed in the far UV moving into the
optical regime. Several ingredients (stabilizers) are under
investigation to prevent polymerization and oxidation in the
scintillator (both can affect the attenuation length of the
scintillator) and promising results were achieved using
organophosphorous compounds. The choice of stabilizers will be
optimized by additional long-term stability tests of concentrated
CBX-solutions ($>10$~g/$\ell$ Gd) with different additives.

The final scintillator can be tested in a 1/5~th scale mock up of
the Double Chooz detector. More than 100~$\ell$ of Gd-CBX-scintillator
were already produced at MPIK and filled into the acrylic inner
vessel of this detector mock up. Periodic tests of its optical
properties will be performed. The general purpose of this mock up
is to find technical solutions for the construction and
integration of the detectors. In particular, the material
compatibility of the scintillator with the acrylic as well as filling
procedures of the detector are tested. Furthermore purification
methods can be tested with this setup. Another Gd-CBX-scintillator
version was produced by LNGS/INR people and is exposed to the same
acrylic as in the mock up in a 30~$\ell$ vessel.

\subsection{Gamma-Catcher}
The 21.5~m$^3$ volume of the $\gamma$-catcher will also be filled with a
liquid scintillator, but without Gd-loading.
Its purpose is to detect gamma rays resulting from prompt positrons
and gammas from delayed neutron capture from neutrino interactions in
the target volume that escape. This scintillator is also enclosed in an acrylic
vessel. Similar requirements as for the target liquid concerning
material compatibility, density and optical properties have to be
applied.

The $\gamma$-catcher scintillator has to match the density of the
target scintillator which is about 0.80 within 1\%. Since the
Gd-loading does not affect the density of the liquid significantly
the same PXE/dodecane ratio in the $\gamma$-catcher as in the
target would fulfill the density requirements.

\begin{figure}
\begin{center}
\includegraphics[angle=0, width=0.6\textwidth]{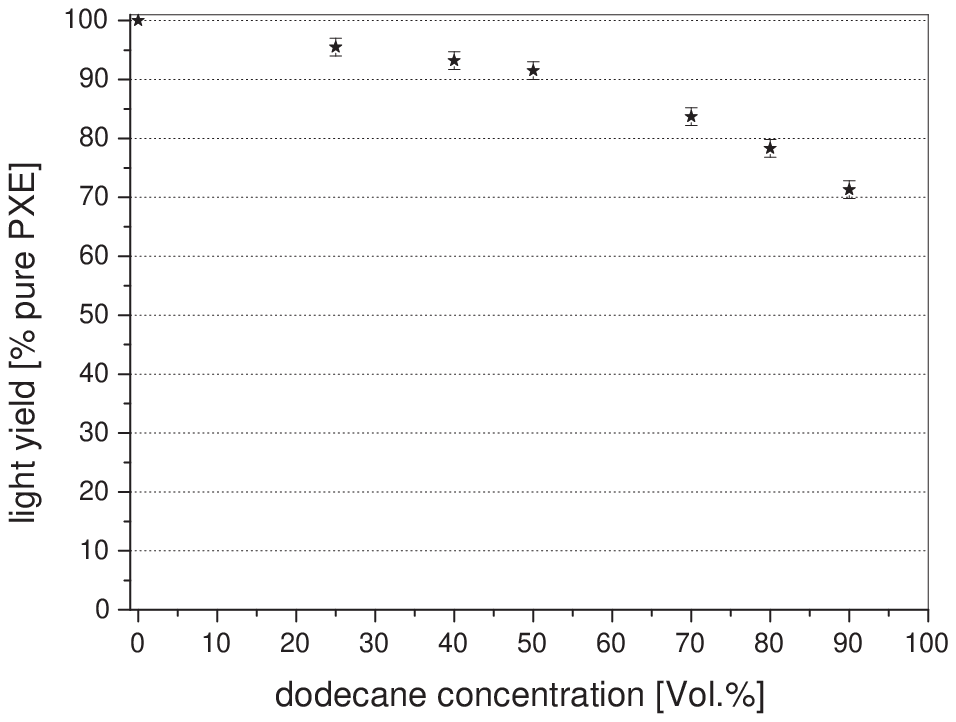}
\caption[LY]{\label{Lightyield}
\it Scintillation light yield of PXE/dodecane mixture
with varying dodecane concentration. The PPO concentration is kept
constant at 6~g/$\ell$.}
\end{center}
\end{figure}

Furthermore, the light yield of the $\gamma$-catcher scintillator
should be similar to the light yield of the Gd-scintillator. There
is some quenching in the Gd-scintillator compared to the unloaded
version and the light yield of the $\gamma$-catcher scintillator
would therefore be $10-20$\% higher in the target. One option
to adjust the light yield is to lower the aromatic fraction in the
$\gamma$-catcher. The dependence of the light yield on the
PXE-concentration is shown in Figure \ref{Lightyield}. On the
other hand a change in the PXE/dodecane ratio implies a change in
the density of the scintillator. This effect can be adjusted by
replacing part of the dodecane by mineral oil having a higher
density. The attenuation length in the $\gamma$-catcher for the
wavelength region of interest is comparable or higher than in the
target and the stability of the optical properties is not viewed
as a problem, since this scintillator has no metal loading.

The $\gamma$-catcher volume of the mock up was filled with about
200~$\ell$ of liquid scintillator. In this test a composition was
chosen to match the density and at the same time the light yield
between the inner two volumes. The scintillator contains 10\%
PXE, 30\% mineral oil and 60\% dodecane. As in the target a
PPO/bis-MSB combination was chosen to shift the light into the
near UV and optical regions.

\subsection{Buffer}
For the buffer volume each detector needs about 100~m$^3$ of
non-scintillating liquid. This volume should shield the active
volume from gamma rays emitted e.g.~by the photomultipliers. Since
there is a thin acrylic vessel between the $\gamma$-catcher and
the buffer this liquid has to match the density of the
scintillators. Additional requirements are material compatibility
with acrylic and the photomultipliers and high transparency in the
wavelength region of the scintillator emission.

It is foreseen to use pure mineral oil in this region. The density
of the mineral oil used in the mock up is slightly too high for the
demands of the Double Chooz detector. Therefore optical pure
mineral oil with lower density has to be found or the mineral oil
has to be mixed with dodecane (density 0.749).

\subsection{Fluid handling and purification}
There will be a scintillator fluid system on-site at the reactor
area and off-site at German institutes. All the mixing
and purification of the liquids will be done off-site. On-site
there will be the filling systems and it is foreseen to have
enough storage area for all detector liquids. All three inner
volumes and maybe the muon inner veto as well have to be filled
simultaneously.

It is planned to purify and mix the target scintillator at MPIK
Heidelberg. The liquid handling system for the target scintillator
has to be made exclusively out of ``Teflon" like material. A prototype
system was already designed and built for the loading of the
mock up. It can be used for the loading, unloading, filtering,
mixing, storage and nitrogen purging of liquids. Additionally, a
column for purification of organic liquids can be attached to the
system. After the Gd-scintillator preparation off-site it will be
transported in dedicated tanks to the Chooz site. The procedure
for the $\gamma$-catcher will be similar. The volume of the
$\gamma$-catcher fits into one typical ISO-container.

On-site there will be filling systems and storage tanks. The
buffer liquid will be stored above ground in about four connected
steel tanks per detector. It would be advantageous to have the
ability to circulate and mix the target liquids in the detectors
to assure that the optical and chemical properties are identical.

\subsection{Material compatibility}
All three inner volumes, target, $\gamma$-catcher and buffer are
in contact with acrylic vessels, therefore the material
compatibility of the respective liquids with the acrylic used in
the experiment is crucial. The compatibility of several
PXE/dodecane mixtures and of both Gd-scintillator versions with
acrylic was tested at the company Roehm in Germany and in Saclay. These
measurements confirmed the improving compatibility with increasing
dodecane concentrations. The PXE/dodecane ratio of 20:80 by volume
mentioned above seems to be the best compromise providing
sufficient material compatibility and scintillator light yield.
The PXE concentration in the target scintillator could be lowered
to about 15\% to improve the safety margin, since the 
overall loss of light would not only be
a few percent (light output decreases, but transparency 
of oil is better than that of PXE). 
No significant difference concerning
compatibility with acrylic was observed between the
Gd-BDK-scintillator and the Gd-CBX-scintillator.

There are also other materials in contact with the scintillators
during the measurement (calibration system), handling and
purification. Steel surfaces should be avoided for the target
scintillator, especially if the Gd-CBX-scintillator would be used.
This system is expected to be surface reactive in contact with
steel and the stability of the scintillator could be affected. In
the production and synthesis of the 110~$\ell$ of Gd-CBX scintillator
for the mock up only PFA, PTFE, PP, PE and glass was used. Only
such materials should be used for the final Double Chooz systems
as well.

\subsection{Scintillator stability}

During the first Chooz experiment, the Gd-loaded scintillator used in
the $\nu$-target showed a fast and unexpected degradation of its
transparency, which has been ascribed to the oxidation by nitrate
ions\cite{bib:chooz}. In Double Chooz the long-term stability of the
target scintillator is of fundamental importance, either to assure
a sufficiently long running time (several years), and to avoid systematics
due to a possible different evolution of the liquids in the two detectors. 

The Heidelberg and Gran Sasso groups of the Double Chooz collaboration
have been producing Gd-loaded scintillator since 2003. A large number 
of samples have undergone extensive
monitoring of their chemical stability, in the laboratories,  where
these scintillators have been synthesized, and especially at Saclay. 

Most of the first R\&D samples have shown some worsening of their
optical clarity within time-scales of several months. These failures
have motivated a careful study of the underlying chemistry. In particular, 
the key role of several factors have been recognized:
the presence of free water in the final scintillator,
initial impurities in the base solvent (PXE), and the possible
polymerizations of the Gd-complex molecules. A successful R\&D on
these items has resulted in the production of a second generation
of Gd-loaded scintillators of remarkably improved optical properties
and chemical stability.

During December 2005 the technological 1/5 Double Chooz mock-up 
has been filled with $\sim$110~$\ell$ of Gd-loaded scintillator
(Gd concentration = 1 g/$\ell$), and $\sim$200~$\ell$ of unloaded scintillator
for the $\gamma$-catcher. Another $\sim$30~$\ell$ Gd-loaded scintillator
sample is being tested in a special tank, made of the same acrylic
selected for the mock-up.  Furthermore, a third formulation is going to be tested
by the same procedure.
The optical
monitoring of all these liquids will assess the long term stability
of the Double Chooz scintillators under experimental conditions that
are as close as possible to those of the full scale experiment. These
tests are therefore intended as a final validation of the Double Chooz
R\&D on scintillators.

\subsubsection{Methods}

The long-term stability of the scintillators developed for Double
Chooz has been investigated by means of spectro-photometric techniques.
The transmission of a collimated light beam through 10 cm of material
is routinely measured with a spectro-photometer. We are thus sensitive
to any chemical evolution of the scintillator leading to an increase
in the absorbency, i.e. to an optical degradation.

For any {}``fresh'' scintillator sample, a 10 cm quartz cell is
filled (V $\sim$100 ml) and the liquid is flushed with $N_{2}$ for
$15^{\prime}$ in order to purge oxygen, which is a potential danger as regards
the chemical stability of the scintillator. The cell is hermetically
sealed through air-tight stoppers and stored in darkness at room temperature.
The transmission T in the wavelength range $300\,\textrm{nm}<\lambda<800\,\textrm{nm}$
is routinely measured, approximately once a month. Since pure quartz
shows very low absorption in the optical wavelengths of interest (the
emission of the secondary fluor is centered around 400-500 nm), we
usually reference our instrument to the transmission T of the beam
in air, defined as T = 100\%.

The above spectro-photometric measurement can be used not only to
monitor the relative changes of the transmission of a sample, but
also to determine the absolute, wavelength-dependent attenuation length.
For the latter case, the effect of the light losses due to reflections
at the air-quartz-liquid and liquid-quartz-air interfaces must be
corrected. This is done by self-referencing the transmission spectra
to the response around $\lambda=600\,\textrm{nm}$. In this region
the transmission is a very flat function of $\lambda$ and it can
be safely assumed that light absorption is negligible (i.e. the absorption
length is infinite). To improve the accuracy of this absolute measurement,
the tiny absorption from the quartz windows is also corrected, by
zeroing the instrument with the response of the empty cell, instead
of air.

The same experimental procedure described above is repeated with a
twin sample stored at elevated temperature, typically $40^{\circ}$C.
This is intended as an accelerated aging test, under the hypothesis
that a change in the temperature influences just the dynamics of the
chemical reactions.
I.e. the chemical evolution at elevated temperature
is the same as that at room temperature, only faster. Experimental
and theoretical arguments exist in favor and disfavor of the above
interpretation of the high-temperature tests (see Section\ref{sec:473}).

Since the beginning of 2006 similar spectro-photometric tests are
performed on the liquids filling the 1/5 mock-up. Samples are taken
from the mock-up approximately every month. Their optical transmission
is measured and compared to that of reference samples taken during
the synthesis and the mock-up filling.

\subsubsection{Results}

\begin{figure}
\begin{center}
\includegraphics[width=.45\textwidth]{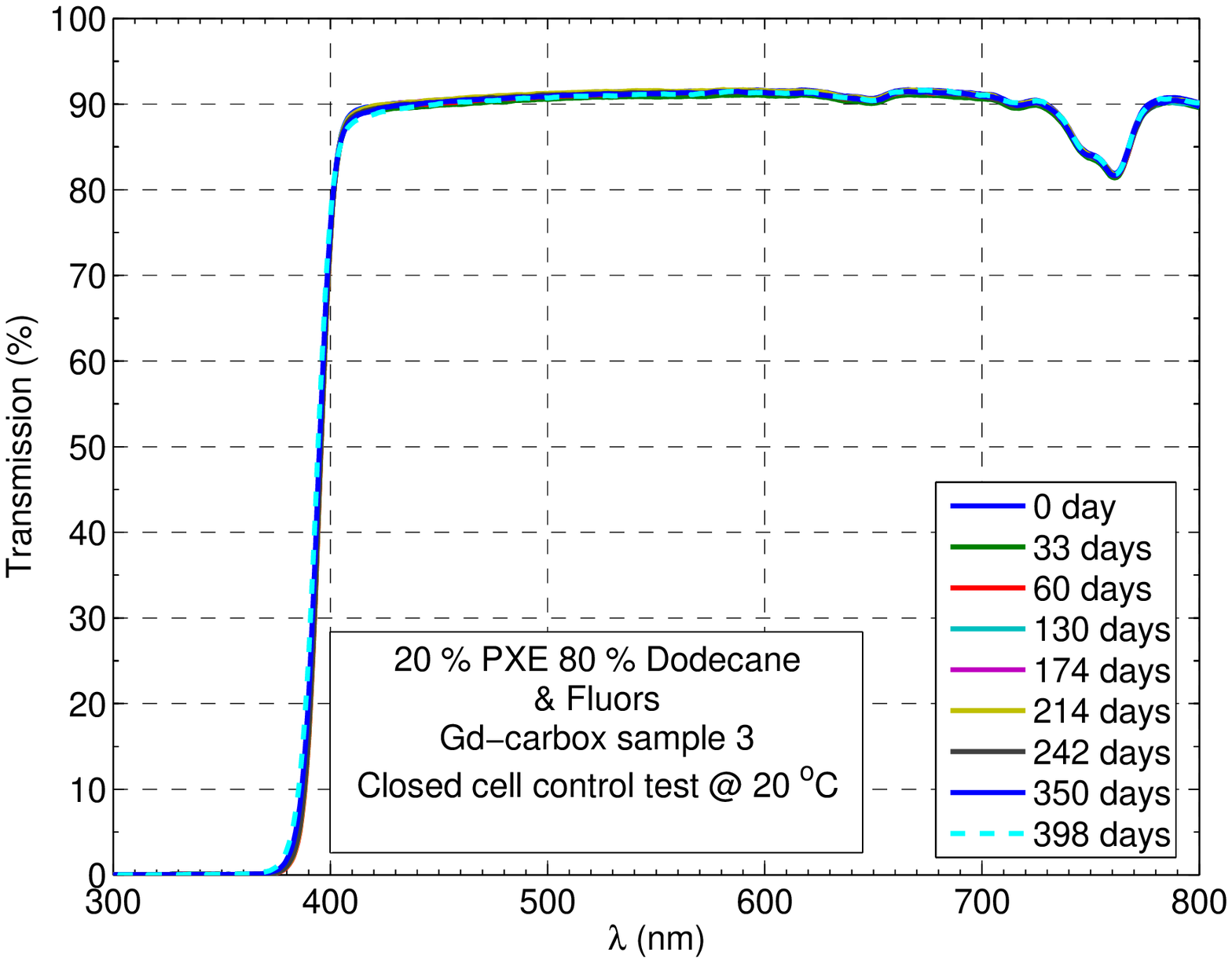}\hfill
\includegraphics[width=.45\textwidth]{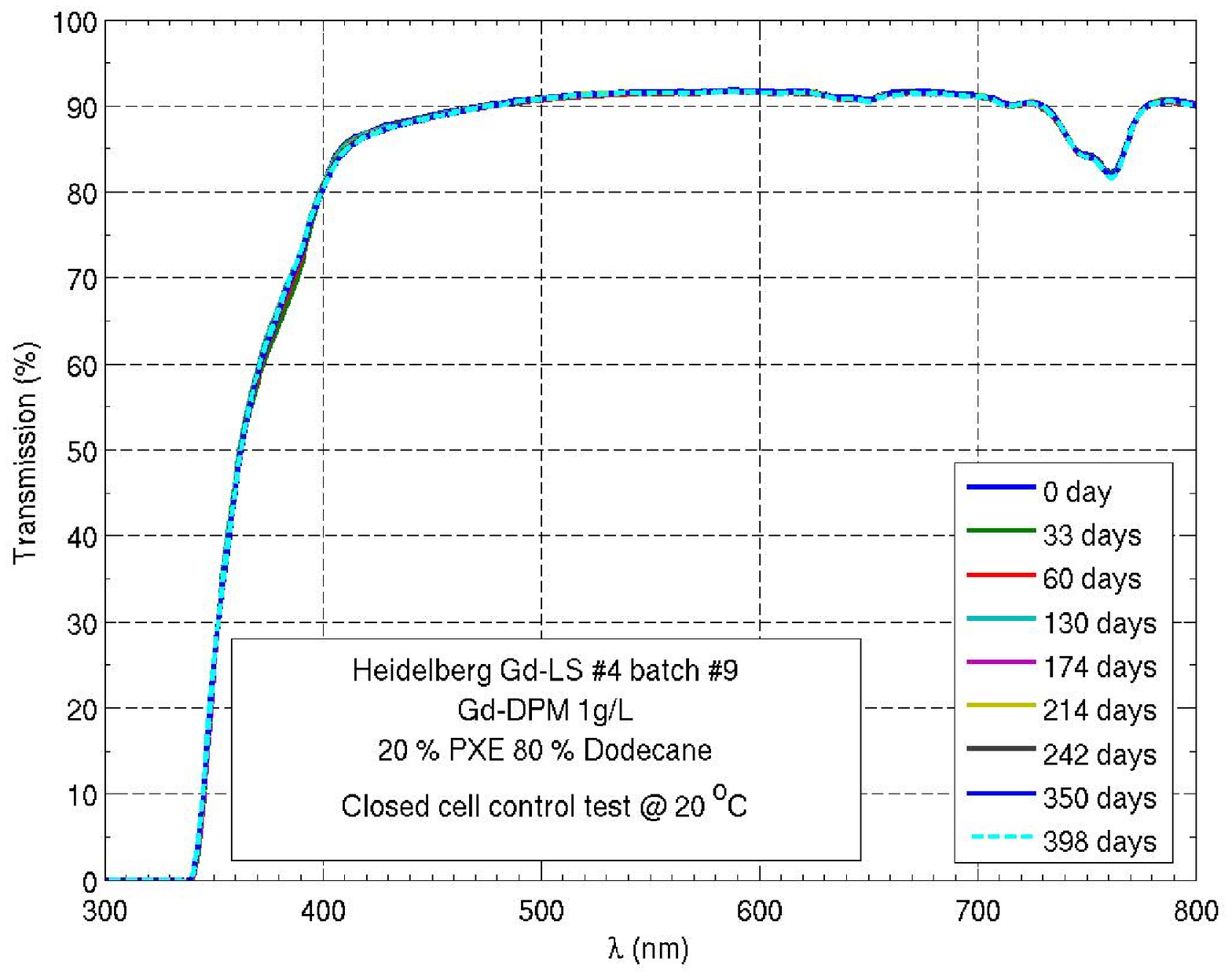}
\end{center}
\caption{{\small \label{fig:Hd41+Gs41 at 20C}Transmission as a function of
the wavelength for two R\&D samples stored at room temperature. Left:
Gd-dpm sample from Heidelberg. Right: Gd-cbx from Gran Sasso. For
both the common base is 20\% PXE + 80\% dodecane. 
The sharp cut-off at $\sim400\,\textrm{nm}$
for the Gd-cbx system is due to the fluors (the Gd-dpm sample shown
here has no fluors dissolved). Each curve corresponds to a scan at
a different elapsed time with respect to the cell filling, from 0
to 242 days.}}
\end{figure}

\begin{figure}
\begin{center}
\includegraphics[width=.5\textwidth]{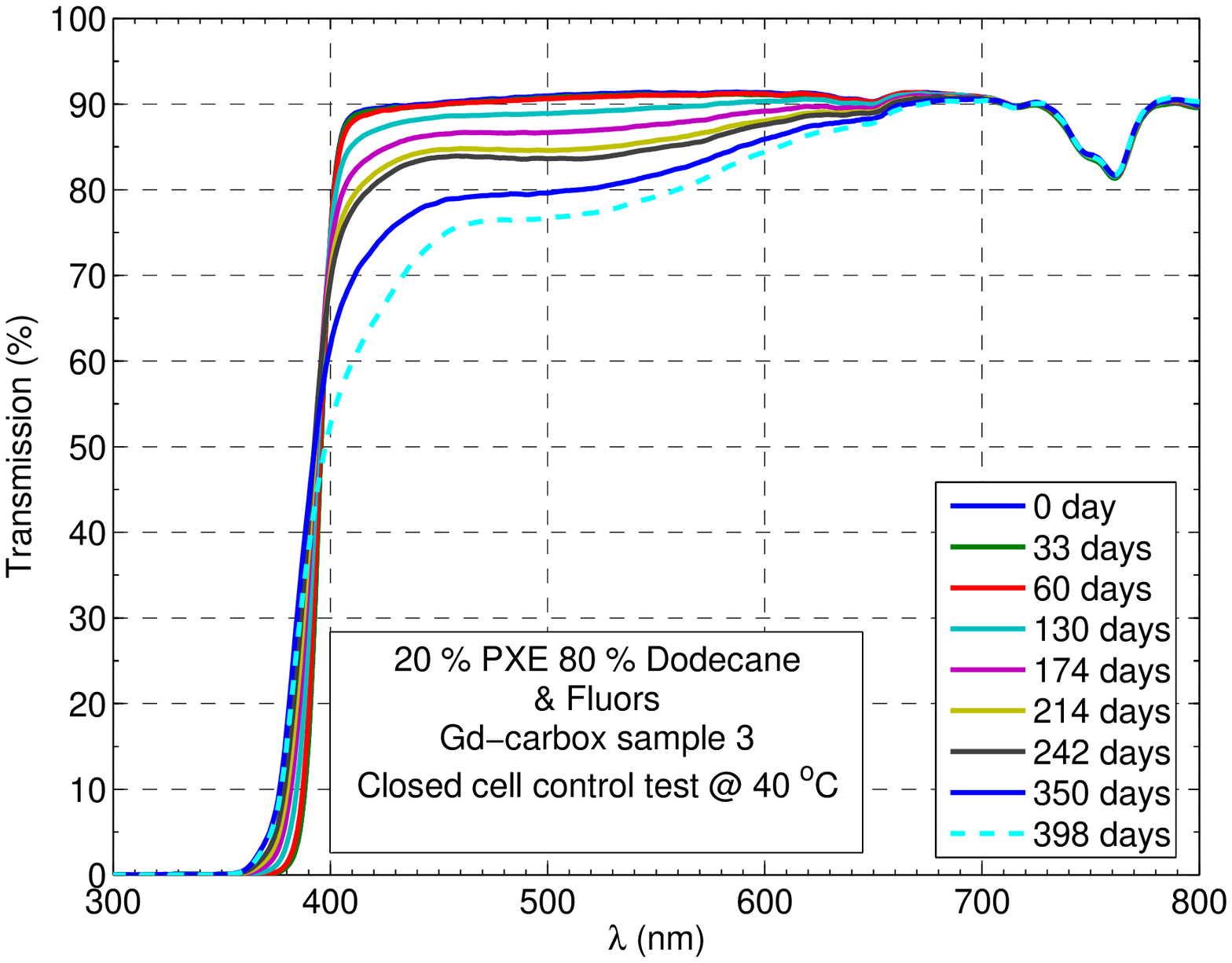}\hfill
\includegraphics[width=.5\textwidth]{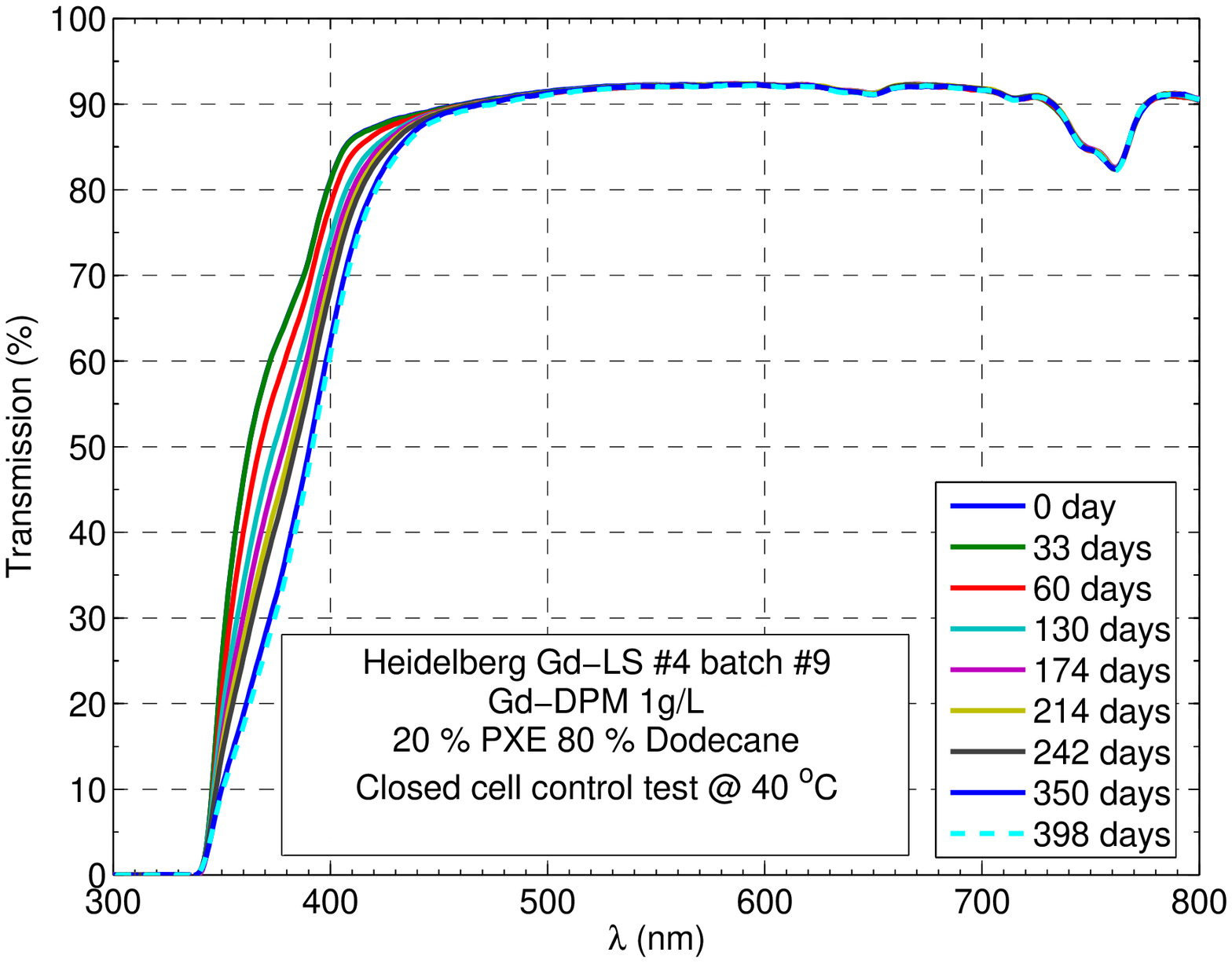}
\end{center}
\caption{{\small \label{fig:Hd42+Gs43 at 40C}Transmission as a function of
the wavelength for the same samples as Figure~\ref{fig:Hd41+Gs41 at 20C},
stored at $40^{\circ}$C. Each curve corresponds to a scan at a different
elapsed time with respect to the cell filling, from 0 to 242 days.}}
\end{figure}

Figure \ref{fig:Hd41+Gs41 at 20C} shows the optical survey of two R\&D
Gd-loaded scintillators synthesized in early 2005. Within the sensitivity
of the instrument, no degradation has been observed for $\sim$1~year
of data-taking.%
 This can be compared to Figure~\ref{fig:Hd42+Gs43 at 40C}, where two
samples from the same batches have been tested at $40^{\circ}$C.
Here a degradation is observed, particularly for the case of the Gd-cbx
system. In the next section this effect will be discussed.

\begin{figure}
\begin{center}\includegraphics[%
  width=0.6\columnwidth]{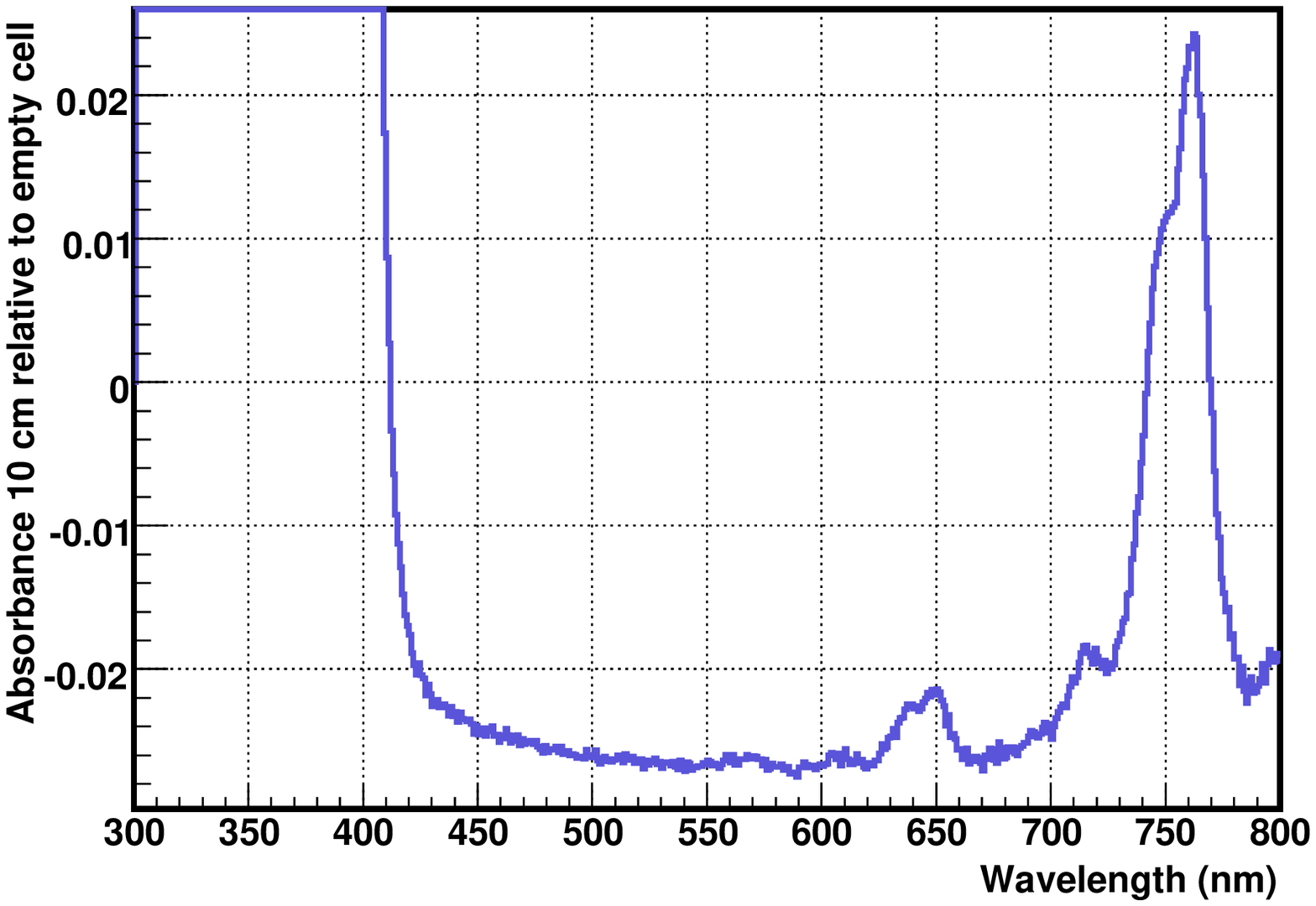}\end{center}

\caption{{\small \label{fig:target_mockup_TA}
Absorbency spectrum for the same liquid (reference
to the empty cell). Due to the uncorrected effect of Fresnel reflection
at surfaces, only the differences with respect to the value at $\lambda\sim600\,\textrm{nm}$
are considered significant.}}
\end{figure}

Regarding the survey of the mock-up liquids, to date only one sample
has been taken from the tanks and analyzed, which happened one
month after filling. The result, shown in 
Figure~\ref{fig:target_mockup_TA},
shows that the scintillator transparency is stable on this time scale

\subsubsection{Discussion}
\label{sec:473}
Spectrophotometry provides a very powerful and precise tool to investigate
the stability of our scintillators. The bin-to-bin error of each measurement
due to the noise of the photo-detector is in all cases negligible,
as it corresponds to a $\gtrsim100\,\textrm{m}$ equivalent attenuation
length (this can be further improved by smoothing the curves, since
one data-point is available per nm in wavelength). Moreover, the 
sample transmission scan over a large wavelength interval provides
diagnostic and self-referencing capabilities. In fact, it is important
to note that no optical degradation has ever been observed in the
wavelength region $\lambda\gtrsim650\,\textrm{nm}$. This gives a
handle to correct the small fluctuations in the offset of each scan
with respect to the others, which are believed to be due to surface
effects and instrumental systematics. This offset-correction is of
the order of a few tenths of percent, at most.

In the case of the first generation of R\&D Gd-loaded scintillators,
there has been a certain consistency between data at room- and elevated
temperature. Whenever an instability appeared at $40^{\circ}$C, it
showed up later with a slower dynamics at $20^{\circ}$C, as well.

On the other hand, it has been
shown that the best formulations developed during 2005 have shown
no degradation at all during $\sim$1~year survey time, while some worsening
of the transmission has been observed at elevated temperature. It
is not possible to correlate the results of the two tests by
defining some {}``acceleration factor''. In general, the Gd-carboxylate
scintillators have shown a worse performance at high temperature.
These systems, however, are expected to be more fragile with respect to
strong temperature variations, since synthesis is based on chemical
equilibrium. The parameters of the synthesis are tuned to drive this
equilibrium toward stability at room temperature, which does not
necessarily imply stability at elevated temperature. In the case of
the beta-diketonate system, instead, the degradation is believed to
be due to the accelerated reactivity of impurities left from the synthesis.
\subsubsection{Simulation: Optical properties}
\label{sec:opticalsimulation}
In Double Chooz -more than in other neutrino detectors - optics 
is a crucial issue. The chemicals required to bring Gd in solution
and stabilize the system have an influence on light generation and
transport processes well beyond the typical cases of binary solvent-fluor
systems. Furthermore, the chemical stability of the liquids are of
major concern and all conceivable scenarios of aging of the liquids
and drift of their optical properties must be carefully evaluated.
Finally, due to the multi-shell structure of the inner detector, where
each volume is filled with a different liquid and separated by acrylic,
processes at the interfaces may have a non-negligible influence on
the detector response. For these reasons, a considerable effort has
been devoted to carefully modeling the detector optics.
\subsubsection*{Scintillators and other liquids}
It has to be noted that the development of the Double Chooz scintillators
has been (and -to some extent- still is) an R\&D project and the best
state-of-the-art formulations have been subject to change. Consequently,
we have decided to build the optics using a micro-physical approach
(see Reference~\cite{bib:lens} for a detailed description of the model). Most
of the optical properties are derived by the actual liquid composition
selected for the simulation (see previous section), and a database
of input properties of the basic ingredients. This database includes
the experimental, wavelength-dependent extinction coefficients, fluorescence
spectra, fluorescence yields, etc. In this way, implementing new liquids
is trivial and moreover the Monte Carlo can provide feed-back for
the fine tuning of the scintillator formulation.

\begin{figure}
\begin{center}\includegraphics[%
  width=0.6\columnwidth,
  keepaspectratio]{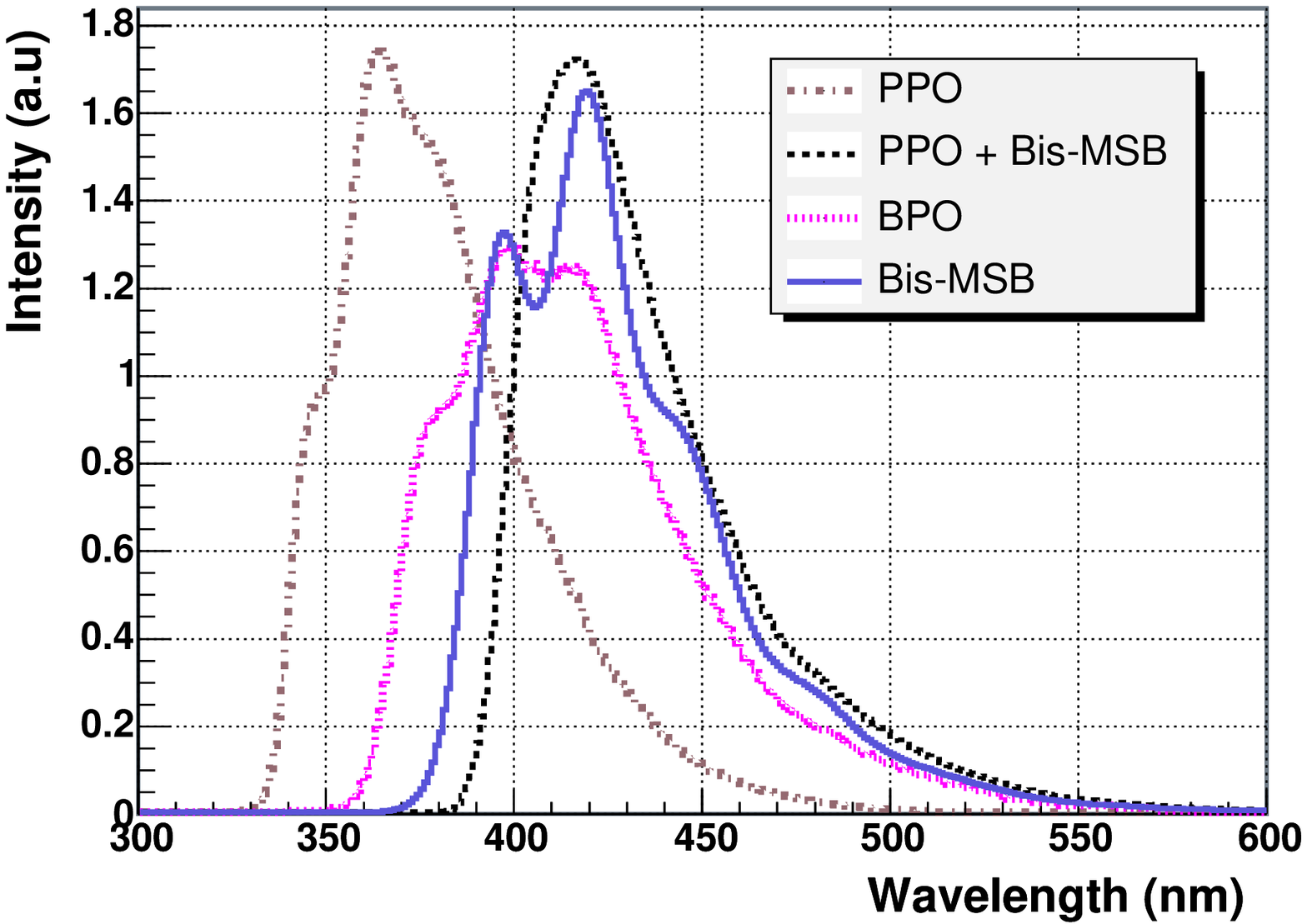}\end{center}

\caption{{\small \label{fig:Emission-spectra}Emission spectra of several
fluors. The primary emission spectrum typically used for the Double
Chooz Monte Carlo is the curve {}``PPO + Bis-MSB''. Re-emission
follows the bis-MSB curve (with the constraint $\lambda_{\textrm{new}}>\lambda_{\textrm{old}}$).
Data from MPIK-Heidelberg.}}
\end{figure}

In DCGLG4sim the choice of fluor and WLS determines the emission and
re-emission scintillation spectra (a-priori different, as opposed
to GEANT4). Figure~\ref{fig:Emission-spectra} shows some relevant examples.
In a system with PPO and bis-MSB, the energy transfer from the primary
to the secondary fluor is very efficient, since the emission spectrum
of PPO matches very well the absorption band of bis-MSB. The resulting
spectrum is essentially that of bis-MSB, peaked around 420 - 430 nm. 

The light yield (LY) is assigned as a free constant material property,
independently for each of the scintillating liquids. The LY of our
scintillators depends on many factors: aromatic/oil ratio, fluor concentration,
Gd-loading. Measurements have been performed in several laboratories of the
collaboration, usually by comparing the response of the investigated
sample to a known energy deposition with the response of a reference
scintillator, whose absolute LY is known. Typical values for our Gd-loaded
scintillators are $\textrm{LY~of}\sim6000-7000\,\textrm{photons/MeV}$. The
formulation of the unloaded $\gamma$-catcher scintillator is tuned to
reduce the LY and achieve a perfect match with the target scintillator.
For the Inner Veto liquid, we consider a 
$\textrm{LY~of}\sim5000\,\textrm{photons/MeV}$. 

\begin{figure}
\begin{center}\includegraphics[%
  width=0.6\columnwidth,
  keepaspectratio]{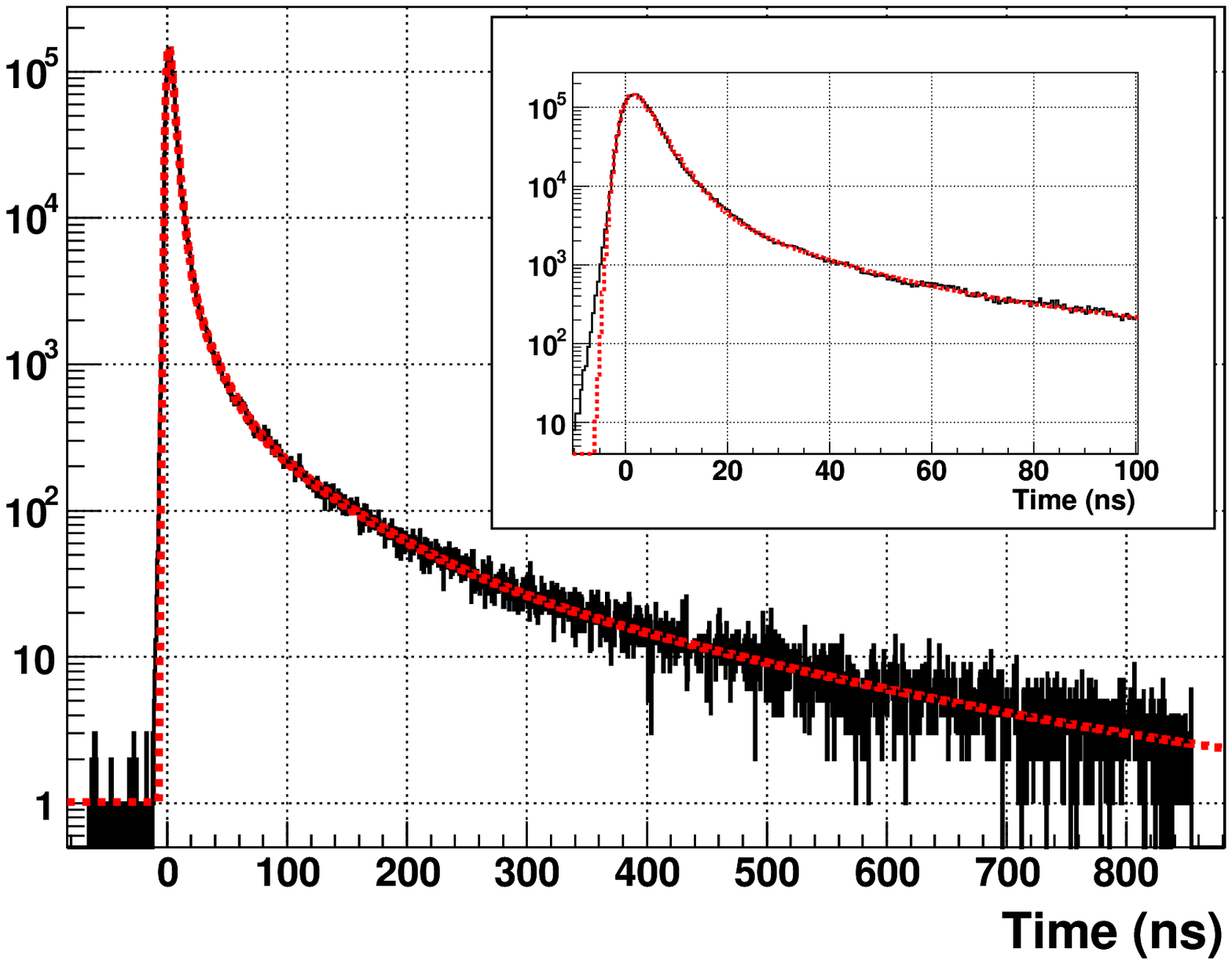}\end{center}

\caption{{\small \label{fig:decay-time-pdf}Experimental time profile for
photon emission after excitation with beta-particles of the PXE-based
scintillator loaded in the CTF of Borexino \cite{PXE_paper}. The
red line is a 4-exponential fit including a convolution with the response
function of the apparatus. The insert shows a magnification of the
first 100 ns. Details of the measurement and analysis in \cite{Tesi-PhD}.}}
\end{figure}

An emission time is assigned to each emitted photon. This is chosen
from a multi-exponential probability density function. So far, the
scintillation time of our scintillators has not been measured;
hence we are currently using the experimental response of pure PXE scintillators
(see Figure~\ref{fig:decay-time-pdf}). The decay constants are given
in Table \ref{tab:Decay-constants}.%
\begin{table}

\caption{{\small \label{tab:Decay-constants}Decay constants ($\tau_{i}$,
in ns) and weights ($q_{i}$, fraction of light emitted with decay constant
$\tau_{i}$)
of the time profile for photon emission after excitation
with beta and alpha particles. Data from Borexino for a PXE/p-Tp(3.0
g/$\ell$)/bis-MSB(20 mg/$\ell$) mixture \cite{PXE_paper}.}}

\begin{center}{\small }\begin{tabular}{c|c|c|c|c||c|c|c|c}
{\small Excitation}&
{\small $\tau_{1}$}&
{\small $\tau_{2}$}&
{\small $\tau_{3}$}&
{\small $\tau_{4}$}&
{\small $q_{1}$}&
{\small $q_{2}$}&
{\small $q_{3}$}&
{\small $q_{4}$}\tabularnewline
\hline
\hline 
{\small beta}&
{\small 3.1}&
{\small 12.4}&
{\small 57.1}&
{\small 185.0}&
{\small 0.788}&
{\small 0.117}&
{\small 0.070}&
{\small 0.025}\tabularnewline
\hline 
{\small alpha}&
{\small 3.1}&
{\small 13.4}&
{\small 56.2}&
{\small 231.6}&
{\small 0.588}&
{\small 0.180}&
{\small 0.157}&
{\small 0.075}\tabularnewline
\end{tabular}\end{center}
\end{table}

\begin{figure}
\begin{center}\includegraphics[%
  width=0.8\columnwidth,
  keepaspectratio]{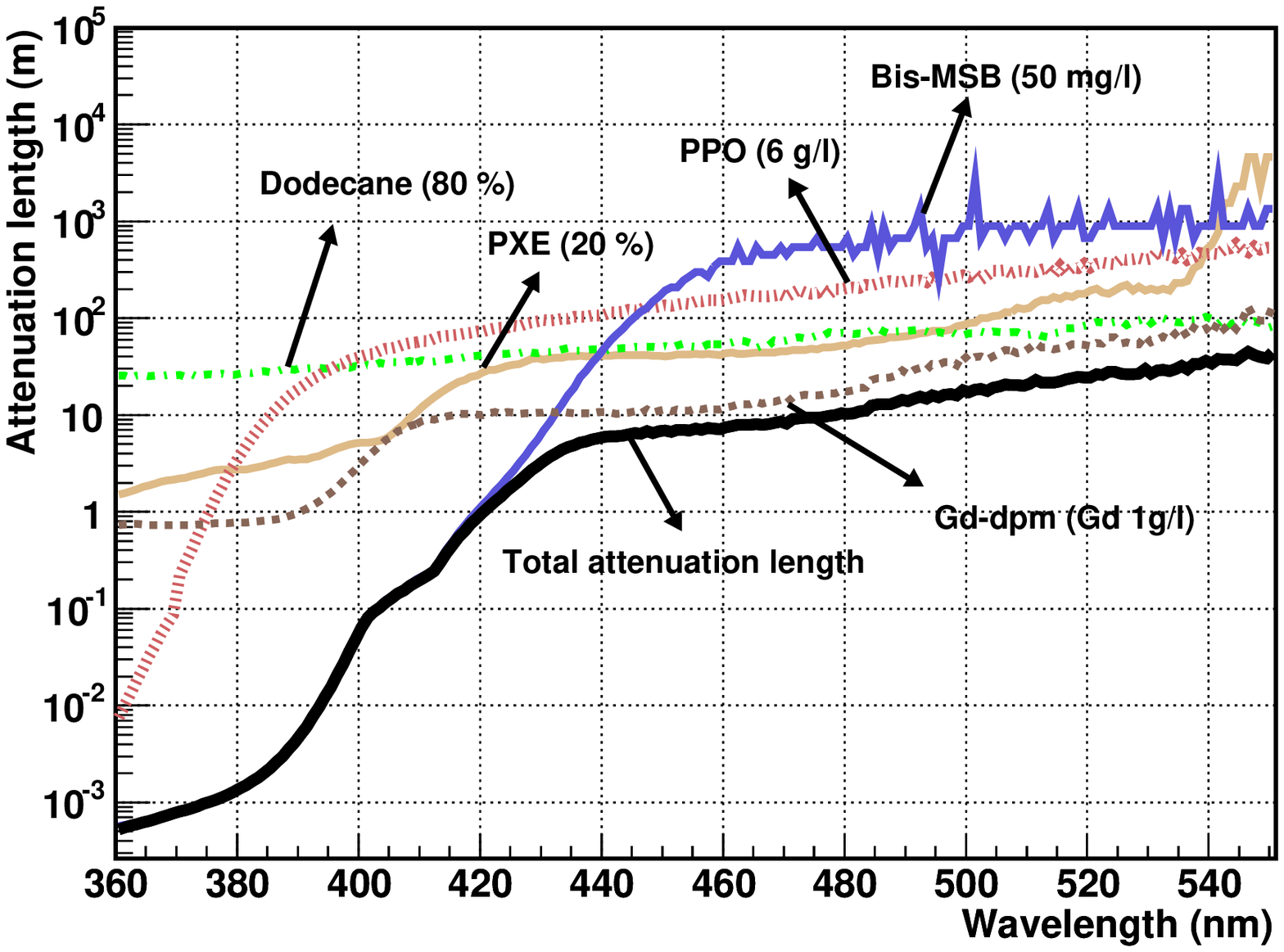}\end{center}

\caption{{\small \label{fig:Attenuation-lengths}Attenuation lengths contributed
from each component of a scintillator suitable for a $\nu$ target for
Double Chooz. All values are calculated for the given concentrations.
Extinction coefficients from MPIK. }}
\end{figure}

Attenuation in the liquids is treated as the sum of three processes:
absorption, wavelength-shift and scattering. 
Global attenuation lengths are calculated by adding the contribution
of all molecular species present in the scintillator cocktail. Spectro-photometric
measurements of the purified single components provide integral information
about the sum of the three processes. By fluorimetric methods it is
possible to disentangle wavelength-shift from the sum of absorption
and scattering. As for the absorption/scattering relative weight,
for the moment we assume that scattering from micro-particles is negligible
(Rayleigh scattering is already included in GEANT4). The above assumption
is conservative, since the possible contribution of the scattering
to the interaction lengths is attributed to absorption. Figure~\ref{fig:Attenuation-lengths}
shows an example of attenuation lengths in a typical Gd-loaded scintillator
formulation, as implemented in DCGLG4sim. The effective interaction
length is given by the composition of all contributions, however it
is important to know the relative weight of each one.

\begin{figure}
\begin{center}\includegraphics[%
  width=0.6\columnwidth,
  keepaspectratio]{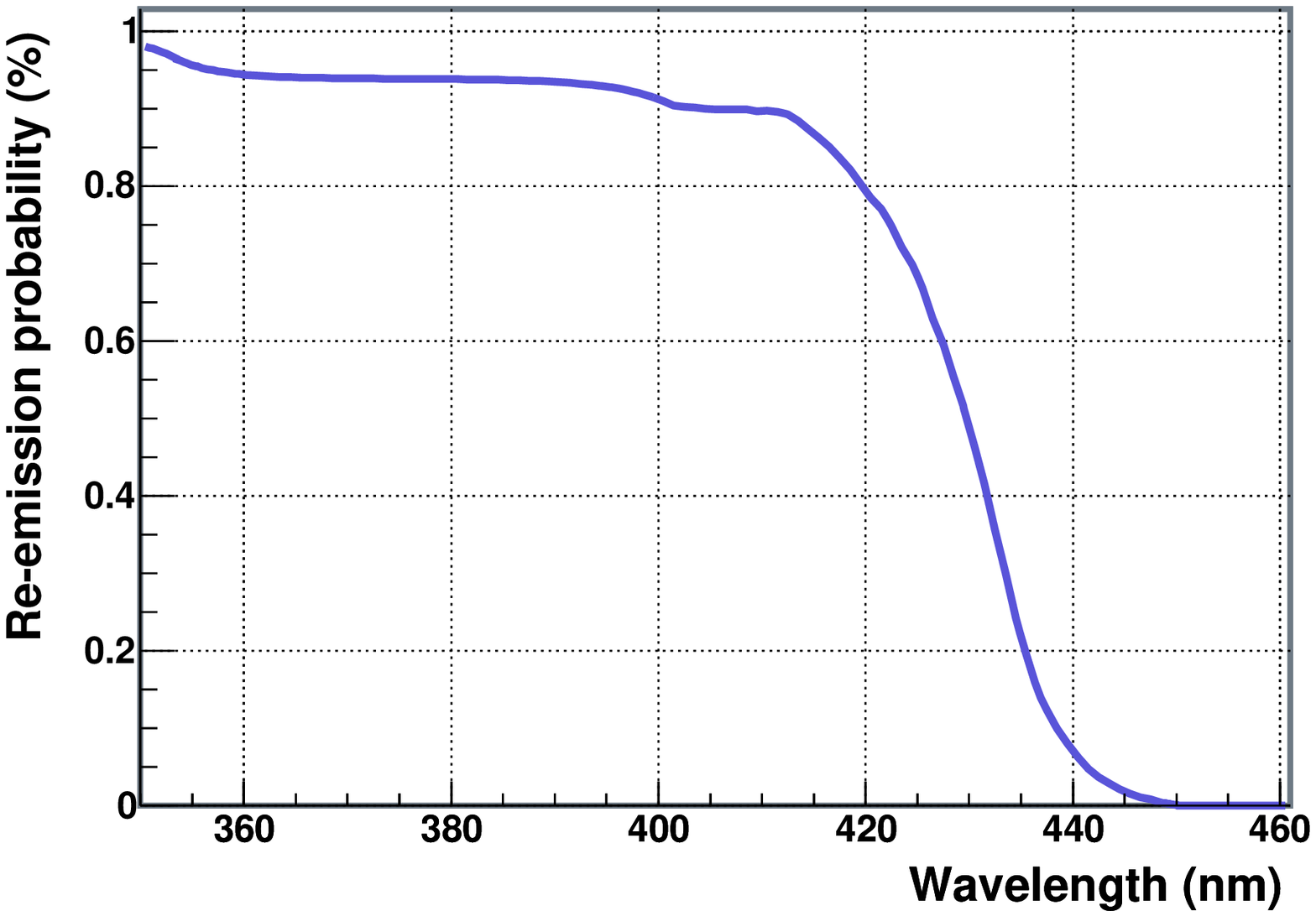}\end{center}

\caption{{\small \label{fig:Re-emission-probability}Re-emission probability
calculated for the scintillator of Figure~\ref{fig:Attenuation-lengths}.
The ordinates represent the probability that an absorption results
in the emission of a new photon, with $\lambda_{\textrm{new}}>\lambda_{\textrm{old}}$.}}
\end{figure}

The re-emission probability is calculated for each wavelength by multiplying
the probability that absorption was due to a fluorescent species (fluor
and WLS) by their wavelength-dependent re-emission probability. The
latter is assumed constant to its literature value up to a wavelength
cut-off that is determined from fluorimetric measurements \cite{bib:lens}.
Figure~\ref{fig:Re-emission-probability} shows the calculation of the
re-emission probability for the scintillator of Figure~\ref{fig:Attenuation-lengths}.

Figure~\ref{fig:Attenuation-lengths} and \ref{fig:Re-emission-probability}
together demonstrate
 that for $\lambda\lesssim420\,\textrm{nm}$, absorption is dominated
by bis-MSB and largely results in a wavelength-shift. This is crucial,
since the PPO emission falls in a wavelength region where absorption
from non-fluorescing impurities is severe (compare Figure~\ref{fig:Emission-spectra}
and \ref{fig:Attenuation-lengths}). In our scintillator, absorption
is most important for $\lambda\gtrsim430\,\textrm{nm}$, where bis-MSB
is no longer an effective wavelength-shifter. In this range we anticipate
an attenuation length of $\sim5-10\,\textrm{m}$. This value can still
be improved by further purification of the Gd-compound and PXE.



%

%
\cleardoublepage
%
%
\section{Photodetection}
\label{sec:photo}
\subsection{Target and Gamma-Catcher}
In this section we present the factors 
considered in the design of the Double Chooz
target array of photomultipliers (PMTs). The requirement is to have 
sufficient coverage and PMT performance to allow the experiment to reach its sensitivity 
goals, while at the same time considering not only the cost of the array, but also the 
fact that PMTs are expected to be one of the major sources of uncorrelated internal 
background due to their relatively high level of radiological contaminants and their 
proximity to the target volume. This is an important point, as accidental correlations 
were roughly one-third of the 
background events recorded by CHOOZ\cite{bib:chooz}.
In addition to standard concerns about PMT performance, Double Chooz also has a unique
requirement that the detector response and hence the light collection be identical in 
two different detectors. This raises concerns about asymmetries in the PMT acceptance 
both due to inherent asymmetry in the dynode structure and due to magnetic fields, 
which might be different at the two detector sites. 
Significant progress has been made to measure this aspect of performance 
in addition to timing, gain, stability and dark noise.
Since the Double Chooz experiment has a 
tight construction schedule without 
 extensive R\&D studies, we are choosing a
 conservative system design and 
use several technical solutions from the previous experiments where members of the Double Chooz 
collaboration were involved. 
\subsubsection{Coverage}

With respect to photodetection, the CHOOZ experiment\cite{bib:chooz}
 can 
be considered as a prototype for each of the two detectors in Double Chooz. Following 
the successful operation of CHOOZ
 with 15\% photocathode coverage, we plan 
a similar photocathode coverage (13.5\%) also using 8-inch photomultiplier 
tubes. With the given detector dimensions such coverage requires 534 8-inch
photomultipliers per detector. The relatively large number of phototubes
guarantees (as confirmed by simulations) that the loss of 
a few tubes per detector
will still 
preserve the identity of each detector response at an 
acceptable level.    
We expect that this number of phototubes will give sufficient energy resolution 
to keep low-energy backgrounds from fluctuating above our threshold. Based on our 
best simulations, we expect the average light collection to be above 180 p.e./MeV 
at the center of the detector increasing by about 
five percent to the periphery of the target.

\subsubsection{Radiopurity Requirements}

The goal for the detector trigger threshold is $\sim$0.5-0.7 MeV (with 
$\>$90\% efficiency at 0.8 MeV and $\geq$~99.97\% at 1.0 MeV)
that will allow high-efficiency detection of antineutrino events 
independent of the systematics of energy-scale non-linearity and 
calibration. In the CHOOZ detector\cite{bib:chooz} the trigger 
threshold for prompt signal detection was chosen at 1.3 MeV. This 
resulted in a raw trigger rate of $\sim$130~Hz that after applying 
software cuts was reduced down to $\sim$65~Hz. Lowering thresholds in 
CHOOZ was limited by the DAQ system. In the Double Chooz detectors 
the single rate situation will be considerably improved by introducing 
passive 17-cm iron shielding around the detector 
(see~Section~\ref{sec:steelshielding}). 
Also, a passive 1062~mm thick
buffer layer of non-scintillating oil (+12-15~mm of acrylics) 
is being introduced between 
the scintillator and the wall on which photomultiplier tubes are 
mounted. Radioactivity of liquid scintillator and buffer oil 
will be controlled by the purification process during the liquid 
scintillator production and at detector filling time. The effect of 
radioimpurities is discussed in more detail in 
Section~\ref{sec:backgrounds}. 
In addition to the above, the careful 
selection and radioassay
of construction materials will help to further reduce 
the single rates.

\par Since borosilicate glass of photomultipliers can contain
 radioimpurities ($^{40}$K, $^{232}$Th, $^{238}$U) and the photomultipliers are 
positioned close to the scintillator volume, particular care 
was given to the selection of PMTs with low concentration of 
radioimpurities in the glass. 
Three leading PMT manufacturers (ETL, 
Hamamatsu, and Photonis/Burle) provided us with samples of 
low-radioactivity PMT glass. Radioassay of these samples 
(Photonis/Burle glass sample is currently being tested) was made 
at the low-counting facility at the University of Alabama and at 
MPI/Heidelberg. The results of these measurements are shown in 
Table~\ref{T:PMT:rad}.
\begin{table}
\begin{center}
\begin{tabular}{l|r|r|r} \hline
source & K(ppm) & U(ppb) & Th(ppb) \\ \hline 
ETL, glass, manufacturer  &  64  &  30 &  30 \\
ETL, glass, Alabama       & 100  &  36 &  21 \\
ETL, glass, MPIK          &  59  &  54 & 156 \\
Hamamatsu, glass, Alabama & 204  & 164 &  66 \\
ETL, dynode, manufacturer & 109  &  35 &  12 \\ \hline
\end{tabular}
\caption{\label{T:PMT:rad} Measurements of PMT radiopurity.}
\end{center}
\end{table}
Radiopurity data for PMT glass from
 ETL and Hamamatsu companies are generally confirmed 
by our radioassay measurements, although there are some discrepancies between 
measurements performed by different groups. It is also possible to have tube to tube variations. 
We currently plan to test non-destructively several 
PMTs from ETL, Hamamatsu, and Photonis to verify and extend these results. 
The current measurements, when input into our U/Th/K/Co event generator and 
detector simulation, result in a background singles rate of less than 10~Hz at a 
threshold of 0.7 MeV. This rate would result in less than 2 events/day above 
1 MeV in our far detector, which is acceptable. Thus the two manufacturers tested 
so far are capable of producing PMTs of sufficient radiopurity for our purposes. 
It is our intent, however, to include radiopurity in the PMT specifications, 
and to perform non-destructive counting on a random sample of delivered PMTs 
to ensure radiopurity quality and consistency. The low-counting facility at 
Alabama is large enough to be able to ``whole body'' count a PMT in this fashion. 
This facility will also be
used to count a reference PMT, which will then 
be crushed and counted again for better accuracy. In this way a conversion factor 
from crushed to whole body counts can be obtained and used to ensure radiopurity
quality during the manufacturing process. Capacitors and other components of the PMT 
base will be screened as well.
The collaboration also has
 access to low background counting facilities in Heidelberg, 
Gran Sasso and Modane. 
\subsubsection{PMT Performance}
Light produced in the liquid scintillator of the
Far and Near targets of Double Chooz 
should be detected by two identical photomultiplier (PMT) systems. The number of 
phototubes (534 per detector) and their geometrical arrangements will be the 
same for both detectors.
With the individual PMT threshold of 0.25 s.p.e. 
(single photo-electrons), one can expect the
following global PMT system performance characteristics:

\begin{itemize}
\item light output and uniformity of detector response: minimum response 
of 180 spe/MeV in the center of the detector increasing by less than 10\% (to 200 spe/MeV) 
at the periphery of the target volume (5\% is the predicted value). This dependence is correctable by 
the reconstruction of the spatial position of the event in the scintillator 
volume;
\item expected accuracy of vertex reconstruction $\sim 9~cm/\sqrt{E(MeV)}$ 
in x, y, z projections folded with particle energy deposition distribution;
\item expected energy resolution $\sim 7.5\%/\sqrt{E(MeV)}$;
\item PMT dark rate is expected to contribute $\sim$ 1 count per second at 
the energy threshold of 50~KeV. We plan the trigger threshold to be 
set at a level 0.7 MeV as determined by radiopurity considerations.
\end{itemize}
\subsubsection{Detailed Specifications}
\label{sec:514}%
Below we discuss parameters of the PMT system that are being used in the design.
These parameters either came from the baseline configuration and set of 
parameters, or were specified by the PMT manufacturers, or have been measured 
by us with the samples of phototubes provided by the manufacturers.
\begin{itemize}
\item 8-inch diameter PMT represents an optimal ratio of glass 
weight to photocathode area (active photocathode diameter 190 mm and 
mass of the phototube is $\sim$1000~g);
\item the total number of PMTs is 534 per detector. On the cylindrical surface they
are arranged as 12 rings of 30 PMTs in each. There are also 87 PMTs on each of 
the bottom and top caps. Figure \ref{PMTlayout} shows the positions of phototubes 
in the detector. The PMTs will be installed in the detector with their axis 
tilted to the center of the target. For non-tilted PMTs in the central plane the 
distance of the apex of PMT from the back supporting wall is 300 mm;
\begin{figure}
\begin{center}
\psfig{file=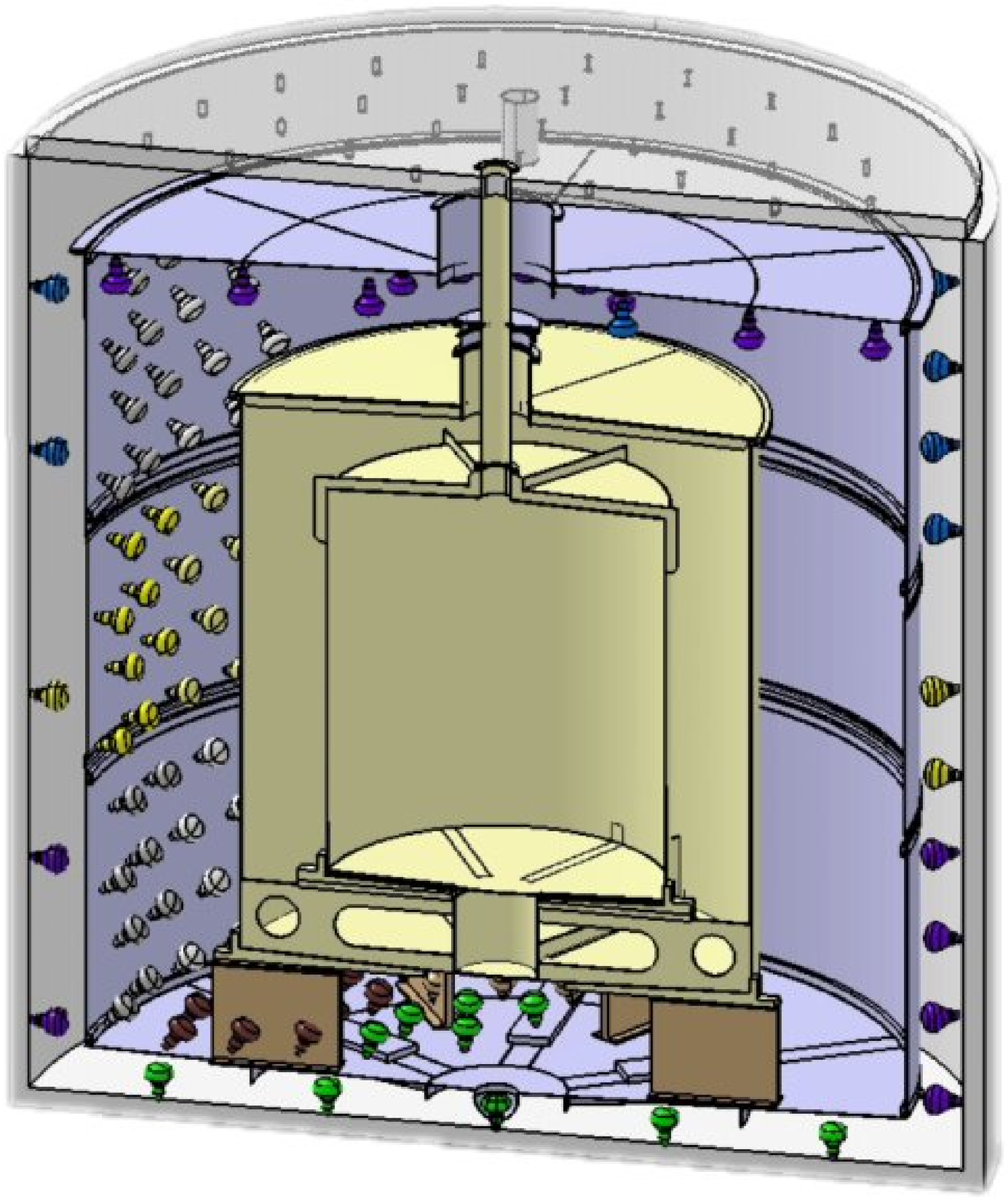,width=0.5\textwidth,clip}
\caption{Target PMT layout in the Double Chooz detector.}
\label{PMTlayout}
\end{center}
\end{figure}
\item from our experience in Chooz, KamLAND, and Super-Kamiokande we expect the possible 
losses of PMT in the detector to be at the level of less than 1\%. The key to low 
PMT mortality rate is comprehensive testing program during PMT installation. 
Total number of PMTs should be large enough to keep the detector uniformity 
constant within $\leq$ 1\% accuracy. For identical response of the two detectors 
inoperable tubes in one detector will require that  the corresponding tube 
in the other detector be disabled in software and possibly removed from the trigger;
\item the total number of PMTs to purchase is 2 $\times$
 550, which will include 3\% spares;
\item for the mechanical support of the PMTs we plan to use the light-weight mounting
developed by the MiniBoone Collaboration. Figure \ref{PMTsupport} shows this 
support structure. During the installation PMTs will be mounted individually 
one-by-one on the walls and bottom/top lids of the stainless buffer vessel.
Construction materials of the support structure (stainless steel and acrylic) 
have been radioassayed to be sure that they produce negligible contribution to the 
detector singles counting rate;

\begin{figure}
\begin{center}
\psfig{file=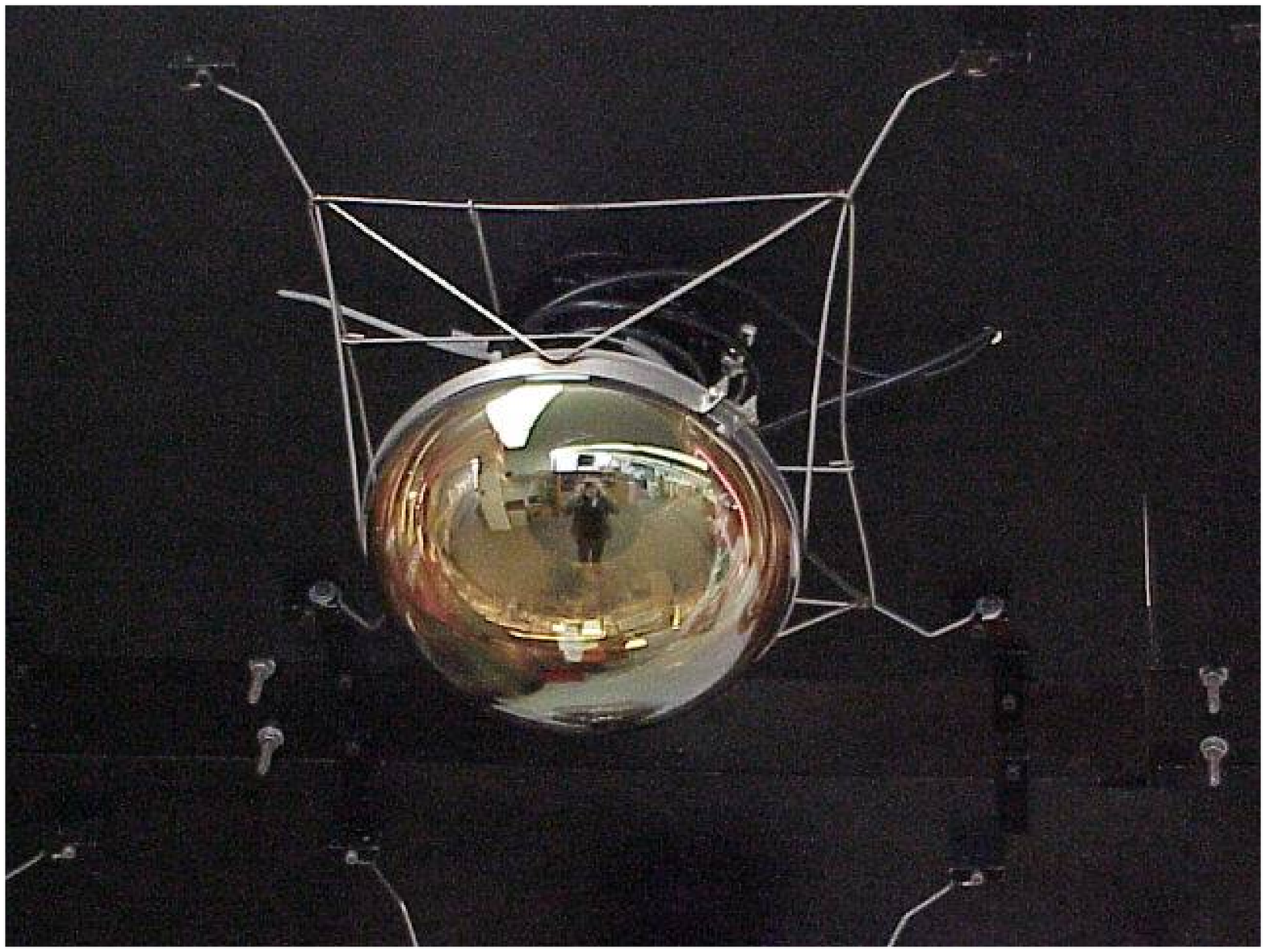,width=0.5\textwidth,clip}
\caption{For support of 8-inch photomultiplier tubes in the Double 
Chooz detector we plan to use the mechanical design of the mini-Boone collaboration, 
but other solutions are also currently being investigated.}
\label{PMTsupport}
\end{center}
\end{figure}

\item positive supply voltage at the 
PMT anode provides the advantage of minimizing 
the dark current and the discharge rate, but introduces RC-coupling in the signal 
processing; 

\item a single cable will be used for supplying high voltage to PMT and for 
signal readout. This minimizes the ground loops noise, cross-talk, and cable cost;

\item low-current PMT voltage dividers (0.1 mA at the nominal voltage) minimizes 
power dissipation in the detector;

\item acrylic + epoxy potted and sealed bases will be constructed by PMT manufacturers;

\item 20-m long oil-resistant RG-303C 50-ohm cable sealed in the PMT base with 
SHV connector at the outer end will be used. Such cables immersed in mineral 
oil are used in the KamLAND detector;  

\item the choice between stainless steel and black sheet 
between the PMTs is currently being discussed. This can 
be decided rather late in the construction schedule;

\item PMT glass resistance to the buffer oil (dodecane) was tested for Hamamatsu and 
ETL tubes (with no effect for equivalent $\geq$ 10 
years of operation).

\end{itemize}

Other PMT parameters important for detector operation (measured with PMT samples
or provided by manufacturers):

\begin{itemize}

\item PMT operation gain is $10^7$;

\item PMT anode is 50-ohm back terminated for robust operation with large signals;

\item quantum efficiency $>20\%$ at 400 nm. As an example Figure~\ref{PMTuniformity}
shows measured non-uniformity of photocathode for Hamamatsu R5912 tube.

\begin{figure}
\begin{center}
\psfig{file=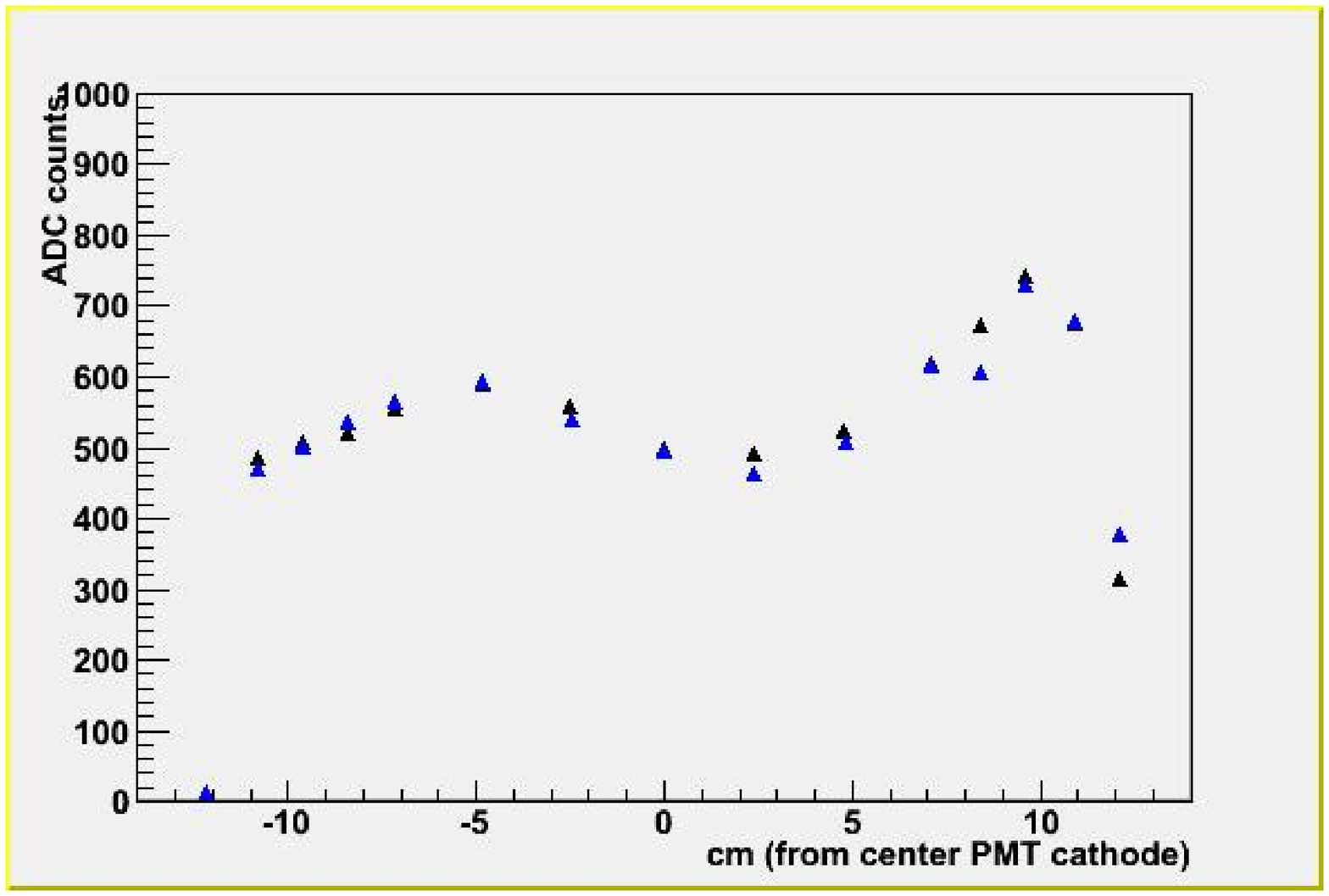,width=0.7\textwidth,clip}
\caption{Example of photocathode non-uniformity for typical Hamamatsu R5912
photomultiplier tube.}
\label{PMTuniformity}
\end{center}
\end{figure}
\item single photoelectron resolution: s.p.e. peak-to-valley $\geq$ 1.5.
Figure \ref{SPEspectra} shows single photoelectron spectra for two 
measured samples of Hamamatsu and ETL phototubes with and without 
discrimination threshold. They demonstrate that low thresholds $\sim$ 
0.25 s.p.e. are practically possible to achieve;
\begin{figure}
\begin{center}
\psfig{file=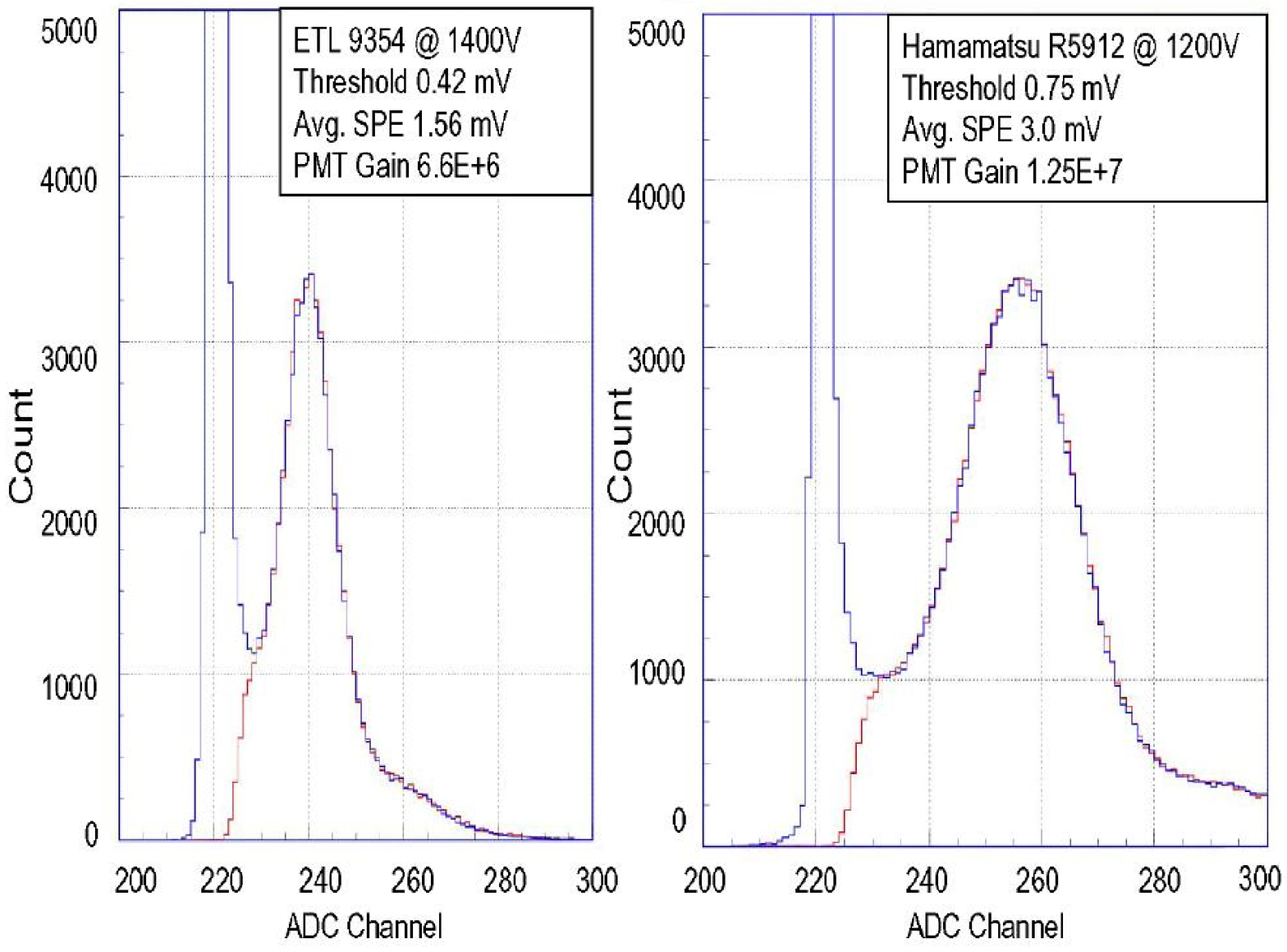,width=0.7\textwidth,clip}
\caption{Untriggered (blue) and self-triggered (red) s.p.e. spectra
of two tested photomultipliers: ETL 9354KB and Hamamatsu R5912 measured with 
low-efficiency LED.}
\label{SPEspectra}
\end{center}
\end{figure}
\item typical dark rate in air specified by manufacturers is 4,000 per second.
Our measurements with 0.25~s.p.e. threshold can reproduce this rate;
however, from our KamLAND experience, we know that 
the dark rate can be higher for 
tubes immersed in the mineral oil and looking at the large scintillator volume.
The nature of this increase in KamLAND is not fully understood, however if we
assume that the dark rate will not be 
scaling with the detector volume and the 
coverage but only with the PMT photocathode area we can conservatively estimate 
the average dark rate per phototube at the 
level of $\sim$10,000-12,000${\rm s^{-1}}$;
\item s.p.e. after-pulses occurring within $\sim$ 60~ns after the main s.p.e. pulse 
are unfortunately present for all tested 8-inch photomultiplier tubes.
They appear generally due to electron backscattering from the first dynode 
with the probability of a
few percent. 8-inch PMT specifications allow 
after-pulses (within 60 ns after the main pulse) at the level of less than 3\% 
and late pulses due to transport of residual ions (within 16 $\mu$s after the 
main pulse) at the level of less than 10\% per initial photoelectron.   
\item with tapered (B-type) 
voltage dividers, all tested phototubes show 
a linear response up to 80-100~mA of anode current
(peak value), which 
satisfies the requirements of Double Chooz, where typical large 
muon signals will produce $\sim$ 60~mA peak value;
\item the typical rise time for the s.p.e. signal (after 20-m of RG-303 cable) is 
$\sim$ 4 ns; FWHM $\sim$ 7 ns; and return-to-the-base time is $\sim$30 ns;
\item transition time spread $\sim$ 2.5 ns FWHM (specified by manufacturers);
\end{itemize}
\subsubsection{Sensitivity to Magnetic Field}
\label{sec:toto}
\begin{figure}
\begin{center}
\psfig{file=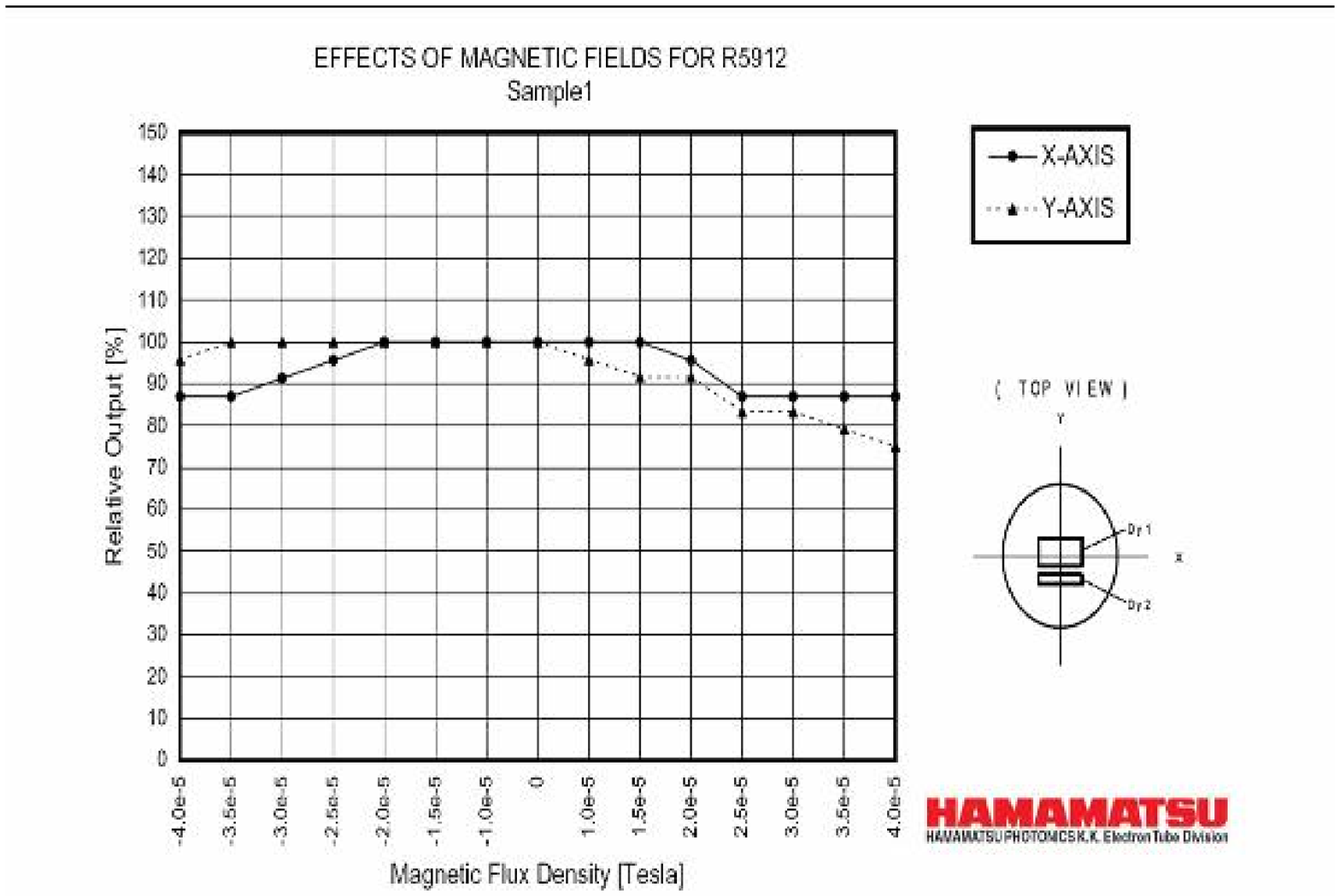,width=5.0in,clip}
\caption{Effect of transversal magnetic field on the 8-inch PMT response.
Measurements provided by Hamamatsu Corporation}
\label{PMTmagnetic}
\end{center}
\end{figure}

PMTs are sensitive to the magnetic field. As an example Figure \ref{PMTmagnetic}
shows the typical effect of the magnetic field transversal to the PMT axis 
on one of the candidate PMTs (Hamamatsu R2512) \cite{hamamag}. It is 
expected that PMTs of other manufacturers will have similar sensitivity to the 
transversal magnetic field.
The magnetic field as known from geomagnetic models \cite{geomag}
has the following components at the Chooz location: 0.43 Gauss Down, 
0.20 Gauss North, 0.003 Gauss West. The unshielded magnetic field transversal 
to the PMT axis can create response difference for the tubes at different 
locations in the detector up to 30\%. This difference in PMT response 
can potentially affect the identity of two Double Chooz detectors and thus 
should be eliminated or reduced.
The presence of 17-cm iron shield around the detector, needed for the 
reduction of radioactivity counts by the scintillator, might result
in additional non-uniform magnetic field inside the detector at the 
place where PMTs are installed. The issue of magnetic shielding is 
addressed in Section~\ref{sec:toto}. 
The required residual transverse magnetic field should not exceed 
0.05~Gauss at any PMT location in the detector.
\subsubsection{PMT Purchase and Delivery}
We have been in contact with three leading PMT manufacturers of 8-inch 
phototubes: Hamamatsu in Japan (R51912), Electron Tubes Limited
in England (ETL 9354KB), and Photonis/Burle in France/US (XP1806). 
All three companies produce 8-inch photomultiplier tubes with 
low-radioactive borosilicate glass and with similar performance 
characteristics. All three companies provided us with samples of 
their tubes for performance evaluation as well as with samples 
of low-radioactive glass for radioassay. Some initial results of tube 
measurements are shown in the previous subsections. We also asked 
manufacturers to integrate potted bases and oil-resistant cables
in the construction of the deliverable product, so we could 
enlist their help in the R\&D effort, save on construction time, 
and guarantee more robust operation of the tubes. Also, we plan to 
specify the type of the voltage divider optimized by us for Double 
Chooz operation which the manufacturer will build from high radiopurity 
components and seal in the PMT/base assembly.
A final decision on the particular manufacturer product will be made 
after verifying the compliance with required level of radiopurity,
detailed performance comparison, and {\em price}.
All three manufacturers have agreed on the following production schedule:
response to the bid specifications: within one month; preproduction
time for the first 110 PMT batch delivery: 4 months; after that monthly
production rate of 110 phototube. Thus, first 550 tubes needed for 
assembly of the Far detector will be produced and delivered in 8 
months after signing the contract. It is important that the proposed rate 
is convenient to all manufacturers and does not saturate their production
capability. We are negotiating with companies the possibility of separating 
the production into two groups of 550 pieces each with the gap in
production of $\sim$ 6 months to reduce the high load on the funding
profile.
%
%
\subsubsection{Acceptance and Characterization Tests}

We believe that the key to reliable and lasting operation of the 
large PMT array system is in comprehensive tests prior to installation 
and commissioning of the phototubes. We plan several levels of tests
performed at different locations and times.

First we agreed with manufacturers that they will provide us with 
certificate of acceptance measurements made at the factory.
It will typically include: cathode and anode luminous sensitivity,
cathode blue sensitivity, anode dark current and dark counting rate,  
supply voltage at the gain $10^7$, peak-to-valley ratio for single
photoelectrons, and transit time spread.

At the Chooz site we plan to rent a warehouse 
with area approx. 3,000 ${\rm m}^{2}$
that will be converted to the clean room facility. There PMTs will be 
unpacked, tested, assembled, and stored. The rate of characterization and 
acceptance tests ($\sim$5 per day) should match the rate of PMT delivery 
(110 per month). These tests will include burning-in time of at least 24 hours 
for each phototube at the voltage providing gain of $\sim$ $10^7$ followed 
by measurements of major performance parameters of each phototube. 
These parameters will include: 
single photoelectron peak vs high-voltage, 
dark rate vs high-voltage, 
photocathode quantum efficiency relative to the reference,
s.p.e. spectrum at nominal gain $10^7$ and peak-to-valley ratio,
measurement of linearity at nominal gain $10^7$,
measurement of transition time spread at nominal gain $10^7$,
pedestal noise RMS,
viewing the PMT signals with an oscilloscope.

The test station will be equipped with two sets of black boxes each 
for 6 PMT positions, with laser- and LED-based light sources, 
and with a VME or a CAMAC-based DAQ system, digital scope and a computer. 
6 tubes simultaneously will be burn-in or tested at a time alternatively 
in two stations. All tests will be performed by a group of two people 
who besides performing the measurements will also maintain the database 
of all PMT measurements and installation parameters. These two people 
should stay at Chooz for extended period time to provide consistent and 
responsible test operation.

The purpose of the characterization/acceptance test will be to obtain 
a set of individual parameters for each tube: operating voltage for nominal 
gain, slope of gain vs high-voltage, s.p.e. spectrum,
 dark rate, transition 
time spread, and linearity which will be included in the data base to
characterize the detector performance. These data can  
later be used in the  
analysis and/or in the detector simulations. Furthermore, in a first stage 
of the experiment, we will need to group PMTs into groups with similar 
high-voltage parameters since four or eight tubes in a group will share
 the common power supply channel.

Selective acceptance radiopurity tests: 1-2 randomly selected tubes 
from each monthly delivery batch will be assayed for radiopurity
(in non-destructive tests). If K, Th, or U content will exceed the specified 
level of contaminations, then other 10 randomly selected tubes from the 
same batch will be radioassayed. If more than 3 out of 10 PMTs will exceed 
the specified contamination level, the whole batch will be returned to
the manufacturer.

\subsubsection{Mechanical Assembly and Cleaning}

The mechanical support system could be adopted from the Mini-BooNE 
design shown in Figure~\ref{PMTsupport}. It is light and can be made 
with radio-pure materials. The PMT circular grip can be firmly adjusted 
during the assembly of the support structure. It should provide 
a reliable hold of the phototube in all possible positions relative 
to the direction of the buoyant force.
The assembly of PMTs and mechanical structures will be performed in 
a clean room next to the characterization/acceptance tests at the Chooz 
site not far from the detectors. Two trained technician will be 
involved. They will clean the PMT, the cable and all the mechanical 
parts with alcohol prior to
and after the assembly, will tighten the PMT grip 
ring with a torque controlled tool and will seal the whole assembly 
in a hermetic plastic bag. Sealed assemblies will be stored and later 
transported to the detector site for installation. Sealing bags will be 
removed inside the clean area of the detector. Assembly production 
(of $\sim$ 25 PMTs per day) will go slightly ahead and in parallel with 
the PMT installation in the detector.
\subsubsection{Installation in the Detector, testing and commissioning}
According to the collaboration plans the installation of the PMTs in a detector 
will require coordinated work of a team of 12 people and should be 
performed within one month (installation of $\sim$25 PMTs per day).
The work will include PMT transportation to the detector site and
opening of the sealed bags. Before the installation each tube will undergo a 
short functionality test with HV in a special dark box. Tubes will be 
bolted to the angle rib and welded to the vessel wall. The position of the tube, 
its serial number, and cable label will be then entered in the computer
database. Cables from several neighboring tubes will be arranged 
in bundles of 16 cables. These bundles will be routed to the 
cable exit tubes on the top of the detector rim through pre-installed
Teflon conduit tubes. This cabling procedure has been successfully used 
in the KamLAND detector.
The phototube installation will start from scaffoldings on the top 
of the cylindrical part of the detector and will proceed by patches 
down and around the cylinder. Photomultipliers on the bottom floor will 
be installed next with cable routed through pre-installed conduit tubes. 
PMTs on the upper lid of the detector will be mounted under the lifted 
lid. Bundles of PMT cables from the upper lids will be routed through 
the conduit tubes attached to the lid to the rim of the detector where 
the cable exit tubes are located. Cable bundles from top lid PMTs will 
loosely pass through the side exit tubes with significant slack. These 
cable bundles will be pulled out to their final extension when the lid 
will be closing.
\subsubsection*{Functional test during installation}
Newly installed PMTs will be tested daily
one-by-one (during the 
evening shifts) for functionality and correct cable 
numbering to the position assignment. The final test with LEDs
 in the central 
position of the detector at the nominal PMT voltages will be 
accomplished upon the installation of all PMTs in the detector
and before the installation of the central acrylic vessels. For this 
test all PMTs will be connected to the HV distribution system 
(HV-signal split boxes) and tested one-by-one with a
dedicated DAQ 
station verifying the functionality of each channel and the 
correspondence between the HV channel number and the signal output.
Daily functional tests during the installation will be performed by
members of the
installation crew.
\subsubsection*{PMT System Commissioning}
After the installation of the acrylic vessels, with the whole detector being closed 
and sealed, and before it is filled with scintillator and oil, a 
commissioning test will be performed. For this test the final 
HV-system and readout electronics need to be installed and connected. 
All tubes will be powered and held under nominal high-voltage 
for 2-3 days. Then, at several high-voltages close to the nominal HV
corresponding to PMT gain $10^7$ a discriminator threshold scan for 
every tube will be performed to establish the threshold corresponding 
to 0.25 spe. Together with that the s.p.e. spectra will be recorded and 
the dark rate for each PMT will be measured. The final choice of the 
high-voltage values will be made by equalizing the response of the PMTs in 
terms of single photoelectron response.
After filling the detector with scintillator and oil the commissioning 
test described above will be repeated. 
The final goal of the commissioning 
test will be to provide the detector with the set of high-voltage values 
corresponding to equal response within $\sim$ 1-2\% to single 
photoelectrons. Note, that relative quantum efficiencies (averaged over 
the whole photocathode area) of the tubes were to be measured during 
the earlier characterization tests. The commissioning of the PMT system will 
be performed by a group of 2-4 PMT group members together with other 
collaborators responsible for electronics and DAQ operation.
\subsubsection*{PMT Documentation Database}
The information from all results of PMT tests, history of each PMT performance,
geometrical position in the detector, cable label, HV-channel, electronic 
channel etc. will be recorded and stored in the PMT database. Created as
html data structure it should be accessible through 
the Internet and should allow
extraction of formatted parameter tables that can be used in data analysis 
and for Monte Carlo simulations.
\subsubsection{Optics of the PMTs}
\begin{figure}
\begin{center}\includegraphics[%
  width=0.40\columnwidth,
  keepaspectratio]{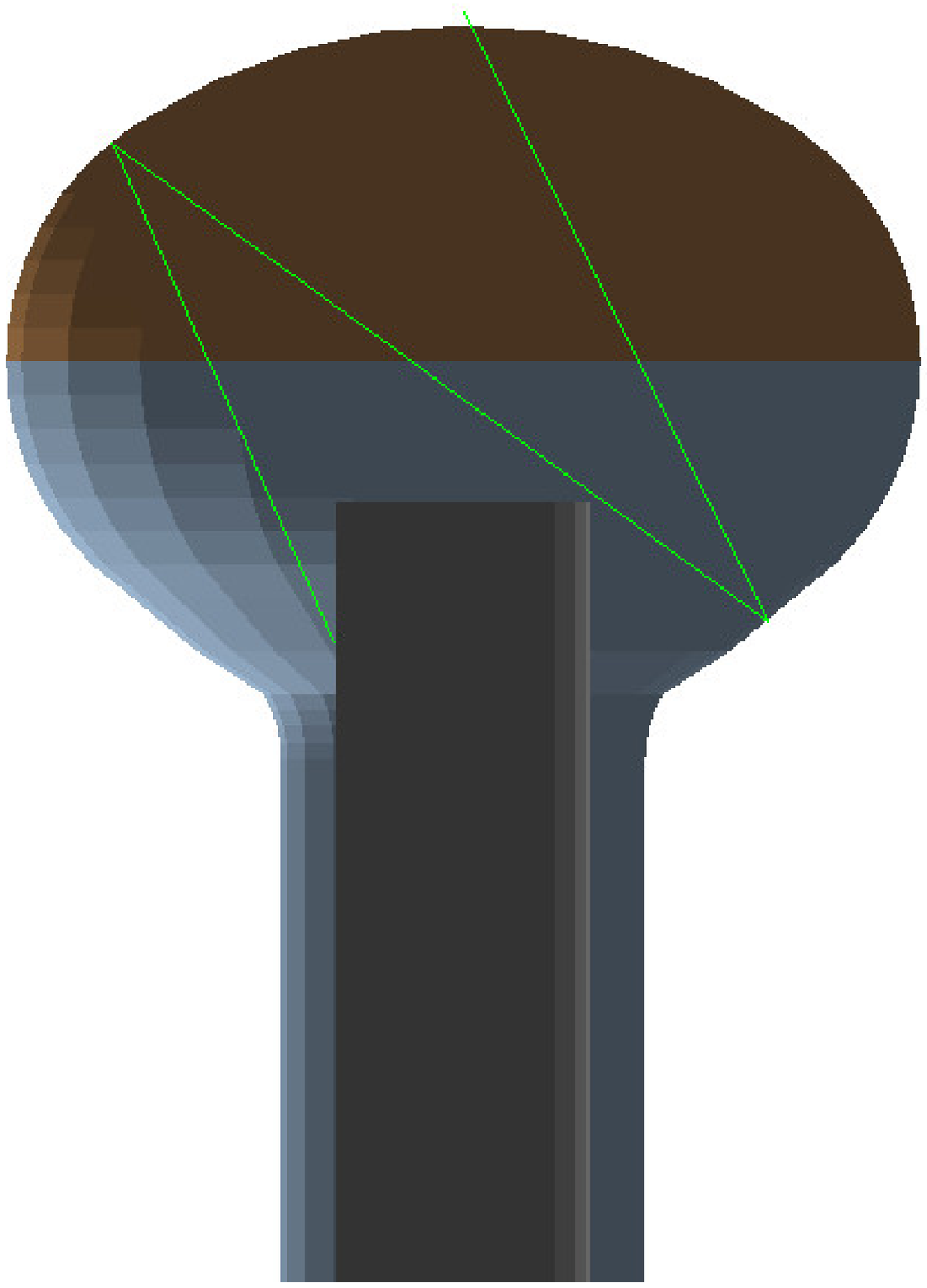}
\end{center}
\caption{{\small \label{fig:Photon-tracking-PMT}Photon tracking in GLG4sim.
The optical model calculates the probability for detection, dissipation,
reflection, and transmission at each interface.}}
\end{figure}

The optics of the PMTs in our Monte Carlo
is managed by the program GLG4sim, through the class \emph{GLG4simPMTOpticalModel}.
This implements the full model of the thin absorbing photocathode
with complex, wavelength dependent refractive index. The model and
its parameters are given in \cite{PMT_Optics}. It is based on experimental
data on the optics of PMTs equipped with the same bialkali photocathode
as the Double Chooz PMTs and leads to the prediction of the probability
of photo-conversion, dissipation, reflection and transmission, for
each crossing of the photocathode. These probabilities are dependent
on the photon wavelength and impinging angle, and on the refractive
index of the medium where the PMT is immersed (e.g., mineral oil).
Photons reflected or transmitted are further tracked, until they are
either lost or detected (Figure~\ref{fig:Photon-tracking-PMT}). 

\subsubsection*{Other materials}

\begin{figure}
\begin{center}\includegraphics[%
  width=0.60\columnwidth]{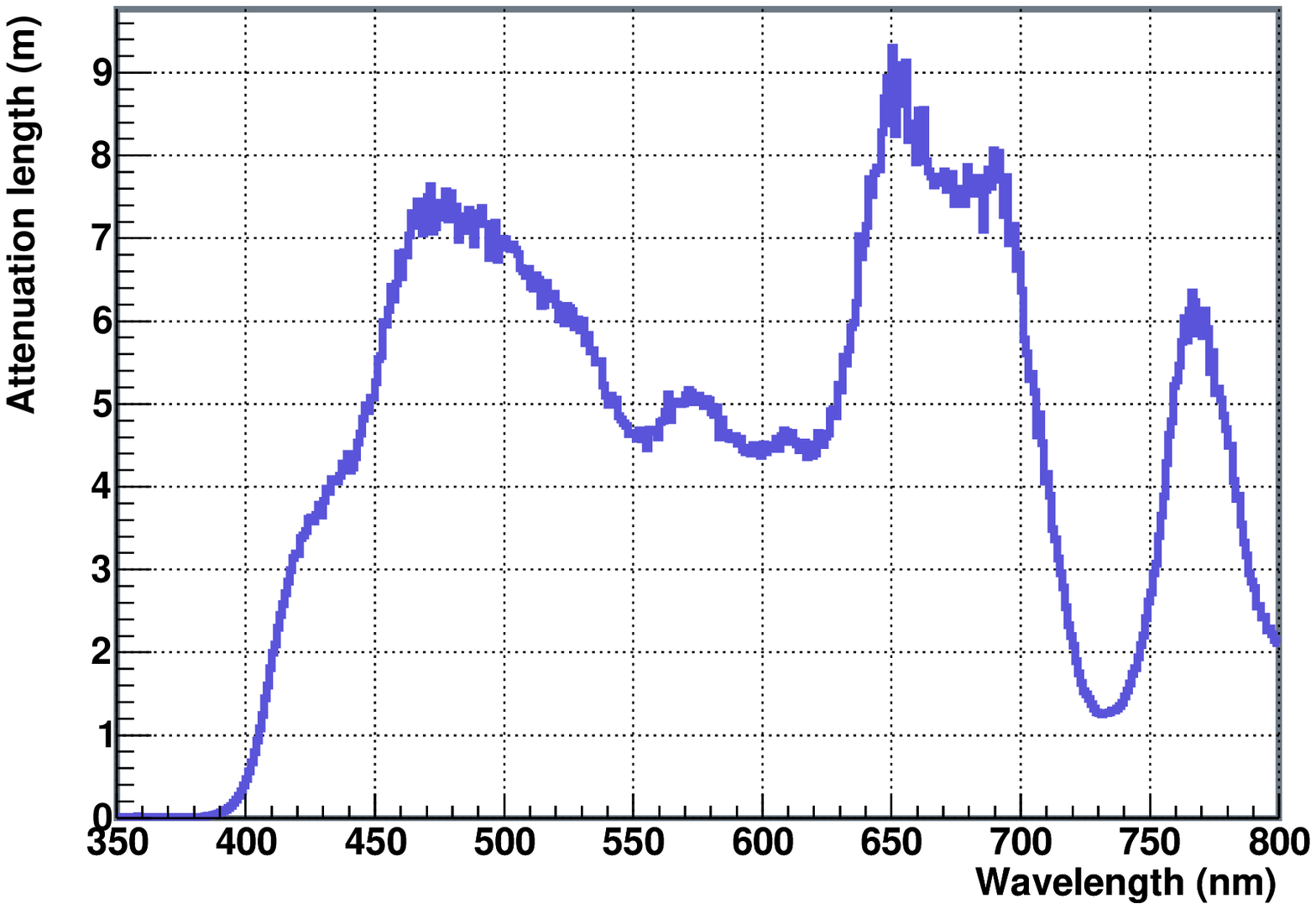}\end{center}
\caption{{\small \label{fig:Att-length-acrylics}Attenuation length for light
in the acrylic selected for the Double Chooz mock up. Data from CEA-Saclay.}}
\end{figure}

The double acrylic vessel containing the scintillators of the target and
$\gamma$-catcher is an optical interface between three regions, each
filled with a different liquid. For the time being, we assume that
the refractive indices of all liquids match perfectly. This is 
anticipated
to be very close to the actual case, since the target and 
$\gamma$-catcher will have the same or 
a very similar PXE/dodecane ratio, and the mineral
oil refractive index should also be in the same range as
 the PXE/dodecane
mixture, since the density will be similar. Acrylics are expected to
have a slightly higher index. For the moment we consider
$\textrm{n}\simeq1.46$ and $\textrm{n}\simeq1.49$, for liquids and
acrylic respectively. This slight mismatch causes some Fresnel reflection
at the liquid-acrylic interface, especially for grazing incidence.
Acrylic is also a potential light absorber/scatterer. The attenuation
lengths in acrylic have been measured in a 20~cm long sample of the
same material chosen for the Double Chooz Mock up. Results are given
in Figure~\ref{fig:Att-length-acrylics}. In the wavelength region of
interest the photon mean free path is $\gtrsim3\,\textrm{m}$.
Another important optical interface is the buffer tank. Several scenarios
are envisaged, and the Double Chooz Monte Carlo must include them
all: blackened walls, raw stainless steel, Tivek coating, etc... The
final decision will be a compromise optimizing light collection and
spatial reconstruction. At the buffer tank an optical interface is
implemented, which is based on the GLISUR model (from GEANT3). This
means that the reflective properties are driven by two effective parameters:
a reflection and 'polish' coefficients. The latter determines
the amount of isotropic diffusive reflection admixed to mirror reflection. 
\subsection{Inner Veto}
Double Chooz will operate two detectors at a 
depth of 300 m.w.e. and 75-100 m.w.e., resp. Due 
to these shallow depths, the rate of cosmic muons 
that penetrate the shielding and enter the detectors is high, 
in particular for the near detector.  In order to reach 
a signal-to-background ratio sufficient for the desired accuracy, 
backgrounds at both detector locations have to be 
well understood and controlled. The dedicated muon identification 
system that is described here will be 
used at both detectors to reject background events that 
are induced by cosmic muons. The muon system is an 
active detector based on liquid scintillator, surrounding the inner 
volumes of the experiment with 4$\pi$ coverage. 
\par Backgrounds for the 
delayed coincidence events can be split into two main branches. 
In accidental uncorrelated background events, the neutrino 
signal is mimicked by two independent processes which happen 
at the same time.
Correlated background events mimic 
the neutrino signal by a prompt and delayed energy deposition that 
is caused by the same initial particle. Cosmic muons are 
capable of producing such correlated background in different ways: 
They can produce fast neutrons via spallation in the surrounding rock, 
which in turn can cross the Veto and Buffer and deposit 1-8 MeV in the 
$\gamma$-catcher  or target scintillator before being captured 
by Gadolinium; negatively charged muons can be captured in 
the detector, followed by neutron emission; also, muon spallation 
on $^{12}$C can produce $^8$He, $^9$Li and $^{11}$Li. These isotopes decay 
on a timescale of 100\,ms for $^8$He and $^9$Li and few ms 
for $^{11}$Li and can produce $\beta n$ cascades. As the
muon rate is of the order of 30\,Hz (far) and 1000\,Hz (near), 
it is not possible to veto these events without unacceptable 
dead times for both detectors. Therefore, a solid 
understanding of these processes and their rates is mandatory.    

Although the veto identification system is not a hardware veto 
in the sense that it suppresses any trigger on a hardware level,
 we will call it "Inner Veto" in this proposal. The part of the 
dataset taken with this system is used to suppress muon induced 
backgrounds during offline-analysis. Analysis, classification 
and finally rejection of events is possible on the basis of 
the data acquired by the veto photomultipliers. Moreover, it 
provides data for further research, e.g. in particle shower 
characteristics. To monitor stability and for timing calibration 
purposes, light flashers will be installed in various positions 
of the veto volume. 

Compared to a water Cerenkov veto, the use of a scintillating liquid yields a factor of 30 more photoelectrons per typical muon event and grants the possibility of identifying incoming fast neutrons by their charged secondaries. 
\begin{figure}[hbtp]
	\centering
	\includegraphics[width=0.4\textwidth]{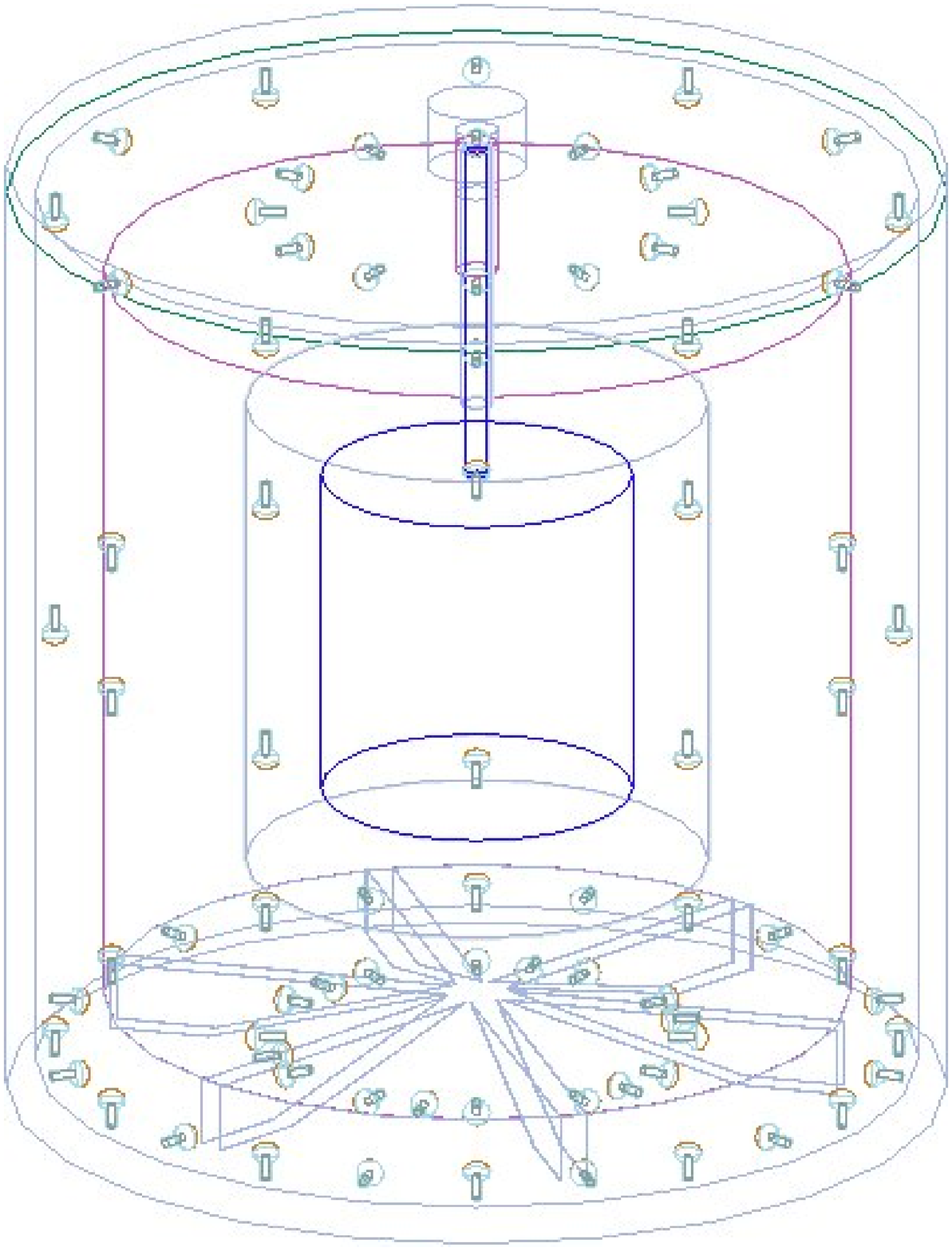}
	\caption{\label{fig:iv_78pms}Detector with inner veto photomultipliers}
\end{figure}
The detection of light created by charged 
particles in the scintillator is done with 8" low 
background photomultiplier tubes. The number and 
the arrangement of these PMTs was optimized using 
the Monte Carlo simulations that are described below. 
The standard layout that evolved from these simulations 
is an arrangement of 78 PMTs in five groups (Figure~\ref{fig:iv_78pms}):
\begin{itemize}
\item{a ring of 12 PMTs on top of the buffer vessel, pointing horizontally, alternating inward and outward,}
\item{a ring of 12 PMTs attached 
to the side wall, 30~cm below 
the lid, pointing alternatively inward and down,}
\item{a ring of 12 PMTs attached to the side wall, in the center, pointing alternatively up and down,}
\item{a ring of 24 PMTs on the bottom of the veto, pointing 
alternatively inward and up,}
\item{and a ring of 18 PMTs on the bottom of the veto, pointing alternatively inward and outward.}
\end{itemize}
The support structure below the buffer vessel demands a higher number of PMTs below the vessel than above.
\begin{figure}[htbp]
	\centering
		\includegraphics[width=0.6\textwidth]{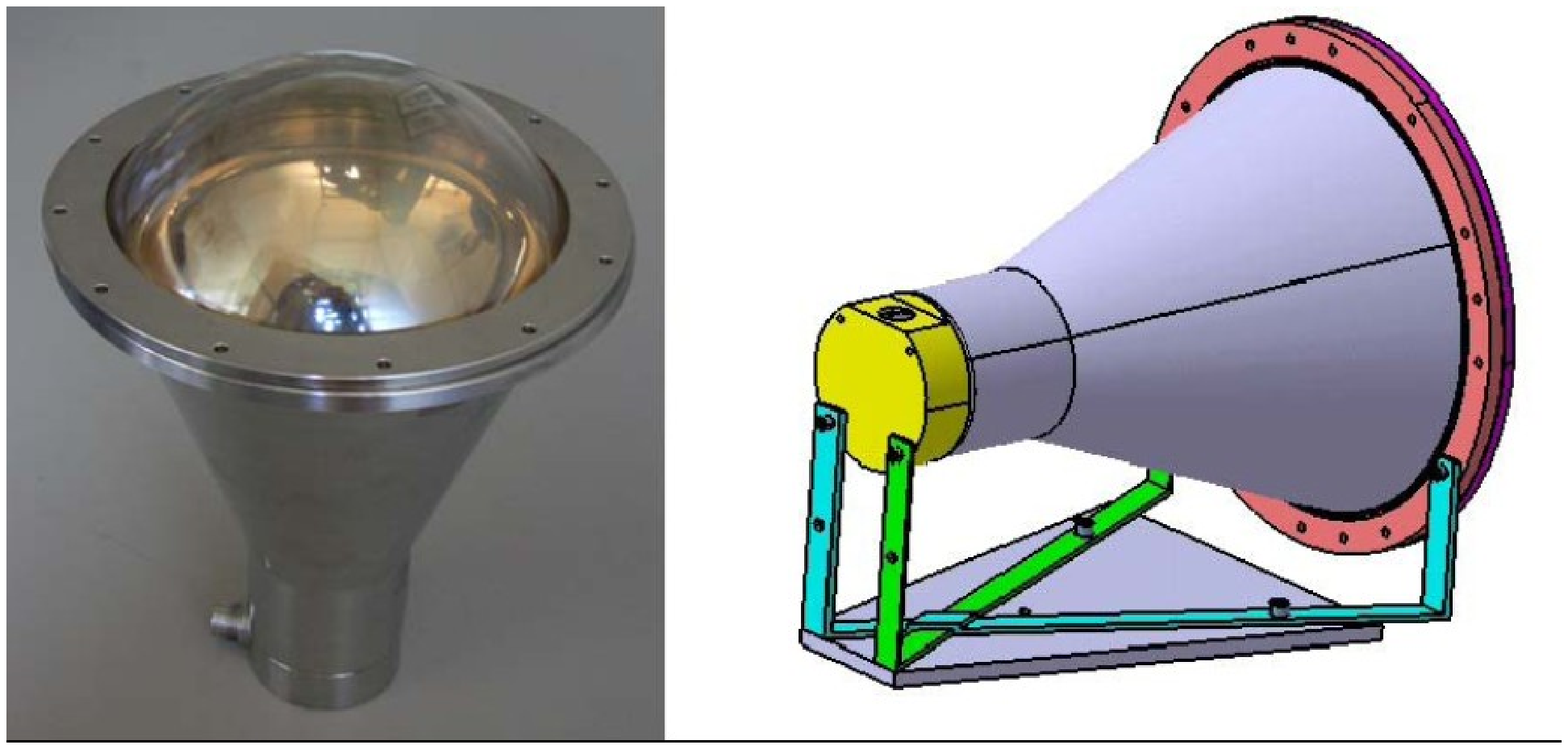}
	\caption{\label{fig:iv_pmt3}PMT encapsulation and mounting 
(Drawing P. Guillou$\rm{\ddot{e}}$t)}
\end{figure}
The PMTs are of the type 9354KB of Electron Tubes Ltd. 
(or a comparable type of another manufacturer). The active 
diameter of the blue-green sensitive photocathode is 190 mm. 
The PMTs can be encapsulated in a stainless steel chassis 
(Figure~\ref{fig:iv_pmt3}), where the Borexino encapsulation design with PET pressure membrane can be used with small modifications. Encapsulating the PMTs has the advantage that no leaks in the vicinity of the connector and/or the voltage divider can occur and mechanical forces on the PM glass tube are minimized. In case of a broken tube, the risk of contamination of the veto liquid and impact on other tubes is reduced. An additional $\mu$-metal shielding inside the steel chassis suppresses the magnetic field of the Earth. An alternative PMT encapsulation design similar to the inner detector is under investigation.

With 78 PMTs of this type, the effective coverage amounts to 0.6$\%$ of the total surface area of the veto detector. To increase the light collection, almost all surfaces in the veto detector will be painted white or covered with a reflective coating (diffuse reflection). Compared to a stainless steel surface, this improves reflectivity by more than a factor of two. Where applicable, TiO$_2$ based white paint will be spray-painted onto the steel surfaces after welding the pieces together. The side walls of the veto cylinder, the floor, the bottom side of the lid and parts of the vessel support can be treated in this way. Because of the limited space between buffer and veto vessel, it is not feasible for the buffer vessel which has to be welded on site. For this reason, highly reflective sheets of Tyvek (e.g. type 1073B or similar) will be attached to it after the three parts of the buffer are welded together. Tyvek was tested and used in Super-Kamiokande, Borexino and in the
  KamLAND experiment. Its diffuse reflectivity for visible light exceeds 90\%.
With an estimated optical attenuation length of several meters, multiple reflections are highly probable, increasing the light collection efficiency by roughly one order of magnitude. This is especially important for the detection of proton recoils.

Figure~\ref{fig:iv_side} illustrates the simulated light collection inside the veto region (for details on the simulations see below). The plot shows the average number of hit veto PMTs for a 5 MeV positron generated at random positions inside the veto. This energy corresponds to the projected energy threshold of the veto system. The lower light collection efficiency in the region below the buffer vessel is due to the support structure that creates shadow effects there.

\subsection{Simulations}
Extensive Monte Carlo simulations were performed to ensure the feasibility 
of the experiment and optimize the detector layout. In case of the muon veto, 
our simulations addressed general design questions (impact of changing the veto thickness on the detection capability, for instance) as well as more detailed questions like the position of single photomultipliers among the support structure on the bottom of the detector. Special attention was given to the performance of the veto in identifying and classifying muon and neutron events and the resulting 
background rates. Muon tracking possibilities were investigated as well. Preliminary results indicate that in many cases, at least
a rough track reconstruction sufficient for classification purposes is 
possible.
The parameters modified during the simulations were the dimensions of the veto 
vessel itself, the optical surface of the veto vessel, and the number and 
placement of the photomultipliers. 

The Double Chooz Monte Carlo simulation is 
based on the Geant4.7.0p1 simulation 
framework provided by 
CERN\cite{Geant4},
GLG4sim\cite{GLG4sim},
which is a generic liquid scintillator neutrino experiment package for 
GEANT4, and a dedicated extension 
to this package called DCGLG4sim. In addition, the muon propagation 
tools MUSIC\cite{bib:music}
and MUSUN\cite{bib:musun}
were used to obtain muon spectra for the detector 
sites, as they are shown in plots \ref{fig:iv_far_mu_energy} and 
\ref{fig:iv_near_mu_energy}. 

\begin{figure}[hbtp]
	\centering
		\includegraphics[width=0.60\textwidth]{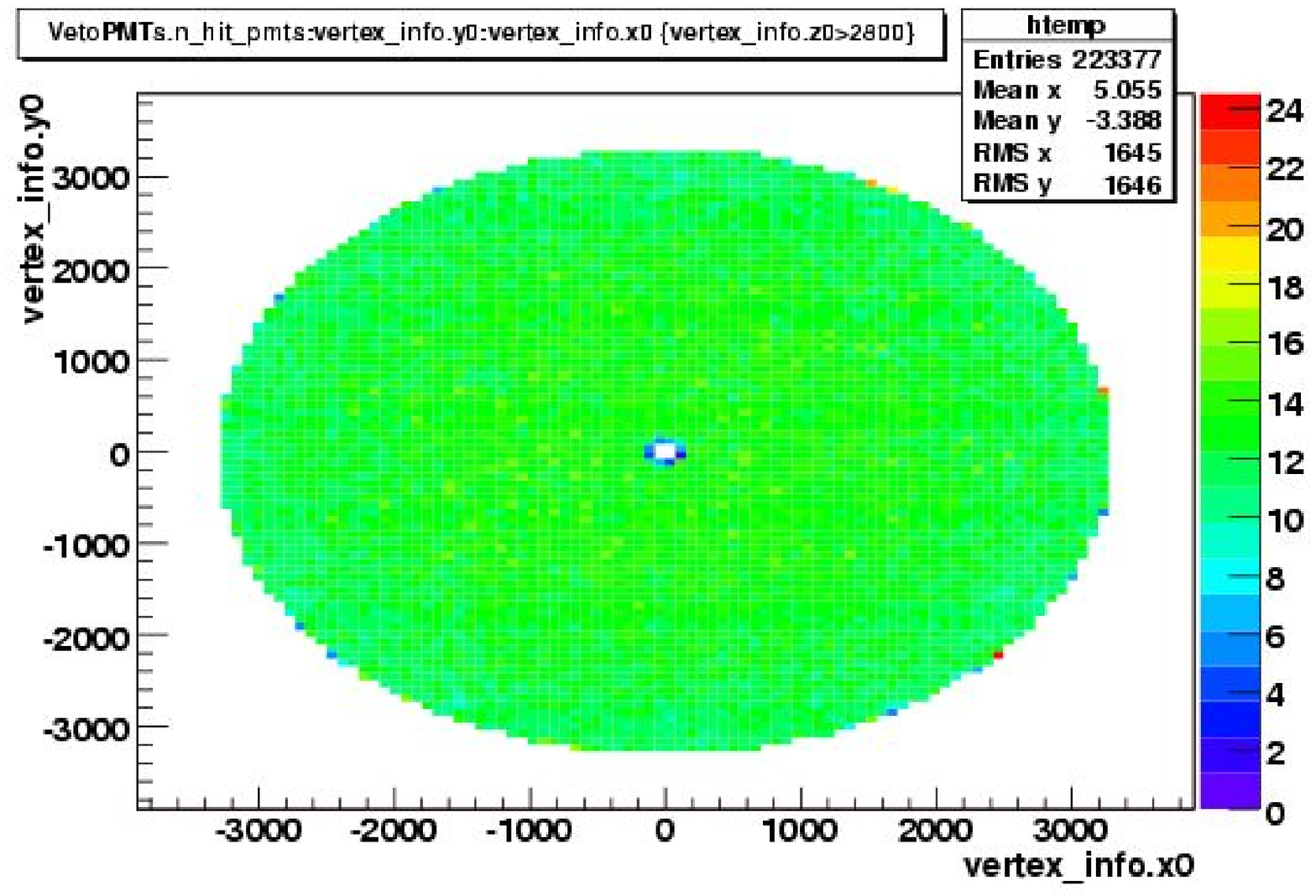}
		\caption{\label{fig:iv_top}5 MeV positrons started at a random point in the inner veto above the buffer vessel show a homogeneous detection multiplicity. The ring in the center corresponds to the neck.}
\end{figure}

\begin{figure}[hbtp]
	\centering
		\includegraphics[width=0.60\textwidth]{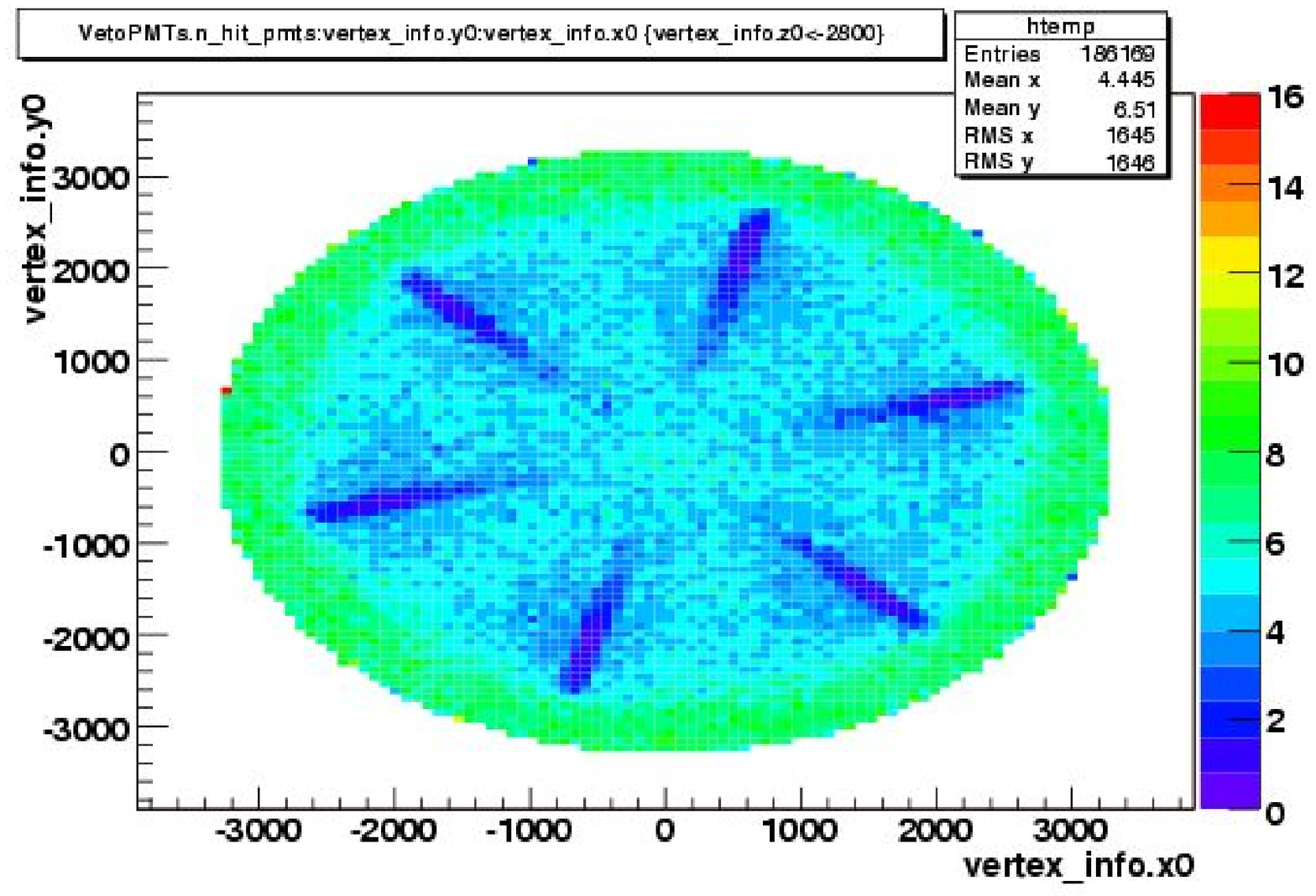}
		\caption{\label{fig:iv_bottom}The support structure at the bottom of the veto strongly deteriorates photon propagation, leading to a position dependent multiplicity.}
\end{figure}

\begin{figure}[hbtp]
	\centering
		\includegraphics[width=0.60\textwidth]{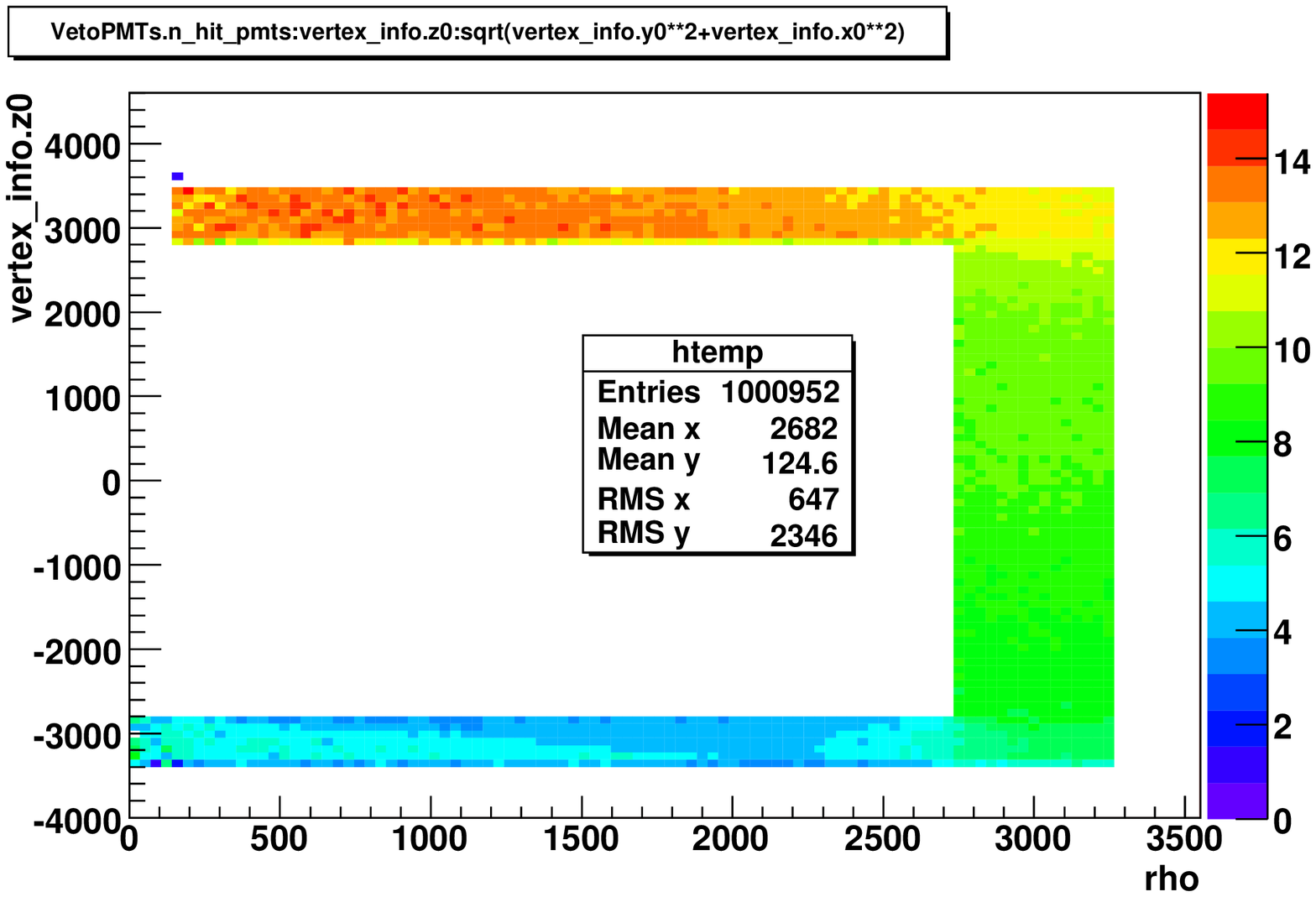}
		\caption{\label{fig:iv_side}This side view shows the number of hit PMTs depending on the position of energy deposition. The effect of the support structure below the veto tank is visible.}
\end{figure}

\begin{figure}[hbtp]
	\centering
		\includegraphics[width=0.70\textwidth]{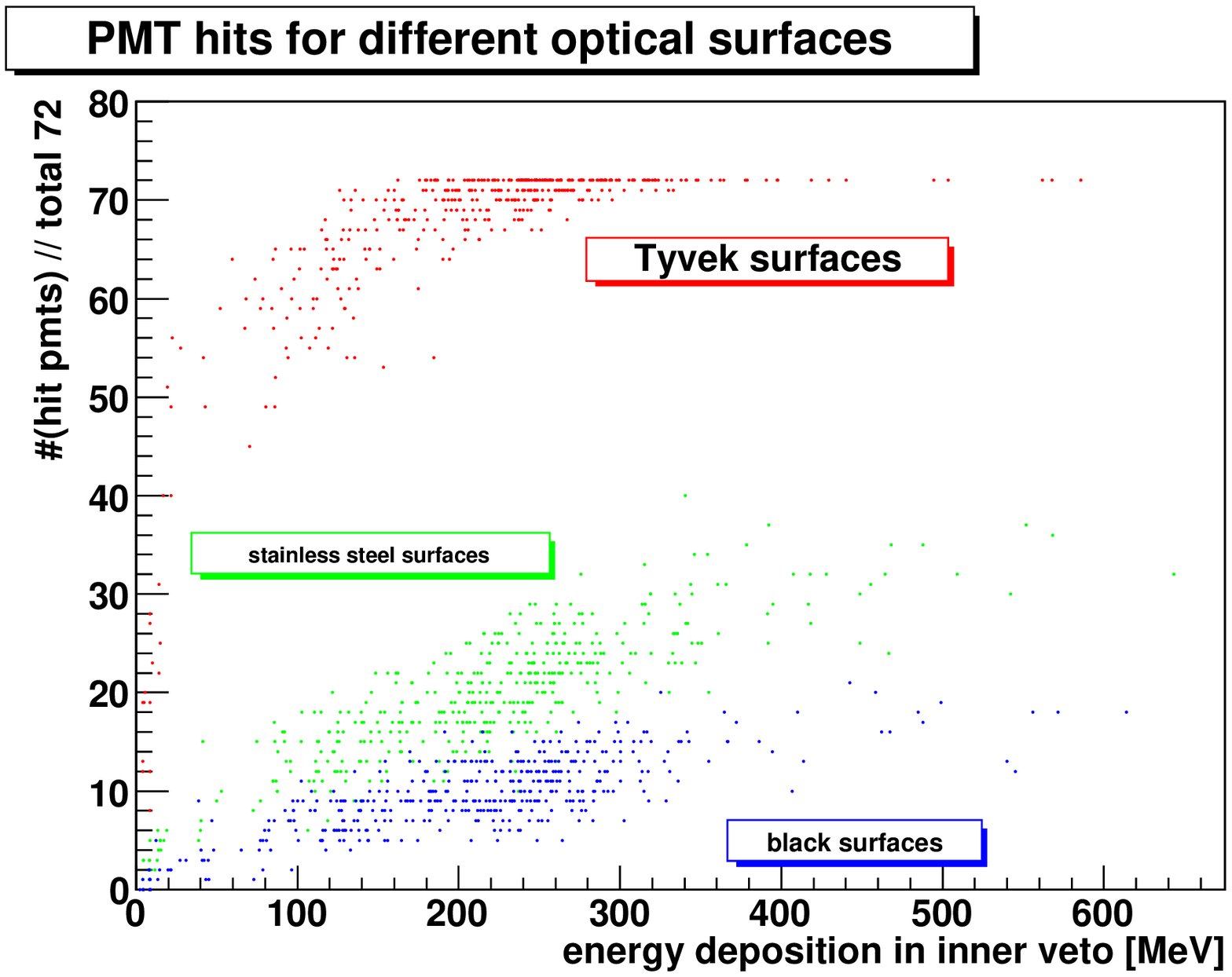}
	\caption{\label{fig:iv_tyvek}
	Comparison of different veto surfaces, using a previous PMT layout. The effect of high reflectivity surfaces is only weakly dependent on the details of the PMT distribution and equally valid for the updated 78 PMT geometry.}
\end{figure}

Choosing a sensible multiplier start distribution, local energy depositions of 5 MeV
by positrons were uniformly distributed throughout the veto volume to test for locations 
where only few photomultipliers would register the scintillation light. By moving 
and adding photomultipliers, then performing the same simulation, and iterating the 
process, the multiplier distribution was optimized to give an homogeneous response 
regardless of the place of energy deposition. Figures~\ref{fig:iv_top}, 
\ref{fig:iv_bottom} and \ref{fig:iv_side}
show maps of the detector, where the color represents the average number of 
photomultipliers hit if an energy deposition in this 
part of the detector happens. 
In parallel, the effect of changing veto thickness and optical surfaces of the veto 
volume were investigated. As the total volume of the far detector is limited by the 
existing pit, there is some motivation for decreasing the veto thickness to increase the target size. It could be shown that reducing the veto 
thickness from 60~cm down to 50~cm does not significantly lower the muon identification power of the veto. 
Simulations indicate that due to the high reflectivity surfaces mentioned previously,
six times as many photons are detected compared to stainless steel surfaces, as shown in 
Figure \ref{fig:iv_tyvek}. Furthermore, the improved reflectivity makes it possible to achieve sufficient light collection with only one ring of side PMTs and facilitates the installation of the buffer vessel, which will be
welded together inside the detector pit. With the usage of Winston cones, the side PMTs might even be unnecessary. Research and simulation work concerning this are in progress.
	  

Using the information gathered with these simulations, the veto detector layout was defined. This allowed starting high statistics runs with either
muons or neutrons as primary particles and analyzing their output regarding both the rejection power of the muon veto, as well as its classification capabilities. 

To estimate the veto performance, two different definitions of \lq\lq rejection power\rq\rq
are useful. We will call the term
\begin{equation}
R_{\mu} = \frac{\rm{detected\,muons\,with\,energy\,deposition\,in\,the\,detector}}{\rm{all\,muons\,with\,energy\,deposition\,in\,the\,detector}}
\end{equation}
the muon rejection power, while
\begin{equation}
R_{c} = \frac{\rm{correlated\,events\,due\,to\,muons\,recognized\,in\,muon\,veto}}{\rm{all\,correlated\,events\,due\,to\,muons}}
\end{equation}
will be called the muon correlated event rejection power.

Both quantities require the definition of a \lq\lq detected event \rq\rq in the inner veto.
Several conditions can be imposed for such an event. One can demand a certain number of hit photomultipliers (i.e. a multiplicity threshold), in a typical time window of a few tens of ns. In addition, a threshold on the number of photoelectrons that are required for a valid photomultiplier hit is applied. Some of the muon events are seen only by few multipliers due to geometry reasons. Therefore, they have a low multiplicity, but possibly high individual PMT signals and can be identified by imposing a threshold for the sum of all collected signals. More sophisticated conditions divide the PMTs in different groups (e.g., rings, sectors, ...) and use the group signals in addition to individual and summed data to decide whether and which kind of a muon event has happened.

At the same time, there are additional practical considerations. Depending on the expected dark rate of the PMTs 
(about 4 kHz for the ETL tubes), a very low multiplicity condition may result in a high false trigger rate. 
High gain, which will be favorable for neutron detection, might result in fast saturation for muon events with long tracks in the veto.

\begin{figure}[hbtp]
	\centering
		\includegraphics[width=0.80\textwidth]{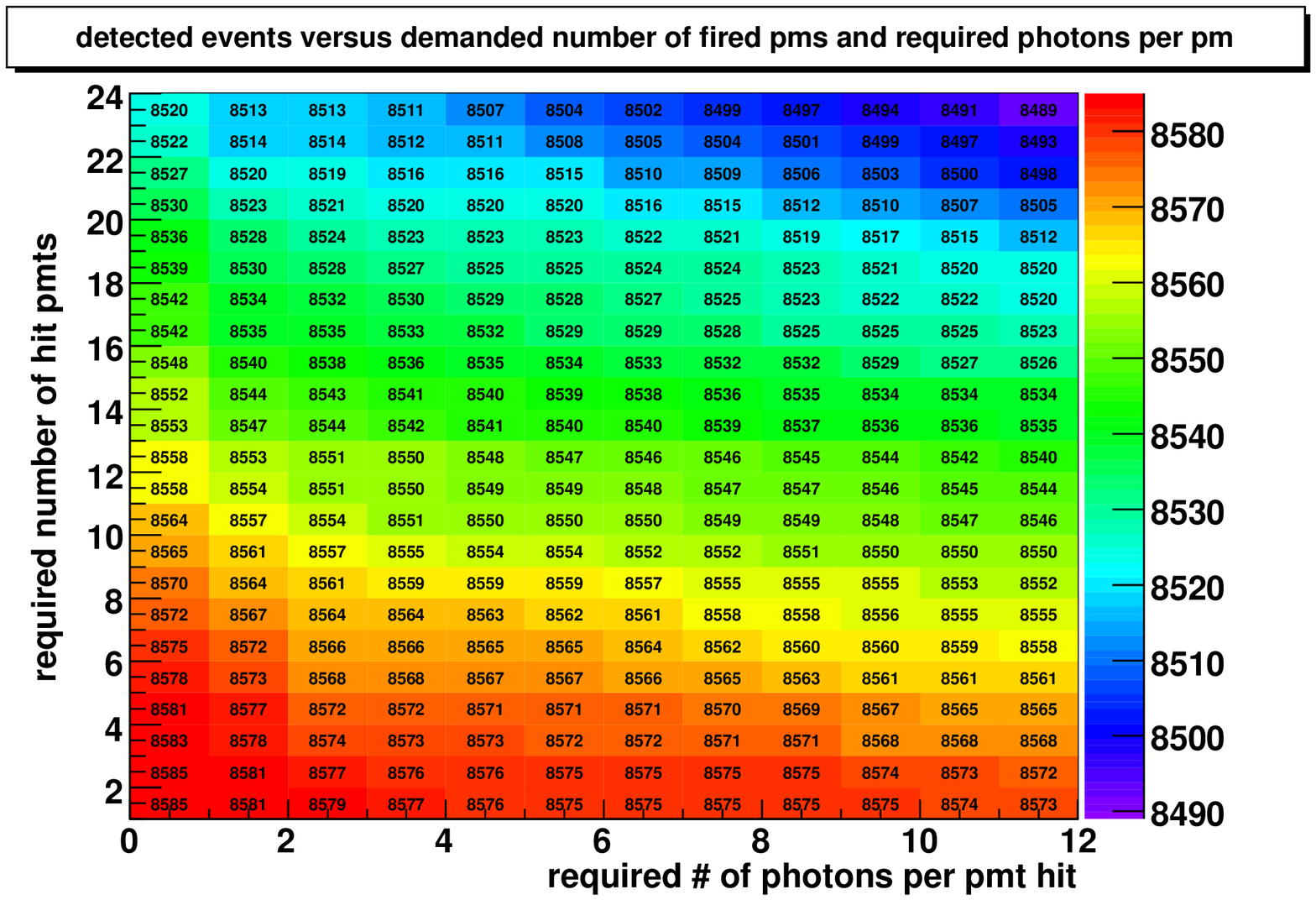}
	\caption{\label{fig:iv_muon_identification_zoomed}
	Muon identification probability plot for a geometry with 60 PMTs. In total, 10000 muons were simulated, 8640 of which generated an energy deposition in the veto or detector. A restriction to events with multiplicity of at least 15 and a PMT threshold corresponding to 5 p.e. would results in 98.8\% muon rejection power without threshold on the energy deposition in the inner detector. The remaining 1.2\% of events are due to very low energy deposition inside. Already a threshold of 10 keV increases this value to 99.5\%.}
\end{figure}

In a simple scheme to get a multiplicity-optimized trigger condition, one just checks the scaling of the rejection power
for variations of both multiplicity and required p.e. per PMT, and chooses conservative settings to compensate
for simulation uncertainties. Figure~\ref{fig:iv_muon_identification_zoomed} gives the number of identified muons as a function of multiplicity threshold and PMT threshold for a sample of 8646 muons.

To determine the muon correlated event rejection power $R_c$, we impose the following restrictions for a correlated event: there has 
to be an prompt visible energy deposition of 1-8\,MeV in the target or 
$\gamma$-catcher, followed by a visible energy deposition
of about 8\,MeV up to 200$\,\mu s$ later. In the current largest sample of muon events (about 100.000 events) with full scintillation,
for all 30 events where these conditions are fulfilled, there is an identified muon in the veto, even when demanding more than 25 triggered PMTs and at least 15 p.e. per PMT. 
For reliable statistics, it is clear that a larger number of events has to be simulated. Currently, simulations are limited by
calculation speed, as a single muon event with full scintillation processes takes about 30 seconds. To increase simulation speed,
two main ideas are investigated. Simulation of secondary particles for events where the primary muon already deposits
large amounts of energy in the veto can be skipped, spending full simulation time only on events where the identification
might depend on secondary particle energy deposition. A huge speedup (on the order of factor 100) 
can be achieved if it is possible to omit scintillation
simulation in favor of simple energy deposition simulation. This might be justified if we can generate a map of the veto volume 
which links energy deposition in a certain region to a PMT hit pattern.

\begin{figure}[hbtp]
	\centering
		\includegraphics[width=0.60\textwidth]{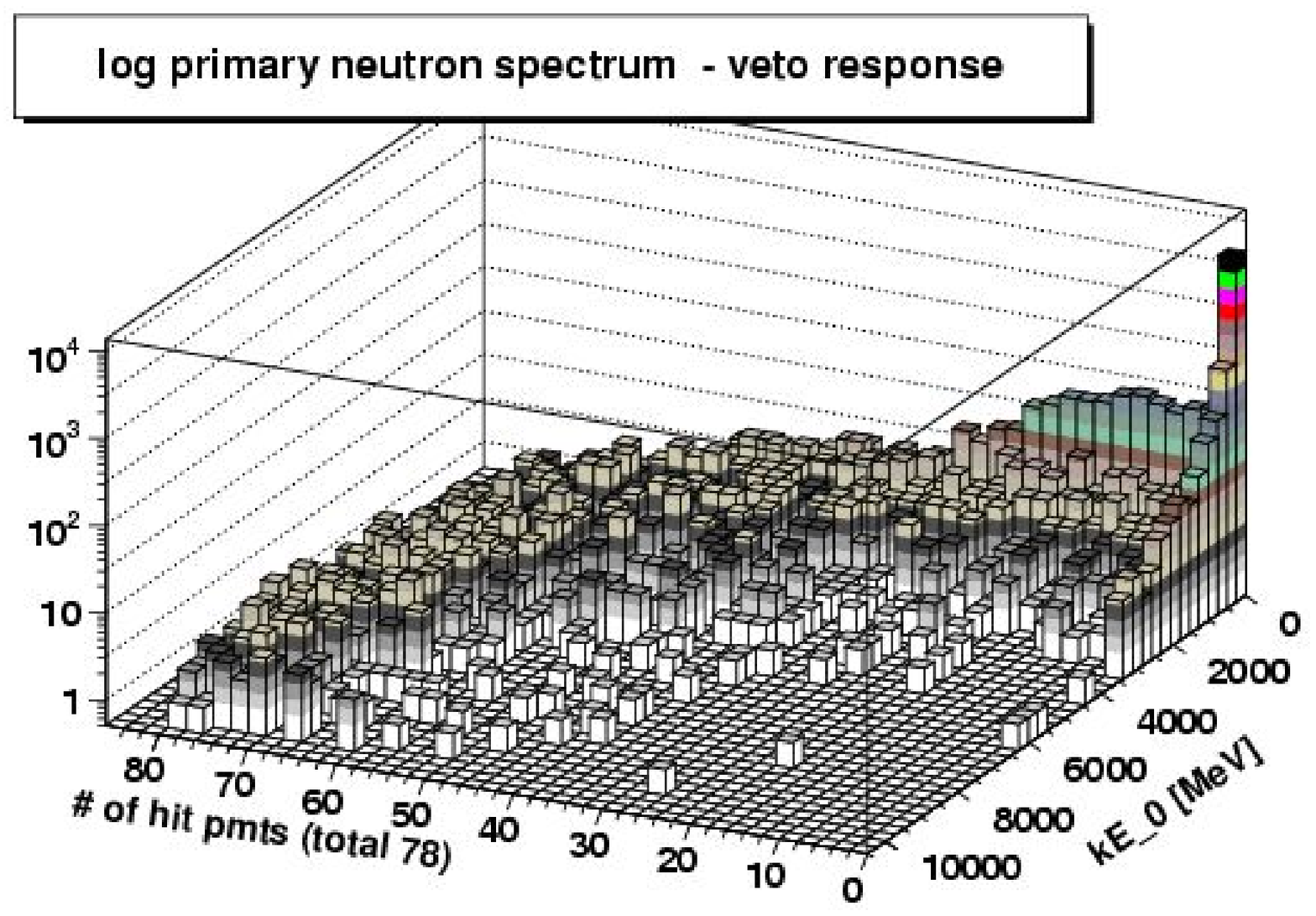}
	\caption{\label{fig:iv_log_neutrons}Neutrons with kinetic energies ($kE_0$) exceeding 3 GeV create large amounts of charged secondaries, resulting
	in strong photon signals.}
\end{figure}

\begin{figure}[hbtp]
	\centering
		\includegraphics[width=0.60\textwidth]{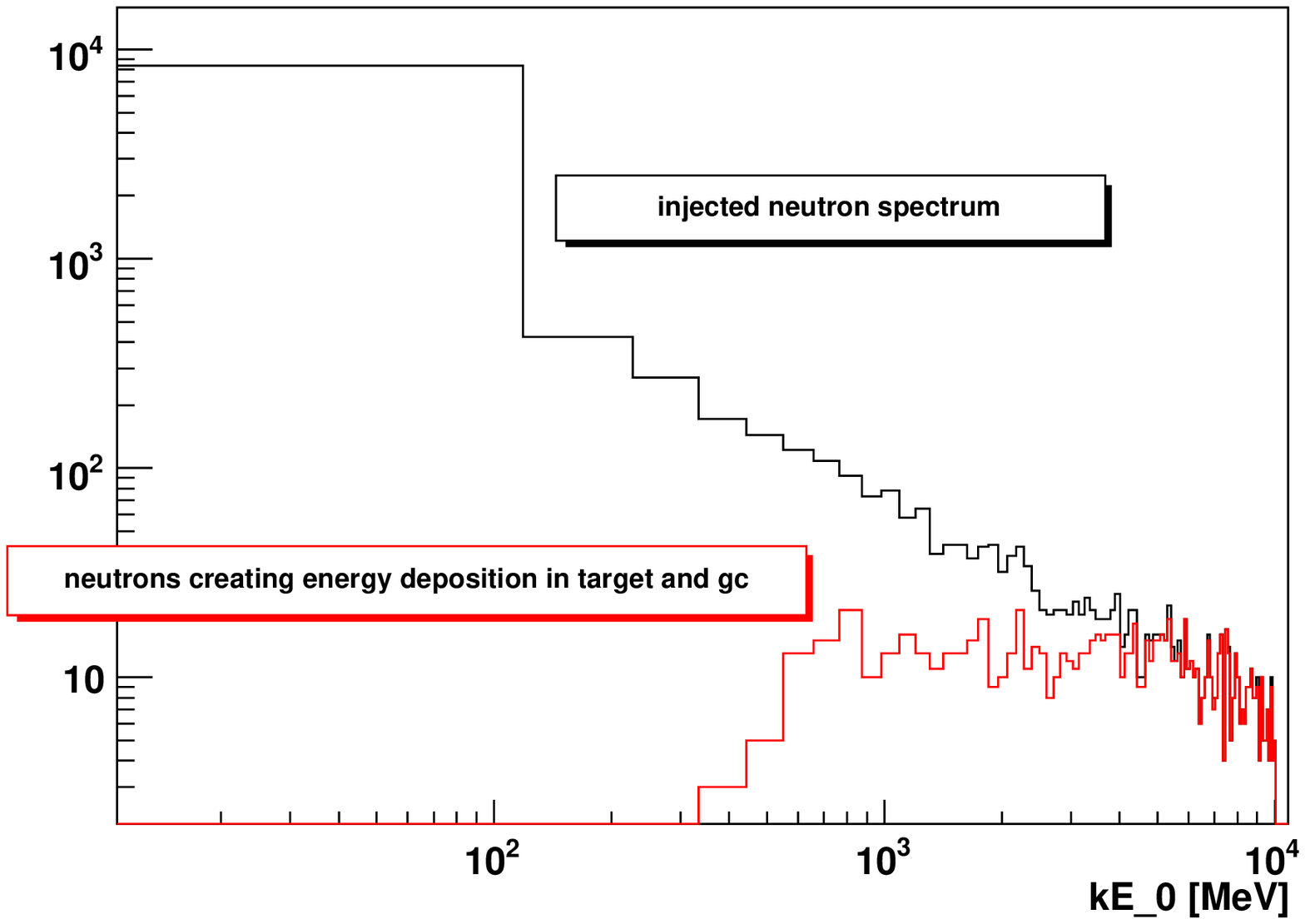}
		\caption{\label{fig:iv_log_neutrons_range}Low energy 
neutrons have a very low probability to reach the target or 
$\gamma$-catcher  if they are created outside the veto.}
\end{figure}

To explore the feasibility of rejecting neutrons in particular, first simulations were done with a logarithmic energy 
distribution to check the veto response to neutrons of all energies with comparable 
statistics. The result is shown in Figure~\ref{fig:iv_log_neutrons}. As can be seen, high energy neutrons are generating
large PMT signals and therefore have a high detection probability. Low energy neutrons, on the other hand,
have only a slight probability of reaching the veto, as their range is limited (see Figure~\ref{fig:iv_log_neutrons_range}). 
Most dangerous are neutrons in the medium energy range, as they can reach the inner vessels without necessarily 
producing strong veto signals.

\begin{figure}[hbtp]
	\centering
		\includegraphics[width=0.60\textwidth]{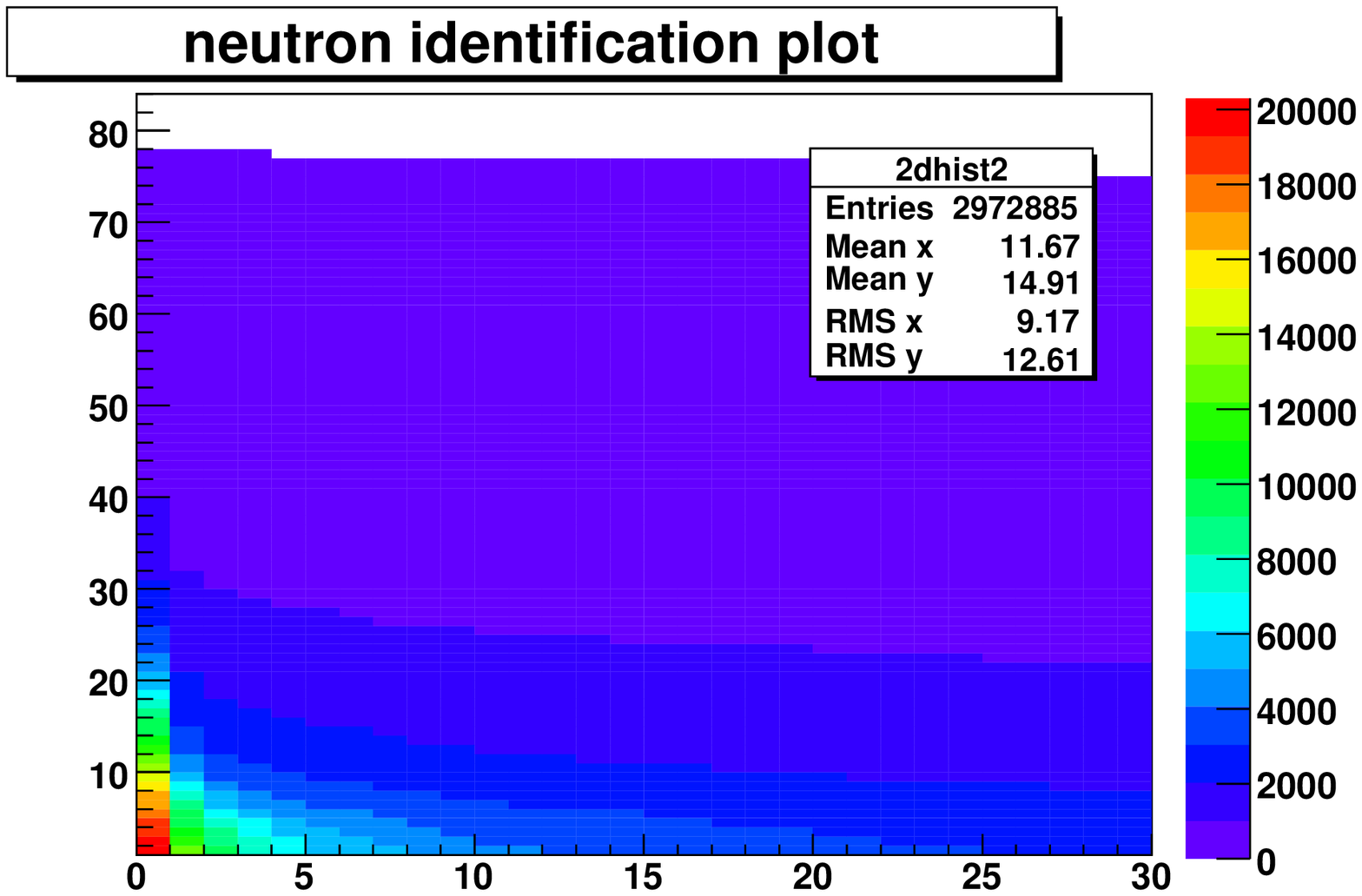}
	\caption{\label{fig:iv_far_neutrons_identification}22185 simulated neutrons out of 6800 lead to a energy deposition in detector and/or veto. For most neutron events, only one p.e. per PMT is registered. Imposing single p.e. threshold for the PMTs and a multiplicity threshold of 8, this results in a neutron rejection power of 71\%.}
\end{figure}

\begin{figure}[hbtp]
	\centering
		\includegraphics[width=0.60\textwidth]{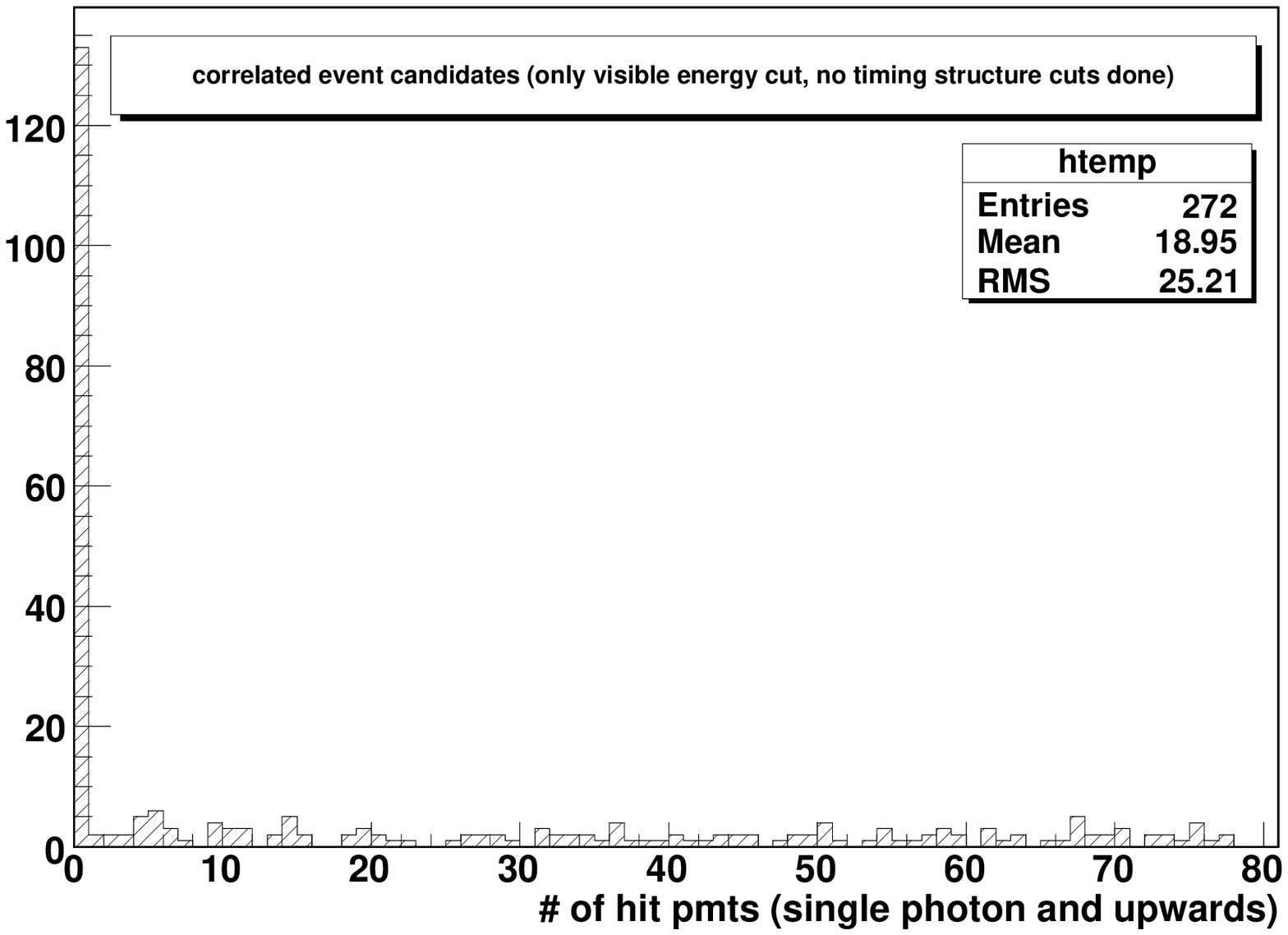}
	\caption{\label{fig:iv_correlated_candidates}Starting from neutrons 
as primary particles, events with sufficient energy deposition for 
a correlated event in the target and/or 
$\gamma$-catcher  are selected. About 30-50
	percent of these are detected in the muon veto in this scenario.}
\end{figure}

By using only the energy spectrum of the secondary neutrons and injecting them 
horizontally into the detector, it is possible to simulate a worst case scenario, 
where a neutron created right outside of the detector with no accompanying shower 
particles covers the minimum distance in the veto on its way to the inner vessels. 
The resulting rejection power can be regarded as a lower limit for the general 
case. 
Figure~\ref{fig:iv_far_neutrons_identification} shows the 
same plot for the neutron identification as 
Figure~\ref{fig:iv_muon_identification_zoomed} for muons. 
Comparison immediately shows neutron identification to be 
much more ambitious, due to the far lower amount of light 
generated. The neutron rejection power will therefore be 
significantly lower than the muon
rejection power, as expected. 
Rejection power for candidates for neutron correlated events are shown in Figure~\ref{fig:iv_correlated_candidates}. Performing only
a minimum energy deposition cut for target and 
$\gamma$-catcher  of 10\,keV, one selects 272 neutrons out of 68000 injected. Of these,
 roughly one half does not show up in the veto at all. About 30-50 percent are detected in the muon veto, i.e. the thresholds conditions are fulfilled.

Another topic that was investigated with the simulation code was classification of events. Classification could be useful in two different situations. There will be an online
classification scheme to decide how much information for a given event is to be saved. A
second and more thorough classification can be achieved in the offline analysis of the data, where
one has complete access to data registered both before and after the event under consideration.
An important aspect of classification is track reconstruction.
The scheme envisioned for track reconstruction starts with the first 
significant energy deposition identified with the photomultipliers. The velocity of muons 
is comparable to c, which gives us a time window in the ns range in which to expect 
signals from other multipliers, varying with their distance from the point of first 
detection. Crossing muons will produce such delayed signals and therefore allow 
classification. Events where no such delayed multiplier signal is found can either be 
stopped muon events or correspond to a muon track cutting through only a short section of 
the veto. By reconstructing the position of the first signal, some stopped muon events 
can be identified without doubt, for instance if the first signal indicates a position 
inside the inner top multiplier ring. A careful analysis of simulated events allows 
probability statements for the other cases.
To incorporate some ordering structure for both simulation and experimental output, the following 
event classification scheme can be applied:
\begin{itemize}
\item Muons can cross the detector, either only through the veto or through inner 
vessels as well. These events will be called \lq\lq veto crossing muon event\rq\rq or 
\lq\lq inner crossing muon event\rq\rq, respectively. 
\item Muons that enter, but do not leave the detector (due to capture or 
decay) will generate \lq\lq stopped muon events\rq\rq. As muon capture 
is an important source of hard-to-veto background, the veto classification capability 
for this kind of event is of great importance. 
\item If a muon event is accompanied by a large 
amount of charged secondary particles, we will speak of a \lq\lq showering muon event\rq\rq. 
\item A
showering muon missing the detector can lead to a \lq\lq partially contained shower event\rq\rq, 
where a sizable amount of secondaries are registered without the primary muon. 
\item Other 
event classes comprise \lq\lq high energy photon events\rq\rq, 
\item \lq\lq internal decay events\rq\rq resulting 
from alpha, beta or gamma emitters residing in the detector material, 
\item and \lq\lq neutron events\rq\rq, which are detected via recoil protons. 
\end{itemize}
The simulations performed are currently analyzed in aspect to both to the rejection power for muon and neutron 
events, as well as its classification capabilities. 

\begin{center}
\begin{table}[hbtp]
		\begin{tabular}{cc}
			\includegraphics[width=6.0cm]{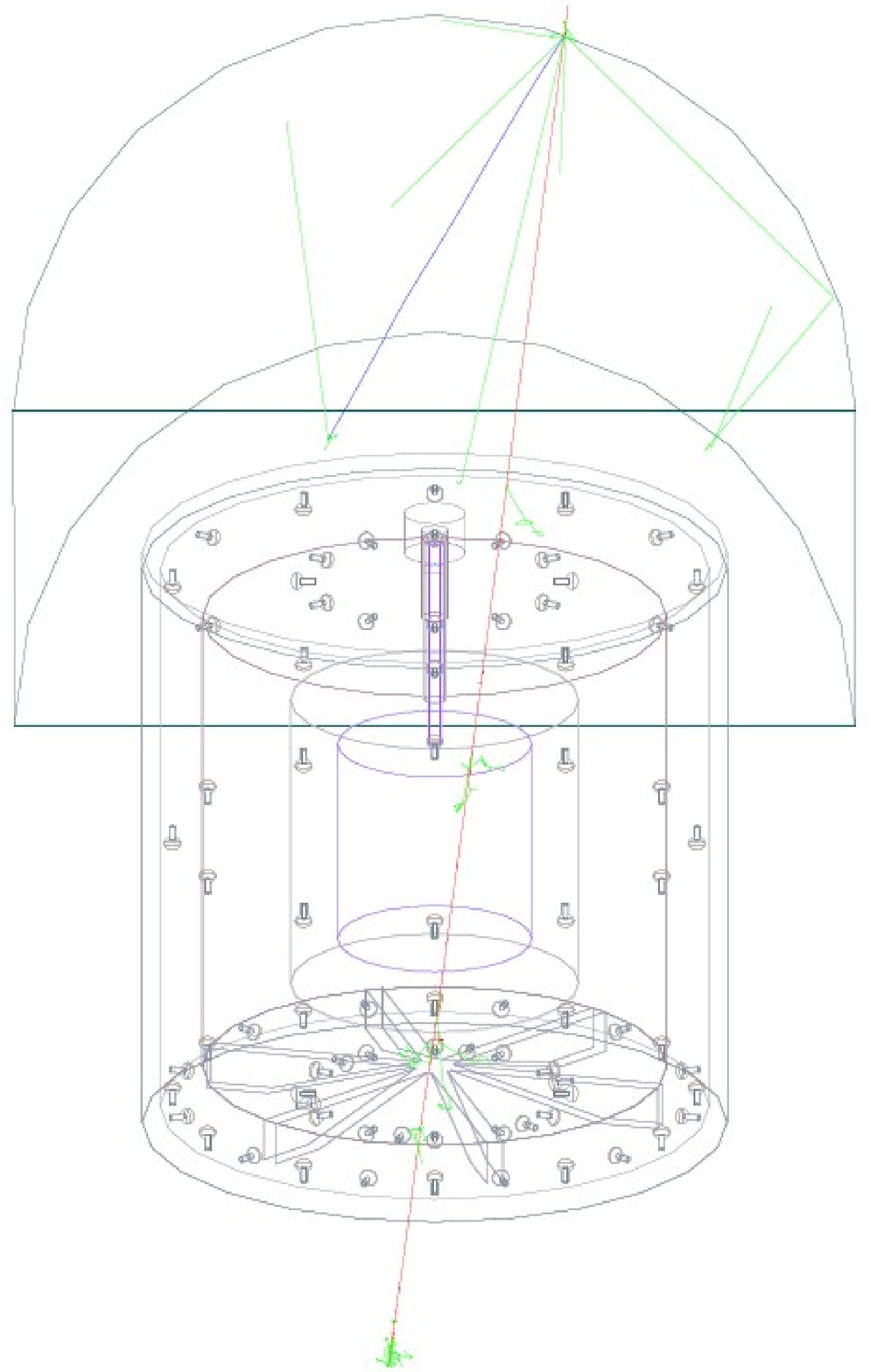} & 
\includegraphics[width=6.0cm]{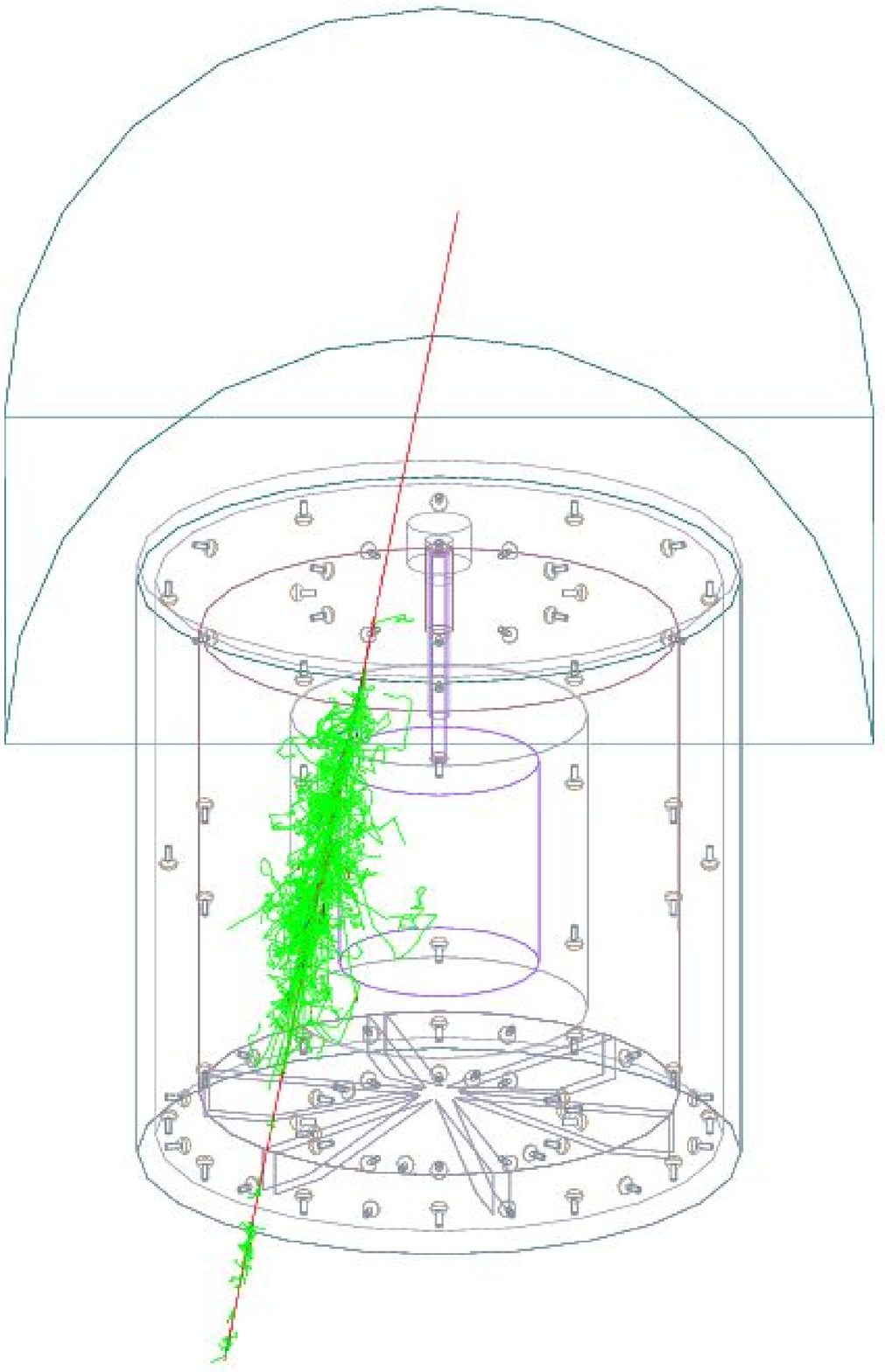}\\
			\includegraphics[width=6.0cm]{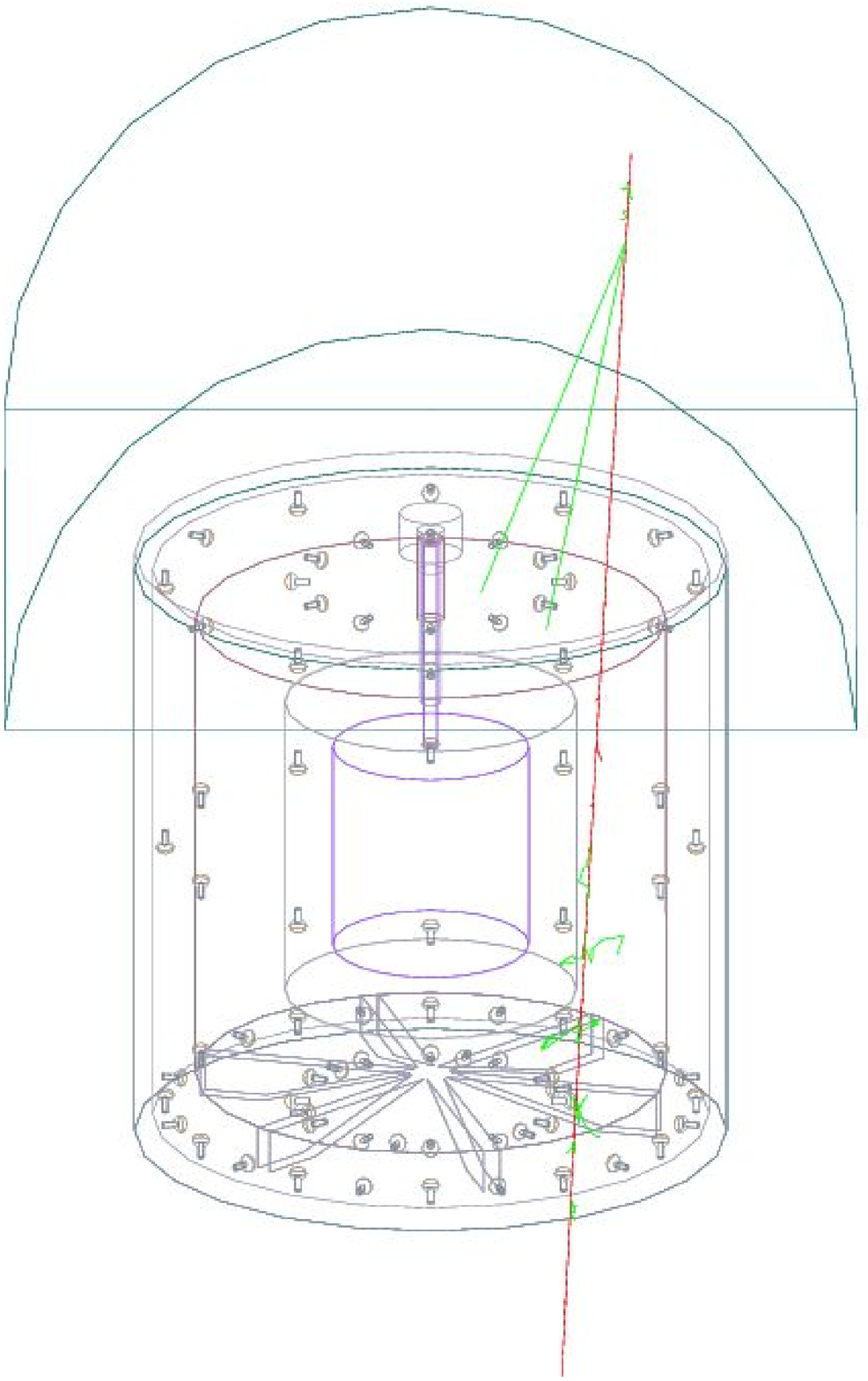} & 
\includegraphics[width=6.0cm]{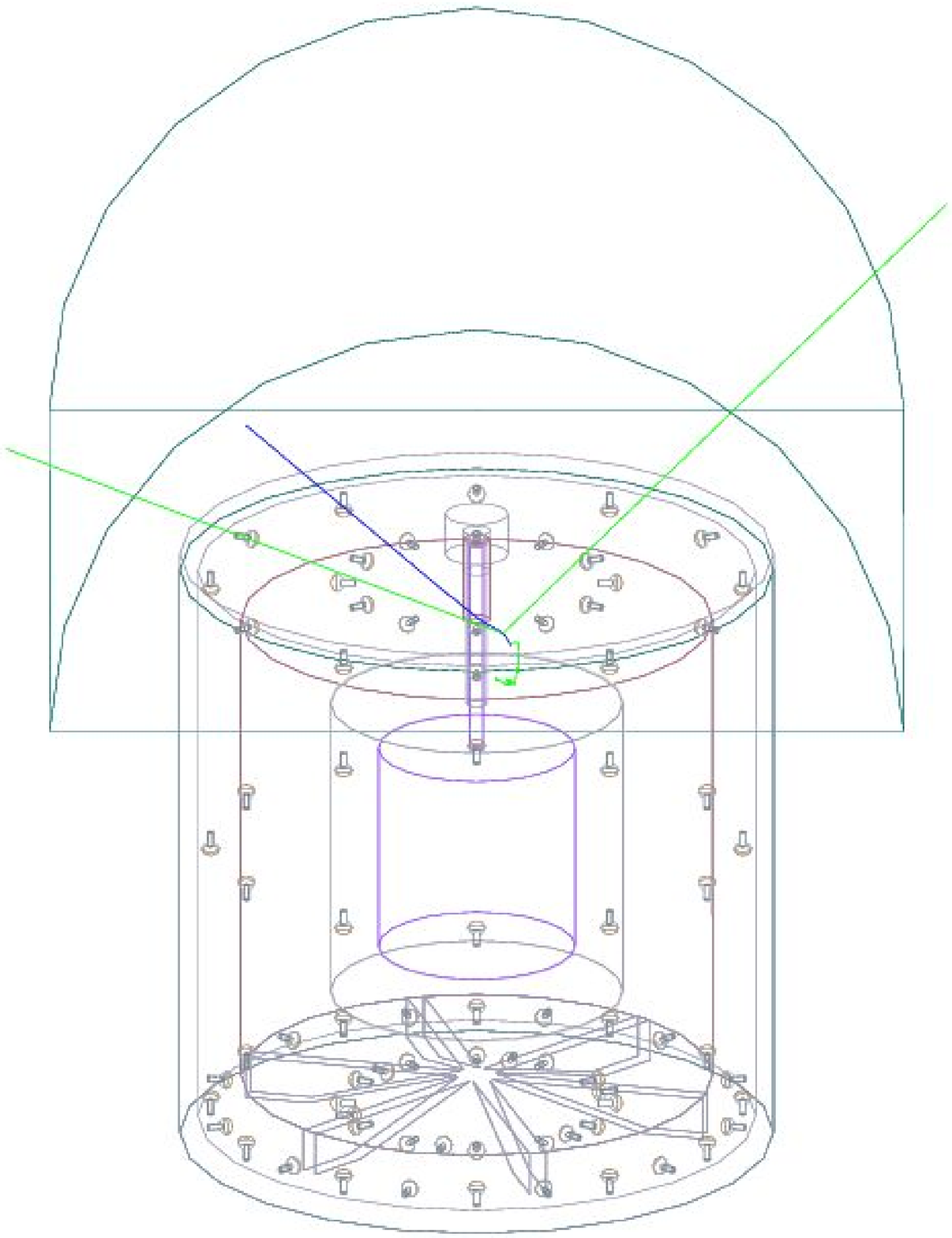}\\
		\end{tabular}
\label{fig:iv_events}
\vspace{0.3cm}
\caption{Typical muon events (red: $\mu^-$, blue: $\mu^+$): 
a through-going muon (upper left), 
a shower event next to the target (upper right), 
neutron creation near to the target (lower left) and a 
stopped muon (lower right).}
\end{table}
\end{center}
\subsection{Magnetics}
\subsubsection{Sensitivity to Magnetic Fields}
It is a well-known fact that large-area PMTs are subject 
to significant variation in collection efficiency due to 
magnetic fields on the order of the Earth's field - 
i.e. a few 100 mG. For most applications this may not be important
as such variations can be removed or reduced by calibrations. For 
detectors using very large PMTs such as Super-Kamiokande
and KamLAND, compensating coils are used to reduce the 
field below 100 mG (the specification for Super-K). In the
case of Double Chooz, the effects of magnetic fields are 
potentially more serious for two reasons:

\begin{itemize}
\item Since it is necessary to build two identical detectors, differences in the magnetic environment between the
Near and Far detectors can lead to non-negligible systematic differences between them
\item The necessity of reducing the singles rates from rock gammas to reduce accidental correlated events
leads to including a relatively thick steel shield (17 cm or more) on the outside of the detector, roughly 65 cm from
the first dynode of the PMTs. This mild steel
could have a substantial permanent magnetic dipole moment. In addition, there could be significant differences
in this moment between the Near and Far detector.
\end{itemize}
Several possible magnetic field mitigation schemes have been investigated:
\begin{itemize}
\item Replace 20-cm iron shield with 10-cm lead shield and use 
Helmholtz coils for Earth mag. field compensation. Replacement of the steel shield with 
lead after the construction is not practical. This option was investigated and
found to be rather expensive;
\item Global mu-metal shielding of PMTs. A mu-metal shield could be installed around the
buffer vessel to reduce the magnetic fields inside the whole detector to acceptable levels. A shielding
engineering company (Amuneal Manufacturing) was called in to make a preliminary design and costing
for such a shield. This option, while feasible, turns out to be also quite expensive;
\item Provide individual magnetic shielding for every PMT in the system.
This shielding might be a simple mu-metal cylinder extending the length 
of PMT for about 10-15 cm above the PMT apex. Such shielding will 
restrict PMT angular acceptance and would have to be compensated for by installation of Winston cones.
Construction of these hybrid cones and accompanying complicated PMT supports represents 
significant additional cost to PMT system.
%
\item The effects of the steel shield could be reduced by measuring the overall dipole moment
of individual pieces (which are in the shape of elongated rectangles) and assembling the shield in
such a way that adjacent pieces have opposite directions. Particularly high-field pieces can be
``de-permed'' in a similar fashion to how Navy ships are demagnetized, a common practice for over 60 years.
Compensating coils can then be installed on the outside of the detector to trim the resultant magnetic field
down to acceptable, near-identical levels in the Near and Far detectors. 
\end{itemize}
The last option listed above is now our baseline one, as it is the least expensive and has a relatively
small impact on the construction schedule.
\subsubsection{Size of the Effect}
\begin{figure}[ht]
\centerline{\includegraphics[width=0.5\textwidth]{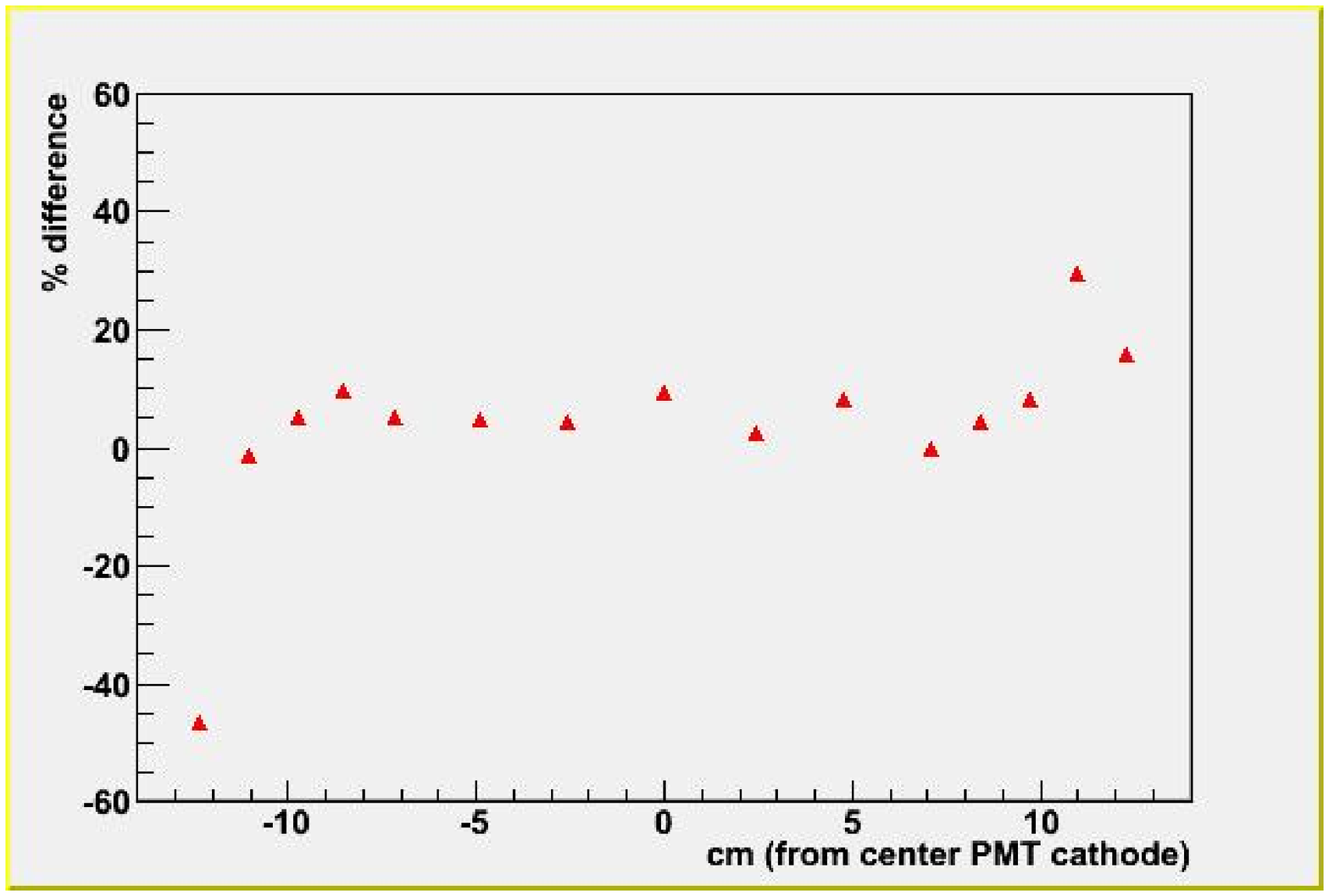}\includegraphics[width=0.5\textwidth]{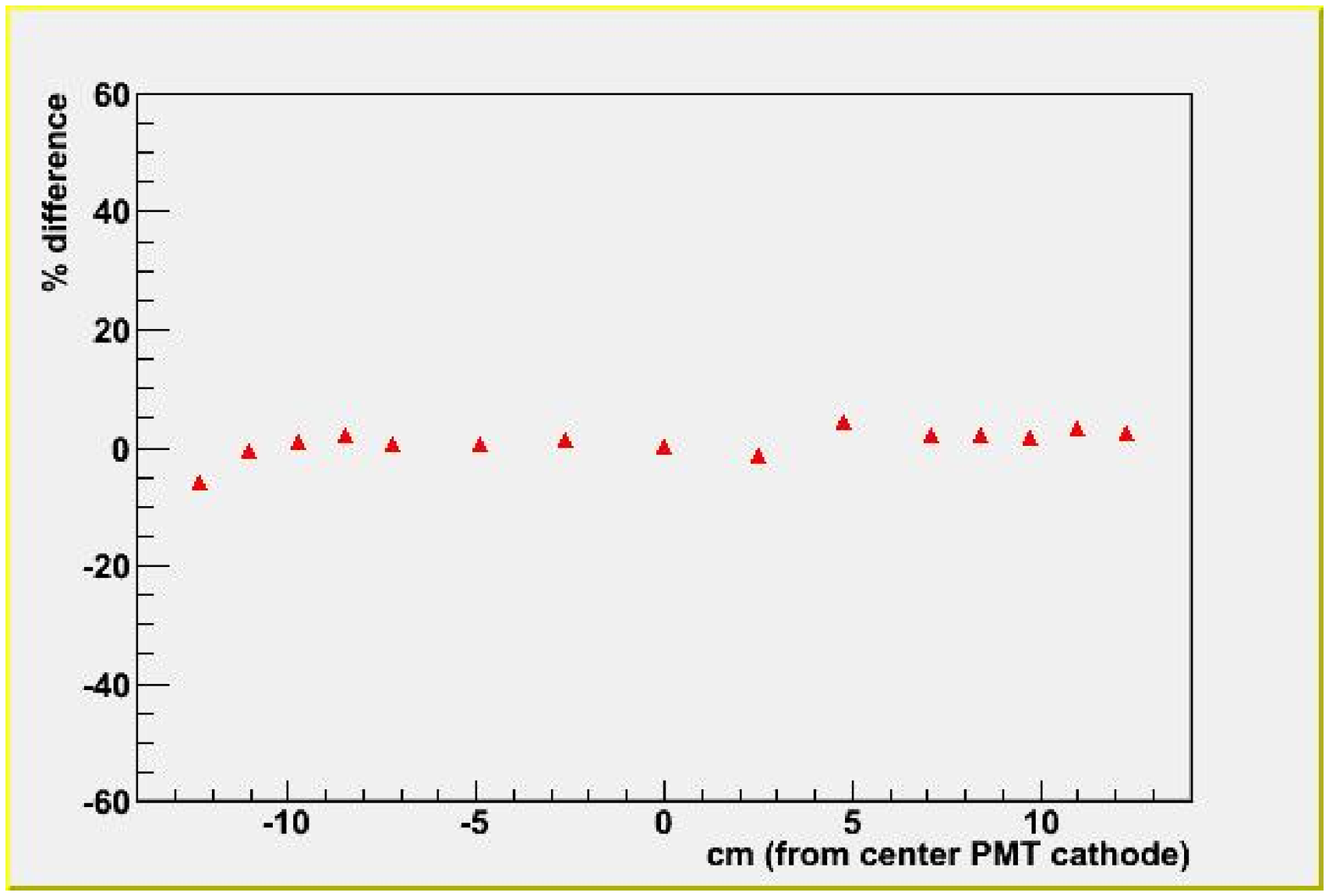}}
\centerline{\includegraphics[width=0.5\textwidth]{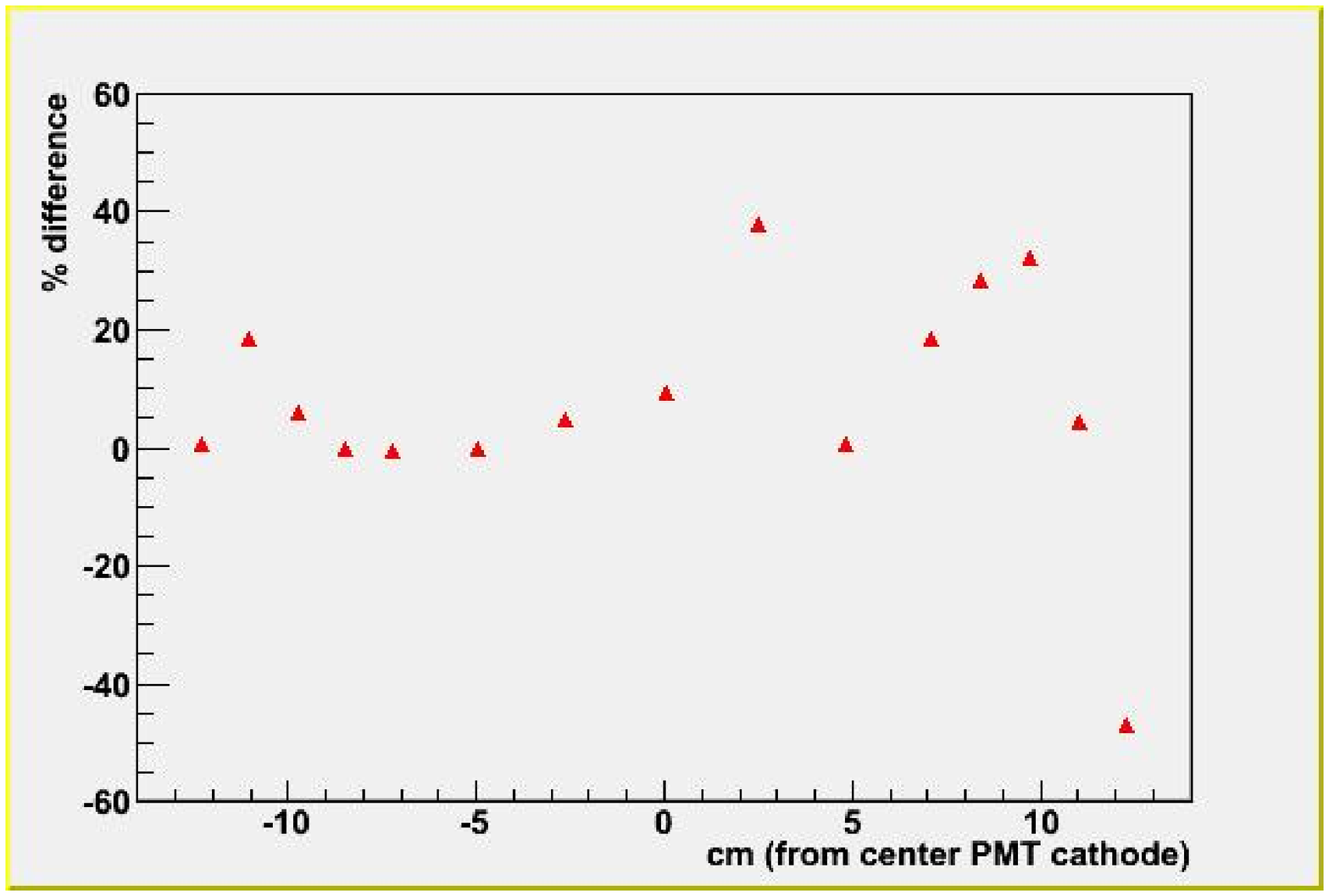}\includegraphics[width=0.5\textwidth]{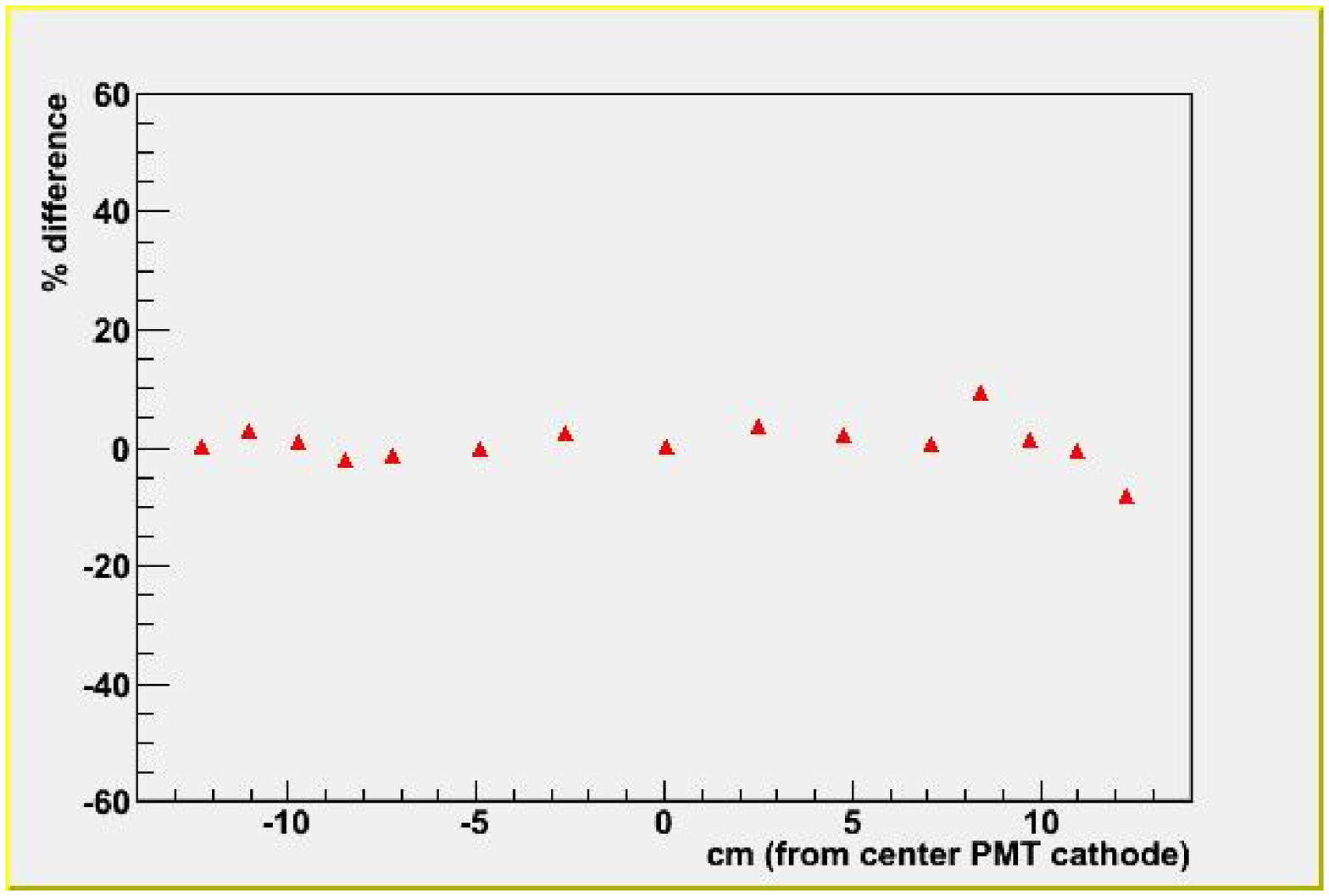}}
\caption{Percent variation in collection efficiency caused by the rotation through ninety degrees of an 8 inch PMT as
a function of position across the face. Top (Bottom) is along the x-axis (y-axis). The right hand figure
shows the same rotation inside a mu-metal shield. The x- and y-axes are defined as in the previous figure.}
\label{F:LSU_R5912}
\end{figure}

The relative collection efficiency of a Hamamatsu R5912 8 inch PMT is shown in 
Figure~\ref{PMTmagnetic} for a PMT with front face illumination from a point source
on the central axis of the PMT (provided courtesy of Hamamatsu 
Photonics). \footnote{Although we use
a Hamamatsu PMT to illustrate the point, the magnetic characteristics of all candidate PMTs
is now under identical evaluation by laboratories inside the Double Chooz Collaboration. It is not
expected that there will be significant differences between candidate tubes.} The x-axis shows the
magnitude of the magnetic field applied along the x- and y-axis. Note that $5\times10^{-5} T$ $\sim 500 mG$ is about
the magnitude of the earth's field at the Chooz site. These data imply that if no action is taken then there would
be a roughly 10\% asymmetry introduced into the detector due to the Earth's field alone.\\
%
%
More detailed measurements made on a Super-K Hamamatsu R5912 by our collaboration indicate that there can also
be bias based on the position the photon strikes the PMT
cathode, adding further complication.  Using the same coordinate
axis as in Figure~\ref{PMTmagnetic}, Figure~\ref{F:LSU_R5912} shows 
the variation in sensitivity across
the front face of an R5912 which is sitting vertically in the Earth's field and rotated 90 degrees about the
vertical axis. These tests were done at LSU, where
the Earth's field is very similar to that expected at Chooz. Vertical/Horizontal is 426/242 mG for Baton Rouge
and 439/200 mG for Chooz. Thus this figure corresponds to a transverse field of about $2.4\times 10^{-5}$ T across
the face. The orientation of the x- and y-axis is the same as 
Figure~\ref{PMTmagnetic}. The right hand
figure shows the same test done inside a mu-metal coated dark box, where the field is reduced to less than one-tenth
of the ambient value. These figures confirm significant variation across the PMT face, especially near the edges
where the path of the electron is very sensitive to the transverse field. They also show that reducing the field
by a factor of ten essentially does away with this problem. In practice, the Super-Kamiokande collaboration
found that a factor of five reduction was sufficient even for 20 inch PMTs for reliably being able to calibrate
out any asymmetries. For Double Chooz we adopt this tested number as our design goal, which with our smaller
PMTs is a conservative number.\\

\subsubsection{De-perming Experiments}

We have determined the typical fields to be expected from mild steel of the type to be used for the Double
Chooz gamma shield using a portable Hall probe and a 10x20x90~cm steel bar sample.
For these measurements, the steel sample was elevated 1 meter above the floor on a wooden table near the
center of a large open room. At 5 cm from the center of the sample the field is on the order of the Earth's
field, about 500 mG total. This falls to roughly 350 mG at 10 cm. At 50 cm the field due to sample is not detectable
against the background field. Fitting the measured data to a single value of the magnetization ($\vec{M}$)
magnetic dipole moment density, fits the data poorly, so the actual situation is more complicated. There may be
steel pieces in which fields are abnormally high, or in which the
field is in an unusual direction. In this case, it may not be possible to use neighbors aligned oppositely to
cancel the long-range field to the desired level and it will be necessary to demagnetize (i.e. ``deperm'')
the steel piece on site. This is done by using a strong applied magnetic field and driving the steel into
saturation along one axis. The current is then reversed and the steel is polarized in the opposite direction, but
with a field roughly 10\% smaller. This alternating-current process continues with the applied field being
smoothly brought to near zero, essentially walking the steel down the hysteresis curve to near magnetization.\\

\begin{figure}[ht]
\centerline{\includegraphics[scale=0.50]{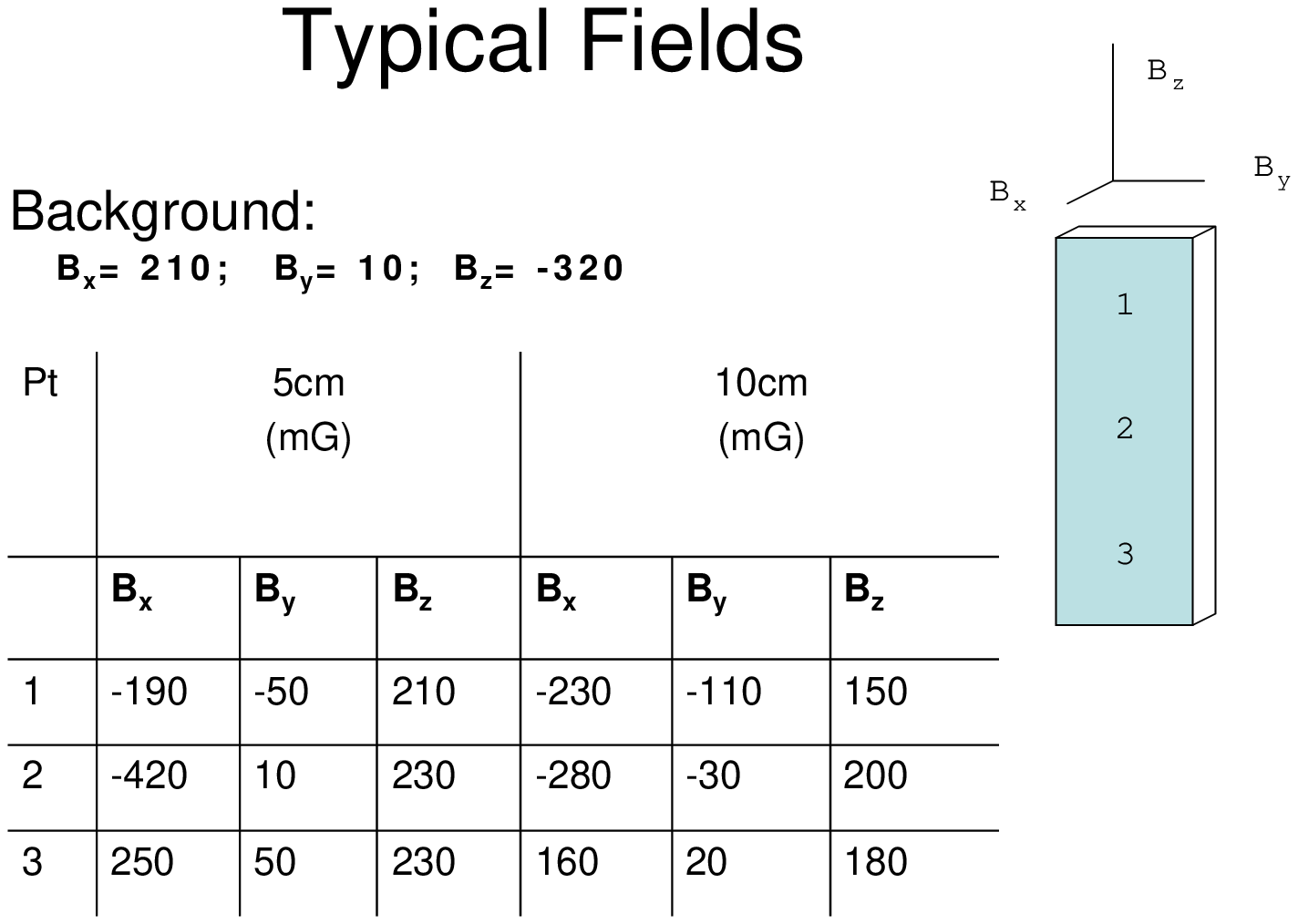}}
\caption{The fields measured near a sample piece of carbon steel using a Hall probe. Background from the
Earth's field is subtracted. Uncertainty is roughly $\pm 50$ $mG$.}
\label{F:Typical_Fields}
\end{figure}

Experiments have been performed with small pieces of carbon steel in which the steel was successfully
demagnetized to the level of the Earth's field using an applied magnetic field of about $10^{4}$ A/m. The
process takes less than one hour. For large steel pieces, this implies that for a winding density of roughly
250 turns/m, that a current of approximately 40 amps will be required. This is possible with commercially-available
power supplies and with standard high-current cables. We intend to next try this procedure on out 90 cm sample,
which is roughly the thickness we plan to use for shielding pieces.\\
\par If successful, a deperming facility will be set up either at APC or the Chooz site. This facility will
have a crane and solenoid capable of handling large shield pieces. It is envisioned that a few pieces
per day could be depermed in this manner. Field measurements on individual pieces will be used send pieces
for deperming on an ``as-needed'' basis.\\
\subsubsection{Compensating Coils and Detectors}
It is expected that the magnetic environment of the detector will be a relatively slow-varying field
due to the sum of the Earth's field and the residual field from the detector shield. In order to reduce
this field to below 100 mG and to trim the fields in the Near and Far detector to as similar a value as
possible compensating coils are planned for Double Chooz. Simulations show that currents of around 50 amps will be required
with 5 coils in the vertical and 5 coils in the horizontal directions. Each coil (total 15) should be separately adjustable
to minimize PMT transverse fields. It is planned to install a system of triaxial magnetometers inside the detector to
allow for field trimming to make the Near and Far magnetic environments as similar as possible.

%

%
\cleardoublepage
%
%
%
\section{Trigger, Electronics, and DAQ}
There are several major tasks that the Double Chooz electronics
will undertake: processing signals from the detector
and distributing them to acquisition systems; forming low-level
triggers; and monitoring the stability and performance of the
detector.

Each of these tasks are addressed in one or more electronics
subsystems, to be implemented with a mix of commercial and 
custom electronics.   The custom electronics is currently 
under development, and the status and plans for these parts
of the Double Chooz electronics will be a major focus for this
document.
The subsystems will be discussed in the order in which signals
pass from the detector to where they are digitized and stored. 
\subsection{High Voltage System}
A research quality high voltage system will be specified and constructed 
for the near and far detectors of Double Chooz to provide clean, stable 
power for the PMTs and outer veto gas chambers. Optimally this system would 
have one HV channel controlling each PMT or gas chamber but to reduce 
the cost of the system, we plan to gang detector channels.  A future 
upgrade could return to the single channel concept.  The ganged configuration 
connects one HV channel to 8 PMTs or 20 gas chambers.  It requires 
us to group PMTs based on their gain curve so that only a narrow range 
of HV values are required within a gang.  To provide uniform detector response, 
fine adjustment of the HV to individual PMTs will be done with Zener 
diodes on the HV splitter boards.  
A single cable design for PMT connections will be used, so both HV 
and signal are carried on the same coaxial cable (RG-303).  
\subsubsection{System Design}
The HV system will be comprised of two parts, one for the near and one for 
the far detector. A single vendor will be chosen to provide the 
main high voltage components for the entire system.  One type 
of mainframe (crate) and a minimum number of module types 
will be used throughout the system.  Common software will be 
written to meet the controls, monitoring and safety requirements as 
described elsewhere in this document.  
To date, the three primary candidate-vendors are CAEN 
from Italy, Connecticut-based Universal Voltronics (UV) and 
ISEG from Germany. 
The baseline detector design calls for 93 and 84 channels 
of high voltage in the near and far detectors, respectively. 
\subsubsection{Requirements and Specifications}
The baseline PMT design requires +2.5~kVDC at 0.3~mA per tube.  
Ganging 8 such tubes results in a total current of 2.4~mA/HV channel.  
The outer veto gas chambers are planned to be +3~KVDC at~0.1 mA.  Ganging
20 such chambers creates a load of 2~mA/ch.  Ganging may need 
to be adjusted depending on vendor product capabilities.

\begin{itemize}
\item Ripple and noise -- under full crate/module load these need to be 
below a level such that the effect on any signal is less than 1~mV peak-to-peak.
\item Voltage set/monitor resolution - this
will be at the level of 1~V or better.
\item Maximum voltage -- there must be a 
hardware provision that allows the setting 
of the maximum output voltage on a module or
crate basis.  This value shall be readback via 
software.  Software settable controls shall provide the 
ability to set software limits for the voltage settings.
\item Current trip set/monitor resolution -- it must be 1~$\mu$A or better.
\item  Voltage Ramp up/down -- it shall be in the range of approximately 10 
to 500~Volts/sec, programmable.
\item  Operating range -- it is approximately 0~C to 40~C, dry atmosphere.
\item HV outputs shall be floating, such that the crate ground is independent 
of the return for the HV channels.
\item  AC power supply -- it shall be approximately 220~VAC, 50~Hz.
\item  Polarity -- both positive and negative polarity modules are required.
\item  Communications -- complete control of the HV system 
must be able to be accomplished via a standard 
TCP/IP protocol based Ethernet connection.  Provisions to control 
individual crates from either a front panel or a 
front panel connection must be provided.
\item  Electrical Connections -- the HV connections must be through 
either standard SHV connectors or an HV certified multi-pin connector 
of proven reliability in this application.
\end{itemize}
\subsubsection{Control and Monitoring Software}
The software to run both detector sites will run on 
a single PC located at the far site, as it 
will be installed first.  All HV crates will be controlled 
via an Ethernet connection.  The type of software needed will 
depend somewhat on which manufacturer is chosen to supply the 
HV system.  The basic specifications however are the same regardless 
of which vendor is chosen.  The software will access and download to 
the crates a carefully controlled configuration file containing 
all channel voltages, trip current settings and limits.  
A crate polling routine will monitor and 
record voltages, currents and temperatures.  Alarms will be 
sent out to the operators as needed should tolerances be 
exceeded. The slow controls system will also monitor 
temperatures, etc. but on a more global scale.  
\subsubsection{Uninterruptible Power Supply}
A UPS will be used to prevent damage to the HV system 
components as well as the PMTs and gas chambers resulting from 
AC line fluctuations, brownouts or blackouts.  A hold time 
of 5-15 minutes is planned, depending on estimated AC interruptions 
and UPS system cost.  A double conversion will be used as 
only a true online, double-conversion UPS can provide pure, 
full-time sine wave AC output free of surges, voltage 
fluctuations and line noise.  It actively converts raw input 
from AC to DC, then back to sine wave AC output 
with enhanced protection from harmonic distortion, fast 
electrical impulses and other hard-to-solve power problems 
not addressed by other UPS types. 
\subsubsection{Cabling}
Cables from the HV modules to the 
HV splitter boards will be provided.  
The type and quantity of the cables will 
be determined after the vendor is selected 
as they may use different connector/cable types.
\subsubsection{Evaluation Phase}
This phase is to test and verify the manufacturers' specification 
claims by performing measurements on actual production HV 
modules and crates supplied by the vendors to the 
University of Tennessee.  These units will be on loan to 
us and will be returned to the vendor at the conclusion of 
the testing.  No direct cost is involved, other than shipping charges.  
Another aspect of this phase is to evaluate the software coding 
requirements for each vendor and to estimate the time and 
expertise required for the task.  The deliverables from this phase 
of the development work will be HV and UPS requirements and 
specifications documents suitable for use in submitting RFQs.

%
%
\def\units#1{\hbox{$\,{\rm #1}$}}

\subsection{Electronics}
\label{sec:electronics}
\subsubsection{HV Splitter}
\begin{figure}
\begin{center}
\includegraphics[width=4in]{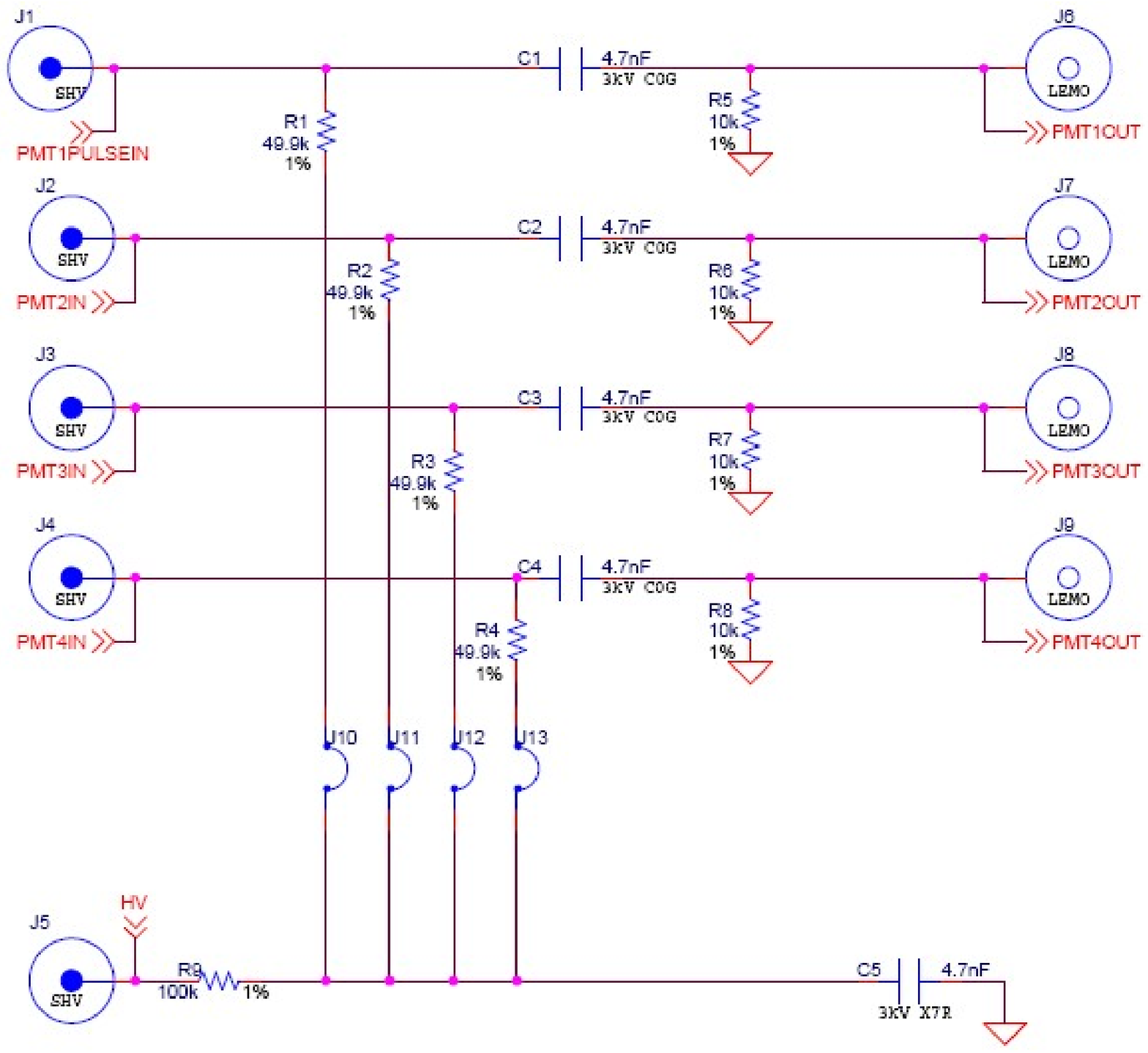}
\caption{Schematic for a four-channel HV Splitter prototype. This 
schematic was used both for a prototype circuit board and to simulate
the performance of the HV Splitter. The jumpers in the schematic 
can be replaced with Zener diodes to allow channel-to-channel 
variation in high voltage.}
\label{fig:electronics_HVS_schem}
\end{center}
\end{figure}

\begin{figure}
\begin{center}
\includegraphics[width=0.5\textwidth]{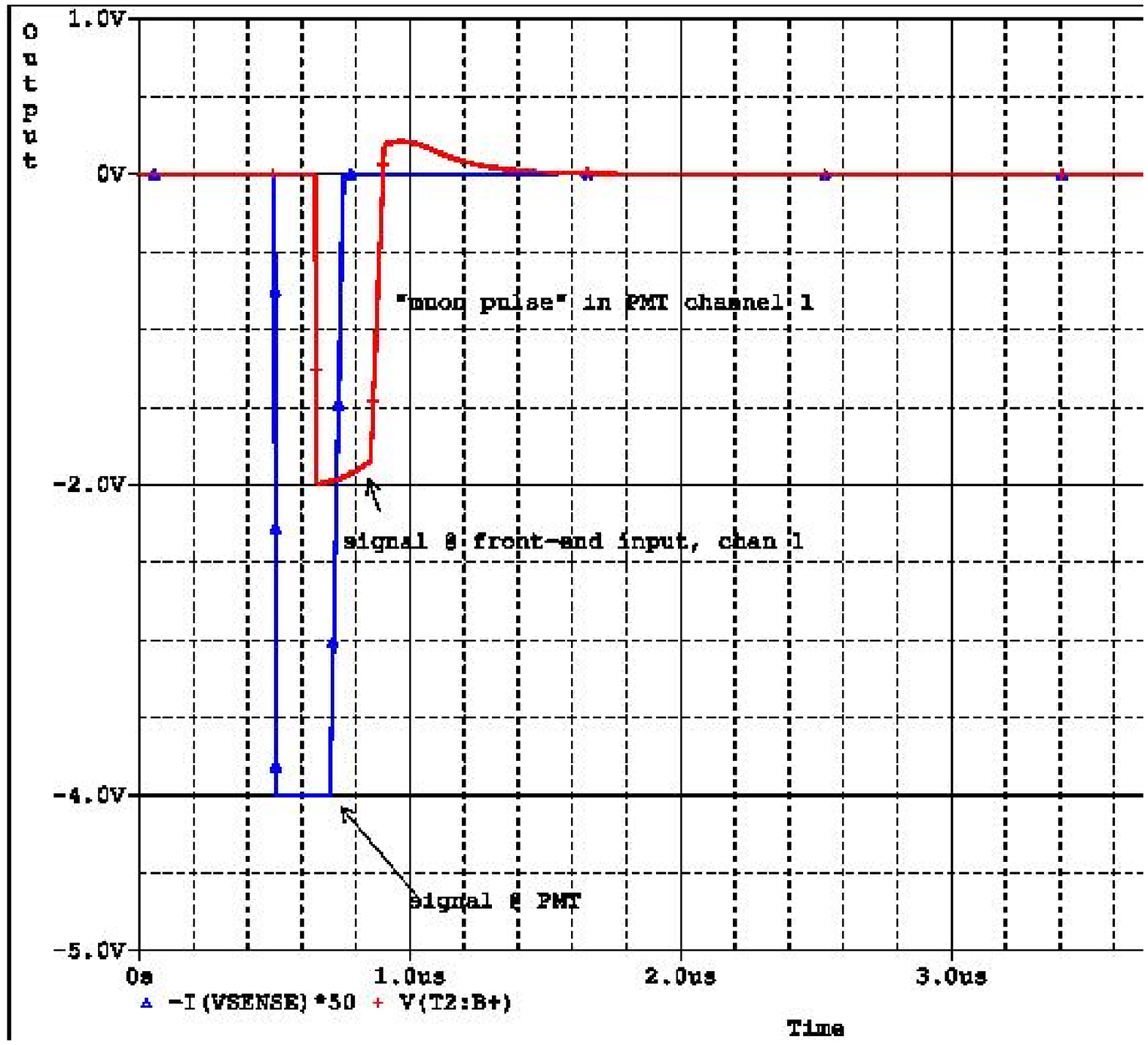}
\caption{Simulated performance of HV Splitter with a square-pulse
input. The use of a square-pulse emphasizes the differentiating
effect of the HV blocking capacitance.}
\label{fig:electronics_HVS_pulseshape}
\end{center}
\end{figure}

%
%
For the inner detector, the PMT signals will first pass through
High-Voltage (HV) Splitters.  Double Chooz will use a single cable
for each PMT, carrying both the HV and the PMT signals, 
for reasons of cost, for minimizing the dead-volume in the
detector and the feed-through area as cables exit the detector,
and to minimize the effects of ground loops between HV and signal
cables.   In all of this, it must be noted that the PMT signals are
expected to be rather small ($\sim$2~\units{mV}), so noise must be
kept well below the millivolt level.
It is the primary job of the HV splitter to decouple the signals coming out
of the PMTs from the HV supplies.  In addition, the HV Splitters
will terminate the PMT cable transmission line to minimize pulse
reflections, distribute HV among several PMTs per HV channel to
reduce the HV cost, and provide filtering of the HV to reduce
noise that comes from the HV power supplies. 
Figure~\ref{fig:electronics_HVS_schem}
shows a schematic for a simple four-channel prototype of the
HV Splitter that has been fabricated, which has the
basic features that are expected to be used in the final HV Splitter
design.
While it is unlikely that any of the Double Chooz PMTs will
be operated above 2\units{kV}, the specs from Hamamatsu indicate a
maximum voltage around 2.4\units{kV}.  As a result, the HV Splitter
is designed with components that are rated for a minimum of
2.5\units{kV}, while most are rated at 3\units{kV}. 
Figure~\ref{fig:electronics_HVS_pulseshape}
shows the simulated performance of the HV Splitter with 
a square-pulse input, with a back-terminated PMT and
100\units{ns} of cable between the PMT and the HV Splitter.
For this kind of `RC' termination, required because of the
need for HV decoupling, one can expect 3--5\%
signal reflections.  Having termination at both the HV
Splitter (where the terminating impedance is the output cable)
and the PMT base reduces the reflections to well below
1\%.  An alternative termination scheme, similar to that
used on the Borexino experiment~\cite{bib:borexino}, was
examined and rejected because the presence of parasitic capacitance
in inductors required the use of significantly smaller
decoupling capacitors, with correspondingly more signal
distortion. 
This prototype was designed to test the scheme of having one
HV channel supply four PMTs; the final design will most
likely have a HV channel supply between two and eight channels.
This scheme depends on being able to group PMTs of similar
nominal voltage, however the requirements are not particularly
difficult to meet: with PMT nominal voltages ranging from 1.2\units{kV}
to 2\units{kV}, on average one can find groups of 8 PMTs within a 10\units{V}
interval.

\begin{figure}
\begin{center}
\includegraphics[height=0.5\textwidth,angle=0]{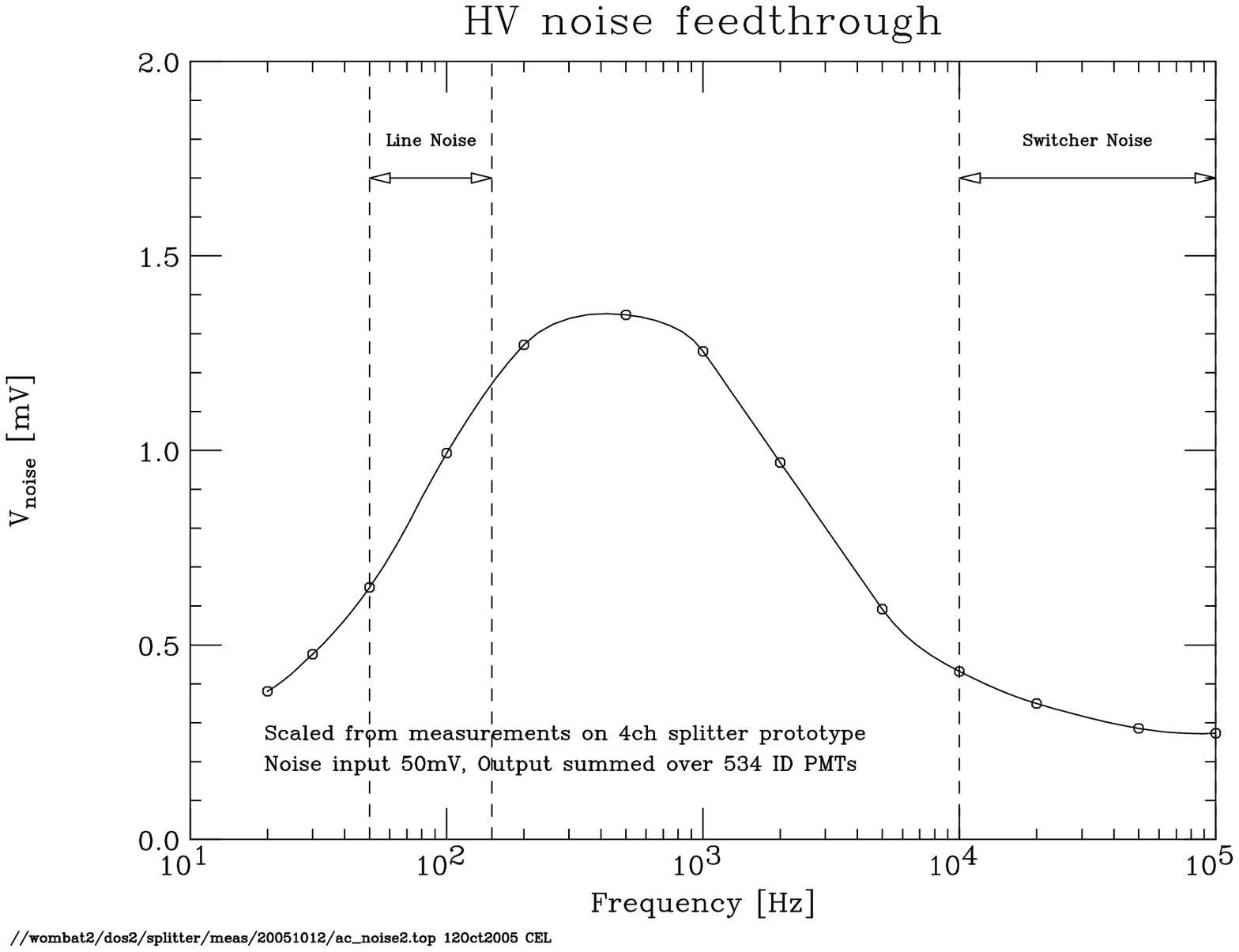}
\caption{Measured feed through of noise on HV supply to 
PMT signals as a function of noise frequency. The measurements
have been scaled to simulate the effect of 50\units{mV} 
of power supply noise on summed PMT signals.  The threshold at
which synchronous noise would overcome the L1 trigger threshold
is approximately 100\units{mV}.}
\label{fig:electronics-HVS-ACnoise}

\end{center}
\end{figure}

The prototype includes jumpers (shown on the schematic as J10
through J13) that can be replaced with Zener diodes, so that
the PMTs on a single HV channel can have a range of supply
voltages.  This would allow PMT-to-PMT voltage difference to
be more like 50\units{V} if needed. 
Noise from the HV power supply, like ground-loop pickup, would
be synchronous across many PMT channels. The summation of the
PMT channels to produce energy triggers means that serious 
attention must be paid to these noise issues. 
The series resistors and decoupling capacitor (C5) on the HV supply
line has the effect of reducing any noise on the HV supply line to
acceptable levels.  For one of the HV supply options the spec for
noise is 50\units{mV}, however the frequency of this noise is not
specified.  It is expected to be either line noise (50\units{Hz})
or noise from switching power supplies (20--100\units{kHz}). A
HV supply unit is being obtained by the University of Tennessee
to measure and characterize this noise.
 Figure~\ref{fig:electronics-HVS-ACnoise}
shows the measured amplitude of sine-wave noise at the signal 
output of the HV Splitter, from a 50\units{mV} 
amplitude noise source on the HV power supply, where resulting
output has been multiplied by the number of ID PMT channels to
simulate the `worst-case' noise scenario.  As can be seen, the
result is less than $\sim$1.5\units{mV}, which is considerably
less than the $\sim$100\units{mV}
summed signal equivalent to a `trigger threshold' event. 
Another concern with ganging PMTs on HV channels is the possibility
of cross-talk between the channels.  In a detector of the
Double Chooz design, one expects cross-talk not to be a serious
concern: the PMTs are looking at the same events, and should 
have similar signals; there simply won't be cases where one
PMT gets a large (muon-like) signal, while an adjacent PMT only
gets single photoelectrons.  However, the simulation of the HV Splitter
shows essentially zero cross-talk ($<$50\units{\mu V} for a 4\units{V}
input pulse) between channels as a result of
ganging PMTs on a HV supply.  
One can still have capacitively- or
inductively-couple cross-talk between channels that is not apparent in
the simulation.  Tests will be conducted at every stage of the HV Splitter
(and other downstream electronics) to measure, and if necessary mitigate,
any cross-talk, but it must be noted that such cross-talk would be 
present regardless of the level of ganging of PMTs on the HV supplies. 
\subsubsection{Front-End Electronics}
\begin{figure}
\begin{center}
\includegraphics[width=0.6\textwidth]{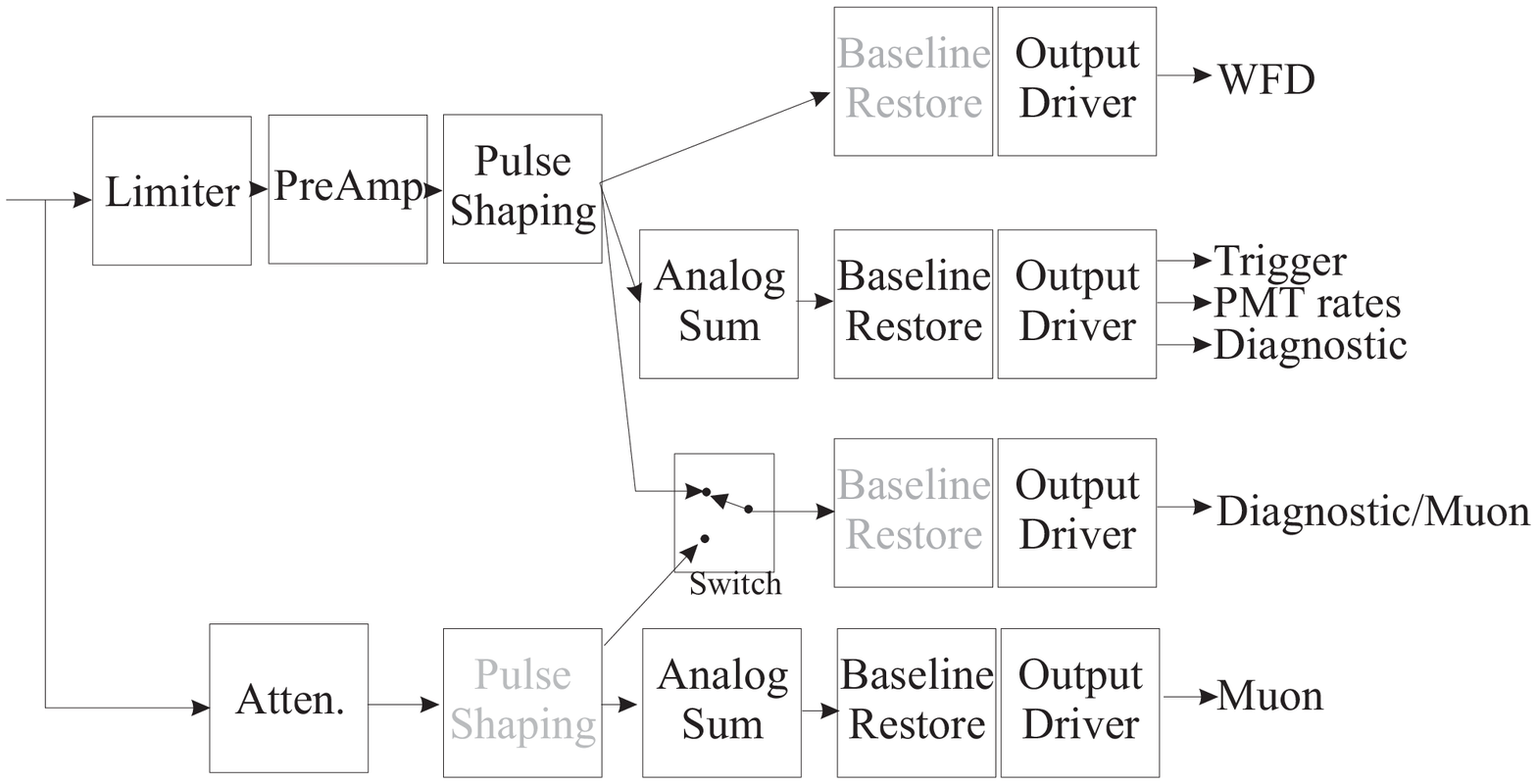}
\caption{Block diagram of Double Chooz analog front-end functions. This
design will also be used for performing analog summation for the L1
trigger.
The `switch' in the diagram will be implemented with a zero-ohm
surface-mount jumper.}
\label{fig:electronics_FE_block}
\end{center}
\end{figure}

The Front-End subsystem receives PMT signals from the HV Splitter, 
performs amplification, pulse-shape and baseline corrections, and
distributes the signals to downstream electronics 
(see Fig.~\ref{fig:electronics_FE_block}). 
The input
signals to the Front-End will be quite small (millivolt level), so 
the entire Front-End will be a pure-analog design to avoid any 
digital noise leaking into the signals. 
\begin{itemize}
\item Amplification: total gain in the range of 10--15, so that 
single photoelectron signals are 20--30\units{mV} amplitude. 
\item Calibration input: single analog input per Front-End module;
the calibration pulse will be fed into all of the channel inputs.
\item Channel disable: RF-relays used to disable input channels
under computer control.  This provides a more robust 
disable than `chip disable' features on the amplifiers. 
The digital logic that controls the channel disable is
completely quiescent during normal data-taking, and is a
simple open-collector TTL bus driven from a single VME
output register per detector. 
\item Pulse shaping: to restore pulse shape from cable attenuation
and HV decoupling capacitor. 
\item Baseline restoration: we will avoid some of the baseline-shift
problems caused by large (muon) signals and capacitive
coupling, by using DC coupling wherever possible.  The
amplifiers needed for high-speed and low-noise
performance, however, tend to have
unacceptably large DC offsets.  To avoid this, we
are developing
baseline restoration circuitry to remove the offsets.
The first full-speed prototype, built and tested in September 2005,
used active feedback of
low-frequencies components in the signals (including DC) to
remove the offsets and improve the rejection of line
frequency pick-up. This prototype was completely successful
at removing DC offsets, but was somewhat more complex than
desirable, and required some minor modification to have
acceptable performance for saturating pulses.  

A new prototype is currently being produced for testing the
front-end channel disable and summation functions. This new
prototype will 
also test a different (`feed-forward') and simpler technique to cancel 
DC offsets. Results from SPICE simulation of the new circuit
are very encouraging, but tests of the prototype will be needed
before a final decision.

\item Pulse summation: a simple analog sum of the input
signals is produced, for use in trigger and
diagnostic systems. 
\item Muon/Trigger channel:  a separate attenuated path
through the Front-End is provided for getting information
about background muon events.  This attenuation helps
to avoid any electronic limitation of the muon signals,
at the expense of low-amplitude signal information.  We
will have the option (by means of zero-ohm surface-mount
jumpers on the Front-End board) of having individual
muon outputs for each input, or of having only a summed
muon output.  Note that the circuitry for the muon
channel is also used in the Level-1 Trigger implementation.
\end{itemize}
The implementation plan is to first test 
and optimize individual functions
of the Front-End (summation, baseline restoration, etc.) 
using specific prototype circuit boards,
then to produce a `few channel' prototype that includes the
full range of circuitry for a limited number of channels, before
proceeding to a final design of the Front-End module. 
\subsubsection{Inner Veto Electronics}
Since - in the current planning - the 78 veto PMTs are considered a part of the detector just as the inner PMTs, the general organization of electronics is very similar to that of the inner photodetection system. High voltage and signal are fed to/from the PMT with a single coax cable. The cables are running through the veto liquid up to the top and to a HV splitter box. The trigger condition relies on the number of hit PMTs, the analog sum and for more sophisticated trigger scenarios on the hit pattern. Inner Veto events can trigger the DAQ of the inner detector, and vice versa. The integration in the DAQ of the experiment will be done on two levels: one the hand using the trigger boards of the experiment on a hardware level, and also in the software that reads, processes and stores the FADC data for all channels .

\subsubsection{Outer Veto Electronics}
The electronics design is based on providing a digital output from the
modules.  A single discriminator threshold will be applied to each of 
the 24 chambers within a module.  To provide this functionality, we 
foresee a single printed circuit board which will be placed on one end of 
each module.  The board will connect to the threaded brass connections 
provided by the mechanical construction of each tube.  A single high-voltage 
connection will supply all chambers in that module.  The analog signal from 
each wire will be read-out and immediately put into a discriminator which 
operates in time-over-threshold mode with a 10MHz clock (i.e. digitized every 
100~ns). This discriminated signal will be passed into an FPGA on the front-end 
card which will compile the output from all 24 chambers into a single 24-bit 
data word.  When a trigger condition is satisfied, a 24-bit timestamp will be 
appended to each 24-bit encoded data word and it will be sent through a 
hot-link connection to the back-end VME system.  We foresee the following 
three trigger conditions:
1)	Simple Coincidence: Since any muon will have greater than a 99\% 
chance of causing a signal in at least 2 out of 3 chambers within a single 
module, this will provide the most effective method for reducing the random 
singles rate while maintaining a very high efficiency of muon detection.  In 
order to avoid missing events which may span a clock boundary, we propose to 
use a simple sliding window of two clock cycles which will be used to define 
a coincidence. 
2)	Long Pulse Trigger: A showering muon may have multiple particles 
interacting in a single chamber over a period of time up to 1-2$\mu$s.  
For that 
reason, the data for a channel which is above threshold for contiguous 
clock-cycles will be stored.  If the simple coincidence described above is 
satisfied at any time during the contiguous period, all digits for the 
contiguous period will be sent to the back-end system.
3)	 Diagnostic Trigger: For calibration or debugging purposes, a special 
trigger will be available.  In this configuration, if any of the 24 channels 
are above threshold, the 24-bit data word will be triggered and sent to the 
back-end system.  Since this will have a high rate, this is foreseen as a 
configurable trigger condition.

The back-end is expected to be based on a single VME 9U crate at each detector 
location.  Communication between the front-end and back-end will be done 
serially with the use of fiber-optic cables, to reduce the possibility of 
ground loops.  The VME modules will consist of independent 
data channels for each input channel.  The data will have address and control 
codes appended and will be stored in the VME processor for collection by the
main data acquisition system.  A synchronized timing system will be 
provided that allows the correlation of events from the outer veto
system with those recorded by the main detector to within 50~ns.

\subsubsection{Toward the Level-1 Trigger}
A detector of a geometry where an inner
scintillator volume is surrounded by PMTs distributed uniformly
on a surface, will have the property that the total amount of
light collected by the PMTs is directly proportional to the
energy deposition, in the absence of light attenuation. 
For Double Chooz the distribution of PMTs is not perfectly 
uniform, however it is close enough to uniformity to make total light
a useful proxy for event energy. In addition,
the small size of Double Chooz makes light attenuation only
a small correction, negligible for the purposes of creating
a trigger. 
As a result, a simple and natural {\it energy} trigger for
Double Chooz is simply to sum up the phototube signals from the
inner detector, and use a discriminator to set an energy level
for the trigger. 
In practice, the summation of so many PMTs means that any
common-mode noise (ground-loop noise, etc.) has to be
very stringently suppressed, since it will be greatly 
amplified in the summation.  This motivates much of the
efforts in upstream electronics systems to avoid or 
ameliorate noise inputs. 
The trigger summation occurs in several stages; the first
stage is in the Front-End modules, where all of the input
signals handled by that module are summed and provided on
a single output. 
Additional levels of summation are required. 
Details can be found in section \ref{sec:trigger}.

%
%
%
\subsection{Trigger and Timing}
\label{sec:trigger}
%
\subsubsection{Concepts}
The trigger and timing system has to provide a highly efficient trigger to 
the Double Chooz experiment for neutrino events as well as for several types 
of background events. 
The trigger has to be reliable, i.e.\ trigger failures must be rare and easily
detectable, and the trigger efficiency must be measurable with sub-percent 
precision. 
The trigger system will also distribute a common clock signal to the 
experiment and provide a time stamp for all events. 

A two-stage trigger system is foreseen. The level-1 trigger is based on 
the analogue signals from the detector which are discriminated and analyzed
in FPGAs (Field-Programmable Gate Arrays). A positive Level-1 decision triggers
the readout of the detector. The readout is followed by a level-2 software 
trigger. 
This section covers the level-1 trigger only. 

There will be independent trigger systems for the far and near detector
with independent clocks. 
The clocks will be synchronized between the two detectors through either a 
cable connecting the two triggers or through GPS systems. 
The frequency of the clocks in both detectors must be identical. 
Synchronization of the phases is not required.

The trigger will receive input from the PMTs of the inner target and the inner
veto. 
The signals of 16 PMTs will be pre-summed in the electronics of the Front-End 
modules (see sect.\ \ref{sec:electronics}). The trigger only 
receives the signal sums. 
Input from the outer veto and the calibration systems can also be processed 
(for example from the laser system sec.\ \ref{sec:laser}). 
It is possible, but not foreseen, to apply veto conditions on the trigger
level. 
Instead an event in the veto detectors will be read out as well as a neutrino 
candidate following it. It is left to the offline analysis to reject 
cosmogenic events.
The signals of the groups of PMTs are discriminated individually to provide 
an approximate multiplicity of the event. In parallel the total signal sums 
from the inner target and inner veto are derived and discriminated on several
levels. The trigger will allow conditions on the total energy and 
multiplicity, or any combination of these, to be applied. 
Special care will be taken to 
avoid the amplification of baseline-shifts, common-mode noise and saturation
through the summation.

To keep the trigger conditions flexible, FPGAs will be used to apply the trigger
conditions to the discriminated signals. For redundancy,
 two independent 
trigger boards will be used for the inner target, each based on only half
of the groups of PMTs. This guaranties that hardware failures do not result in 
the loss of events and can be easily detected.
A third trigger board will handle the inner veto and
additional trigger inputs. A careful simulation will determine the initial 
trigger conditions. They can be adapted once experience with the real detector
is available. The following list gives initial ideas on some of the necessary 
trigger conditions:
\begin{itemize}
  \item{\bf Inner Target Neutrino Trigger A}\\
       Approximately 0.5 MeV energy threshold on the total energy in the
       inner target and at least 2 groups of PMTs above noise (to exclude
       single sparking PMTs from triggering). The trigger is based on the
       first half of PMTs only (groups A).
  \item{\bf Inner Target Neutrino Trigger B}\\
       Same as above, but based on the second half of PMTs (groups B).
  \item{\bf Low Energy Neutrino Trigger A + B} \\
       This neutrino is the same as above, but with a lower energy threshold. 
       It's rate will be too high and therefore only a
prescaled selection of events
       will create a real trigger. The energy threshold and 
the prescale factor will be adjusted
based on the detector noise. This trigger allows the threshold 
       behavior of the main neutrino trigger to be studied.  
  \item{\bf Inner Target Muon Trigger A + B}\\
       A muon will create a signal in the inner target that fires all groups 
       of PMTs. The trigger will be based on multiplicity only. 
  \item{\bf Inner Target Muon Trigger A + B -- PMT with low gains --} \\
       Some PMT set with a low gain could be installed to better detect and tag 
       showering muons.
  \item{\bf Inner Veto Muon Trigger}\\
       A trigger based on energy threshold and minimum multiplicity in the 
       inner veto.
  \item{\bf Inner Veto Low Energy}\\
       A trigger on the inner veto with a lower threshold to study the 
       trigger efficiencies. Scaled-down in rate as above. 
  \item{\bf Random Trigger}\\
       A trigger with an adjustable frequency triggering the readout at 
       fixed time intervals. The events will be uncorrelated with any
       physics process in the detector (therefore random) and will allow
detector noise to be studied.
\end{itemize}


\subsubsection{Trigger Outline}
\begin{figure}[bth]
 \includegraphics[width=18cm]{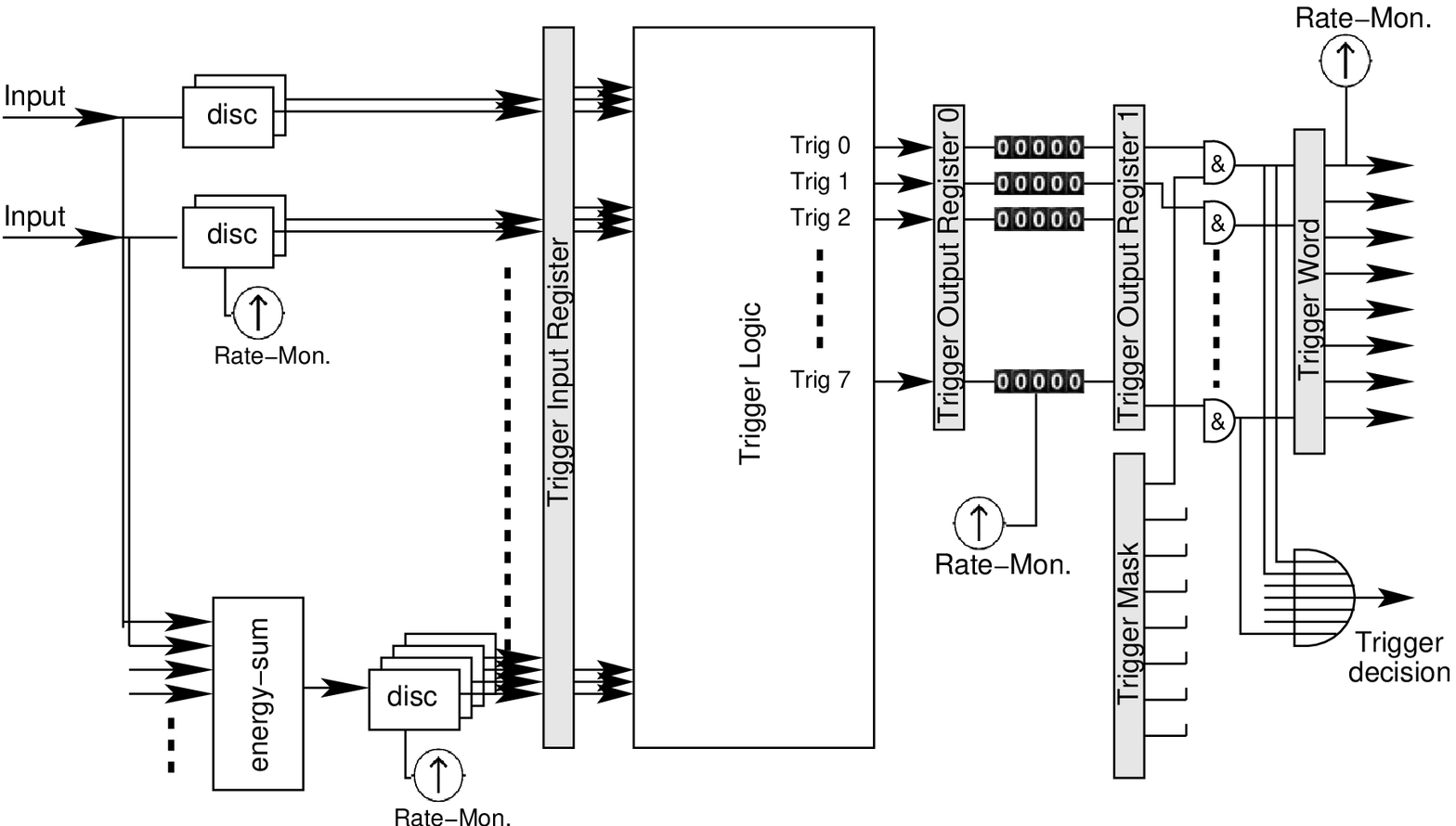}
 \caption{Schematic layout of the trigger. Input from the front-end electronics
          on the left, output to the data acquisition on the right. Shaded 
          rectangles are registers that will be read out. The discriminators 
          are labeled 'disc'.}
 \label{fig:trigger} 
\end{figure}

A schematic view of the trigger is shown in fig.\ \ref{fig:trigger}. 
There will be three such units (trigger boards) for each detector. 
Two handling the inner target with half of the PMTs each and the third one 
handling veto and calibration.
A global trigger word is combined from the three trigger boards on the master
board and the trigger decision is formed as a logic 'OR' of all triggers.

Input signals from groups of PMTs are received and discriminated on two 
different, programmable thresholds. The energy sum is derived and discriminated 
on four levels. 
For the veto and calibration triggers the energy sum will be modified and 
input for logical signals will be provided.
The output rate of the discriminators is available for monitoring. 
The status of the discriminators is stored in the input register
and read out in case of an event. 
The trigger logic is fully programmable and may combine any input signals. 
Up to 8 different triggers can be created in each board. 
The output of the trigger logic is stored in the output register 0. 
A scaler for each trigger allows the reduction of its rate
by accepting only every 
n.th trigger, where n is programmable. 
After down-scaling, the triggers are combined with a mask that allows 
individual triggers to be suppressed.
The result is transmitted to the master board.

\subsubsection{Clock and Timing}
The trigger and timing system will provide a 62.5 MHz clock to the whole
detector. For this purpose a free running oscillator will be placed on the 
master board to allow stand-alone running. A possibility of
 synchronizing the 
clock to an external reference is foreseen. With a cable connecting the
far and the near detector the two cocks could be synchronized. 

A separate timing module is under discussion that receives the clock and an 
absolute time from the GPS system. With such a timing module both detectors
could be synchronized to GPS. It would also allow an absolute time stamp for
each event, should this be necessary for physics. 

The trigger will have the possibility to receive a busy signal from each 
subdetector should it not be ready for data taking. 

\subsection{Data Acquisition System}
\label{sec:daq}
The Double Chooz Data Acquisition System aims to record neutrino interactions
observed through the annihilation of a positron and
the delayed capture of a neutron by a Gadolinium nucleus. 
In addition, many other events are also considered 	worth
reading out, most of which are induced by cosmic rays. 
The event rate amounts will be around $600$~Hz in the
near detector.

The main principle of the designed system is to digitize and store
on mass storage every event occurring in the detector which deposits
more than $0.5$ MeV in the inner cylinder (Target + Gamma Catcher)
or $5$ MeV in the Inner Veto. The digitization of the photomultipliers
will be performed by Waveform Digitizers, with
one channel per photomultiplier. This has been chosen 
in view of recording pulse shapes, for the absence of dead time
and the simplicity of operation.
\subsection{Main Features of The Waveform Digitizer module}
\label{sec:daqWfd}
The Waveform Digitizers are built from 8-bit Flash-ADCs operated at
$500$~MHz, static random access memory, and a smart memory controller,
which allows uninterrupted digitization and read-out without dead
time.
The digitizer is currently under co-development at AstroParticle and
Cosmology Laboratory (APC) and Costruzioni Apparecchiature Elettroniche 
Nucleari (CAEN). It
will be commercially available as a NIM module with USB or optical
fiber data link and as a VME64x module. Double Chooz will use the
VME module which will house 4 channels in a 1-U-wide device. A NIM/USB
prototype is currently under test. Important parts of the design have already
been validated but the firmware is still under development. The transition
from the NIM/USB model to the VME one will be straightforward, due
to the modular design. The production of modules for Double Chooz
will start at the end of 2006.
\subsubsection{Operation of the Waveform Digitizer}
\begin{figure}
\begin{center}
 \includegraphics[scale=0.6]{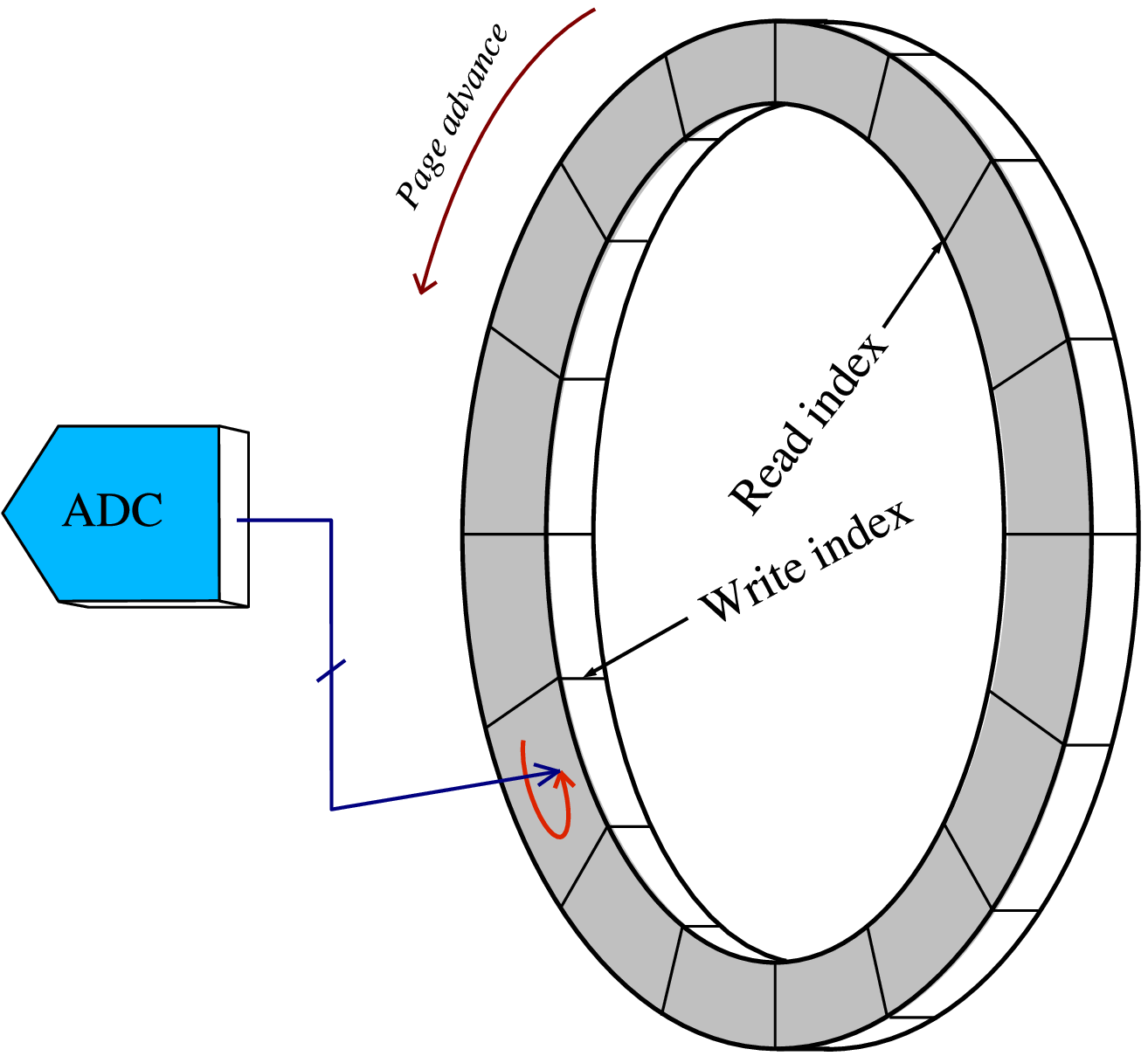}
\caption{Operation of the Waveform Digitizer as a FIFO\label{fig:rotating-buffer}}
\end{center}
\end{figure}
The four channels of the Waveform Digitizer module (WFD) 
are operated with one $62.5$~MHz clock, which can be generated 
onboard or taken
from a front-panel input. In Double Chooz, the WFDs will all use
the same clock, delivered by the trigger system. In this way,
all channels of all modules are synchronized.
From the $62.5$~MHz clock, a $500$~MHz one is generated onboard by
a Phased Lock Loop (PLL) to clock the flash-ADCs.  Triggers are accepted at the
$62.5$~MHz clock rate.
The onboard memory may be accessed through the VMEbus without disturbing
the digitization, like with a dual-port memory. The memory controller
manages the flash-ADC data in the following way, illustrated by Fig.
\ref{fig:rotating-buffer}:

\begin{itemize}
\item The ADC samples are continuously stored in a $1024$~Bytes rotating
buffer, thus keeping a history of the past $2048$~ns. The memory is
divided in $1024$ such rotating buffers (or  pages). 
\item On the first $62.5$ MHz clock after detection of a trigger, the
last ADC sample is written to the current rotating buffer; subsequent
samples are written to the next buffer. The time difference between
this trigger and the previous one is stored, along with the value
of a 16-bit word read from the front panel. The time elapsed between
triggers is later used to search for coincidences; the 16-bit word
is used to read the event number and various flags generated by the
trigger system. Rotating buffer, time difference and 16-bit word constitute
an event for the waveform digitizer system. 
\end{itemize}
The Waveform Digitizer's trigger logic and the computer in charge
of the read-out keep synchronized by the mean of two registers:

\begin{itemize}
\item The Write Index indicates the buffer currently in use to store
flash-ADC data. It is incremented at each trigger. 
\item The Read Index indicates the first buffer that cannot be re-used
yet because it has not been read. This index is manipulated through
the VMEbus. When the Write Index is just below the Read
Index, the trigger is disabled, causing dead-time; therefore the
absence of dead time depends on the ability of the VME master to smoothly
read the incoming data. 
\end{itemize}
This synchronization mechanism is exactly the one of a FIFO, where
the elements are the events, with a capacity of 1023 events, but with
the additional feature that the full content of the FIFO is exposed
to the read-out system, not only the most ancient event.
\subsubsection{VME layout of the Read-out}
\label{sec:daqReadout}
The Waveform Digitizer can handle a variety of VME access modes, ranging
from D16/D32 random access to BLT64, 2eVME and 2eSST which allows
data transfer rates from $160$ to $320$~MByte/s. The modules will
be plugged into VME crates (7 for the PMTs of the inner
cylinder), with 19 WFDs and a VME master in each crate. The VME masters
will be interconnected by a 1 Gbit/s Ethernet private network to a
computer in charge of event building. Each of them will perform the
read-out of its own crate. 
%
\subsubsection{Second Level Trigger and Data Reduction}
The principle of the Data Acquisition of Double Chooz is to record
all single events, leaving to the offline analysis the task of deciding
what to do with them. However the amount of data available in the
Waveform Digitizers is too big to be easily managed offline.

For each event, a $2~\mu$s record is available. This is much more
than the duration of an event, ($<200$~ns). Considering a duration
of $200$~ns for each event and a trigger rate of $600~{\rm s^{-1}}$, the
total data flow would be around $2.6$~TB/d for only the pulse shapes
of the target PMTs, which is still too big. Actually, in the near
detector, most of the events will be cosmic muons crossing the Inner
Veto and missing the Target. For these events, 
only the starting and ending times need
to be recorded, while for neutrino candidates, the whole pulse
shape could be recorded for all channels. For other events, various
data reduction schemes can be considered if the events can be sorted
online.

The second level trigger, implemented at the heart of the read-out
software, addresses this goal, based on the following pieces of information: 

\begin{itemize}
\item the trigger flags provided by the various discriminators, 
\item the time information recorded by the Waveform Digitizers. 
\end{itemize}
Combinations of these informations lead to fast identification of
coincident low energy events (neutrino candidates), low energy singles
(gammas), and cosmic rays crossing the inner cylinder or not. No event
will be rejected online; the second level trigger will only decide
the size and format of the data. A possible data reduction strategy
is detailed below.

\subsubsection{A possible Data Reduction Strategy}

A possible example of data reduction
strategy is presented in table~\ref{tab:reduc},
where the following event categories have been considered: 

\begin{itemize}
\item Coincidences: two energy depositions above the 0.5 MeV threshold
in the target+$\gamma$-catcher volume of the detector, 
within a time window ($\sim$200~$\mu$s)
chosen to be larger than the one applied for the 
offline identification of neutrino
candidates. Besides neutrino interaction events, correlated backgrounds
and accidental coincidences will be retained. The full available information
on pulse shape will be stored for such events. 
\item Gammas: a single energy deposition above 0.5 MeV and below
50 MeV in the target+$\gamma$-catcher volume of the detector. The rate 
of such events
will be about 10 Hz, dominated by gammas. Storing the full waveform
information would lead to a huge data flow ($\sim70$~GByte/d):
one possibility is to record the pulse starting time for each PMT
and the waveform for groups of PMTs. The full information could be
stored for singles depositing an energy larger than 6 MeV, which include
neutrons, occurring with a much lower rate ($\sim$10$^{-2}$ s$^{-1}$). 
\item Muons: a veto window will be applied in the offline analysis
 after each muon
crossing the detector. For such events, it will be sufficient to record
the global timing, together with the minimal information in the event
header. 
\item Muons through target: muons interacting in the detector can
give origin to radioactive ions which decay with rather long lifetimes
($3\tau\sim0.6$ s) mimicking neutrino interactions. Whenever a coincidence
occurs, all the muons crossing the target volume one second before
will be recorded. The large deposited energy will cause all PMTs in
the central region to be saturated and the only data which may be
used is the starting and ending
times of each pulse. This, combined with the timing
and charge in Veto PMTs, can provide the necessary handle for a rough
reconstruction of the muon direction and to distinguish muons stopping
or showering inside the detectors from the ones crossing it. There
is margin to store more details on the Veto PMTs, such as pulse rise time
or charge in a given time window, or the charge and even full waveform
for few dedicated PMTs with lower gain (non saturated) in the target
region. 
\end{itemize}
The final strategy will be defined after detailed simulation studies
and on the basis of experience during the first year of data taking
with the far detector alone, where the rate of singles and coincidences
will be the same as in the near detector while the cosmic muon flux
will be lower by about a factor of 10.

\begin{table}
\begin{center}\begin{tabular}{|l|l|c|l|c|}
\hline 
Category &
 Identif. &
 Rate &
 Stored data &
 Data flow \tabularnewline
&
 criteria &
 ($Hz$) &
&
 ($GByte/d$) \tabularnewline
\hline
Coincidences &
 $2\times E_{T}>0.5$&
 $0.1$&
 Full &
 $2$\tabularnewline
&
 in $<$200 $\mu$s &
&
&
\tabularnewline
\hline
Singles &
 $0.5<E_{T}<50$&
 $10$&
 $t+Q$ for all PMTs &
\tabularnewline
&
 \&$E_{V}<5$&
&
 +$WF$ per groups &
 $5$\tabularnewline
 - neutrons &
 $E_{T}>6$&
 $0.01$&
 Full &
\tabularnewline
\hline
Muons &
 Inner Veto $\mu$ trigger &
 $600$&
 global time &
 $1.5$\tabularnewline
\hline
Muons through target &
 $E_{T}>50$&
 $20$&
 $t$ for all PMTs &
 $2$.5\tabularnewline
 1 s before coincidence &
 (\&Inner Veto $\mu$ trigger) &
&
 $Q$ for Veto PMTs &
 \tabularnewline
\hline
\end{tabular}\end{center}
\caption{\label{tab:reduc}
A possible example of data selection on the basis of event sorting
operated by the second-level trigger. The event categories are described
in the text. Measured energy in the target volume and in the Veto
are denoted by ${\rm E_{T}}$ and ${\rm E_{V}}$ respectively and are 
always expressed
in MeV. The event rates are calculated for the Near Detector}
\end{table}
\subsubsection{Dead time}
The Waveform Digitizers behave like FIFOs; they do not incur any dead
time as long as the read out system (VMEbus, processors, network,
mass storage) can sustain the data rate. The designed system is able
to sustain the data flow through the VMEbus and the Ethernet network;
therefore the full read out process is free of dead time. Although
the online data reduction aims mostly at making data easily manageable
offline, as a side effects it keeps the data flow even farther below
the capability of the system. In case, by some unforeseen effect,
the digitizers would miss a trigger, this would be detected by the
online program which would stop immediately the acquisition, to avoid
corrupting the data already taken with a dead time uncertainty.

A general discussion of the dead time and the induced systematic uncertainty
can be found on page~\ref{'dead'}.
\subsection{Slow Monitoring}
\newcommand{\onewire}{1-Wire\textsuperscript{\textregistered}}
\newcommand{\ibutton}{iButton\textsuperscript{\textregistered}}
A slow monitoring and control system is required to control systematic
effects that could impact the experiment, to allow automated scans of
parameters such as thresholds and high voltages, and to provide
alarms, warnings, and diagnostic information to the experiment
operators.  The quantities to be monitored and controlled include
temperatures and voltages in electronics, experimental hall
environmental conditions, line voltages, liquid levels and
temperatures, gas system pressures, radon concentrations, photo-tube
high voltages, and discriminator settings and rates.  Most of these
functions can be accomplished using ``\onewire'' devices from Dallas
Semiconductor \cite{one-wire-webpage}.  The high voltage and
discriminator subsystems will have their own control and readback
hardware.  All slow monitoring and control systems will use the same
database and history log software.  A computer in each experimental
hall will monitor the local 1-wire bus and a local radon monitor,
acquire any data provided by other subsystems, record the data, and
make the data available via the local internet connection.
\subsubsection{Monitoring via 1-Wire interface}
\begin{figure}
\begin{center}
\includegraphics[width=\textwidth]{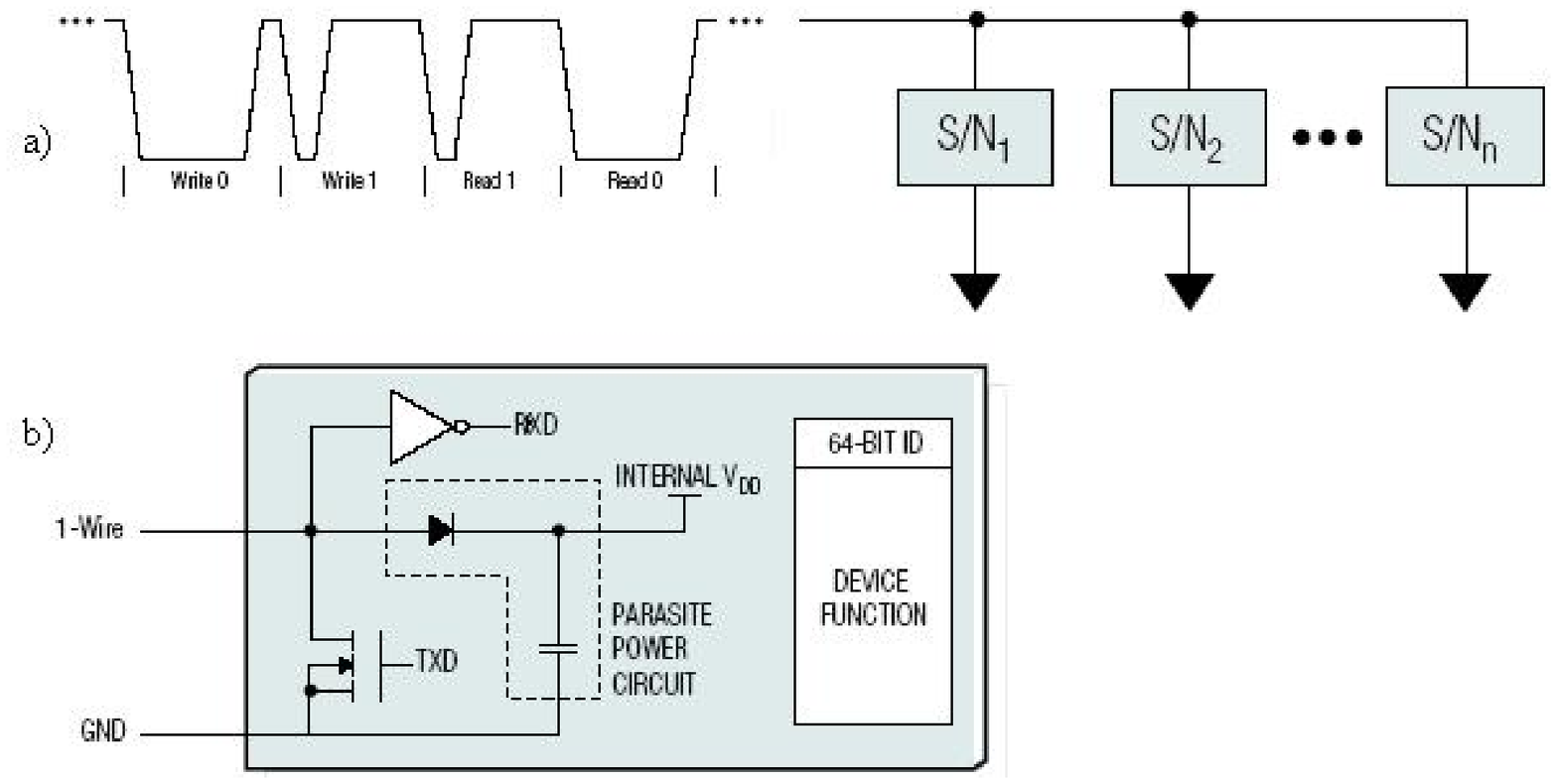}
\end{center}
\caption{The \onewire\ bus: (a) control, readback, and power provided
to multiple devices over a ``single'' wire; (b) parasite power circuit
captures power during high period of \onewire\ waveform.  (Adapted from
figures by Maxim IC / Dallas Semiconductor.)}
\label{Fig:onewirefigs}
\end{figure}

\begin{figure}
\begin{center}
\includegraphics[width=0.7\textwidth]{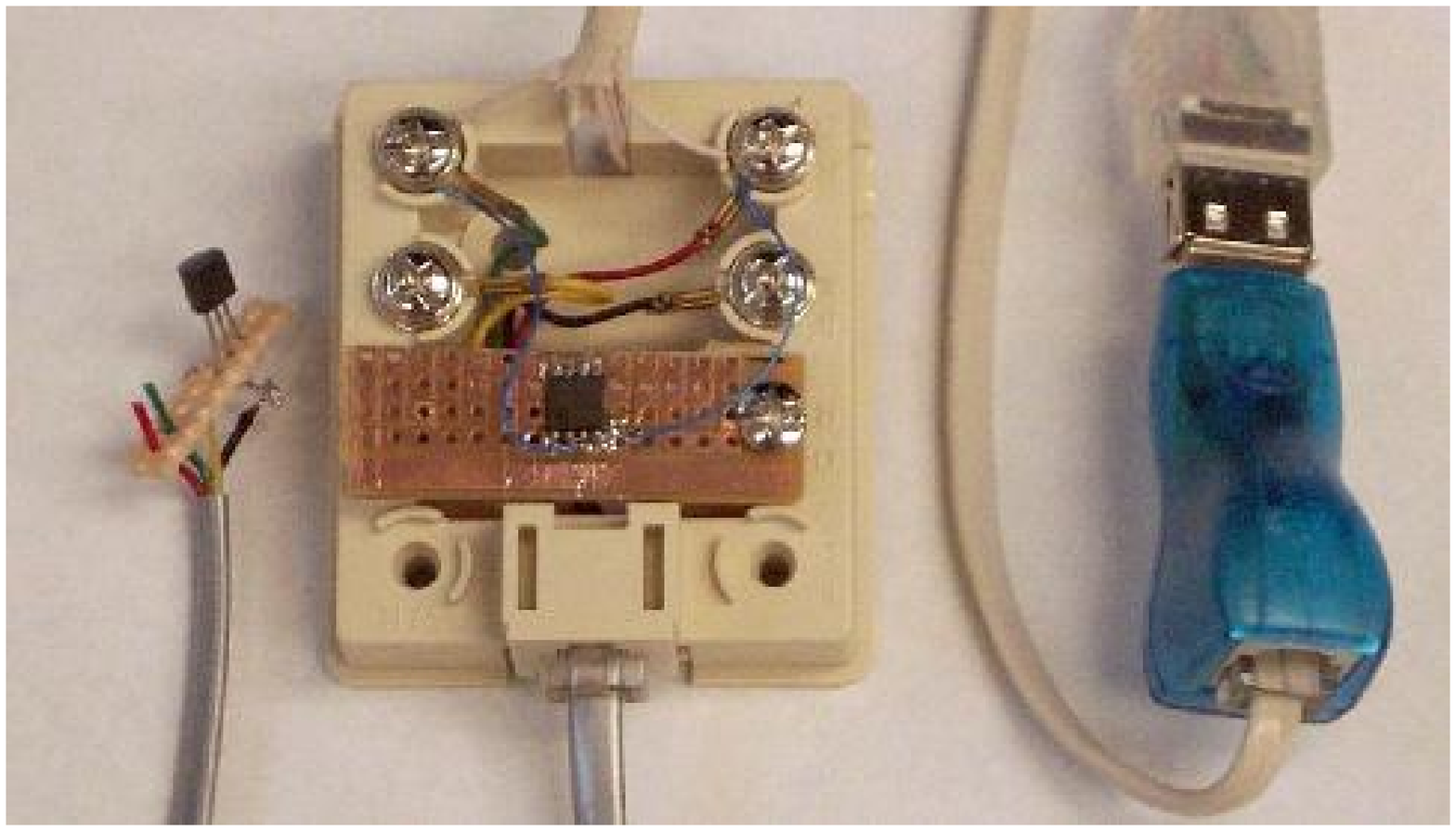}
\end{center}
\caption{Examples of 1-Wire devices. From left to 
right: digital thermometer in 
TO-92 package; 4-channel ADC on prototype board, mounted in RJ-11 
modular phone jack box; USB to 1-Wire interface.}
\label{Fig:packagephotos}
\end{figure}

The ``\onewire'' line of semiconductors from Maxim IC / Dallas
Semiconductor use a simple interface bus that supplies control,
readback, and power to an arbitrary number of devices over a single
twisted-pair connection~\cite{one-wire-webpage}, 
see~Figure~\ref{Fig:onewirefigs}. A variety of sensor and control
functions are available in traditional IC packages and
stainless-steel-clad ``\ibutton s'', see Figure~\ref{Fig:packagephotos}.
Each device has a unique, factory-lasered and tested 64-bit
registration number used to provide device identification on the bus
and to assure device traceability.  Some devices are available in
individually calibrated NIST-traceable packages.  The features of low
cost, multidrop capability, unmistakable device ID, and versatility
make this an attractive choice for implementing the slow
instrumentation and control system.
Implementation details and expected performance for several subsystems
are given here.
\subsubsection*{Crate and card temperatures and voltages:} 
Temperature and voltage monitoring can be included in any custom-built 
electronics at
 a component cost of only a few dollars per device using DS18S20 and
 DS2450 chips and low-cost modular connectors to connect to the
 \onewire\ bus.  Note in addition to the temperature and voltage
 functions, the unique ID on each chip provides automatic tracking of
 any card swaps.  Trivial custom boards containing only these
 components can be used to monitor temperature and bus voltages on
 crates which otherwise contain no custom-built electronics.

\begin{figure}
\begin{center}
\includegraphics[width=\textwidth]{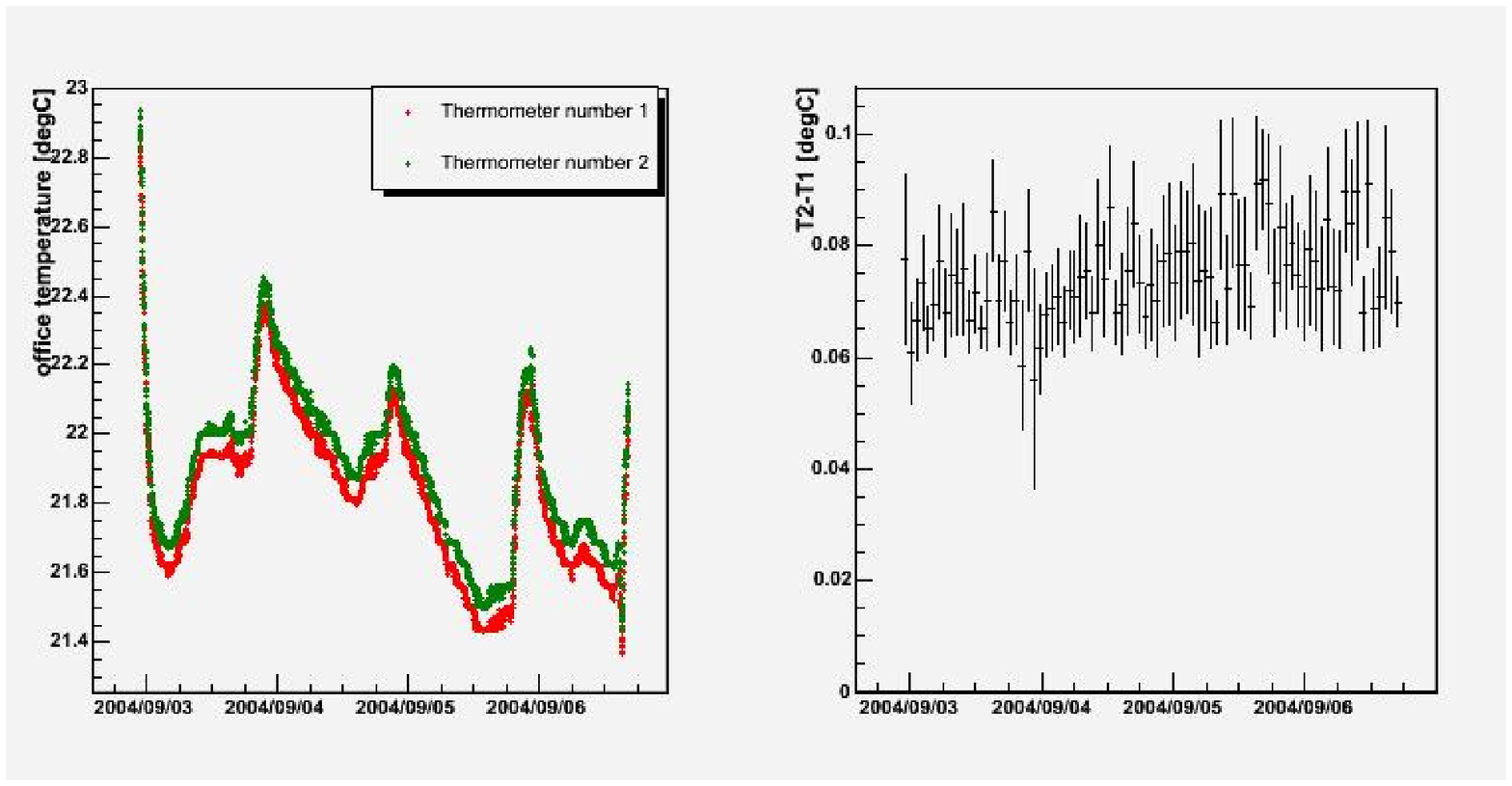}
\end{center}
\caption{Data from three day test of several \onewire\ samples: (a)
 temperature in an office at Kansas State University vs.\ time as read
 by two adjacent DS18S20 thermometer chips; (b) difference
 in reading of the two chips.}
\label{Fig:threedaytest}
\end{figure}

 In these chips, digitization of temperature and voltage is initiated
 by an explicit ``convert'' command from the bus master.  Temperature
 conversion takes 900~ms, and digitization of the four 12-bit channels
 of the DS2450 takes less than 4~ms total.  During conversion, the bus
 master may communicate with other devices if the chips have an
 external source of power; if a chip is powered parasitically from the
 bus, then the master must maintain the bus level high throughout the
 conversion.  Testing of samples provided by Maxim IC / Dallas
 Semiconductor confirm that multiple devices can maintain their
 internal state while all are powered parasitically from the same bus.
 Figure \ref{Fig:threedaytest} shows data from a three day period
 during which two DS18S20 thermometers were sampled once a second by a
 program written in Java.  The two thermometer chips were mounted in
 direct contact with each other, and recorded the same temperature to
 within a small fraction of a degree.  In this test, the thermometers
 were located about two meters from the bus master, and another
 \onewire\ device was connected on the same bus about three meters
 further downstream.  No failures or interruptions occurred during
 this period.
\subsubsection*{Experimental hall environment:} 
The DS1923 ``Hygrochron''
 \ibutton\ looks ideal for monitoring temperature and humidity.  In
 addition to the humidity and temperature functions, this device
 features on-board battery backup and automatic logging to internal
 memory independent of external control.  Power or computer failures
 will not interrupt the temperature and humidity record.  This
 \ibutton\ is designed for tracking sensitive products during shipment
 or other handling.  Each DS1923 is individually calibrated and NIST
 traceable.  Another \ibutton, the DS1922, provides similar
 functionality without the humidity function.

 Monitoring of barometric pressure and other environmental factors is
 easily achieved using the DS2450 ADC and one or more external
 transducers.  (Note: A complete \onewire\ weather station is even
 available \cite{one-wire-weather}.)  Another interesting
 ``environmental'' condition to monitor is ambient light level in the
 experimental area: many systematic effects in past and present
 neutrino experiments have been attributed, correctly or incorrectly,
 to electrical or optical noise introduced by lighting, and a simple
 phototransistor addresses the issue handily.  The phototransistor
 technique also can be used to monitor status LEDs on devices which
 lack electrical status outputs.  AC line voltage is also easily
 monitored using a \onewire\ ADC and a trivial circuit.

\subsubsection*{Liquid levels and temperatures, and gas pressures:} 
Transducers should be provided for monitoring important aspects of the detector
 such as scintillator and buffer oil levels, temperatures, and
 pressures in any gas systems used.  Transducers should produce
 voltages in the 0 to 5 V range for maximum compatibility with the
 DS2450 ADC.  The further specification, purchase, and installation of
 such transducers 
are the responsibility of the respective subsystems.

\subsubsection*{Simple controls:} 
The DS2890 is a \onewire\ digitally controlled
 potentiometer.  It can be used to provide slow control for simple
 servos, power supplies, or other devices controllable by an external
 analog signal.  At present, there is no definite plan to use this
 capability, although the discriminator levels could possibly be
 controlled in this way.  Support for slow control as well as
 monitoring should be provided in the software for maximum
 flexibility.

\subsubsection{Radon monitoring}

Professional continuous radon monitors have become readily available
and relatively inexpensive.  An example is Sun Nuclear's Model 1027
\cite{SunNuc1027}.  Each experimental hall will have at least one
radon monitor read out by the slow control PC.  The data will be
stored and made available via the same interface used for all slow
monitor data.



\subsubsection{Interface to other subsystems}


Some hardware subsystems may have important slow monitor that cannot
be made available on the \onewire\ interface or the serial ports of
the slow monitor computers.  Examples may include the clean room
particle counters, the high voltage power supplies, and the
discriminator circuitry in the trigger system.  In such a case, either
the hardware itself or a computer which monitors and controls it
should make the data available via network TCP connection.
``Virtual'' monitor data, such as capture time, event rates, or other
quantities determined by online analysis, could also be recorded by
this mechanism.  The software on the master slow monitor computers
will poll these external servers and make all slow monitor data
available in a common framework.  This is preferable to each subsystem
providing a separate data interface.  In the common framework,
systematic correlations may be studied among any variables.  Support
for control functions and synchronization with externally controlled
devices should be provided in the software to allow for scans of
controlled parameters such as high voltages and threshold levels.  The
common framework will allow dependent variables observed in one
subsystem, {\it e.g.}, discriminator rates, to be easily correlated
with parameters monitored or controlled by some other subsystem, {\it
e.g.}, high voltage.


%
\cleardoublepage
%
%
\section{Calibration}
\subsection{Calibration Goals}
\label{sec:calibration}
The two main tasks for which calibration is 
critical are the estimate of the inverse beta
decay detection efficiency and the determination of the 
energy scales  for positrons, gammas,
and neutrons.  Over time,
the relative detection efficiency between the near and far detectors 
should be known with
an uncertainty less than 0.5\%   
 including 
all deadtime effects.  
With regard to 
energy scales, 
the tolerable uncertainty depends on the level and 
nature of backgrounds.  We have adopted 1\% as the 
maximum uncertainty on the absolute
energy scales for gammas and positrons
because it is realistically achievable; 
the corresponding
relative uncertainties between the near and far detectors 
should then be much less than 1\% 
 since
the same energy-scale determination 
methods will be used 
 for both detectors.   

For the neutron energy scale, a less stringent requirement 
suffices: the uncertainty in the estimate of the  
visible neutron kinetic energy should be limited to 20\%.  
This relaxed requirement reflects the large signal-to-background ratio 
expected for each detector as well as the mild
 energy dependence of the fast-neutron background  across 
the reactor-neutrino energy range.
Our approach in specifying a system to achieve these goals builds 
on our extensive experience in calibrating the Palo Verde, CHOOZ, 
Super-Kamiokande, KamLAND, and MiniBooNE
detectors.

The uncertainty budget for the relative detection efficiency is taken to be 
(a) 0.2\% from the energy cut (``6 MeV cut'') 
applied to select neutron candidates,
(b) 0.1\% from the cut on the neutron
 capture time, (c) 0.25\% from deadtime, and (d)
$<$0.2\% from the requirement (if used)  of spatial 
correlation between the prompt and
delayed subevents.
  
 The uncertainty in the absolute energy scale receives contributions from the 
 uncertainties in extrapolating the energy response of the scintillator from 
gammas to
positrons and uncertainties in the corrections associated with 
event position.  These
uncertainties combined must be controlled to 1\% or better.   

In developing the design of
the calibration program described below, it has been assumed that the
contributions to the absolute positron energy-scale uncertainty are
controlled to
(a) 0.7\% from energy response extrapolation and (b) 0.7\% from position
corrections.

It is practically unavoidable that calibrations probe the detectors 
under different conditions
than do the signals and backgrounds of interest.    
It is likely not feasible
to build and deploy 
inverse beta decay sources, 
and even the detector responses 
to positrons and neutrons separately cannot be directly measured.
For example, the typical energies of neutrons produced in inverse 
beta decay induced
by reactor neutrinos 
are tens of keV, whereas the energies of neutrons produced by 
available calibration sources lie above 1 MeV.   
Moreover, the calibration process 
introduces structures into the detector 
that are not present during normal data taking.
It is therefore necessary to use detector simulation to estimate the detector response
to inverse beta decay and relevant backgrounds; 
the role of calibration is 
then to provide the information 
needed for tuning and checking the simulation to the required accuracy.   Included in
this process is correcting the calibration data for the effect of 
structures introduced into
the detectors for calibration.

In addition to providing measurements of 
detector response 
that will control its
uncertainties within the levels stated above,  the calibration program must 
provide
information 
that can be used to check assumptions about other aspects of detector
performance and 
 to carry out analyses using different sets of selection criteria or
 alternate methods of background subtraction.
  Specifically, the scope of calibration 
  will also include the following:
\begin{enumerate}
\item A check that the trigger is fully efficient for inverse beta decay events.
\item Calibration of the efficiency of a spatial correlation cut to 
within 0.2\%.
\item Measurement of the detector response to  
neutrons, for modeling the background 
due to fast neutrons.
\end{enumerate}
\subsection{Calibration Sources and Deployment}
We will employ three types of calibration sources: 
gamma sources, neutron sources, and light flashers. 
\subsubsection{Gamma Sources}
Gamma sources will be used 
for the following purposes:
\begin{enumerate}
\item Precisely measure the response of scintillator (target and 
$\gamma$-catcher) to gammas  
from well below inverse beta decay 
threshold to at least $\approx$5 MeV.   Spanning this 
range and beyond will also
facilitate understanding the 
roles of quenching and Cerenkov radiation, uncertainties in which
propagate through to the uncertainties in the 
positron energy scale derived from the gamma
energy scale.
\item Measure light transport properties (absorption, re-emission, 
speed of light) of the liquids in 
the target, $\gamma$-catcher, and buffer.
\item Measure PMT $t_0$'s.
\item Check relative efficiency of far and near detectors with respect to trigger, data
acquisition, event reconstruction, and event selection.
\item Global monitoring of detector stability.
\end{enumerate}

The set of gamma sources to be used is enumerated in 
Table~\ref{tab:gamma-sources}.
\begin{table}\caption{Gamma sources to be
used in calibrating Double Chooz}\label{tab:gamma-sources}
\begin{center}
\begin{tabular} {|l|c|c|c|}
\hline
& Source & $E_\gamma$ (MeV)& Half--life \\
\hline
1 & $^{203}$Hg & 0.289 & 46.6~d  \\ 
2 & $^{137}$Cs & 0.667 & 30.1~y  \\
3 & $^{68}$Ge & 0.511 + 0.511 & 271~d  \\
4 & $^{60}$Co & 1.333 + 1.173 & 5.27~y \\
5 & $n$ capture on H & 2.223 & \\
6 & $n$ capture on C & 4.94 & \\
\hline
\end{tabular}
\end{center}
\end{table}

The sources will be encapsulated and stored so that the same physical sources
can be deployed in all volumes in both detectors.    
Gamma sources 2--4 could be encapsulated in a single package. 
The activities of at least one of the 
sources $^{68}$Ge, $^{60}$Co, or $^{137}$Cs will be
accurately known (1--2\%) to check absolute efficiencies.    The 
activity of each isotope (and the 
total activity of any combined sources) 
should be well above the expected background trigger 
rate but low enough to avoid large detector deadtimes.    Over the course 
of the experiment, 
new sources may need to be prepared for the isotopes with 
half--lives much shorter than a year.
For the convenience of global monitoring and other 
calibration tasks that may be carried out more frequently, 
sources may also be provided for use
in only one detector or sub-volumes within a detector.
There will be significant continuum background due to 
$n$ capture on Gd underlying
the  4.94\,MeV line and hence it 
is likely to be visible only for long source deployments.    
The feasibility of deploying other
high-energy gamma sources 
is being explored.

\begin{figure}[hbtp]
\begin{center}
\includegraphics [width = 0.40\textwidth] {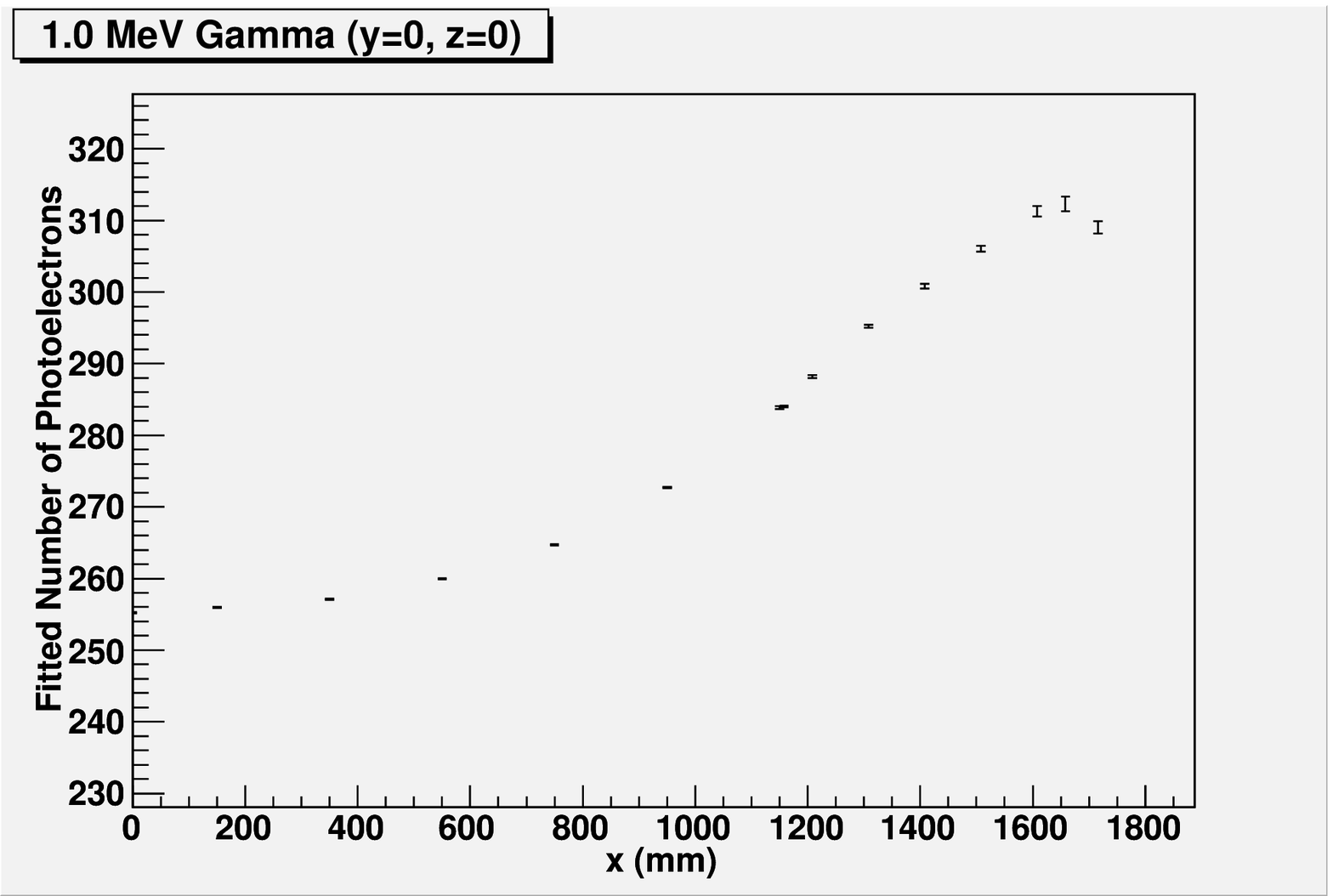} 
\includegraphics [width = 0.40\textwidth] {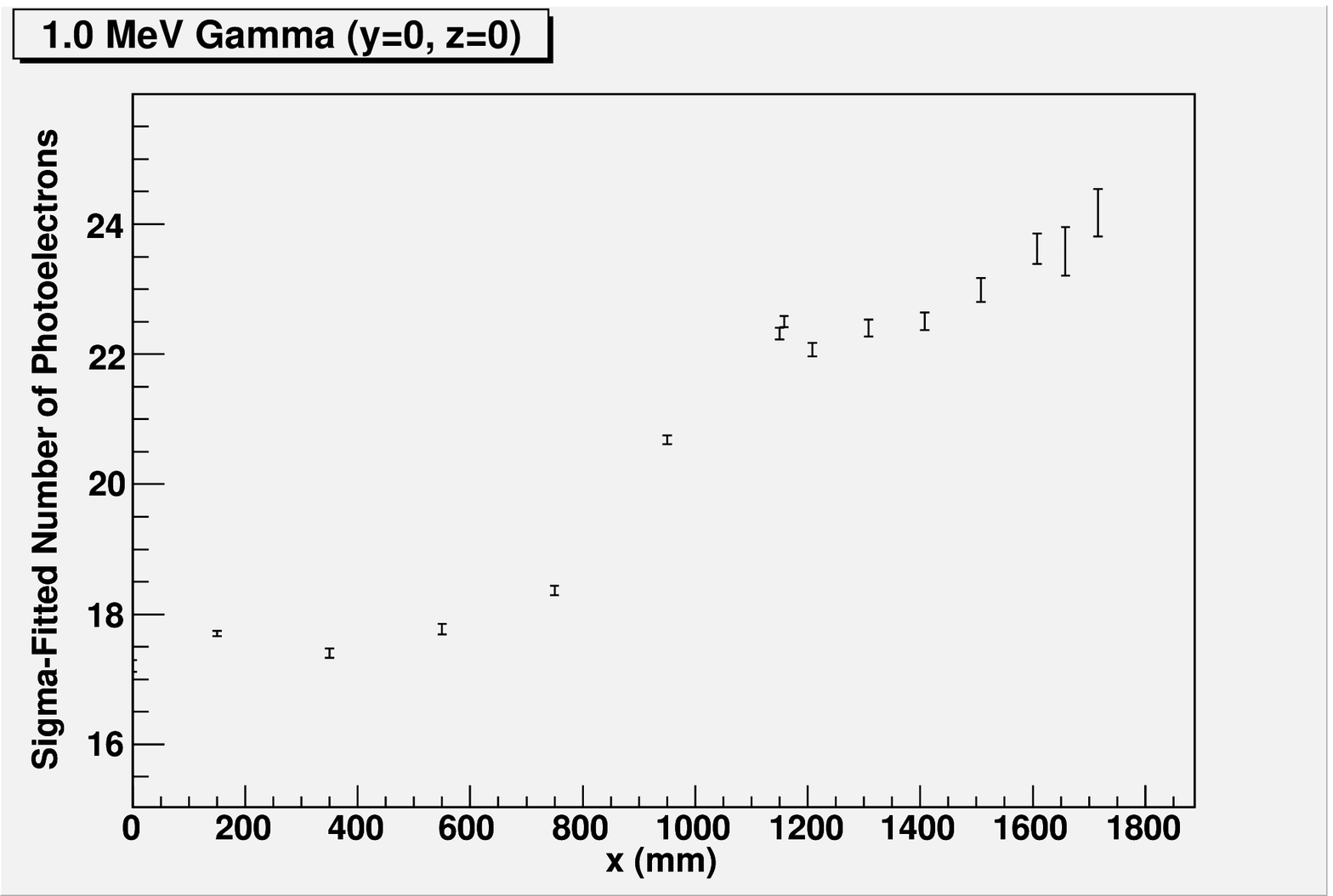}
\end{center}
\caption{Center of Gaussian fitted 
to peak in photoelectron distribution versus
Center (left) and sigma (right) of Gaussian fits
 to peak in photoelectron distribution versus
distance along the $x$-axis for 1 MeV gammas.}
\label{xvar}
\end{figure}

Detector simulations have been carried out to estimate how the 
detector response
to gammas 
will vary with position.    The results are typified by  what is shown in
Figure \ref{xvar} for samples of 1~MeV gammas generated along the 
$x$-axis.    
The strongest variations occur near the target--$\gamma$-catcher
boundary and across the $\gamma$-catcher.   To avoid unacceptable uncertainties
in energy reconstruction, the deployment system will have the capability to
position  gamma sources  near the 
target--$\gamma$-catcher boundary and at points within the $\gamma$-catcher.  
Furthermore,
to 
ensure that the uncertainty in the energy reconstruction due to position 
uncertainty
is less than 0.7\%, the systematic uncertainty in 
source position must be
less than 1.5~cm.  The particular choice of deployment positions
in the $\gamma$-catcher will be further discussed in 
connection with calibrations with neutron
sources. 

\subsubsection{Neutron Sources}
Neutron sources will be used 
for the following purposes:
\begin{enumerate}
\item Measure the relative neutron detection efficiency between the 
two detectors. 
\item Measure the absolute neutron detection efficiency at selected 
points in each detector.
Such a measurement is required to help check the accuracy 
of the neutron simulation and
for analysis of 
data for which there is only one detector, most notably the far
detector data taken before the near detector is turned on.
\item Measure the energy response 
of the target and $\gamma$-catcher to neutrons, 
to enable accurate simulation of
the prompt energy spectrum of fast neutron events. 
\item Gauge the relative deadtime  
of the two detectors to a source of correlated events.
\end{enumerate}

As with the gamma sources, the neutron sources will be 
encapsulated and stored so that the same sources
 can be deployed in both detectors. To measure the 
absolute efficiency, deployment of a source
 instrumented with a dedicated device to tag 
neutron emission, such as
the one used in the CHOOZ experiment, is planned.  Using 
the detector itself to tag neutron 
emission, by registering the prompt
 energy deposited by accompanying gammas,  is under study.
Am--Be, for example) will overlap
In order to measure the energy response of
the detector to neutrons, the energy distribution of the 
neutrons emitted by each source
must be well known and corrections must be made for 
the energy deposited by 
any gammas accompanying the neutron emission.

Detector simulation has been carried out at the 
photoelectron multiplicity and 
timing level to  understand
where neutron sources should be deployed and the degree 
of corrections that 
would have to be applied to the efficiencies measured 
with Cf--252 or Am--Be 
sources to obtain the efficiency for neutrons produced by 
inverse beta decay.
Monoenergetic neutrons were
generated uniformly throughout the target, uniformly throughout the 
$\gamma$-catcher, or at specific points.  Several 
energies were simulated: 0.01 MeV as
a characteristic energy for reactor neutrino events, 2 MeV 
as the approximate energy
for Cf--252 neutrons, and 4 MeV, which is close 
to the mean energy for Am--Be neutrons.
For each event, photoelectrons were classified as 
prompt or delayed, depending on
whether they were registered less or more 
than 200~ns from the start of the event.  The
mean time of the delayed photoelectrons was 
used as the capture time.  Neutron
selection cuts were defined based on the total 
number of photoelectrons detected and
the neutron capture time.  

\begin{figure}[hbtp]
\begin{center}
\includegraphics [width = 0.4\textwidth] {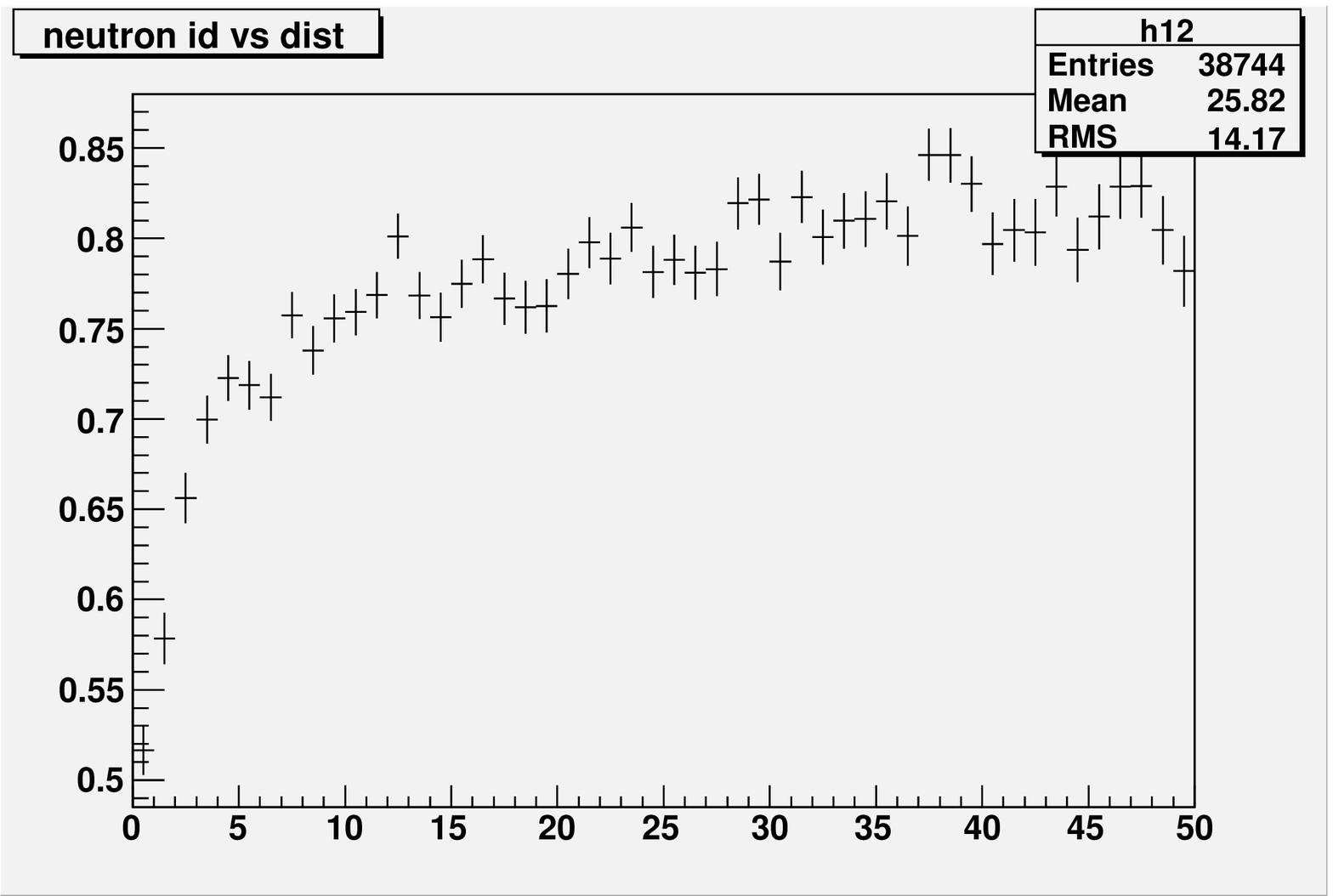}
\includegraphics [width = 0.4\textwidth] {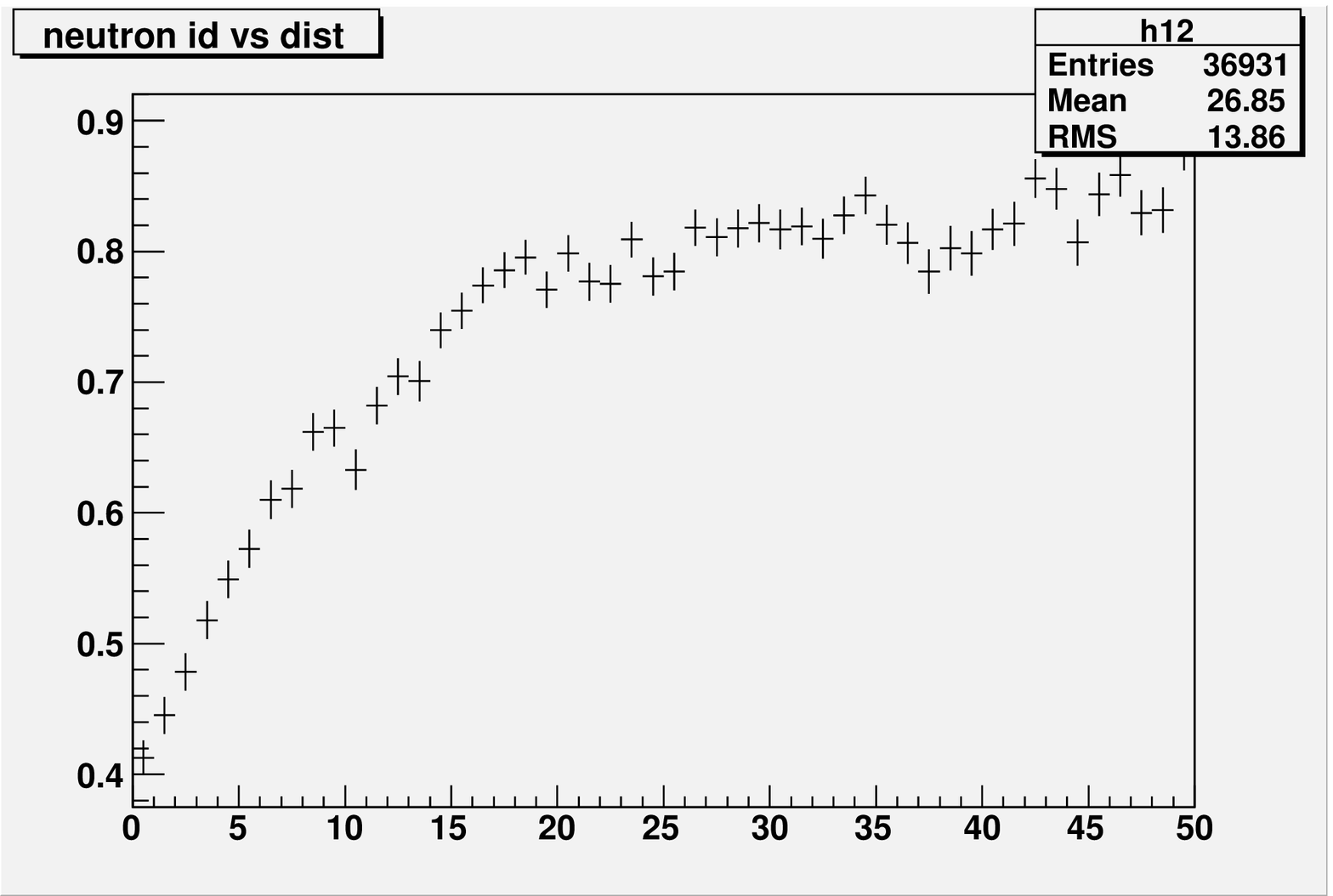}
\end{center}
\caption{Neutron detection efficiency versus 
distance (in cm) between neutron creation
point and target--$\gamma$-catcher boundary 
for neutrons generated in the target:   
(left) 0.01~MeV neutrons; (right) 4~MeV.}
\label{neff}
\end{figure}

The following points can be drawn from the simulations 
performed to date:
\begin{itemize}
\item The neutron detection efficiency varies significantly throughout 
the target, hence
the calibration system should have the
capability to deploy neutron sources throughout
the target volume.  Figure \ref{neff} shows the neutron 
detection efficiency as a function
of distance to the target--$\gamma$-catcher boundary for 
neutrons generated uniformly
in the target.
\item  Efficiency loss 
due to neutrons produced in the target escaping into the 
$\gamma$-catcher is 
termed ``spill-out'' while efficiency gain 
due to neutrons 
produced in the $\gamma$-catcher being captured on 
Gd in the target is termed 
``spill-in''. 
For neutron sources such as Am--Be and Cf--252, spill-in and 
spill-out are separately
large effects (
$\approx$5\%) but appear 
largely to cancel, as would be expected.   
However, since (a) the neutron
detection efficiency is to be estimated with
an uncertainty of a fraction of a percent and (b) the range 
of distance over which spill-in and spill-out occur is much greater (
$\approx$20~cm) for Am--Be and Cf--252 sources than it
is (
$\approx$5~cm) for neutrons from inverse beta decay,  deployment of neutron
sources 
is planned both within the $\gamma$-catcher and in the target near
the 
target--$\gamma$-catcher boundary, 
so as to 
measure this effect at representative 
points and help tune/check the MC simulation
of neutron transport and capture.
\item  The neutron detection efficiency depends on 
the energy distribution of the neutrons.
The differences in neutron detection efficiency between neutron
sources and inverse beta decay are large enough that corrections 
must 
be applied in order
to satisfy the requirements 
on 
uncertainty in 
detection efficiency.  
Multiple sources 
(Cf--252 and Am--Be), with different neutron energy spectra, 
will be deployed so as to measure and 
reduce 
this uncertainty.   The possibility is also being explored of deploying a 
low-energy neutron source, 
in order to reduce the uncertainty in 
transferring 
efficiency measurements made at higher
neutron energies to the energy range of neutrons produced by reactor neutrinos.
\end{itemize}

Detector simulation was also used to estimate the relative uncertainty in 
energy scale between the near and far detectors that could be tolerated while maintaining 
an uncertainty of 
0.2\% in 
neutron detection efficiency due to the ``6 MeV'' cut.   The
energy-scale uncertainty so obtained is 100~keV.

\subsubsection{Light Flashers}
\label{sec:laser}
Fast lasers of several 
wavelengths will 
also be used to calibrate the detector.
The main purpose is for vertex reconstruction, which depends 
critically on light transport
and PMT timing.  In principle,  vertex reconstruction could 
be calibrated with
gamma sources alone, but laser calibrations have the 
advantages of (a) providing
signals of precisely known timing and amplitudes and 
(b) probing the response of the detector to
specific wavelengths.  Moreover, the laser systems can 
be designed to provide a
wide range of well-known intensities, so that laser 
calibrations can be used to
help interpolate the detector response between gamma 
calibrations and extrapolate it
beyond the range of signal strength covered by 
gamma calibrations.  The
capability of lasers to provide very  strong 
signals is useful in understanding the 
response of the detector to cosmic events, many of 
which excite the detector in a regime
where the PMT response is expected to be 
significantly nonlinear.     We have adopted the following
as requirements for the laser calibration system:

\begin{itemize}
\item The laser light output must 
either be highly isotropic (preferred) or its
anisotropy must be 
precisely known: this greatly simplifies the use of laser 
calibration data for
light transport and energy calibration studies.   
\item The width of the laser pulse must be 
small compared to the scintillator
decay times so that the distribution of photon arrival times 
at the PMTs does not
depend strongly on the details of the laser 
pulse timing.  Furthermore, the timing
of the leading edge of the laser pulse (before attenuation filtering) 
must be known with a precision 
better than 1 ns so that studies of PMT timing  can be carried out
using as a timing reference a trigger pulse derived from the 
leading edge of the 
laser pulse.
\item At least two widely separated wavelengths must be 
provided: one that excites
the scintillator to produce pulses with the same 
spectral and timing characteristics as
those due to 
charged particles, and the other whose 
absorption length in
scintillator is very long so that calibrations of
 PMT timing can be carried out with minimum
complications due to re-emission and scattering. 
\item The dynamic range of the laser system must 
extend from much less than one
photoelectron per PMT on average to several 
hundred photoelectrons per PMT. 
\item Fiber assemblies and diffusers must be 
selected so as to ensure that light is injected at
precisely known points and the self-shadowing of 
the system is either negligible or 
adequately known.
\end{itemize}

Because of (a) the overlapping capabilities of 
calibrations with light flashers 
and those 
with radioactive sources and (b) the expectation that measurements
of intrinsic PMT gain and timing characteristics will not
depend significantly on the position of the source, 
deployment of the light flasher sources only on a vertical 
line through the target is
sufficient.  
For the sake of simplicity, the symmetry axis of the detector is 
preferred, but deployment along another vertical line would also 
be 
suitable.

\subsection{Calibration Deployment}
\subsubsection{Introduction}
The purpose of the calibration deployment system is to 
deploy calibration sources into the
target and $\gamma$-catcher regions.    
The calibration sources,  the motivation for using
them, and basic deployment requirements have 
been described above.
 The deployment systems utilized by the near and 
far detectors will be identical.

The deployment system must
be designed to accommodate 
gamma sources,
terminated fibers illuminated by external lasers, and 
neutron sources (untagged and tagged).  The characteristic
dimensions of these source will range from a few mm 
to a few cm, and their masses from a few tens 
to a few hundreds of grams.    
The calibration system
must be capable of positioning sources
at representative points in the target and
$\gamma$-catcher with an uncertainty less than 1.5~cm.

The materials and geometry of the deployment system must be chosen to minimize
uncertainties in the corrections for shadowing and absorption.   
Detector simulation
studies have been carried out to set the maximum dimensions that can be
used and still meet the energy-calibration uncertainty requirements.
Each material used in the calibration system must be 
compatible with all elements
of the detector environment to which it is exposed.  
Furthermore, measures must be taken to
protect the detector from radio--contamination by the calibration systems.
The process of deploying the calibration
system must not affect detector performance.  The setup of the
calibration deployment system should not be awkward or time-consuming, hence 
calibrations that 
are carried out frequently will be largely automated.

We next describe 
the deployment system in three parts: 
deployment 
methods, 
detector interface, and 
control systems.

\subsubsection{Deployment Methods}
 
 The methods of source deployment for the target 
region will differ 
 from that for 
 the $\gamma$-catcher region because of the different 
geometries and different calibration
 requirements; therefore, they are discussed separately.
 \vskip 0.2in
 \noindent{\underbar{{Target}}}
 
\begin{figure}[hbtp]
\begin{center}
\includegraphics [width = .95\textwidth] {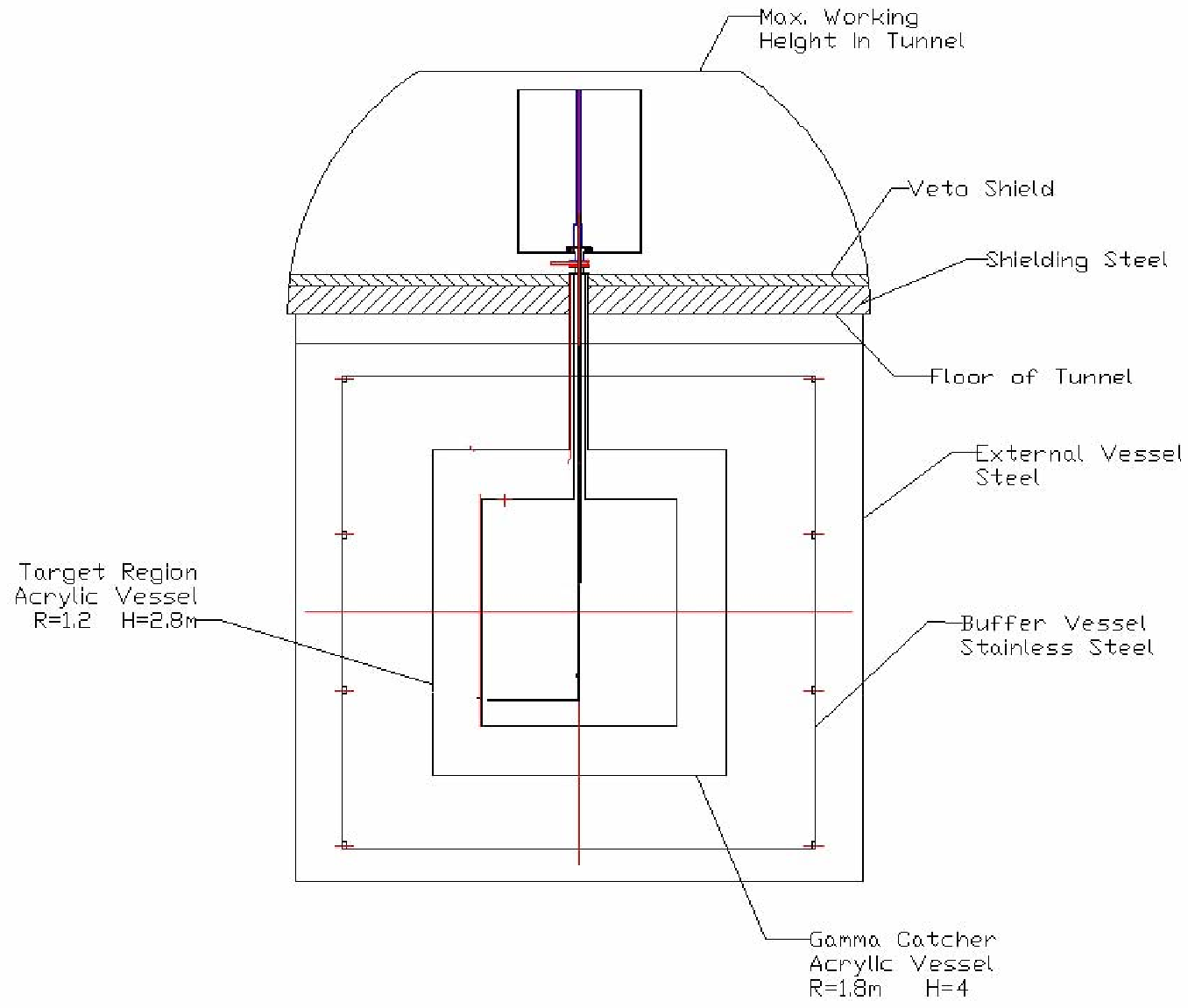}
\end{center}
\caption{Schematic of the articulated arm system for 
deploying calibration source
throughout the target region. }
\label{artarm}
\end{figure}

Because we 
intend to use the entire 
volume of the target,
the 
calibration sources must be 
deployable throughout the 
target  region.
This will be accomplished using an articulated arm
shown schematically 
 in Figure \ref{artarm}.
 The articulated arm is comprised of a telescoping vertical 
shaft, supported from the
 calibration interface above the detector, with a 
fixed-length arm pivoting at the lower end.  By adjusting the length and
 azimuthal
angle of the telescoping shaft and the angular 
position of the arm,  a source attached 
to the end of the arm can be
deployed at any position within a cylindrical volume.
During calibration, an operator will 
attach a 
calibration source to the source holder, deploy
the source into the target at the desired positions, 
and then retract the source to the
detector interface.    
A cable and winch system (``fish-line'') will also be installed for
frequent deployment
of sources along the target $z$-axis. 

Those components of the deployment system that 
come into direct contact with the
scintillator will be checked for material compatibility.  
samples for changes in 
Components that will not come into direct contact with the 
scintillator but that 
will be 
exposed to its vapors  must 
also be tested.   
\vskip 0.2in
\noindent{\underbar{{$\gamma$-catcher}}}

The $\gamma$-catcher requires its own calibration because 
its light yield and properties of
light and neutron transport will likely differ from those 
of the target.    These differences 
can be measured using gamma and untagged neutron sources.  

\begin{figure}[hbtp]
\begin{center}
\includegraphics [width = 0.5\textwidth] {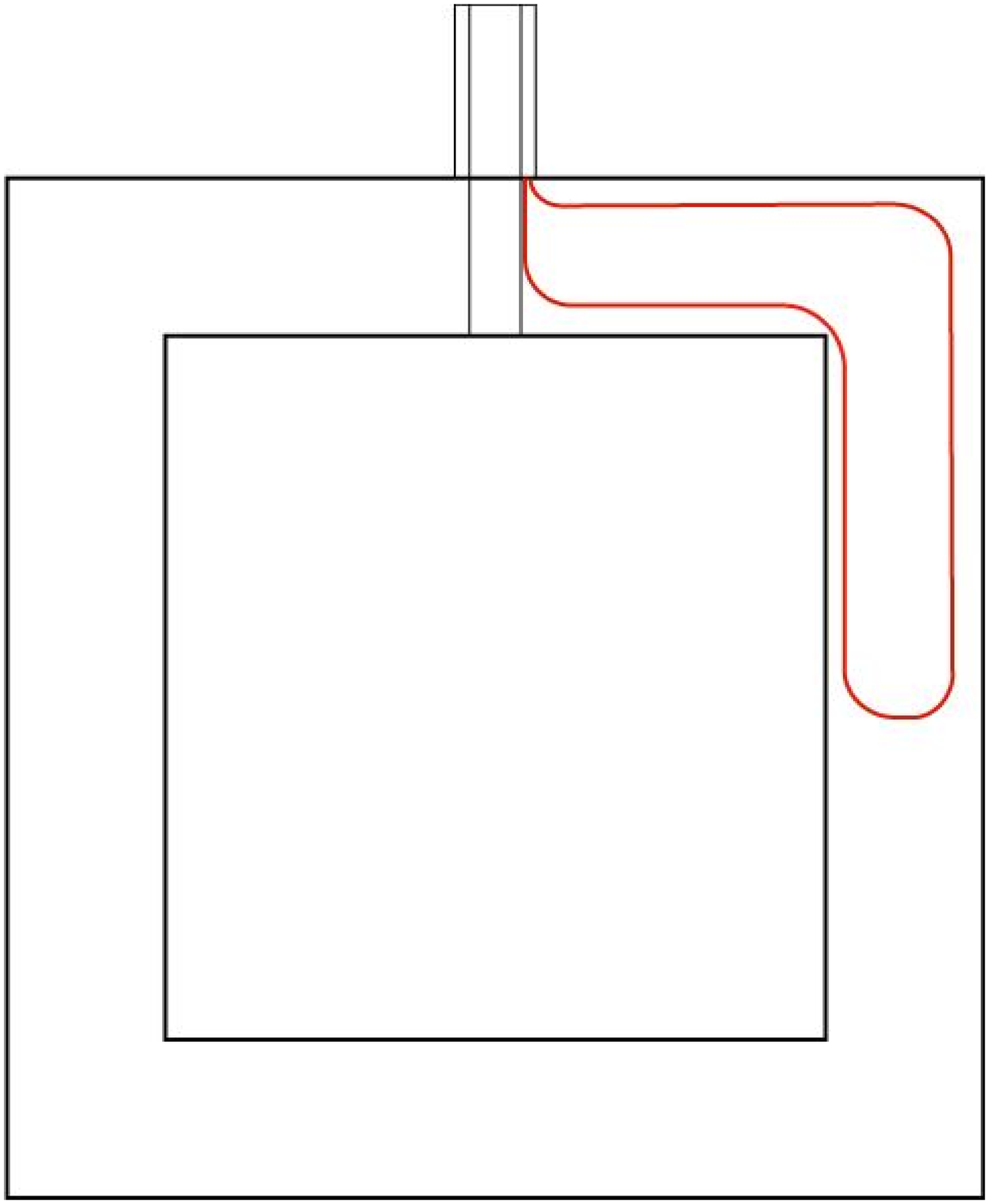} 
\end{center}
\caption{A possible guide-tube path for deployment of sources in the 
$\gamma$-catcher}
\label{gt}
\end{figure}

A guide tube system will be used to deploy calibration 
sources in the $\gamma$-catcher.  The tubes must be small in diameter
in order to 
avoid shadowing of scintillator light and to 
minimize dead material and absorption.  The tubes will 
run into the
$\gamma$-catcher from a manifold in the 
detector interface.   Figure \ref{gt} shows a 
path that a guide tube, designed
to have 
two accessible ends, 
could take in order to allow sources to be positioned at
representative points within the $\gamma$-catcher.
The calibration sources will 
be attached to
a wire and pushed through the tube. 
The position of the source can be determined from the length
of wire inserted into the tube and an accurate survey of the 
guide-tube geometry.   

The number of guide tube(s) to be deployed in the 
$\gamma$-catcher is being studied.
The connection of the wire to the sources must be 
such that the same sources can
be deployed in the $\gamma$-catcher as in 
the target.  As with the deployment system for
the target, all components will 
have to be checked for material
compatibility before final selection and
carefully cleaned before installation into the experiment.

\subsubsection{Detector Interface}

The detector interface is the region that 
has access to the target and 
$\gamma$-catcher and
that 
can be accessed from the outside to introduce 
and remove calibration sources.  
To prevent contamination of the detector, the 
detector interface will incorporate a glovebox for source 
manipulation and an airlock through which sources can be 
inserted and removed.
The volume of the glovebox 
must be large enough 
that sources can be
easily manipulated and the deployment system can be 
assembled and disassembled safely and easily as needed.
The glovebox must be connected to an N$_2$ gas system and purged so 
that it has the same atmosphere each time 
it is opened
to the detector. 
avoid introducing backgrounds.    
The
glovebox will be equipped with radon and oxygen 
monitors for flagging leaks and monitoring
the progress of purging.
While the
detector interface is open to the detector, it must 
also be light tight, which may
mean that the operator must view the interior of 
the glovebox using infrared illumination
and cameras.   Feedthroughs for
laser fibers and control and power cables must be hermetic.  
To bring sources into
the detector interface, a transfer box is needed: sources 
are placed into this box through
an external door, then the transfer box is purged, 
after which the operator opens an
internal door and brings the source inside the detector interface.  

\subsubsection{Control Systems}
 
 All control and sensor channels for the deployment systems 
must be interfaced to
 a computer system for 
 purposes of automation and 
 monitoring.  Although 
 it should be possible to monitor the state 
of the deployment system remotely, for
 safety and convenience, the computer system 
 controlling and 
 monitoring
 the deployment system should be installed near the detector 
 interface.
 Interlocks,
 both hardware and software, must be implemented to 
ensure that the detector is
 not opened to light, that the door 
 between the detector interface and the detector
 is not closed while sources are being deployed, etc.  
 Full exercise of the control program
 and interlock system will be part of the testing of 
the deployment systems.

\subsection{Calibration and Monitoring Plan}

During the commissioning of each detector, it is assumed that a light flasher
will be deployed at the center of the detector for debugging and for
adjusting PMT gains.  
Once commissioning is complete and detector operation
has stabilized, a full calibration program of the 
detector will be carried out using 
gamma, neutron, and light flasher sources deployed at 
representative points
in the target and $\gamma$-catcher.

After the first full calibration, the process of  monitoring begins.
The purpose of the monitoring process is to provide the data needed to
 accurately
interpolate detector performance between full calibrations and to help
determine when the detector performance has changed to the degree that
a full calibration is needed to maintain the required accuracy.
On a regular basis (initially at least once per week), a gamma source and a
neutron
source will be deployed both in the target (along the $z$-axis using the
``fish-line'' system) and in the
$\gamma$-catcher.  Once both detectors are
operational, monitoring will be carried out on the same schedule for both
detectors, with at least one gamma source and one neutron source deployed
in both detectors.  
In addition to monitoring with sources, analyses looking at
natural backgrounds should be developed to 
provide information on
detector stability on a day-by-day basis.    The frequency of monitoring
with sources can 
be adjusted once the collaboration has experience with
the actual stability of the detectors and the irreducible impact of the
monitoring
process itself on detector stability.

Full calibrations are to be carried out (a) when the detector 
performance has changed
to the degree that monitoring data alone, combined with 
the previous full calibration,
no longer suffices to meet the calibration goals of the detector,
(b) after a major
change to the detector, or 
(c) when the maximum interval between full calibrations as set by
the collaboration has been reached.    To the greatest extent 
practicable, full calibrations of 
the near and far detectors will follow the same schedule.

%
\cleardoublepage
\providecommand{\boldsymbol}[1]{\mbox{\boldmath $#1$}}

\providecommand{\tabularnewline}{\\}
\section{Simulation and Software}
It is not unusual that the design and engineering of a new experiment is
heavily aided by computer simulations.  Rightfully speaking,
a large portion of the Double Chooz proposal is assisted by simulation
studies in such a way that simulated results are fully integrated into all of
the essential parts of the entire document.  The role of the
simulation studies is best appreciated in their proper context so that there is
no need to reiterate simulation results already explained elsewhere in the
proposal.  However there are interesting facts related to simulation
and software that are not already mentioned in other parts of the proposal
and are most suited in a
chapter dedicated for a panoramic view of the architecture and
philosophy of software development in Double Chooz.  This chapter outlines
the basic concepts of the software tools used in detector R\&D and data
analysis for the Double Chooz experiment.

\subsection{Software and Simulation Goals} 
The main goals of simulation studies can be divided into two major categories:
(1)~R\&D including engineering and detector designs, and (2)~data analysis
and physics discoveries.  The first category constitutes the main emphasis
of the studies during the proposal writing phase of the project while the
second category typically continues until the end of the project.

\subsubsection*{Simulation Requirements}
One of the key purposes of simulations is 
to optimize the detector design.
The optimization is done with respect to background sensitivity, detector
resolution and cost.  The simulation is expected to faithfully model the 
gamma ray and nuclear properties of the detector, particle transport, 
backgrounds
and shielding studies.  It must correctly represent the optical properties of
the scintillator, acrylic, photo-tubes and detector walls.  Possible distortion
from electronic effects must be included.

\subsubsection*{Software Acceptance Tests}
Reliability and stability are the goals of Double Chooz software.  A lot
of efforts are
invested in cross-checking the simulation results with existing experimental
data as well as the detection of programming bugs such as inability to compile,
runtime crashes and memory leaks.  Nominal stability tests and detailed
comparisons with observations are made to insure that the physics model
accurately represents the Double Chooz environment.

\subsubsection*{Comparison with Prior Measurements}
There are existing experimental results on scintillation light yield,
attenuation, emission spectra, absorption spectra and surface optical
properties that are compared with the simulation.  Nuclear and electromagnetic
properties such as photon attenuation length, neutron absorption gammas and
capture time are also checked against the known properties of the materials.
For example, the simulated angular distributions of the muon background at the
far site using \texttt{MUSIC} has been checked with the experimental results
of a 1994 cosmic ray measurement on location and are shown to be in good
agreement~\cite{tang:music}.

\subsubsection*{Comparison with CHOOZ Data}
We are fortunate that a middle baseline neutrino experiment CHOOZ has already
been carried out at the far detector site using similar 
technology~\cite{bib:chooz}.  Data
from the first CHOOZ experiment for both reactor on and reactor off running
are available for Double Chooz.  This has helped
in tuning the neutron response and in modeling the background.

\subsubsection*{Quality Control}
All software packages are subject to rigid quality criteria for reliability,
accuracy and practicality.  Double Chooz software packages are designed
to help the collaborators to be productive in physics research.  Documentation
is actively maintained which  includes operating instructions, summaries of
test runs and descriptions of the algorithms employed.

The Double Chooz software packages are centralized 
in the CVS repository at Lyon
(code named CC-IN2P3).  This repository is easily accessed via the internet
for 
all members of  this international collaboration.  It permits version control
so that certified codes can be used in production, while new codes are being
developed.  If problems are encountered during quality testing, the prior
version of the code can be retrieved from the archive.  Documentation on the
accessibility of Double Chooz software packages is also an
important part of the quality control program.
\subsection{Software Tools}
Double Chooz programming tools are built on a collection of scientific
software packages such as \texttt{GEANT}, \texttt{FLUKA}, \texttt{MCNP},
\texttt{MUSIC} and \texttt{ROOT}.  There are enough overlaps in functionality
among these software packages that they provide a means for the
necessary cross checks for the physics goals, engineering and design of the
experiment.  For
example, a recent study has shown that \texttt{GEANT4} produces fewer
muon-induced spallation neutrons than \texttt{FLUKA} above 100~GeV by almost
a factor of 2 in some materials~\cite{araujo:2005}.  In a high precision
experiment such as a reactor $\theta_{13}$ measurement, variation in
simulated results by a factor of 2 from various software packages tends to
make a difference.  For this reason, efforts are made in the collaboration to
maintain some level of redundancy and repetition as cross checks
to ensure the reliability of the engineering design of the detectors.

\subsubsection*{GEANT 4} 
\label{sec:glg4}
The primary tool for modeling detector response for Double Chooz is
\texttt{GEANT4}~\cite{Geant4} and more particularly an extension of it called
\texttt{GLG4sim}~\cite{GLG4sim} which is a general purpose detector
simulation package derived from a special purpose counterpart called
\texttt{KLG4sim} used by the KamLAND collaboration.  The full
acronym of the Double Chooz version of \texttt{GLG4sim}
is \texttt{DCGLG4sim} which is usually referred to as \texttt{DCG4sim}
or simply \texttt{DCsim} for short.  \texttt{DCsim} simulates all the
relevant physics processes that occur in a liquid scintillator detector.
Generation of primary events is accomplished by using either external event
generators or codes already built into \texttt{GEANT4}, \texttt{GLG4sim}, or
\texttt{DCsim}.  Electronics is simulated by a separate software
package called \texttt{esim}~\cite{bib:esim} which has no inherent
dependence on \texttt{GEANT4}, \texttt{GLG4sim} or \texttt{DCsim}.

With a few exceptions, standard \texttt{GEANT4} toolkits are used to
implement essentially all physical processes
involving the passage of particles with energies above a few keV
through the materials in the detector.  A combination of standard
\texttt{GEANT4} plus \texttt{GLG4sim} extensions are used to simulate
optical photon physics, from optical photon production to
photoelectron emission at the surface of the photocathode. Double
Chooz collaborators from Saclay have made critical contributions 
to this code~\cite{MANDsimMotta}.  
Neutron thermalization in \texttt{GEANT4} is implemented 
in the module \texttt{NeutronHP} using
version 3.7 of the \texttt{GEANT4} neutron data libraries 
and supplementary
data files supplied by the author of 
\texttt{NeutronHP}.
Neutron capture on all isotopes except Gd is handled by \texttt{NeutronHP},
while capture on Gd is calculated using a code written by Double Chooz
collaborators from Subatech~\cite{MANDsimZbiri}.  Fast neutron 
production can be simulated using
\texttt{GEANT4} modules for leptohadron processes or external
packages ({\it e.g.} \texttt{FLUKA}~\cite{bib:fluka}) 
with the products fed to \texttt{DCsim} as externally 
generated primary events.
\subsubsection*{GEANT 3} 
A \texttt{GEANT3}-based simulation package has been 
written mainly for the
modeling the detector response and the 
calculation of internal and
external radioactive backgrounds. This simulation 
includes all the known
materials and their geometrical characteristics, 
including a database of multiple measurements of the U,
Th, K, and Co concentrations of typical materials.  Some examples of the
typical materials are steel, acrylic, PMT glass, and PMT dynode structures.
Shield design against external gammas and the assessment of the
effect of PMT radioactivity are aided by this simulation.

The package contains a U, Th, K, Co event generator that has incorporated all
of the known beta transitions and gamma lines of
these decays.  The generator includes the appropriate multiple
gamma-gamma correlations and branching ratios for
each of the decays.  The high speed performance of the package makes it
possible to simulate several billion nuclear decays 
and to track the daughter
particles to assist the design of the detector.

\subsubsection*{MCNP} 
\texttt{MCNP}~\cite{bib:mcnp} (Monte Carlo N-Particle code) is developed
and owned by Los Alamos National Lab.  It is used to study nuclear processes,
in particular thermal neutron transport and capture in the studies of
non-invasive calibration of the gamma-catcher.  This calibration method uses
neutron activation of sources within the gamma-catcher.  \texttt{MCNP} is used
to study neutron diffusion.

\subsubsection*{FLUKA} 
Low energy neutrons in \texttt{FLUKA} are simulated by 
a multi-group strategy.
Below the energy threshold (typically $E_n<19.6$~MeV), neutron energy is
divided into multiple (typically 72) groups.  The transition probability
from one group to the next
during neutron transport is computed through the down-scattering matrix.  Above
the energy threshold, continuous cross sections are used. 
The usage of \texttt{FLUKA}\cite{bib:fluka} at Double Chooz is mostly limited
to the estimation of the muon-induced neutron
rate at the near detector site.  Studies~\cite{wang} show that
\texttt{FLUKA} reproduces observations of underground neutrons very accurately
so that it is an indispensable tool to model fast neutrons.
The muon flux is defined at the surface~\cite{bib:reyna} 
and \texttt{FLUKA} is used to propagate muons and neutrons to the
new laboratory depth through standard rock.  The studies performed
with \texttt{FLUKA} are used particularly to optimize designs for
the near laboratory and the outer veto system.

\subsubsection*{MUSIC} 
\label{sec:music}
\texttt{MUSIC} (Muon Simulation Code)~\cite{bib:music} 
is a \texttt{FORTRAN} subroutine that
calculates muon energy loss and angular deviation in a material.  The default
cross section files that come with the \texttt{MUSIC} package are calculated
for standard rock only.  In the present application, the material used in the
simulation is defined by the real Ardennes
Mountain rock profile that has a mean density, atomic number, atomic
mass and radiation length of $\langle Z\rangle=11.8$, $\langle A \rangle=24.1$,
$\rho=2.81\,\rm g/cm^3$ and $\lambda=23.3\,\rm g/cm^2$ respectively.  The
approximate chemical composition of the Ardennes rock is 58\% $\rm SiO_2$,
19\% $\rm Al_2O_3$, 17\% $\rm FeO$, 4\% $\rm MgO$ and 2\% $\rm K_2O$ in
elemental percentages.  The above information of the rock composition is used
in the calculation of
the cross section files used with \texttt{MUSIC} for this specific case.
\texttt{MUSIC} is well tested and its results are shown to agree with
experimental data~\cite{tang:music}.  The advantage
of \texttt{MUSIC} is its modularity that gives it the flexibility to be easily
integrated into more complicated simulations involving non-standard geometry
that are not defined in other simulation packages such as \texttt{FLUKA} and
\texttt{GEANT}.  This is particularly useful when simulating
underground muon overburden with a non-flat topographic mountain profile.
Since there is sufficient information specific to the Ardennes Hill that
is built into the present \texttt{MUSIC} simulation, the package is sometimes
referred to as \texttt{DCMUSIC} (\texttt{MUSIC} for Double Chooz)
by the collaborators.

\begin{figure}
\centering
\includegraphics[scale=0.45]{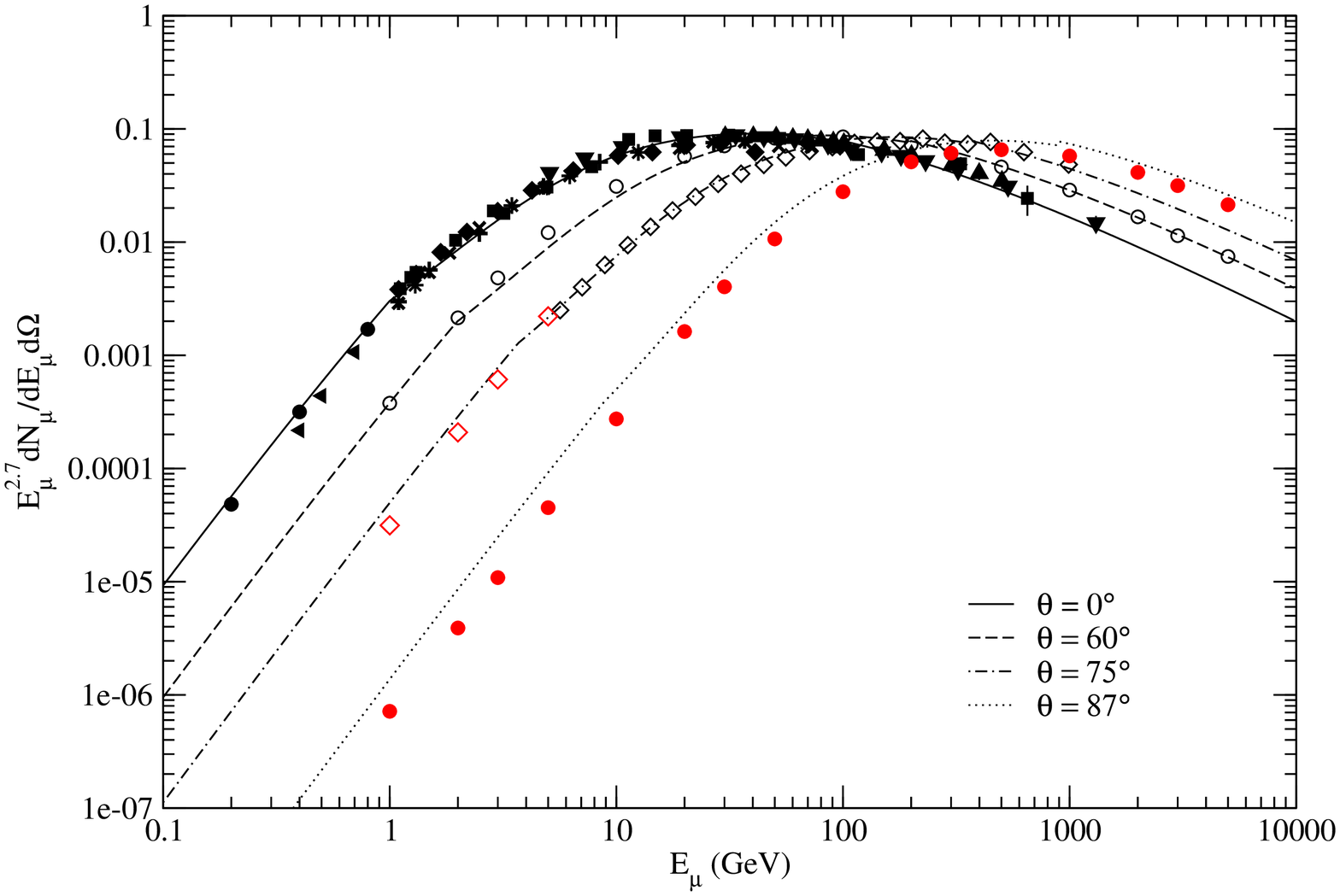}
\caption{\label{gaif}
Fits of the modified Gaisser parameterization to experimental 
data in the low
energy regime between 
$\theta=0$ and $\theta=87^\circ$.  The experimental data
are taken from References~\cite{gaisser}--\cite{hayman}.}
\end{figure}

The standard Gaisser parameterization for surface muons is not accurate in
the low energy regime.  Typical $\theta_{13}$ reactor experiments
place the near detector close to the reactors and subsequently close to
ground surface so that an accurate parameterization of low energy surface muons
is needed in this situation.  In the case of Double Chooz,
the far detector is sufficiently close to the steep shallow hill side that
a good description of the low energy surface muons is also needed.  For
these reasons, the low energy part of the Gaisser parameterization is
modified.  The fits of the modified parameterization to experimental data
are illustrated in Fig.~\ref{gaif}.

\begin{figure}
\begin{center}
\includegraphics[scale=0.45]{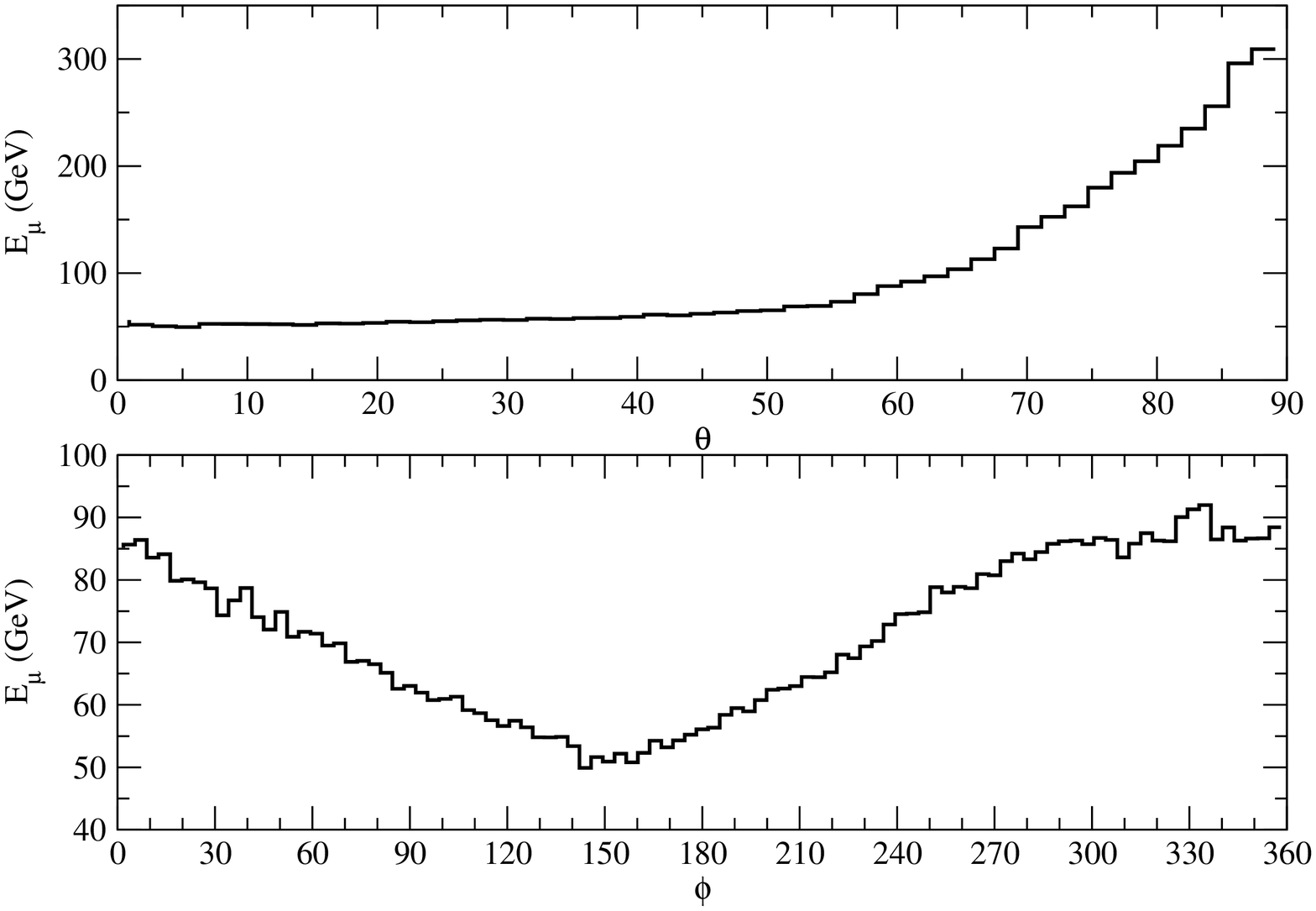}
\caption{\label{ept1c}
Average muon energy $E_\mu$ versus $\theta$ and $\phi$ at the Double Chooz
far site.  The total number of simulated events is $5\times10^6$.}
\end{center}
\end{figure}

\texttt{DCMUSIC} computes average observables such as
average final muon energy, flux and rate as well as various distributions.
The integrated muon intensity and average energy are defined as
\[
J_\mu=\int_\Omega d\Omega \,\int^\infty_0 dE_{\mu0}\,
P(E_{\mu0},X,\theta^\star,\phi)\,
\frac{dN_{\mu0}(E_{\mu0},\cos\theta^\star)}{dE_{\mu0}\, d\Omega},
\]
\[
\langle E_\mu\rangle=\frac{1}{J_\mu}
\int_S d\Omega \,\int^\infty_0 dE_{\mu0}\,
\frac{dN_{\mu0}(E_{\mu0},\cos\theta^\star)}{dE_{\mu0}\, d\Omega}
\int^\infty_0 dE_\mu\,E_\mu\,P(E_\mu,E_{\mu0},X,\theta^\star,\phi),
\]
where $dN_{\mu0}(E_{\mu0},\cos\theta^\star)/dE_{\mu0}\, d\Omega$ is the
modified Gaisser parameterization.
The probability function $P(E_{\mu0},X,\theta^\star,\phi)$ defines
the survival probability of a muon with initial energy $E_{\mu0}$ traversing
a slant depth $X$ from the zenith angle $\theta^\star$ and the azimuthal angle
$\phi$.  In order to simulate the integrals accurately, the solid angle
covering the mountain profile from the perspective of the detector must be
sampled uniformly.  In fact initial muon energy $E_{\mu0}$, cosine of the
zenith angle $\cos\theta$ and the azimuthal angle $\phi$ are generated
uniformly in the present algorithm.  Muons with uniformly generated energies
and zenith angles are propagated through the Ardennes rock along slant
depths that are specified by uniformly generated values of
$\theta$ and $\phi$.  At the
end, all the surviving muons are multiplied by the corresponding values
of the modified Gaisser function as weights, added up, divided by the
total number of throws and then multiplied by the proper range of integration.
The result is a very accurate and efficient method of computing average muon
energy and integrated intensity.  Furthermore, the simulated muon data can
be easily binned to produce all kind of distributions.  An example is
the average muon energy per angle as shown in Fig.~\ref{ept1c}.
The muon rate in a detector can be calculated as
\[
R_\mu=\int d{\mathbf A}\cdot\hat r\,
\int_\Omega d\Omega \,\int^\infty_0 dE_{\mu0}\,
P(E_{\mu0},X,\theta^\star,\phi)\,
\frac{dN_{\mu0}(E_{\mu0},\cos\theta^\star)}{dE_{\mu0}\, d\Omega},
\label{micro}
\]
where $d\mathbf A$ is a differential area element along the the detector wall
and $\hat r$ is a unit vector along the muon line of sight.  The simulated
results of muon flux, average energy and rate at the far site are
$J_\mu=6.23\times10^{-5}$~$\rm cm^2\,s^{-1}$, $E_\mu=60.2$~GeV and
$R_\mu=45.4$~Hz respectively.  It is sometimes assumed that final muons
inside an underground lab are mostly vertical.  On the contrary, it is found
from the \texttt{DCMUSIC} simulation that a substantial fraction of the muons
reach the far lab from the shallow hill side at an angle around $45^\circ$.
This information is useful not just for Double Chooz but the optimization of
inner veto designs of future underground detectors.

\begin{figure}
\begin{center}
\includegraphics[scale=0.45]{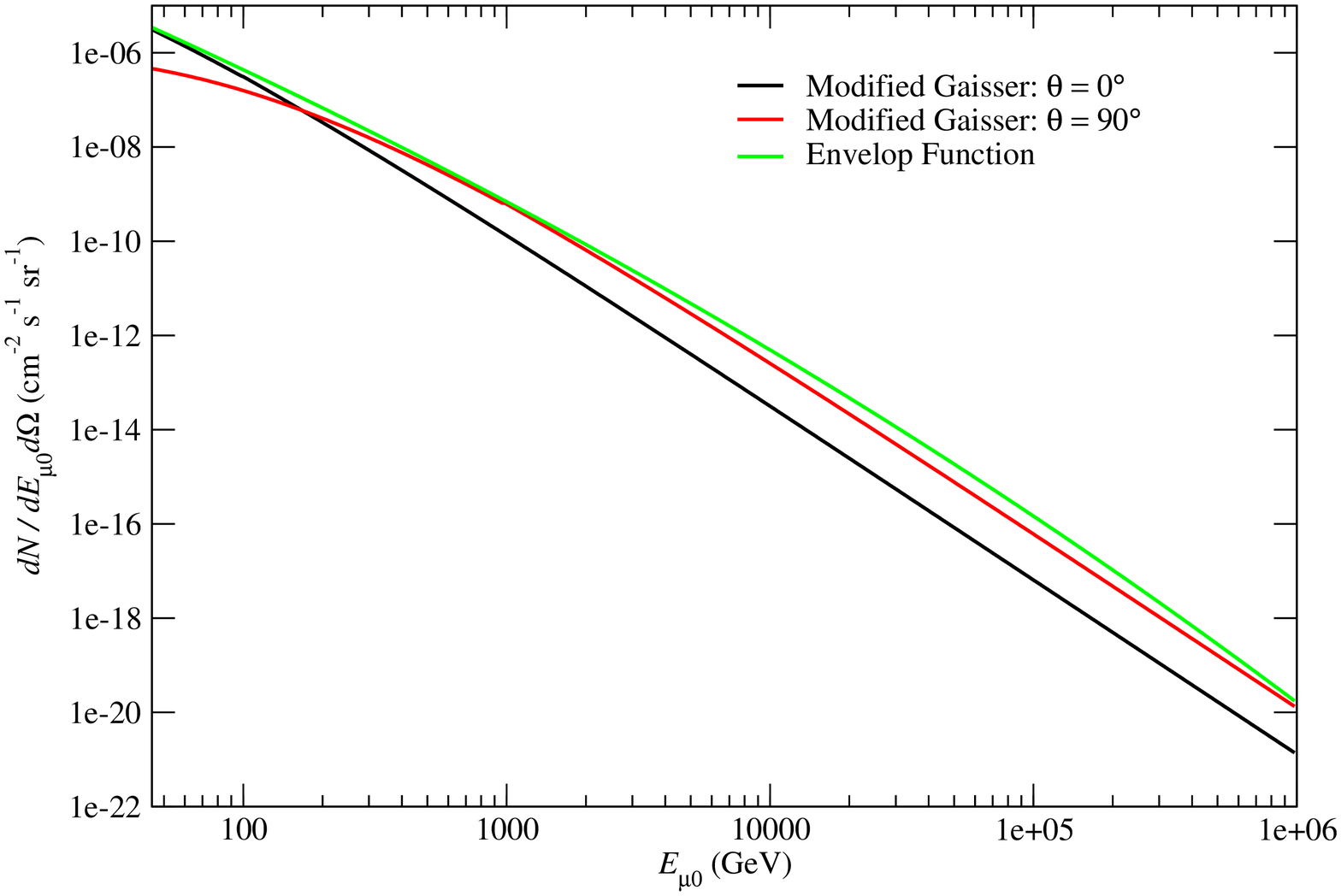}
\caption{\label{env}
A ketch of the envelop function $f(E_{\mu0})$ in relation to the modified
Gaisser function at $\theta=0^\circ,\,90^\circ$ for
$45\le E_{\mu0}\le10^6$~GeV.}
\end{center}
\end{figure}

Uniform generation is a very accurate and efficient method to compute
average observables but not the most straightforward way to generate
muon events as inputs for \texttt{FLUKA} and \texttt{GEANT}.  To this end,
generation according to the surface muon distribution is still the preferred
method.  The traditional approach of generation according to a distribution
is to randomly simulate a value between the maximum and the minimum of the
distribution and then try to invert the distribution by trial-and-error.  This
technique usually fails in the case of a sharply peaked distribution
because inversion by trial-and-error tends toward an infinite loop because
randomly generated trial solutions tend to miss the sharp peak.
Fortunately a trick is found to envelop the sharply
peaked distribution with a function that defines a narrower range of values to
try during the inversion process.  The envelop function is
\[
f(E)=4\times10^{16}\exp\left(-42E^{0.05}\right)
\]
Fig.~\ref{env} illustrates the quality of the functional envelop.
The result is a very efficient inversion
algorithm despite the sharpness of the modified Gaisser distribution.
At the Double Chooz far site, the minimum
rock slant depth is about 93~m so that initial muons of energy less
45~GeV are not expected to survive so that the simulation can begin
at that energy.

The strength of \texttt{DCMUSIC} is concentrated on the simulations of muon
overburden underneath complicated mountain profiles because they cannot
be easily constructed by standard software packages.  In this sense,
\texttt{DCMUSIC} is useful mostly in the case of the far detector.
The near detector on the other hand is located underneath a flat overburden
so that either
\texttt{FLUKA} or \texttt{GEANT} can easily define a simple topographic
geometry.  Other details on muon background simulations can be found
in section~\ref{sec:musim}.

\subsubsection{Event Generators} 
Members of the Double Chooz collaboration have developed a fairly complete
neutrino event generator.  The code includes the source neutrino distribution
of two extended reactor cores with built-in
time variation due to fuel evolution.  Neutrinos are propagated point to point
from their creation in the reactor to their interaction in the near or far
detector.  Oscillation effects are included in the propagation.  Interactions
can occur in any portion of the detector, including the target.

\subsubsection{Nuclear Transport Properties} 
The nuclear properties of the simulated detector materials have been checked
against measured and tabulated values.  For photon and neutron
transport, the attenuation lengths and inelastic properties
are consistent with observation.  Efforts have been
made to simulate the multiple photon cascades from neutron
capture on Gadolinium.  This process is either not modeled or is
poorly implemented in many standard simulation packages, such as
\texttt{GEANT}.
The simulation has been carefully tuned to give a neutron capture
cascade consistent with that observed with the first CHOOZ
experiment.

\subsection{Vertex and Energy Reconstruction} 
The event reconstruction developed for Double Chooz is based on a maximum
likelihood algorithm that makes use of all information available in any
given event.
An event is fully characterized by the four-vertex $(x,y,z,t)$ in the
coordinate system of the detector, direction $(\phi,\theta)$, and energy $(E)$.
Thus, for any given event defined by the set of parameters $\vec\alpha$,
\[ \vec\alpha=(x,y,z,t,\phi,\theta,E), \]
the likelihood for measuring a set of PMT charges $(q_i)$ and times $(t_i)$ in
the Double Chooz detector is the product over the individual charge and time
likelihoods at the PMTs:
\[ {\cal L}_{event} = \prod_{i=1}^{N_{PMTs}} {\cal L}_q(q_i;\vec\alpha)
                                             {\cal L}_t(t_i;\vec\alpha). \]
Reversing the meaning of the likelihood function, ${\cal L}_{event}$ is the
likelihood that the event is characterized by the set $\vec\alpha$ given the
set of measured charges $(q_i)$ and times $(t_i)$.
Maximizing the event likelihood ${\cal L}_{event}$ (or equivalently minimizing
$- \ln {\cal L}_{event}$) with respect to $\vec\alpha$ determines the optimal
set of event parameters.

Given the scintillator concentration in the target volume of Double Chooz, one
can safely assume that events emit effectively only isotropic scintillation
light of strength $\Phi$ (photons per steradian), which is proportional to the
event energy $E$.
The average number of photoelectrons (PEs), $\mu_i$, expected at a PMT of
quantum efficiency $\varepsilon_i$, at a distance $r_i$ from the event vertex,
and subtending a solid angle $\Omega_i$ is given by
\[ \mu_i = \varepsilon_i \, \Omega_i\,\Phi\,\exp(-r_i/\lambda). \]
For distances $r_i$ much larger than the PMT radius $R$, $r_i \gg R$, the solid
angle is defined by
\[ \Omega_i = \frac{\pi R^2}{r_i^2}\,f(\cos\eta_i), \]
where $f(\cos\eta)$ is the angular response function of the PMT, i.e., the PMT
efficiency as a function of the angle of incidence of the light with respect to
its normal, $\eta_i$.
Although the scintillation attenuation length $\lambda$ and the individual
quantum efficiencies of the PMTs are wave-length dependent, only average,
effective values are used in this approach -- which provide a very good
description of the light model.
All reconstruction parameters (attenuation lengths, solid angles and quantum
efficiencies) are determined self-consistently from control data samples.

The charge likelihood ${\cal L}_q(q;\vec\alpha)$ for any given PMT is directly
obtained from the (normalized) probability of measuring a charge $q$ for
a predicted value $\mu$, ${\cal P}(q;\mu)$, since $\mu$ itself depends on the
set of event parameters $\vec\alpha$.
The negative charge log-likelihood look-up tables can be obtained in two
different ways:
\begin{enumerate}
\item[(a)]
Assuming that the probability of measuring $n$ PEs at a given PMT at which
one predicts a charge $\mu$ in the presence of the light source is governed by
Poisson statistics,
\[ P(n;\mu) = \frac{1}{n!} \, e^{-\mu} \mu^n, \]
the probability of measuring a charge $q$ for the predicted value $\mu$ is
given by
\[ {\cal P}(q;\mu) = \sum_{n=0}^\infty P(q;n) P(n;\mu), \]
where the $P(q;n)$ functions are the charge response functions of the PMTs,
{\it i.e.} the probability of measuring a charge $q$ given a number of PEs $n$.
The single PE response function, $P(q;1)$, can be obtained directly from low
intensity laser calibration runs, while the higher ones $(n>1)$ can be obtained
from $P(q;1)$ by multiple random sampling.
\item[(b)]
Alternatively, the entire two-dimensional ${\cal P}(q;\mu)$ distribution
can be obtained from laser calibration runs which cover a wide range of light
intensities.
This method is independent on the underlying photon statistics at the PMTs, and
can be also cross-checked with similar measurements derived from other data
sets.
\end{enumerate}

\begin{figure}
\begin{center}
\includegraphics[bb=20 260 550 550, width=0.60\textwidth]
                {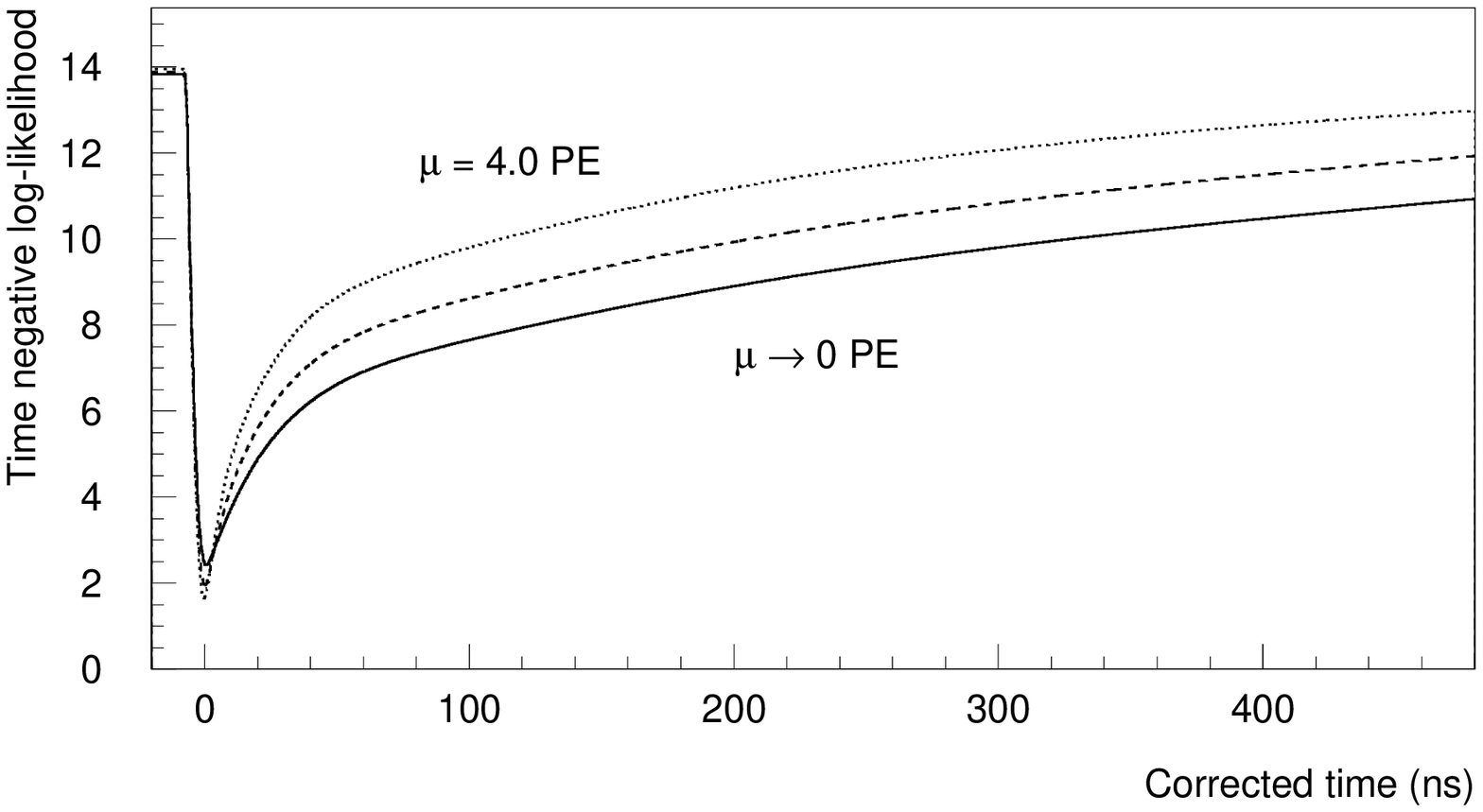}
\end{center}
\caption{Negative time log-likelihood functions for $\mu=0$, $2$, and
         $4\,\mbox{PE}$.}
\label{fig:is-1}
\end{figure}

The time likelihood for any {\em hit} PMT is a function of both the corrected
time, $t_{corr}^{(i)}$, defined as:
\[ t_{corr}^{(i)} = t_i - t - \frac{r_i}{c_n}, \]
(where $t_i$ is the measured time at the PMT, $t$ is the event time, $r_i$ is
the distance from the event vertex to the PMT, and $c_n$ is the speed of light
in the medium), and the predicted charge at that particular PMT, $\mu_i$.
Several such distributions are illustrated in Figure~\ref{fig:is-1} for
$\mu=0$, $\mu=2$, and $\mu=4\,\mbox{PE}$.
In essence they are obtained through folding a sum of exponential decays
(typical for scintillation light) with a Gaussian representing the PMT jitter,
taken to be $\sigma_t=1.2\,\mbox{ns}$ throughout these simulations.

\begin{figure}
\begin{center}
\includegraphics[bb=20 30 550 550, width=0.55\textwidth]
                {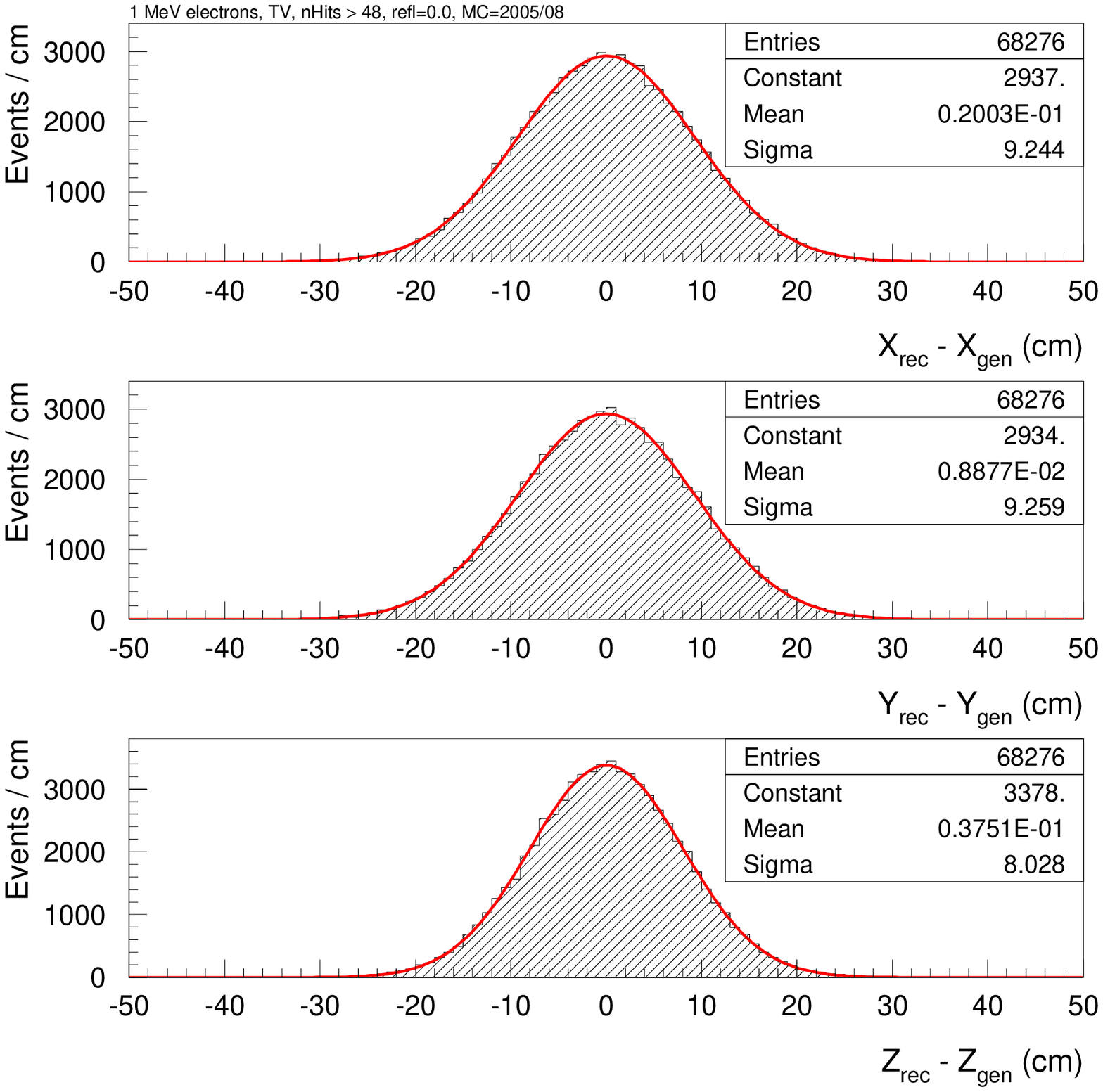}
\end{center}
\caption{Position reconstruction accuracy for $1\,\mbox{MeV}$ electrons
         generated uniformly throughout the target volume.}
\label{fig:is-2}
\end{figure}

The position reconstruction performance of this algorithm is shown in
Figure~\ref{fig:is-2} for 1-MeV electrons generated uniformly throughout the
target volume of Double Chooz.
The accuracy is about $8.8\,\mbox{cm}$ in each direction, corresponding to
a spatial position resolution of $15.3\,\mbox{cm}$, without any systematic
pulls in any direction.

\begin{figure}
\begin{center}
\includegraphics[bb=20 30 550 550, width=0.55\textwidth]
                {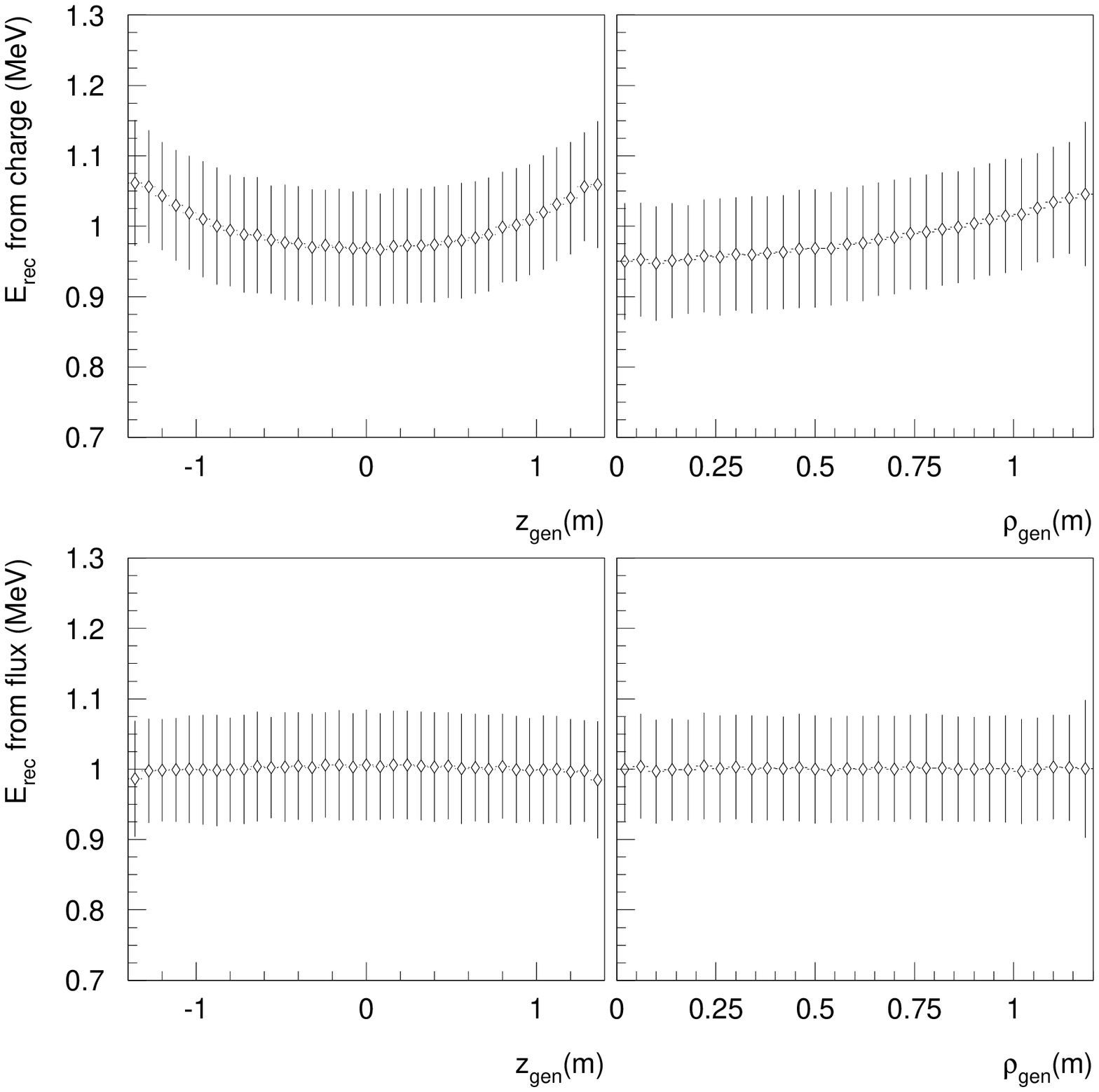}
\end{center}
\caption{Reconstructed energy versus generated $z$ and $\rho$ (radial distance)
         as obtained from the total visible charge (top) and fitted flux
         (bottom) -- both without any position corrections.}
\label{fig:is-3}
\end{figure}

The energy reconstruction can be obtained by two different methods:
either through the total visible charge with corrections as a function of
$\rho=\sqrt{x^2+y^2}$ and $z$,
\[ E = Q\,f(\rho,z), \]
given the cylindrical geometry of the detector,
or directly through the fitted flux $\Phi$, after a simple rescaling,
\[ E = c_E\,\Phi. \]
The proportionality constant $c_E$ can be obtained from known radioactive
calibration sources.
The variation of the reconstructed event energy as obtained from the total
visible charge is shown in Figure~\ref{fig:is-3}, as obtained from the same
sample of 1-MeV electrons.
The same figure also illustrates the fact that the reconstructed energy as
obtained directly from the flux is essentially independent on the event
position in the detector.
The energy resolution obtained through this method is $7.4\%$.

The reconstruction performance has also been investigated in the case in which
the detector walls are painted white (instead of black), in order to enhance
the reflections.
In turn, these would increase the overall light level in the detector, leading
to a better energy resolution and a more uniform charge response.
For a reflectivity coefficient of $0.4$, the average total visible charge
increased by about 38\% in the presence of raw stainless steel walls, {\it i.e.},
the energy resolution is expected to improve by a factor of approximately
$\sqrt{1.38} = 1.17$.
At the same time one expects the position resolution to be slightly degraded,
as a direct consequence of the broadening of the timing distributions by
additional reflections.
With the new set of time log-likelihood functions, we find that the position
reconstruction has an overall spatial position resolution of $15.3\,\mbox{cm}$,
which happens to be identical to the performance of the detector with
non-reflective walls.
The negative effect introduced by the broadening of the timing distributions
is counter-balanced by the increase in the number of hit PMTs.
The energy reconstruction obtained from the fitted flux yields a resolution of
$6.4\%$, in good agreement with the expected improvement calculated from the
total amount of light.
The optical model in this case is slightly different than that in the case of
no reflections, with a different effective attenuation length and a solid
angle $\Omega$ that scales differently than $1/r^2$.

\subsection{Calibration}
\subsubsection{Light Sources} 
The primary purpose of the laser calibration system is to quantify and
monitor pertinent properties of each individual PMT, such as PMT gain, relative
quantum efficiency, pulse-height versus photoelectron linearity, and
timing.
Other functions of the system include the measurement and monitoring of
the effective attenuation length of the scintillator over the lifetime of
the experiment, the determination of the charge and time likelihood
functions as needed by the reconstruction algorithms, as well as
{\it in-situ} measurements of the position reconstruction accuracy.

Single PE response, relative quantum efficiencies, and gain
calibrations can be obtained from low intensity, low frequency background
runs using the central laser light source.
Special calibration runs with high intensity light levels via the central
light source determine uniquely the time offsets of the PMTs.
Time slewing corrections are determined from laser calibration runs
covering all intensities and all light sources.
These data sets will also be of essence in determining the charge and time
likelihood functions.

  \subsection{Resources}
        
    The offline computing environment of Double Chooz is defined by a
    number of constraints:  smooth transition between the two stages of the 
    experiment (far detector only first and then both the far and near
    detectors together), issues of
    the data flow, on-site storage and analysis capabilities,
    networking and the data transport, off-site long term storage
    capabilities and data analysis.
 \subsubsection{On site resources}

\begin{enumerate}
 \item[(a)]\underline{Initial stage: Far detector only}

      During this first stage, site huts will be installed at the
   exit of the tunnel equipped with a user room for the computing
   and storage equipments.  Databases supported by reliable mass
   storage devices (RAID drives) guarantee a few weeks of raw data
   buffering for calibration data.  A cluster of workstations
   are used for offline processes on-site analysis of the calibration
   runs, and first level raw data analysis if needed.

      Internet connections will be configured via
   a regional network to the nearest university, the French 
   national IN2P3 network PHYNET and the French university network
   RENATER.  Raw data will be transfered to the IN2P3 Computing Center
   using file transfer protocols such as bbfFTP and gridFTP optimized for
   the transfer of large data files over the internet.

      A special connection utilizing the Chooz village internet 
   infrastructure will be provided to Chateau de L'Aviette for the
   remote control and monitoring of the detector, on site alarms, and all   
   things necessary to shift workers.
 
\item[(b)]\underline{Final stage: far and near detector}  

     When both detectors will be ready to run at the same time, the user
    room will be moved closer to the near detector site, together with
    the workstations and storage facilities.
    A high bandwidth fiber optic or wireless link will connect the far site 
    and the new user room.

       The rate of data flow projected in the final configuration is
    of the order of 15-25 GB/d (see the DAQ summary paragraph).  Like
    the initial stage of the experiment, a few weeks of raw data storage
    capacity and databases are planned on site for calibration runs.
    This operation conserves the RAID drives (a few TBs) used during the
    first stage of the experiment.

       The local network to Chateau de L'Aviette and the internet
    connection in the new user room will stay the same. 
\end{enumerate}
      
  \subsubsection{Off-site resources}

     The IN2P3 Computing Center (CC-IN2P3) in Lyon (France) will be used as 
    a data repository (raw data and processed data) for
    off-line analysis, as it is already used for the CVS code repository.

       Based on the data flow rate foreseen in the final configuration, 
    the total amount of raw data accumulated during the life time of the 
    experiment should be of the order of few tens of TB taking 
    into account an usual value of duty cycle of the order of 70\%. 
    CC-IN2P3 is equipped with the HPSS (High Performance Storage System) 
    designed to cope with the increasing storage capacity
    demand coming from several experiments, and will easily handle the full
    set of Double Chooz data over its life time. HPSS gives its users
    a transparent and fast access to data as though if they were on disk. A
    bookkeeping database for the data files is probably needed.  Some
    simulation data files larger than 100~MB can also be stored in HPSS.
 
       User accounts will be granted to all Double Chooz Collaboration
    members who want to use the computing facility consisting of 
    more than 1300 processors running under UNIX (mainly Scientific Linux
    and SL3).  CC-IN2P3 will provide general assistance to Double Chooz users,
    including the use of site-licensed softwares, technical support and
    24-hour operators.

\subsubsection{RoSS}

\texttt{RoSS} (Read-Out System Simulation)
is a package that relies on MC data generated with 
\texttt{DCsim} to simulate the response of the readout system of the 
Double Chooz detectors. It holds all the knowledge 
about the readout system elements (simplistically regarded as a set 
of gains and threshold with time-windows) so that we can characterize 
their effect to the physics responses to be measured by 
Double Chooz to high accuracy.

\texttt{DCsim} simulates all the relevant scintillator-light interactions
between the point of creation in the liquid scintillator and the point of
detection on the PMT photocathode. 
It outputs the PEs (photo-electrons) from each PMT to calculate 
the energy deposition in the detector. \texttt{RoSS} takes each
PE simulated by \texttt{DCsim} as an input, whose only information is the
time of creation on the 
photocathode.  All effects associated to the quantum 
efficiency of the photocathode are calculated by \texttt{DCsim}, while all 
processes converting each PE into a sampled digitized electrical 
signal are simulated by \texttt{RoSS}.  In order to accomplish these
functionalities, \texttt{RoSS} includes responses in the DAQ system (PMT and 
electronics), time and charge modulations (bandwidth, 
linearity and fluctuations) in the entire analogue-to-digital conversion
process.  Response correlations both in time (PMT
after-pulsing) and across channels (crosstalk) are integrated into the
simulation by default, along with the trigger logic and electronics 
(discriminators) behaviors. As part of the 
trigger, Double Chooz relies on level-2 software-online-trigger
to perform some event-type classification necessary to 
cope with the expected DAQ rates.

Conceptually, the design philosophy of \texttt{RoSS} relies on as little
simulation of  "micro-physics" (behind the readout system response)
as possible. 
Instead, it utilizes empirical responses curves obtained from 
measurements in dedicated test-benches. This approach has, at least, 
two major advantages:  (1) The simplification of the implementation 
since empirical response PDFs (probability distribution functions)
integrate inclusively over all effects 
that matter, and (2) this approach avoids the introduction of readout element 
models that are difficult to test or tune (if feasible at all) and do 
not necessarily improve the accuracy of the overall performance of 
the simulation for the scope of Double Chooz while enlarges the 
number of unnecessary degrees of freedom of the simulation. This approach is 
expected to model the necessary complexity adequately in the final 
MC versus DATA comparisons. Therefore, a possible oversimplification of the 
simulation can be tackle when motivation is found data-driven. The only one
external dependence of \texttt{RoSS} beyond \texttt{DOGS} is \texttt{ROOT}.

\subsubsection{DOGS}

\texttt{DOGS} (Double Chooz Offline Group Software)
refers to the whole offline software of Double 
Chooz. It relies on \texttt{ROOT} for I/O functionalities. 
It contains dedicated packages devoted to specialized 
tasks necessary for the offline data analysis.  However, the main 
feature of \texttt{DOGS} is a common interface (or framework) by 
which all packages can interact and/or exchange information and data 
easily. The built-in functionalities carry out data
processing operations such as energy deposition, position reconstruction,
event-type recognition, energy reconstruction (upon calibration), sub-
detectors monitoring and so on. Currently most of this packages beyond the 
common interface are still under construction.

The common interface for data exchange is incorporated into the "Event" 
package. Its function is to wrap data within a groups of smart data-
capsules which are designed for the flexibility of the 
data structure necessary to reuse the same data containers (and 
depending software) throughout the full lifetime of data along the 
offline analysis.  As a critical feature for the remaining \texttt{DOGS} 
software design, once the Event structure is defined, all packages 
know how to exchange information/data independently of the type of 
data treated. Simulated and experimental data can be held within the
data-capsules transparently so that the analysis software cannot distinguish
between simulated and experimental data.  The data-capsules are "smart" in that
they have the capability to save and retrieve their contents from the 
storage area.  This flexibility allows the 
selection of the data to be used at each analysis such that the user 
can configure the run with as little I/O overhead as possible.


%
%
%
\cleardoublepage
%
%
\section{Non proliferation activities}
\label{sec:nonproliferation}
Within the Double Chooz collaboration and at the request of the 
International 
Atomic Energy 
Agency (IAEA), studies are being
conducted to evaluate the interest of 
using antineutrino detection to remotely monitor nuclear power 
station. Indeed, 
the existence of 
an antineutrino signal sensitive to the power and to the isotopic 
composition of a reactor core, as proposed by Mikaelian 
et al.~\cite{Mik77} and as 
demonstrated by the Bugey~\cite{Dec95} 
and Rovno experiments~\cite{Kli94}, could provide 
a means to address certain safeguards applications.
If this method proves to be
 useful, the IAEA may decide that
 any new nuclear power plants should include 
an antineutrino monitor.

The high penetration power of antineutrinos and 
the detection capability might provide 
a means to make remote, non-intrusive 
measurements of plutonium content in reactors~\cite{Ber02}. 
The antineutrino flux and energy spectrum depend upon the 
thermal power and on the 
fissile isotopic composition 
of the reactor fuel. Based on predicted and observed $\beta$ 
spectra, the number of antineutrinos per fission 
from $^{239}$Pu is known to be less than the
 number from $^{235}$U, and the energy 
released larger by 5$\%$. Hence an hypothetical reactor 
able to use only $^{235}$U would induce in a detector an 
antineutrino signal 60$\%$ higher than 
the same reactor producing the same 
amount of energy but burning only $^{239}$Pu (see table). 
This may offer a means to monitor changes in the relative 
amounts of $^{235}$U and $^{239}$Pu 
in the core. If made in conjunction 
with accurate independent measurements of the thermal power
 (with the reactor temperature and the flow rate of 
cooling water), 
antineutrino 
measurements might provide an estimate of the isotopic composition of 
the core, in particular its plutonium inventories. The shape of the 
antineutrino spectrum can 
provide additional information 
about core fissile isotopic composition. 

Because the antineutrino signal from the reactor 
decreases as the square of the distance 
from the reactor to the detector a precise ``remote" measurement 
is really only practical 
at distances of a few tens of meters if one is constrained to "small" detectors of the order 
of few cubic meter in size. Without any extra experimental effort, the near detector of 
the Double Chooz experiment will provide the most important dataset of anti neutrino detected 
($5 \times 10^5$ events per year). The 
antineutrinos energy spectrum recorded at a given 
time will be correlated to the fuel composition and to the thermal power provided by EDF; 
it is expected that individual components
 due to each fissile element ($^{235}$U, $^{239}$Pu) could
 be extracted with some modest precision and serve as a benchmark 
for this techniques.
\begin{table}[htpb]
\centering\begin{tabular}{lrr}
                                 &  $^{235}$U                     &     $^{239}$Pu            \\
\hline
Released Energy per Fission      &  201.7 MeV                     &   210.0 MeV          \\
Mean Energy of $\nu$             &  2.94 MeV                      &   2.84 MeV           \\
$\nu$ per fission $>$1.8 MeV     &  1.92                          &   1.45               \\
Reactor neutrino cross section   &  $\sim 3.2\,10^{-43} \,\, cm^2$ &  
 $\sim 2.76\,10^{-43} \,\, cm^2$ \\  
\hline
\end{tabular} 
\caption{\label{tab:nonprolif1}Main features of fissions elements chains producing electron antineutrinos.}
\end{table}
To fulfill the goal of non-proliferation additional lab tests and theoretical calculations should 
also be performed to more precisely estimate the underlying neutrino spectra of plutonium and uranium 
fission products, especially at high energies. Contributions of decays to excited states of daughter
 nuclei are mandatory to reconstruct the shape of each spectrum. 
As concluded by P. Huber and Th. Schwetz~\cite{Huber:2004xh} 
a reduction 
of the present errors on the anti-neutrino fluxes 
of about a factor of three is necessary to achieve this goal.
\subsection{Experimental effort}
The precise measurement of $\beta$-decay spectra from fission products produced by the 
irradiation of a fissile target can be performed at the high flux reactor at Institut 
Laue Langevin in Grenoble~(ILL). The reactor produces the highest neutron flux in the world. 
The fission rate of a fissile material target placed close to the reactor core is about 
$10^{12}$ per second. It is possible to choose different fissile elements as target in order 
to maximize the yield of the nucleus of interest. Using the LOHENGRIN recoil mass 
spectrometer~\cite{Loh04}, measurement of individual $\beta$-spectra from short lived fission 
products are possible ; in the same irradiation channel, measurements of integral 
$\beta$-spectrum with the Mini-INCA detectors~\cite{Mar06}, could 
be envisaged to study the evolution with time 
of the antineutrino energy spectrum of a nuclear power plant.
\subsubsection{Experiments with Lohengrin}
The LOHENGRIN recoil mass spectrometer offers the possibility 
to measure $\beta$-decays of 
individual fission products or at least of isobar 
fission products. This is done by selecting, 
within the A/q ratio, the fission products released by 
the fissile target located 23~m upstream. 
The fissile target ($^{235}$U, $^{239}$Pu, $^{241}$Pu, 
…) is placed into a neutron flux of $6 \times 10^{14}\,\, 
{\rm n cm^{-2} s^{-1}}$, 
50~cm from the fuel element. Recoil 
fission products are selected with a bipolar magnetic field followed 
by an electrostatic condenser. Before the 
focal plane of the spectrometer, a focusing magnet has 
been added in the last few years
to reduce the dispersion of particles and to increase the count rate 
by a factor~7. At the focal plane, the 
detection of the fragments is done in an ionization chamber. 
At the end of the ionization chamber, the fragments could be 
implanted on a moving tape, and the 
measurement of subsequent~$\beta$ and 
$\gamma$-rays are recorded by a $\beta$-spectrometer and 
Ge-clover detectors, respectively. Coincidences between these
 two quantities could also be made to 
reconstruct the decay scheme of the observed 
fission products or to select one fission product. 
Fragments with half-lives down to 2~$\mu$s can be 
measured, so that nuclei with large $Q_\beta$ 
(above 4~MeV) can be measured. 
The LOHENGRIN experimental 
objectives are to complete existing $\beta$-spectra of individual fission 
products~\cite{Ten89} with new measurements (for the main contributors to 
the detected $\beta$-spectra) 
and to provide integral 
$\beta$-energy spectra for mass-selected fission products 
coming from different 
fissile isotopes ($^{235}$U, 
$^{239}$Pu, $^{241}$Pu). This ambitious experimental 
program is motivated by the fact 
- noted by C. Bemporad~\cite{Bem02} - that 
unknown decays contribute as much as 25$\%$ of the antineutrinos 
at energies $>4$~MeV. Folding the antineutrino energy spectrum 
over the detection cross-section for inverse
 beta decay enhances the contribution of the 
high energy antineutrinos to the total detected flux by a 
factor of about 10 for $E_\nu \, > \, 6$~MeV. Although 
integral spectra measured by Schreckenbach et al.~\cite{Sch85} 
are rather precise, better than 2$\%$ up to 8~MeV, there is 
a disagreement with experimental integral
 spectra made by Tengblad et 
al.~\cite{Ten89}, with important errors: 
5$\%$ at 4~MeV, 11$\%$ at 5~MeV and 20$\%$
 at 8~MeV quoted.
The focus of these experiments will be on 
neutron rich nuclei with yields very different in $^{239}$Pu and
 $^{235}$U fission. In the list: 
$^{86}$Ge,$^{90-92}$Se, $^{94}$Br, $^{96-98}$Kr, 
$^{100}$Rb, $^{100-102}$Sr, 
$^{108-112}$Mo, $^{106-113}$Tc, 
$^{113-115}$Ru $\ldots$ contribute to the 
high energy part of the spectrum and 
have never been measured.
\subsubsection*{Irradiation tests}
\begin{figure}[htb]
\centerline{\epsfig{file=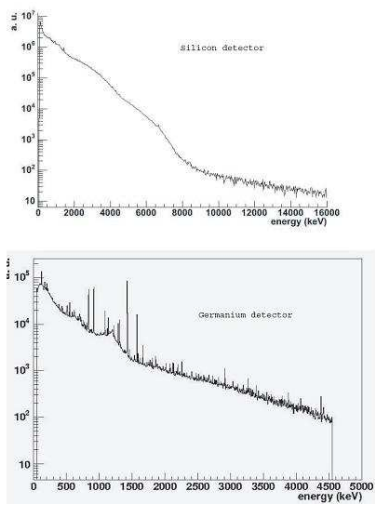,width=0.5\linewidth}}
\caption{Beta energy spectrum (a) recorded with the silicon 
detector corresponding to beta decay 
of fission products with mass A=94. The 
fission products arising from the LOHENGRIN spectrometer were implanted 
on a mylar tape of adjustable velocity in front of the silicon detector.
 The highest velocity was selected
 in order to enhance shorter-lived 
nuclei such as $^{94}$Kr and $^{94}$Br. The gamma energy 
spectrum (b) obtained
 with the germanium detector corresponding 
to the same runs is displayed also.}
\label{fig:np1}
\end{figure}

A test-experiment has been performed during two weeks of the summer
of 2005. The isobaric chains A=90 and
 A=94 were studied. Indeed some isotopes of these 
selected masses possess a high~$Q_\beta$ energy, and 
contribute significantly to the high energy part of 
the antineutrino spectra following $^{235}$U and $^{239}$Pu
 fissions, ($^{99}$Br: $Q_\beta \, = \, $ 10.3~MeV, $^{90}$Rb: 
$Q_\beta$~6.6 MeV, $^{94}$Br: $Q_\beta$ = 13.3~MeV, 
$^{94}$Kr: $Q_\beta$ = 7.7~MeV, $^{94}$Rb: $Q_\beta$ 
= 10.3~MeV).
Moreover they exhibit different fission yields 
after $^{235}$U and $^{239}$Pu fission~\cite{Eng94}. 
The half-lives (from a few tens of ms to years) and 
yields (from $10^{-4}$ to a few per~100~fissions) 
of these nuclei are diverse. The well-known nuclei, 
such as $^{90}$Br, will serve as a test of the experimental  
set-up, while the beta decay of more exotic nuclei 
such as $^{94}$Kr and $^{94}$Br will constitute a test 
case for how far one can reach in the very neutron rich 
region with this experimental device. 
The fission fragments arising from a 
6~mg $^{235}$U target placed in the thermal neutron flux reactor 
(a few $10^{14} \, \, {\rm n cm^{-2} s^{-1}}$) were selected through the 
LOHENGRIN spectrometer. 
The $\beta$-spectrometer was a thick cooled 
Si(Li) detector allowing a $\beta$ detection up to 15 MeV. 
The recorded data (see Figure~\ref{fig:np1}) will also 
validate the simulation described in the following 
section, in particular the evolution over time of 
the isobaric chains beta decay spectra. Indeed, 
thanks to the chopper we could vary the 
implantation durations on the tape, selecting longer or shorter 
lived nuclei, and adjusting the velocity to enhance some nuclei 
with respect to the others, depending on 
their half-life. Moreover the fission rate 
of the $^{235}$U target diminishes quite rapidly, and the target 
was changed during the experiment, so that the simulation of the 
burn up of the target will be tested. 
Because of the huge background due to 
the other nuclear charges within the same mass line, an accurate
 measurement of individual beta spectra is not possible with 
such a set-up. An identification of the atomic
 number of the fission products coming out 
from LOHENGRIN is required. 
Coincidence measurements between $\beta$ and $\gamma$ could 
be used with an increased solid angle but 
the identification of the decaying nucleus 
is not possible when the decay occurs toward the ground state.
 The best solution would be to identify the atomic number of 
the fission products through their 
trajectory in an ionization chamber, and to be 
able to detect the trajectory of the emitted electron. 
The spatial correlation would allow us to determine from 
which fission product the electron comes. 
The track of the electron could be measured 
with a wire chamber, in which the electron would lose 
only a very small part of its energy and the rest of 
its energy would be measured in a silicon detector.
 These improvements of the experimental set-up are 
under study. Experiments on other facilities 
(ISOLDE, GANIL...) could be also performed, 
depending on the nuclei to study. 
\subsubsection*{Integral $\beta$ spectra measurements}
In complement to individual studies on LOHENGRIN, more 
integral studies can be envisaged using the 
so called ``Mini-INCA chamber" at ILL~\cite{Mar06} 
by adding a $\beta$-spectrometer (to be developed).
 The existing $\alpha$- and $\gamma$-spectroscopy station 
is connected to the LOHENGRIN channel and offers the
 possibility to perform irradiations in a quasi 
thermal neutron flux up to 20 times the nominal value
 in a PWR. Moreover, the irradiation can be 
repeated as many time as needed. It offers then the unique
 possibility to characterize the evolution of 
the $\beta$~spectrum as a function of the irradiation time and
 the irradiation cooling. The expected modification 
of the $\beta$~spectrum as a function of the irradiation 
time is connected to the transmutation induced by 
neutron capture of the fissile and fission fragment 
elements. It is thus related to the 
``natural" evolution of the spent-fuel in the reactor. 
The modification of the $\beta$ spectrum as a 
function of the cooling time is connected to the decaying 
chain of the fission products and is then a mean to 
select the emitted fragments by their lifetime. 
This information is important because long-lived fission 
fragments accumulate in the core and after a few 
days mainly contribute to the low energy part 
of the antineutrino-spectra.
Due to the mechanical transfer of the sample from 
the irradiation location to the measurement station 
an irreducible delay time of 30 min is imposed leading 
to the loss of short-lived fragments.
\subsubsection*{Prospect to study fission of $^{238}$U}
The integral beta decay spectrum arising from~$^{238}$U 
fission has never been measured. 
All information relies on theoretical computations~\cite{Vog89}. 
Some experiments could be  envisaged 
using few MeV neutron sources in Europe 
(Van de Graaf in Geel, SINQ in PSI, ALVARES or SAMES
 accelerators at Valduc ...).  Here the total absence of 
experimental data on the $\beta$ emitted in 
the fission of $^{238}$U changes the context of 
this measurement compared to the other isotopes. 
Indeed any integral measurements performed could be 
used to constraint the present theoretical 
estimations of the antineutrino flux produced in 
the fission of $^{238}$U. In any case it seems rather
 difficult to fulfill the goal of a 
determination of the isotopic content from antineutrino
 measurements as long as this important part of the energy 
spectrum is so poorly known.
\subsection{Simulations}
\subsubsection{Simulations of diversion scenarios}
The IAEA recommends the study of specific safeguards scenarios. 
Among its concerns are the confirmation
 of the absence of unrecorded production of fissile 
material in declared reactors and the monitoring of
 the burn-up of a reactor core. The time required 
to manufacture an actual weapon estimated by the IAEA
 (conversion time), for plutonium in partially irradiated 
or spent fuel, lies between 1 and 3 months. 
The significant quantity of Pu to be sought is 8~kg, 
to be compared with the 3~tons of $^{235}$U contained 
in a Pressurized Water Reactor~(PWR) of power 900~MW$_e$ 
enriched to 3$\%$. The small magnitude of the 
desired signal requires a careful feasibility study.

The proliferation scenarios of interest involve different kinds 
of nuclear power plants such as light 
water or heavy water reactors (PWR, BWR, CANDU...), 
it has to include isotope production reactors of 
a few tens of MW$_{th}$, and future reactors (e.g., PBMRs, Gen~IV 
reactors, accelerator-driven sub-critical
 assemblies for transmutation, molten salt reactors). To 
perform these studies, core simulations with 
dedicated Monte Carlo codes should be provided, coupled 
to the simulation of the evolution of the 
antineutrino flux and spectrum over time.

We started a simulation work using the widely used 
particle transport code MCNPX~\cite{Mcn05}, coupled with
 an evolution code solving the Bateman equations for 
the fission products within a package called MURE
 (MCNP Utility for Reactor Evolution)~\cite{Mur05}. This 
package offers a set of tools, interfaced with MCNP
 or MCNPX, that allows us to define easily the geometry of 
a reactor core. In the evolution part, it has access to
the 3834 nuclei possibly considered, the set of 
nuclear data: nuclear masses from ``The 2003 Atomic
 Mass Evaluation" data (National Nuclear Data Center, 
BNL), the NNDC nuclear wallet card and the nuclear
 decay data from the JEF3T library. The nuclear reaction 
data used are taken from the usual nuclear 
databases~\cite{End00}. MURE is perfectly adapted to 
simulate the evolution with time of the composition of
 the fuel, taking into account the physics of a reactor 
core, especially the simulations of neutrons.  We are adapting
 the evolution code in order to be able to simulate the 
antineutrino spectrum and flux, using simple 
Fermi decay as starting point. The constitution of a 
database of more specific beta decay for each fission
 product will be necessary in order to improve the accuracy 
of the computed antineutrino spectra. 

The extended MURE simulation will allow to perform sensitivity 
studies by varying the Pu content of the
 core in the relevant scenarios for the IAEA. By varying 
the reactor power, the possibility to use antineutrinos
 for power monitoring can be evaluated.

The implementation of the simulation is under way. 
Antineutrinos from reactors arise from the beta decay
 of several hundreds of fission products. Each decay occurs 
generally toward several excited states of 
the daughter nucleus, and the antineutrino energy spectra are 
the sum of hundreds of individual beta spectra. 
It is then necessary to know the decay scheme of 
these nuclei (branching ratios and available energies 
toward different final states, shape of the spectrum related 
to the transition type, allowed -Fermi or 
Gamow-Teller or forbidden). Moreover the decay schemes 
of the most unstable isotopes are not well 
determined and sometimes unknown. Preliminary results 
show that nuclei with half-lives lower than 1 s emit
 about 70$\%$ (50$\%$) of the $^{235}$U($^{239}$Pu) antineutrino spectrum 
above 6 MeV. The high energy part of the
 spectrum is the energy region where Pu and U spectra 
differ mostly (see Figure~\ref{fig:np1}). 
The influence of the beta decay of these nuclei on 
the antineutrino spectrum might be dominant too in
 scenarios where rapid changes of the core composition 
are performed, e.g. in CANDU reactors refueled on line. 

The appropriate starting point for this 
scenario is a representative PWR, like the Chooz 
reactors. For this reactor type, simulations of the 
evolution of the antineutrino flux and spectrum over
 time will be provided and compared to the accurate 
measurement provided by the near detector. This should
 provide the precision on the fuel composition of an 
antineutrino detector like Double Chooz and of independent
 thermal power measurements. An interesting point to study 
is at the time of the partial refueling of
 the core, thanks to the fact that reactor like Chooz 
(N4-type) does not use MOX fuel.
\begin{figure}[htb]
\centerline{\epsfig{file=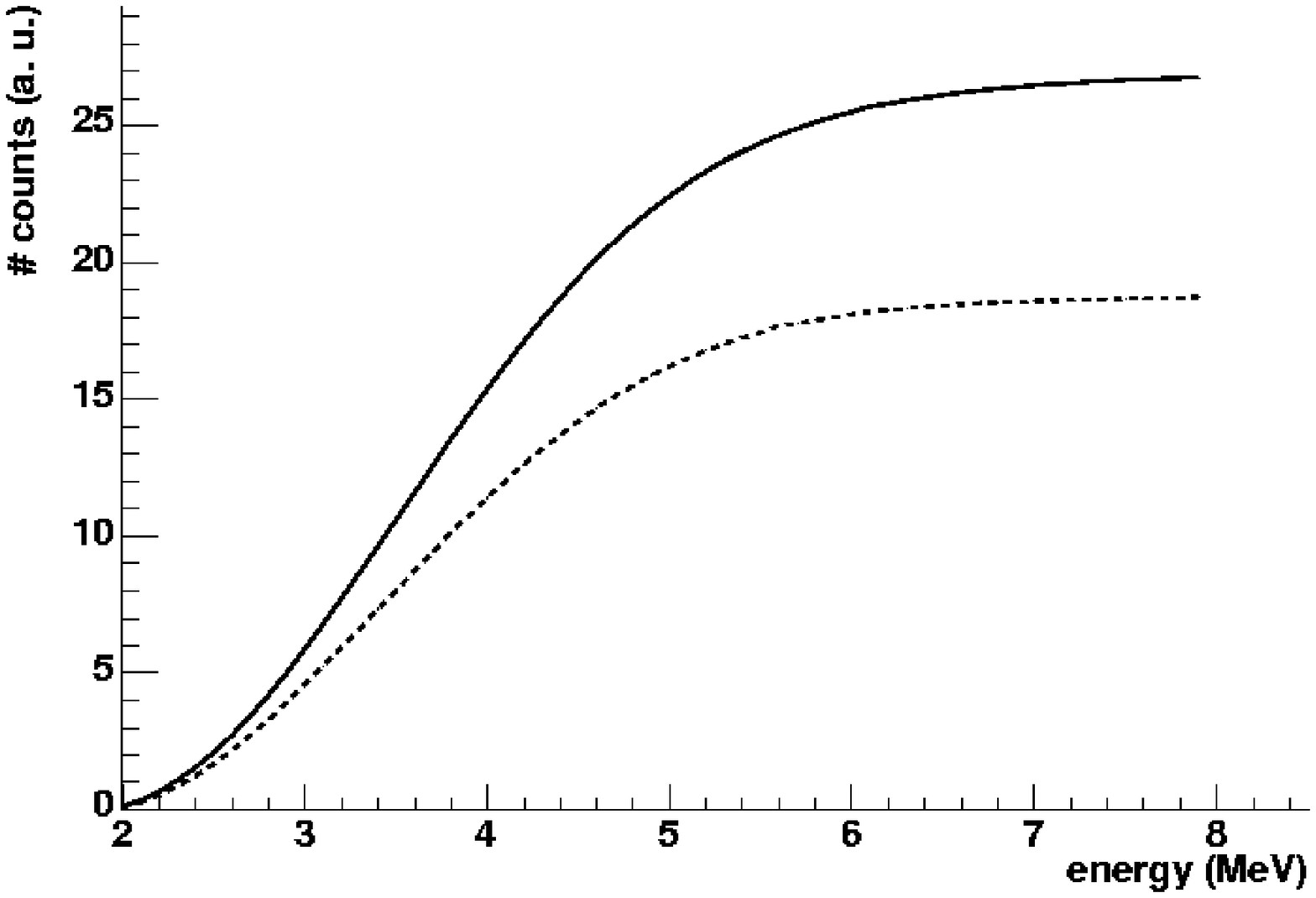,width=0.5\linewidth}}
\caption{Cumulative number of antineutrinos folded over the 
detection cross-section, as a function of 
antineutrino energy for $^{235}$U (plain 
line) and for $^{239}$Pu (dashed line), computed with the MURE program.}
\label{fig:np2}
\end{figure}
In any case the measurement performed by the Double 
Chooz experiment with its near detector will constitute
 the most precise determination of the antineutrino 
emitted by a PWR. In addition the detailed follow-up of
 the evolution of the fuel burn-up, controlled by the 
use of fission chambers, will constitute an excellent
 experimental basis for the above feasibility studies 
of potential monitoring and for bench-marking fuel 
management codes.
\subsection{Experimental effort in the U.S.}
\begin{figure}[htb]
\centerline{\epsfig{file=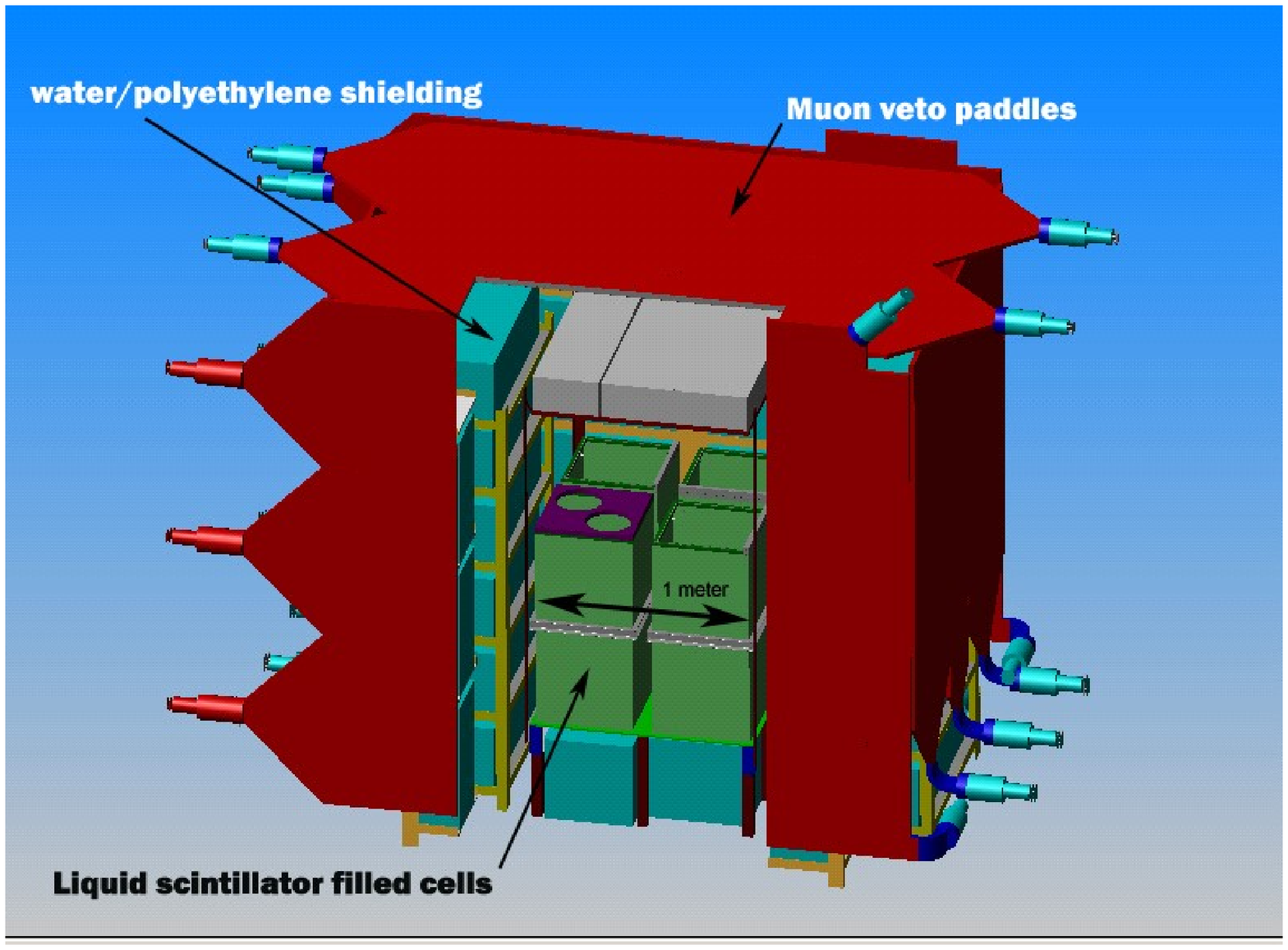,width=0.5\linewidth}}
\caption{A cutaway view of the LLNL/SNL antineutrino 
detector deployed at the San Onofre Nuclear Generating Station.}
\label{fig:npus1}
\end{figure}

\begin{figure}[htb]
\centerline{\epsfig{file=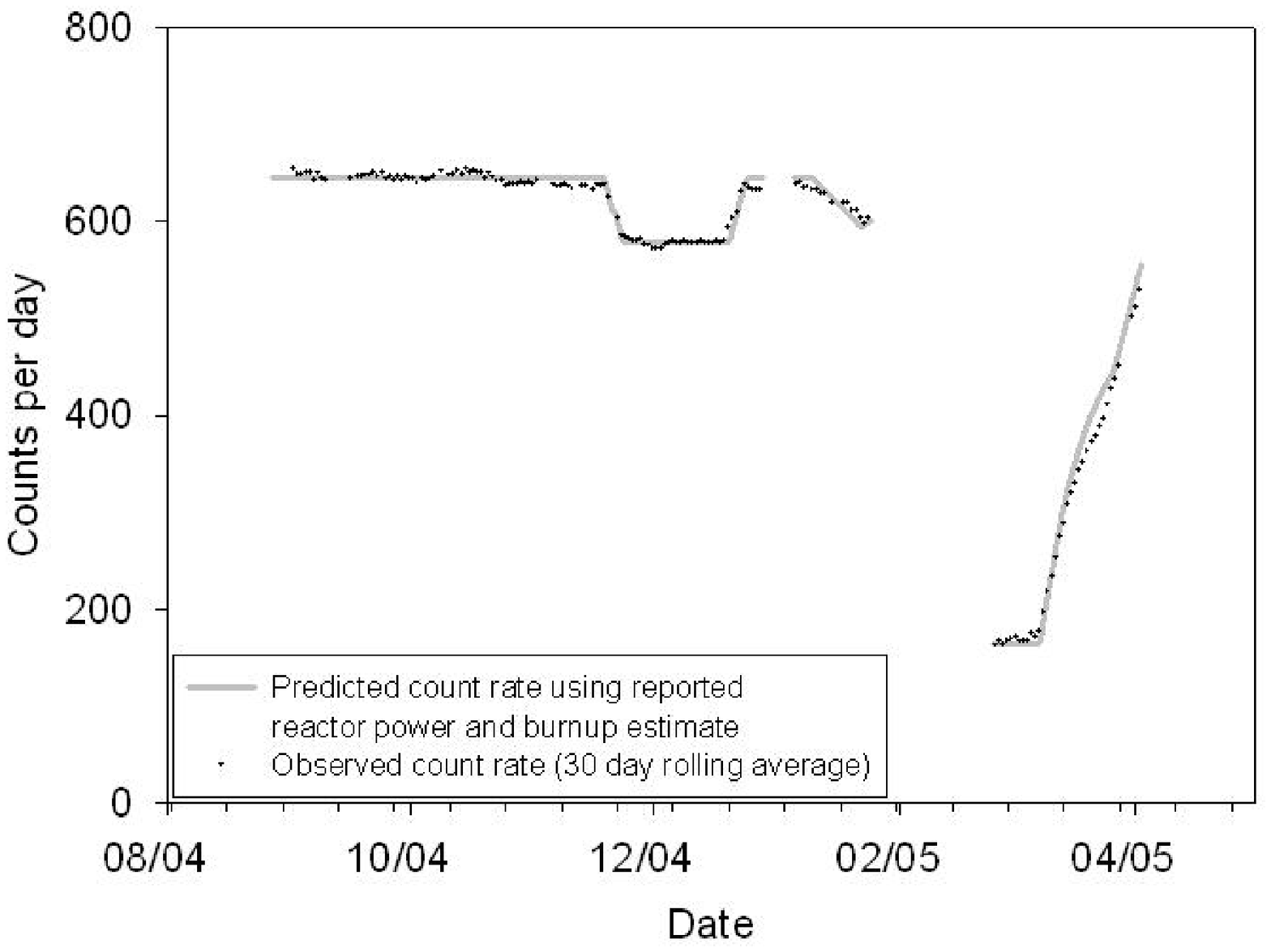,width=0.5\linewidth}}
\caption{Demonstration of reactor power tracking in SONGS.
A 30 day rolling average of observed antineutrino candidate 
events plotted with an estimate of the expected observation calculated 
from publicly reported reactor power data and a 
simple linear approximation of fuel burnup. Data in 
February-March 2005 with a daily count rate below 
200 candidates per day corresponds to a period of 
zero reactor power, and is thus a measure of the 
antineutrino mimicking background. The net antineutrino detection 
rate at full reactor power is thus about 400 antineutrinos per day.}
\label{fig:npus2}
\end{figure}

\begin{figure}[htb]
\centerline{\epsfig{file=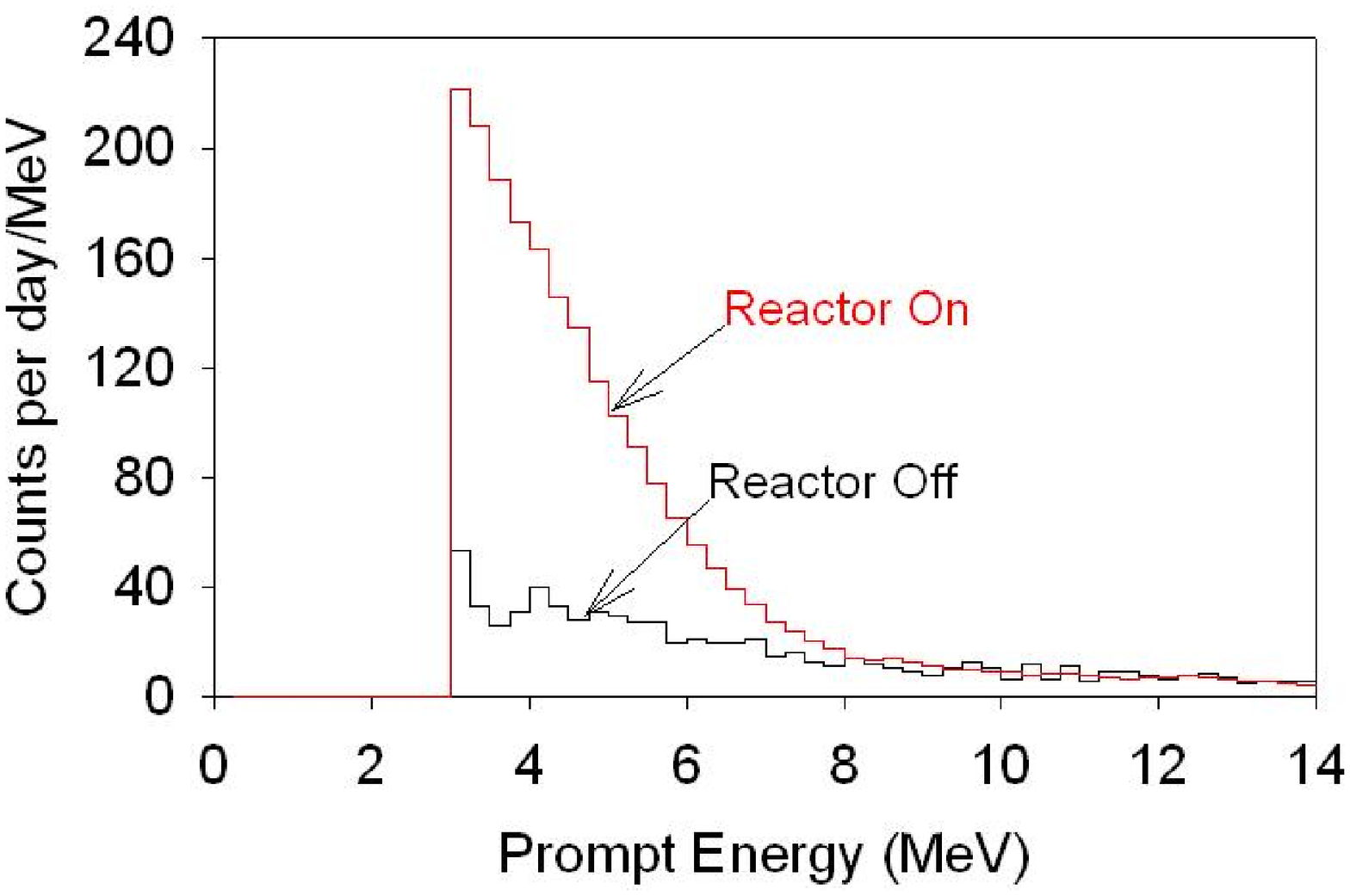,width=0.5\linewidth}}
\caption{Comparison of the measured energy spectra in SONGS with the 
monitored reactor on and off. The different shapes of the 
two spectra conform to expectations and confirm that reactor antineutrinos 
are being detected.
	These measure the energy spectrum of the first event in 
an antineutrino-like event pair. 
The reactor on and off spectra are expected 
to be equivalent above about 9 MeV as is seen.}
\label{fig:npus3}
\end{figure}

The experimental program for development of nonproliferation detectors 
in the United States is led by Lawrence Livermore National Laboratory 
and Sandia National Laboratories. The LLNL/SNL work has 
consisted of installing and operating a prototype detector at the 
3.46 GW$_{th}$ San Onofre Nuclear Generating Station (SONGS) in 
Southern California. The detector, now operating at SONGS at 
a distance of 25 meters from the core, and with an 
overburden of about 20 m.w.e., is shown in Figure~\ref{fig:npus1}. It has 
a one cubic meter active liquid scintillator volume and 
an approximate 2.5 meter $\times$ 3 meter footprint including shielding. 
It consists of a muon veto system for rejecting cosmic ray 
backgrounds, a water/polyethylene shield to reject neutron and 
gamma backgrounds, and a central doped liquid scintillator detector 
which registers antineutrino interactions. As seen in 
Figures~\ref{fig:npus2} and \ref{fig:npus3}, the data clearly 
demonstrates: detection of a high rate of reactor antineutrinos 
with good signal to background; that the rate is consistent 
with what the theory of antineutrino interactions predicts; and 
that changes in reactor power can quickly (within a few hours) 
be detected by tracking the antineutrino rate. The plot of daily 
rate versus time (Figure~\ref{fig:npus2}) also shows a two sigma deviation 
of the antineutrino rate from a constant value over a six month 
period, with the linear reduction in total rate consistent with 
a prediction that includes a fuel burnup estimate. Current effort 
is focused on confirming the indications of 
fuel burnup seen in this data. 
\subsection{Toward a prototype}
If we want to propose to the IAEA a neutrino detector able 
to help in monitoring future nuclear power plants, 
the next step in this effort has to merge the two present approaches:
the Double Chooz approach with a good energy measurement, 
a good signal to noise ratio, but expensive;
and the SONGS approach with a robust, simple, automatic, cheap, 
but with poor antineutrino detection efficiency (about 
10\%), a modest signal to background ratio (3 to 1), 
and poor energy resolution.
We thus are considering a new 
prototype with a size small enough to be installed very close 
to the reactor core (30~meters or so), but using a technique 
able to clearly detect neutrinos. 
Such a prototype will be considered as a 
demonstrator to be shown to the IAEA and at the same time it is 
already usable tool to measure the thermal power.
As an intermediate goal, we can foresee measurements
with this prototype at ILL with its core of roughly 
pure $^{235}$U. It would allow the recording of a very pure 
neutrino signal from $^{235}$U fission only 
followed by the evolution due to burn-up. 
Such a clean experiment would help to calibrate the 
neutrino signal versus the thermal power, and will 
also give some confidence for the simulation effort.
\subsubsection{Complementary studies}
At this stage it would be important to evaluate alternative 
techniques of antineutrino detection
 ($^3$He counters, $^6$Li liquid 
scintillators). Indeed some stability requirements 
(temperature, humidity...) could favor one particular 
technique over another one.
For a realistic device we cannot expect that 
the coverage against external muons will be as 
important as in the case of Double Chooz, hence 
the impact of backgrounds to the
antineutrino signal has to be evaluated carefully.
\subsection{Conclusions}
A realistic diversion (about 10~kg of Pu)
 has an effect on the antineutrino signal which is very 
small. The present precision on the antineutrino 
spectrum emitted in fissions is not  
enough to allow a determination of the isotopic content in
 the core sensitive to such diversion.
On the other hand, the thermal power 
measurement is a less difficult job. Neutrinos will sample the whole
 core, without attenuation, and thus would bring valuable information on 
the power with totally different 
systematics than present methods.
Even if its measurement is not persuasive by 
itself, the operator cannot hide any stops or change of 
power, and in most cases, such a record made with an 
external and independent device, virtually impossible 
to fake, will act as a strong deterrent.
In spite of the uncertainty mentioned 
previously, we see that the most energetic part of the spectrum
offers the best 
possibility to disentangle fission from $^{235}$U and $^{239}$Pu. 
The comparison between the cumulative numbers 
of antineutrinos as a function of antineutrino 
energy detected at low vs. high energy is an efficient 
observable to distinguish pure $^{235}$U and $^{239}$Pu.
The IAEA also seeks the possibility of monitoring large 
spent-fuel elements. For this application, the 
likelihood is that antineutrino detectors could 
only make measurements on large quantities of beta-emitters,
 e.g., several cores' worth of spent fuel. In the time of 
the experiment the discharge of parts of the core
 will happen and the Double Chooz experiment will 
quantify the sensitivity of such monitoring.
More generally the techniques developed for 
the detection of antineutrinos could be applied for the 
monitoring of nuclear activities at the level of a 
country. For example a KamLAND type detector deeply
 submerged off the coast of the country, would offer 
the sensitivity to detect a new underground reactor 
located at several hundreds of kilometers. In that 
respect, progress in detecting media
(Gd doped liquid scintillators) would be of great help.

%
%
%
%
%
%

\end{document}